\documentclass[rmp,twocolumn,showpacs,aps,floatfix,nofootinbib]{revtex4}
\usepackage{graphicx}
\usepackage{epsfig}
\usepackage{tabularx}

\makeindex

\usepackage{relsize}

\newcommand{\gevcc}{GeV/$c^2$}
\newcommand{\epem}{$e^+ e^-$}
\newcommand{\chisq}{$\chi^2$}
\newcommand{\amulo}{a^{\rm had,LO}_{\mu}}
\def\nb{nb}
\def\ev{eV}

\def\gev{GeV}
\def\pipi{\pi^+\pi^-}
\def\Kp {K^+}
\def\Km {K^-}
\def\mumu {\mu^+\mu^-}
\def\KKppch {\Kp\Km\pipi}
\def\ppz {\pi^0\pi^0}
\def\KKppnt {\Kp\Km\ppz}
\def\KKKK {\Kp\Km\Kp\Km}
\def\kst{K^*(892)}

\begin{document}
\title{Hadron Production via \epem\ Collisions
with Initial State Radiation}

\author{V.P.~Druzhinin, S.I.~Eidelman, S.I.~Serednyakov, E.P.~Solodov}
\affiliation{Budker Institute of Nuclear Physics SB RAS,  Novosibirsk, 630090,
  Russia}
\affiliation{Novosibirsk State University, Novosibirsk, 630090, Russia}

\begin{abstract}
A novel method of studying \epem\ annihilation into hadrons
using initial state radiation at $e^+e^-$ colliders is described.
After brief history of the method, its theoretical foundations
are considered. Numerous experiments in which exclusive cross sections of 
\epem\ annihilation into hadrons below the center-of-mass energy of 5
GeV have been measured are presented. Some applications of the results
obtained to fundamental tests of the Standard Model are listed. 
\end{abstract}

\maketitle
\newpage
\tableofcontents
\clearpage
\section{Introduction}

\subsection{Why is low energy \epem\ annihilation interesting?}

\begin{figure}
\includegraphics[width=.4\textwidth]{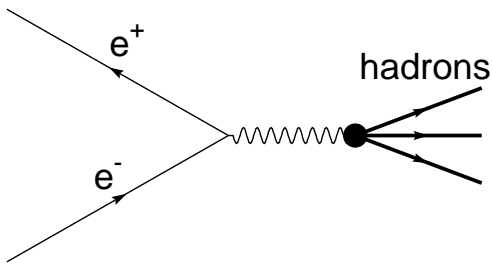}
\caption{The lowest-order Feynman diagram 
describing the process of $e^+e^-$ annihilation into hadrons.
\label{diag1}}
\end{figure}

Studies of low energy \epem\ annihilation into hadrons
are of great interest for theory and have numerous applications.
According to current concepts, \epem\ annihilation into hadrons proceeds
via an intermediate virtual photon which produces a pair of quarks, 
$q\bar{q}$, followed by the hadronization of quarks into observed hadrons. 
This process is described by the lowest-order Feynman diagram shown 
in Fig.~\ref{diag1}. 
When the initial energy of \epem, or equivalently of the
intermediate virtual photon, is large enough, the process of
hadronization is well described by Quantum Chromodynamics 
(QCD). At small energies, lower than 2--3 GeV, 
produced hadrons are relatively soft and intensively interact with
each other forming hadronic resonances. At the moment QCD fails to
describe this  energy region. 
Because of that, it is vitally important to gain sufficient
information from experiment to be used as an input to various 
QCD-based theoretical models. QCD sum rules are an example of how
measurements of total and exclusive cross sections can be used
to extract such fundamental parameters of theory as the strong coupling
constant $\alpha_s$, quark and gluon condensates~\cite{qcdsum}.

Precise knowledge of vacuum polarization effects based on
the total cross section of \epem\ annihilation into hadrons is necessary
to estimate the hadronic contributions to the running fine-structure
constant and thus determine its value at the $Z$ boson mass,
$\alpha(M^2_Z)$, a key component of the high-precision tests of the
Standard Model~\cite{fredold,ej95,bp05,hagi,actis}.
 
Improvement of the precision with which the total cross section 
of \epem\ annihilation into hadrons is known is also needed for
a more accurate estimation of the hadronic contribution to the
muon anomalous magnetic moment since it is one of the crucial limiting
factors in a search for New Physics~\cite{muonold,gourdin,brook}.

There is an important relation between spectral functions in 
\epem\ annihilation into hadrons 
with isospin $I=1$ and corresponding $\tau$ lepton decays based on
conservation of vector current (CVC) and isospin symmetry~\cite{tsai,sakurai}.
While first detailed tests of such relations showed satisfactory
agreement between such spectral functions~\cite{sanda,ei91}, higher accuracy 
reached in both \epem\ and $\tau$ lepton sectors revealed 
possible systematic effects not accounted for 
in the  $e^+e^-$ and/or $\tau$ experiments~\cite{dehz1,dehz2}.
Understanding of these effects is crucial for improving
the accuracy with which the hadronic contributions to the muon
anomalous magnetic moment can be estimated from $\tau$ decays
to two and four pions as was first suggested by~\cite{alem}.   
          
Detailed measurements of the energy dependence of various
exclusive cross sections allow to improve our knowledge of 
vector mesons and look for new states, both of light~\cite{dru07}
and heavy quarks~\cite{rmp08}.

\subsection{Idea of ISR}
\begin{figure}
\includegraphics[width=.4\textwidth]{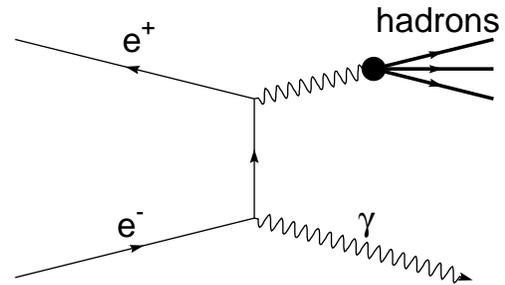}
\caption{The lowest-order Feynman diagram 
describing the initial state radiation process 
$e^+e^- \to \gamma +$ hadrons.
\label{diag2}}
\end{figure}

In \epem\ collider experiments exclusive and total hadronic cross
sections are usually measured by scanning the accessible energy range.
The process of $e^+e^-$ annihilation is accompanied by 
emission of one or several photons from the initial state. The lowest-order
Feynman diagram describing initial-state radiation (ISR) is shown
in Fig.~\ref{diag2}.
The quantity measured directly in the experiment is
the visible cross section
\begin{equation}
\sigma_{\rm vis}=\frac{N}{L}, 
\end{equation}
where $N$ is the number of selected events of the  process 
$e^+e^- \to \mbox{ hadrons}+n\gamma,\;n=0,1,2,\ldots$, 
and $L$ is the integrated luminosity 
of the collider collected at the center-of-mass (c.m.) $e^+e^-$ energy $2E_0$. 
The  visible cross section can be related to the Born cross section 
$\sigma_0$ corresponding to the lowest-order diagram of
Fig.~\ref{diag1} via the integral~\cite{str_func_meth}, providing
the $10^{-3}$ accuracy:
\begin{equation}
\sigma_{\rm vis}=\int\limits_0^{1-m_{\rm min}^2/s} \varepsilon(s,x) 
W(s,x) \sigma_0 (s(1-x))
{\rm d}x,\label{eqn1}
\end{equation}
where $s=4E_0^2$, $x$ is an effective fraction of
the beam energy $E_0$ carried by photons emitted from the initial state,
$m_{\rm min}$ is the minimal possible invariant mass of the
final hadrons, $\varepsilon(s,x)$ is the detection efficiency for
the  process $e^+e^- \to \mbox{ hadrons}+n\gamma$ as a function of $x$ 
and $s$. The so-called radiator function $W(s,x)$ 
taking into account higher-order QED contributions, in particular,
from the diagram in Fig.~\ref{diag2}, is fully calculable in QED~\cite{actis}.
Due to the photon emission from the initial state
the visible cross section depends on the Born cross section at
all energies below the nominal $e^+e^-$ c.m. energy $2E_0$.

In conventional scanning experiments the influence of ISR
is suppressed by the requirements of the energy and momentum balance between
the final hadrons and the initial $e^+e^-$ state. In this case the detection 
efficiency has $x$ dependence close to the step function: 
$\varepsilon(s,x)=\varepsilon_0(s)$ for $x<x_0$, and zero for $x>x_0$.
At small $x_0$, the equation (\ref{eqn1}) can be rewritten:
\begin{equation}
\sigma_{\rm vis}=\varepsilon_0(s)\sigma_0(s)(1+\delta(s)),
\end{equation}
where $1+\delta(s)$ is the radiative correction factor,
which takes into account 
higher-order QED corrections. To calculate this factor it is necessary to 
know $s$ dependence of $\sigma_0$ in the range from $s(1-x_0)$ to 
$s$. For slowly varying cross sections, $\delta$ is about 10\%,
and can be determined with accuracy better than 1\% using existing data on 
the cross section energy dependence.
Thus, in scanning experiments, from the data collected at 
the c.m. energy $\sqrt{s}$  
the cross section $\sigma_0(s)$ is determined directly. 

Another approach is also possible. Equation (\ref{eqn1}) can be 
rewritten in the differential form:
\begin{equation}
\frac{{\rm d}\sigma_{\rm vis}(s,m)}{{\rm d}m} =
\frac{2m}{s} \varepsilon(s,m) W(s,x) \sigma_0(m),
\label{eqn3}
\end{equation}
where we have made a transformation to the variable 
$m=\sqrt{s(1-x)}$, the invariant mass 
of the hadronic system. At non-zero $x$ the dominant 
contribution to the visible cross section comes
from the one-photon ISR (Fig.~\ref{diag2}). With the inclusion of the ISR 
photon momentum into the selection conditions on the energy and momentum 
balance, the non-zero detection efficiency for ISR events can be obtained in
a wide range of the hadronic invariant mass. 
So, from the measurement of the mass spectrum for the process 
$e^+e^- \to \mbox{hadrons}+\gamma$ at fixed c.m. energy $\sqrt{s}$ 
the cross section $\sigma_0(m)$ can be extracted in the 
invariant mass range from threshold to the mass close to $\sqrt{s}$.

The idea of utilizing initial-state radiation from a high-mass state
to explore electron-positron processes at all energies below that state 
was outlined long ago in Refs.~\cite{baier1,baier2}.  A possibility of 
exploiting such 
processes at high luminosity $\phi$- and $B$-factories was discussed in
Refs.~\cite{arbus,kuehn,ivanch,kon99} and motivated studies described in 
this paper.

Analysis of ISR events
at $e^+e^-$-factories provides independent and contiguous measurements of
hadronic cross sections in the low-energy region and also contributes to the
 spectroscopy of low-mass resonances.

\subsection{Calculation of ISR and accuracy\label{CalofISR}}
In the lowest order (Fig.~\ref{diag2}) the probability of the initial-state
radiation of the photon with the energy $xE_0$ and the polar angle $\theta$
is as follows~\cite{Wxs,baier1}:
\begin{eqnarray}
w_0(\theta,x)&=&\frac{\alpha}{\pi x}
\left [
\frac
{(1-x+\frac{x^2}{2})\sin^2{\theta}-\frac{x^2}{2}\sin^4{\theta}}
{\left ( \sin^2{\theta}+\frac{4m_e^2}{s}\cos^2{\theta} \right )^2}
\right.\nonumber\\
&&\left. -\frac{4m_e^2}{s}\frac
{(1-2x)\sin^2{\theta}-x^2\cos^4{\theta}}
{\left ( \sin^2{\theta}+\frac{4m_e^2}{s}\cos^2{\theta} \right )^2}
\right ],
\label{ebm}
\end{eqnarray}
where $\alpha$ is the fine-structure constant, and $m_e$ is the electron mass.

The ISR photon is predominantly emitted at small angles with
respect to the beam axis. 
\begin{figure}
\includegraphics[width=.4\textwidth]{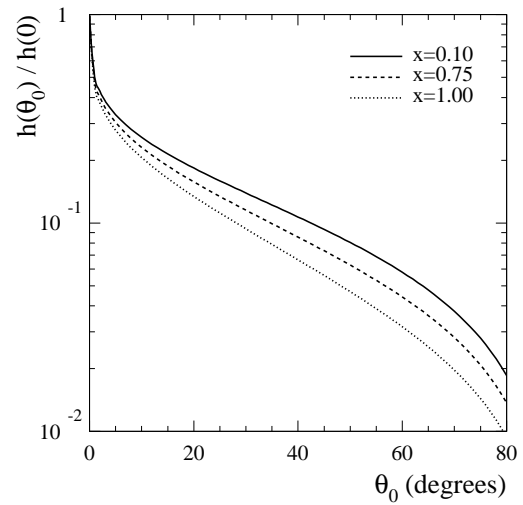}
\caption{The relative probability for the ISR photon to be emitted
into the polar angle range $\theta_0<\theta<180^\circ-\theta_0$ for 
three representative values of $x$.
\label{fig3}}
\end{figure}
In Fig.~\ref{fig3} we present the dependence of the function
$W_0(\theta_0,x)/W_0(0,x)$ on the polar angle limit $\theta_0$, where
\begin{equation}
W_0(\theta_0,x)=\int\limits_{\theta_0}^{\pi-\theta_0} w_0(\theta,x) 
\sin{\theta} \rm{d}\theta.
\end{equation}
The integration is performed for  three values of $x$ at $2E_0=10.58$ GeV,
the c.m. energy of $B$-factories. It can be seen that the  
angular distribution of the ISR photon weakly depends on $x$ and 
that a considerable fraction
of the photons is emitted at large angles. In the next section we will discuss
two approaches to study ISR events, a tagged and untagged one. 
In the tagged approach
the ISR photon should be detected, i.e., emitted at a large angle,
into the fiducial volume of the detector. 
\begin{figure}
\includegraphics[width=.4\textwidth]{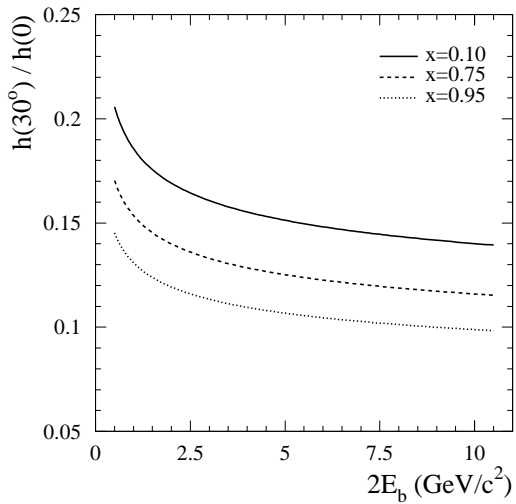}
\caption{The relative probability for the ISR photon to be emitted
into the polar angle range $30^\circ<\theta<150^\circ$ as a function
of the $e^+e^-$ c.m. energy for three representative values of $x$.
\label{fig4}}
\end{figure}
At $B$-factories ($2E_0=10.58$ GeV) about 10\% of high-energy ISR photons 
have $30^\circ<\theta<150^\circ$. This angular range approximately corresponds
to the fiducial volume of the electromagnetic calorimeter of the BABAR
detector.
The fraction of the large-angle ISR increases with decrease of the energy
as shown in Fig.~\ref{fig4}. The compact expressions for $W_0$ can be written
for two practically applicable cases.
For the range of integration $\theta_0<\theta<\pi-\theta_0$, 
$\theta_0\gg m_e/\sqrt{s}$
\begin{eqnarray}
W_0(\theta_0,x)& = &\frac{\alpha}{\pi x}\biggl [ (2-2x+x^2)
\ln{\frac{1+\cos{\theta_0}}{1-\cos{\theta_0}}} \nonumber\\   
&&-x^2\cos{\theta_0}\biggr ].
\label{eq5}
\end{eqnarray}
For the full range of polar angles $0<\theta<\pi$
\begin{equation}
W_0(0,x) = \frac{\alpha}{\pi x}(\ln{\frac{s}{m_e^2}}-1)(2-2x+x^2).
\label{eq6}
\end{equation}

\begin{figure}
\includegraphics[width=.4\textwidth]{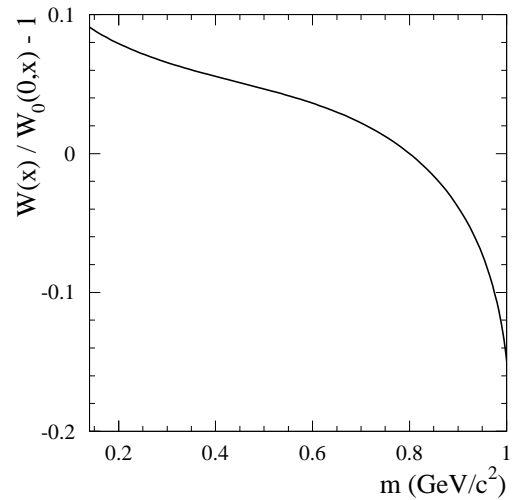}
\caption{The mass ($m=2E_0\sqrt{1-x}$) dependence of the relative difference 
between the radiator function $W(x)$ from Ref.~\cite{str_func_meth} and the 
lowest-order function $W_0(0,x)$ for $2E_0=1.02$ GeV.
\label{fig5}}
\end{figure}

The formulae given above describe ISR processes in the lowest QED order. To 
estimate a contribution of higher-order diagrams (loops and related to
extra photon emission) the function $W(x)$ from Ref.~\cite{str_func_meth}
can be used, which takes into account soft multiphoton emission and 
$\alpha^2$ terms in the leading logarithmic approximation. In this 
approximation
the accuracy $\Delta W/W$ is expected to be better than 1\%.
The relative difference between $W(x)$ and $W_0(0,x)$ as a function of
the invariant mass of the final hadronic system is shown in Fig.~\ref{fig5}
for $2E_0=1.02$ GeV, the c.m. energy of the $\phi$-factory in Frascati. 
It is seen that the radiative correction to the lowest-order radiator function
reaches 15\%.
It should be noted that the size of the radiation correction depends on
experimental conditions. For example, in Ref.~\cite{babrpp} the function 
$W(x)$ is calculated at $2E_0=10.58$ GeV with conditions that the 
highest-energy ISR photon has a polar angle in the range 
$20^\circ<\theta<160^\circ$ and that the invariant mass of the hadronic system 
combined with the ISR photon is greater than 8 GeV/$c^2$. The latter condition
restricts the maximum energy of extra photons emitted from the initial state. 
With these conditions the radiative correction factor 
$1+\delta=W(20^{\circ},x)/W_0(20^\circ,x)$
is close to unity with the maximum deviation $\delta$ of about 2\%.

To provide accuracy better than 1\% required for the measurement of the
exclusive hadronic cross sections at low energies, the calculation of
the radiator function should include the higher-order radiative correction,
in particular, due to emission of extra photons.
Several theoretical papers are devoted to study radiative corrections to ISR 
processes, for example, \cite{arbus,kuehn,khoze1,khoze2,rodr021,czyz03}.
The approaches of Refs.~\cite{kuehn,rodr021,czyz03} allow one
to develop generators of Monte Carlo (MC) events and are used in 
analyses of experimental data.  In Ref.~\cite{kuehn} the photon emission at
large angles only is considered; radiative corrections are calculated 
in the leading
logarithmic approximation with the structure function 
technique~\cite{strfun,strfun1}.
The accuracy of the method is determined by neglecting sub-leading $\alpha^2$
contributions and estimated in Ref.~\cite{rodr01} to be about 1\%.
In Refs.~\cite{rodr021,czyz03} the one-loop corrections and exact matrix
element for emission of two hard photons are calculated. The accuracy
of this next-to-leading order (NLO) calculation is estimated to be about 
0.5\%~\cite{rodr021} due to the higher-order effects.

\subsection{Monte Carlo generators}
The calculation of the radiator function is usually performed by the 
Monte Carlo method. A special computer code referred to as an 
``event generator'' provides events  (sets of the four-momenta of the 
final particles)
distributed over the phase space according to the matrix element squared of
the process under study. The phase space can be restricted by some 
conditions on the angles and energies of the generated ISR photons.
These conditions should be looser than the actual experimental conditions
used for event selection. 

The interaction of the generated particles with the detector and the detector 
response are then simulated. In modern experiments the detector simulation is 
based on the GEANT4~\cite{GEANT4} package. The simulated events are 
reconstructed with the program chain used for experimental data. The detection
efficiency is determined as the ratio of the mass spectrum of simulated events
that passed selection criteria to the spectrum of generated events. 

Most of ISR analyses discussed in this paper are based on two event generators.
Historically, EVA was the first ISR Monte Carlo generator.  
The AfkQed package used in the BABAR experiment at the SLAC
$B$-factory  is a development
of the EVA generator~\cite{kuehn,kuehn2} initially designed to simulate
ISR production of the  $2\pi$ and $4\pi$ final states with 
an ISR photon emitted at large angles. The soft-photon radiation from 
the initial state is 
generated with the structure function method~\cite{strfun,strfun1}. Two extra
photons are emitted in the directions of the initial electron and positron.
The program has a modular structure allowing to implement easily new
hadronic modes. The AfkQed package includes generation of $2\pi$, $3\pi$,
$4\pi$, $5\pi$, $6\pi$, $\eta\pi^+\pi^-$ states, modes with kaons 
$K\bar{K}+ n\pi\;~n=0,1,2,3,4$, and protons $p\bar{p}$, $p\bar{p} 2\pi$.
The generation of the process $e^+e^-\to \mu^+\mu^-\gamma$  is also included
into the AfkQed package. For this process  both  initial- and
final-state radiation (FSR) diagrams and their interference are taken into 
account.  For the charged particles the final-state 
radiation is generated using the PHOTOS package~\cite{PHOTOS}.

The Phokhara event generator is used in the BABAR and Belle experiments at 
the $B$-factories, and in the KLOE experiment at the $\phi$-factory. 
Its latest version 6.1~\cite{Phokhara6} includes generation 
of the $2\pi$, $3\pi$,
$4\pi$, $K\bar{K}$, $p\bar{p}$, and $\Lambda\bar{\Lambda}$ hadronic states,
and the process $e^+e^-\to\mu^+\mu^-\gamma$. The initial-state radiation is 
generated
in NLO~\cite{rodr021,czyz03}, i.e., one or two photons can be emitted by the 
initial electron and positron. The generator can be used for simulation of
both tagged and untagged ISR measurements. For the processes
$e^+e^-\to\mu^+\mu^-\gamma$, $e^+e^-\to\pi^+\pi^-\gamma$, and $e^+e^-\to
K^+K^-\gamma$, NLO FSR radiative corrections are implemented.
In particular, a hard ISR photon can be accompanied by emission of a soft 
photon from the final state.

For all the hadronic states except the two-body $2\pi$ and $K\bar{K}$
as well as  $\pi^+\pi^-\pi^0$, the 
structure of the electromagnetic hadronic current entering  the matrix
element of the process $e^+e^-\to$ hadrons is model dependent and 
the object of a study by itself. This model dependence is the 
second source of the
theoretical uncertainty. For most of the measurements of multihadron 
cross sections its contribution significantly exceeds the 0.5-1.0\%
uncertainty of the radiator function. To estimate the model uncertainty, 
the distributions of hadrons in data are compared to the corresponding
simulated distributions. Usually, the difference between the detection 
efficiency obtained with different models of the hadronic currents is
taken as an estimate of the model uncertainty.
\section{Experimental Techniques}
\subsection {Tagged and untagged ISR\label{taguntag}}
\begin{figure}
\includegraphics[width=.4\textwidth]{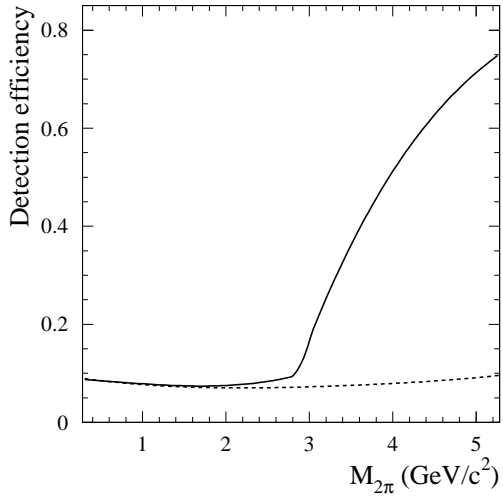}
\caption{The detection efficiency for the process $e^+e^-\to \pi^+\pi^-\gamma$
at $2E_0=10.58$ GeV as a function of the $2\pi$ invariant mass for  
untagged (solid curve) and tagged (dotted curve) ISR photons.
\label{fig6}}
\end{figure}
There are two approaches for studying ISR events. In the first approach,
the untagged one, detection of the ISR photon is not required, 
but all the final 
hadrons must be detected and fully reconstructed. The ISR events 
are selected by the requirement that the recoil mass against the hadronic
system be close to zero. The mass dependence of the detection efficiency
for the process $e^+e^-\to \pi^+\pi^-\gamma$ at $2E_0=10.58$ GeV
is shown in Fig.~\ref{fig6}. The efficiency is calculated with the Phokhara 
event generator in leading-order mode. The detector acceptance for charged
pions is assumed to be limited by the condition $30^\circ<\theta<150^\circ$
which corresponds to the polar angle coverage of the BABAR detector. The solid
curve in Fig.~\ref{fig6} represents the efficiency for the case of untagged 
ISR photons. For two-pion masses below 3 GeV/$c^2$ the detection efficiency
is about 10\% and changes very slowly with mass. At these relatively low  
invariant masses, pions are produced in a narrow cone around the vector 
opposite to the ISR photon momentum and therefore can be
detected only if the ISR photon is emitted at a large angle. The dotted
curve in Fig.~\ref{fig6} represents the detection efficiency for the case
of a tagged ISR photon. The photon polar angle is required to be in the range
from $30^\circ$ to $150^\circ$. It is seen that tagged and untagged
efficiencies are very close in the mass range below 3 GeV/$c^2$. 
For higher masses the small-angle ISR begins to contribute to the untagged
efficiency leading to its rapid increase, whereas the efficiency for the 
case of a tagged ISR photon varies insignificantly.

At $B$-factories the untagged approach is used for measurements of exclusive
cross sections for masses of produced hadronic systems above 3.5 GeV/$c^2$.   
The untagged detection efficiency is very sensitive to the angular 
distributions of the final hadrons. Therefore this approach is suitable for
the measurement of hadronic processes with well defined dynamics, for
example, $e^+e^- \to D\bar{D}$ or $e^+e^- \to D^\ast\bar{D}$.  
For multihadron final states this strong sensitivity to hadron angular
distributions can lead to a sizeable systematic uncertainty of the measurement. 

\begin{figure}
\includegraphics[width=.4\textwidth]{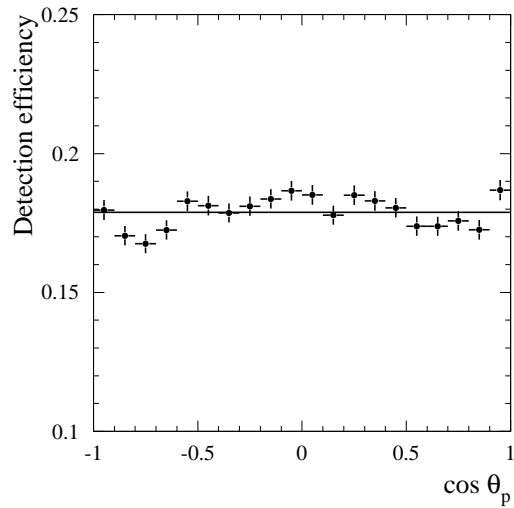}
\caption{ The $\cos{\theta_p}$ dependence of the detection efficiency
for the process $e^+e^- \to p\bar{p}\gamma$~\cite{babrpp}, 
where $\theta_p$ is the proton
angle measured in the $p\bar{p}$ rest frame with respect to the ISR photon 
direction. The horizontal line indicates the detection efficiency averaged 
over $\cos{\theta_p}$.
\label{fig7}}
\end{figure}

All measurements of exclusive cross sections of \epem\ annihilation
into light hadrons at 
$B$-factories were performed using the tagged approach. In contrast to the case 
of untagged ISR, the efficiency for events with a detected photon depends 
weakly on the angular distributions of the final hadrons. As an example,
the angular dependence of the detection efficiency for the process 
$e^+e^-\to p\bar{p} \gamma$~\cite{babrpp} is shown in Fig.~\ref{fig7}, 
where $\theta_p$ is the proton angle in the $p\bar{p}$ rest frame measured 
with respect to the ISR photon direction. This advantage of the tagged ISR 
approach allows one to measure the cross section for 
multihadron final states with a relatively small model uncertainty.

\begin{figure}
\includegraphics[width=.4\textwidth]{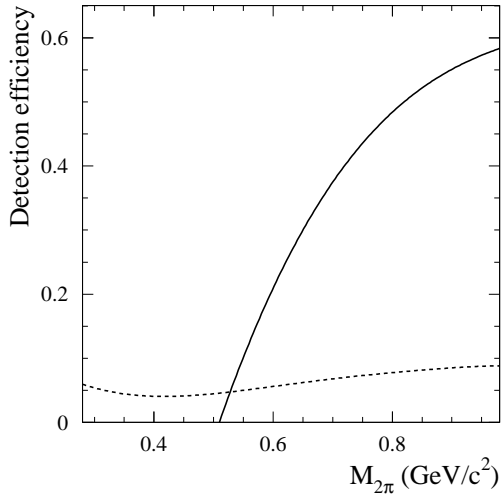}
\caption{The mass dependence of the detection efficiency for the process 
$e^+e^-\to \pi^+\pi^-\gamma$ at $2E_0=1.02$ GeV for two selections,
untagged ($\theta_\gamma<15^\circ$ or $\theta_\gamma>165^\circ$) and 
tagged ($50^\circ < \theta_\gamma <130^\circ$), shown by the solid and
dashed curves, respectively. The pion polar angles range from
$50^\circ$ to $130^\circ$.
\label{fig8}}
\end{figure}

Since ISR photons are emitted predominantly
along the beam axis, in untagged ISR measurements the additional condition 
that $\cos{\theta_\gamma}$ is close to $\pm1$ can be used,
where $\theta_\gamma$ 
is the polar angle of the momentum recoil against the hadronic system in 
the $e^+e^-$ c.m. frame.
In particular, in Refs.~\cite{kloe1,kloe2} the condition $\theta_\gamma
<15^\circ$ or $\theta_\gamma>165^\circ$ is used to select  
$e^+e^-\to \pi^+\pi^-\gamma$ events at the $\phi$-factory. This condition
allows to significantly reduce background from the decay $\phi\to 3\pi$ and
almost completely remove the FSR background, i.e.,  
$e^+e^-\to \pi^+\pi^-\gamma$ events with the photon emitted from
the final state. It should be noted that the FSR contribution
related to radiation by pions is negligible in $B$-factory experiments 
due to the smallness of the pion electromagnetic form 
factor at $s=112$ GeV$^2$. At this energy, the structure-dependent 
contribution, for example, of the processes $e^+e^- \to f_0\gamma$ and
$e^+e^- \to f_2\gamma$ is also expected to be small. Theoretical
estimations for the cross sections of these processes at large $s$ are
absent in literature. An estimate was made for the process 
$e^+e^- \to p\bar{p}$ in Ref.(Aubert,2006a). The FSR contribution
(including a structure-dependent part) was found to be less than 
$10^{-3}$ for the $p\bar{p}$ mass below 4.5~GeV.
The detection efficiency for the process
$e^+e^-\to \pi^+\pi^-\gamma$ at $2E_0=1.02$ GeV with the condition on
$\theta_\gamma$ described above is shown in Fig.~\ref{fig8}. 
The pion polar
angles are required to be in the range $50^\circ$--$130^\circ$. Due to this
restriction the detection efficiency falls rapidly with decreasing 
$2\pi$ mass. The untagged approach was used in Refs.~\cite{kloe1,kloe2}  to
measure the $e^+e^-\to \pi^+\pi^-$ cross section in the mass range from
0.592 to 0.975 GeV. The tagged approach allows one to access the 
near-threshold
mass region. The detection efficiency for $\pi^+\pi^-\gamma$ events with
a detected photon ($50^\circ<\theta_\gamma<130^\circ$) is shown in 
Fig.~\ref{fig8} by the dashed curve.  This selection was also used in the
KLOE experiment~\cite{kloe3,kloe3a} and allowed to reduce the lower mass boundary for
the cross section measurement from 0.592 to 0.316 GeV.

\subsection {Hadronic mass resolution and mass scale calibration\label{masres}}
  The detector resolution on the hadronic invariant mass and the accuracy of the
mass scale calibration are important experimental parameters for the ISR 
cross section measurements.

The mass resolution  $\sigma_m$ is usually determined using MC simulation
as RMS of the $(m_{\rm meas}-m_{\rm true})$ distribution, where $m_{\rm meas}$
and $m_{\rm true}$ are the measured and generated invariant masses,
respectively. The experimental value of the mass resolution can be extracted
from the fit of the measured line shape of a narrow resonance, for example,
$J/\psi$. 

\begin{figure}
\includegraphics[width=.4\textwidth]{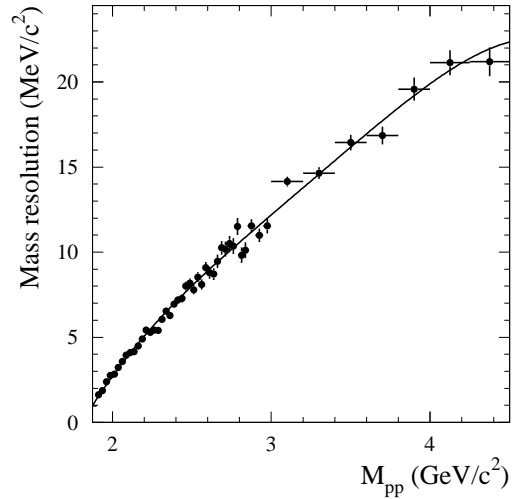}
\caption{The mass dependence of the $p\bar{p}$ mass resolution obtained 
from MC simulation for the process $e^+e^- \to p\bar{p}\gamma$ in 
Ref.~\cite{babrpp}. The curve represents the result of 
a polynomial fit.
\label{resolmass}}
\end{figure}

In general, the invariant mass can be represented as a sum of the 
two terms: $m_{\rm meas}=\Sigma_i m_i + 
\Delta m(\vec{p_1}, \vec{p_2}, {\ldots} )$,
where $m_i$ are masses of stable hadrons produced in the process under
study, and  $\Delta m$ is the term depending on the final particle momenta 
$\vec{p_i}$.  The mass resolution $\sigma_m$ is determined by the precision of
the measurement of the momenta of the charged hadron tracks and photons from 
$\pi^0$ decays. Since $\Sigma_i m_i$ has no sizeable spread, and 
the $\Delta m$ term and its uncertainty are minimal near threshold and grow
with the mass increase, it is expected that $\sigma_m$ also increases
with mass. 
As an example, the mass resolution versus the proton-antiproton mass
for the ISR process  $e^+e^- \to p\bar{p}\gamma$~\cite{babrpp} is shown 
in Fig.\ref{resolmass}. 

At $B$-factories the mass resolution for multihadron systems 
consisting of light quarks
varies from 4--7 MeV/$c^2$ at the mass of 1.5 GeV/$c^2$ to 6--11 
MeV/$c^2$ at 3 GeV/$c^2$; the worse values are for hadron states
with neutral pions. The hadronic cross sections in the mass region between the
$\phi$- and $J/\psi$-meson resonances do not contain structures with 
a width
comparable to the detector resolution. 
The 25-MeV/$c^2$ mass bin was chosen for a study of most of the processes with
light hadrons. With such a bin size the distortion of the mass
spectrum shape because of  resolution effects is small. 
A smaller bin size was used for analyses of the processes 
$e^+e^-\to p\bar{p}\gamma$ and $e^+e^-\to \pi^+\pi^-\gamma$. 
For the former, it is important
to study a near-threshold enhancement in the mass dependence of the proton 
electromagnetic form factor. The good $p\bar{p}$ mass resolution for masses
below 2 GeV/$c^2$ (see Fig.\ref{resolmass}) allows to measure the
cross section in this region with the 5 MeV/c$^2$ mass bin~\cite{babrpp}.
The $e^+e^-\to\pi^+\pi^-$ cross section near the $\rho$-meson peak 
was measured  
in the BABAR experiment with the mass interval of 2 MeV/c$^2$~\cite{babr2pi},
which is significantly smaller than the $\pi^+\pi^-$ mass resolution 
(about 6 MeV/c$^2$ at the $\rho$ peak). The unfolding of resolution 
effects from 
the high-statistics (about one-half million events) mass spectrum was 
performed with the procedure described in Ref.~\cite{unfold}. The procedure
uses a mass-transfer matrix that gives the probability that an event with
true mass in an interval $i$ is reconstructed with $m_{\rm meas}$ in 
interval $j$. The transfer matrix is usually obtained using MC simulation
and corrected to take into account a difference in the resolution
between data and simulation.

The measurement of the $e^+e^-\to\pi^+\pi^-$ cross section at the 
$\phi$-factory~\cite{kloe2} with the KLOE detector was performed with the 
0.01 GeV$^2$ step in the squared mass $s^\prime=m^2_{2\pi}$ corresponding to 
a mass bin width of 6.5 MeV/$c^2$ near the $\rho$ peak. The mass resolution of 
the KLOE detector is about 1.3 MeV/$c^2$ at the $\rho$ mass. The
resolution effects are substantial only in the mass region of the 
$\omega$-$\rho$ interference.
For comparison with theory, these effects were removed by unfolding
the mass spectrum using the Bayesian method~\cite{unfold1}.

For the $J/\psi$ and $\psi(2S)$ produced in ISR processes the observed line 
shapes are fully determined by the detector resolution. In this case  
better mass resolution leads to the larger signal-to-background ratio.
For the process $e^+e^- \to 2(\pi^+\pi^-)\pi^0\gamma$~\cite{babr5pi} 
 in the mass region of  the $J/\psi$ and $\psi(2S)$ mesons discussed
 in Section~V,  the value of the mass resolution 
obtained from the fit to the $J/\psi$ spectrum
is about 9 MeV/c$^2$, in good agreement with MC simulation. 

For the final states containing charmed and charmonium 
mesons ($J/\psi \pi^+\pi^-$, $D\bar{D}$, {\ldots}), the typical resolution
in the 4--5 GeV/$c^2$ mass range is about 5 MeV/$c^2$.  The corresponding
cross sections were measured with the 20-25 MeV/$c^2$ mass bin. For
these final states the influence of the limited mass resolution on the cross
section measurement is negligible.

The precision of the absolute mass scale calibration can be tested by 
comparison of the measured mass values for known resonances with their 
nominal values. For many multihadron states (see Sec. V) the mass calibration
is performed at the $J/\psi$ mass. The difference between the measured and 
nominal~\cite{PDG04} $J/\psi$ masses is found to be less than 1 MeV/$c^2$ 
(see, for example, Refs.~\cite{babr5pi,babrkkpi}). For the $3\pi$ final state
the mass scale shift was determined at the $\omega$- and $\phi$-meson 
masses~\cite{babr3pi}:
$m_\omega-m_\omega^{\rm nominal}=-(0.2\pm0.1)$ MeV/$c^2$ and
$m_\phi-m_\phi^{\rm nominal}=-(0.6\pm0.2)$ MeV/$c^2$. We conclude
that for the measurements of hadronic cross sections at $B$-factories the 
mass scale is defined with a relative accuracy better than or about  
$5\times10^{-4}$.

\subsection {ISR luminosity}
It is clear that a radiation of a hard photon significantly decreases    
the cross section, so the ISR technique can be efficient at
high-luminosity colliders only.
To compare the effectiveness of the ISR method for the measurement of hadronic
cross sections with direct $e^+e^-$ experiments, it is useful to introduce
the concept of ISR luminosity. The mass spectrum for the ISR process
$e^+e^-\to X\gamma $ is expressed in terms of the ISR differential
luminosity $dL/dm$ and the Born cross section for the process 
$e^+e^-\to X$ as
\begin{equation}
\frac{dN}{dm}=\varepsilon(m) (1+\delta(m)) \sigma_0(m) \frac{dL}{dm},
\label{crss}
\end{equation}
where $1+\delta(m)=W(m)/W_0(m)$ is the
radiative correction factor discussed in Sec.~\ref{CalofISR}.  The ISR 
luminosity is proportional to the total integrated luminosity $L$ collected in
an experiment and the lowest-order radiator function given by Eq.(\ref{eq5}) or
Eq.(\ref{eq6}) depending on the angular range used for determination of the 
detection efficiency $\varepsilon(m)$:
\begin{equation}
\frac{{\rm d}L}{{\rm d}m}=W_0(m)\frac{2m}{s} L.
\label{ISRlum}
\end{equation}

\begin{figure}
\includegraphics[width=.45\textwidth]{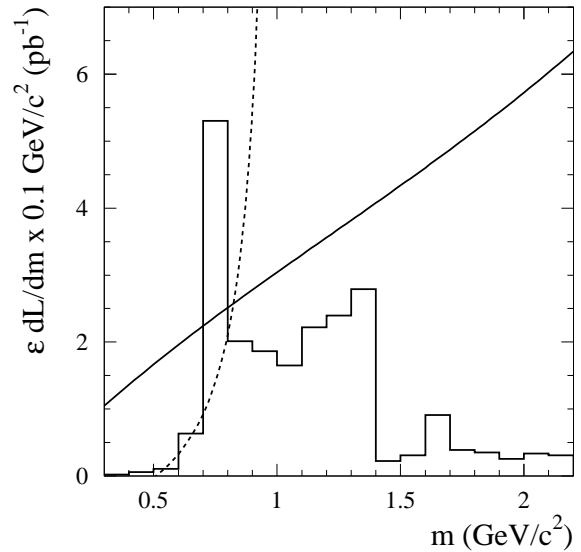}
\caption{The mass dependence of the ISR differential luminosity multiplied
by the detection efficiency.  
The solid curve shows the $\varepsilon {\rm d}L/{\rm d}m$ for the $B$-factory 
($2E_0=10.58$ GeV, $L=500$ fb$^{-1}$, tagged ISR photon), while the
dashed curve shows the same function for the $\phi$-factory 
($2E_0=1.02$ GeV, $L=240$ pb$^{-1}$, untagged ISR photon).
The histogram represents integrated luminosities
collected in direct $e^+e^-$ experiments with the 
SND detector~\cite{snd3pi} at the Novosibirsk VEPP-2M collider (below
1.4 GeV/$c^2$), and with the DM1~\cite{dm16pi} and 
DM2~\cite{dm23pi} detectors at the Orsay DCI 
collider (above 1.4 GeV/$c^2$).
\label{ISRLum1}}
\end{figure}
\begin{figure}
\includegraphics[width=.45\textwidth]{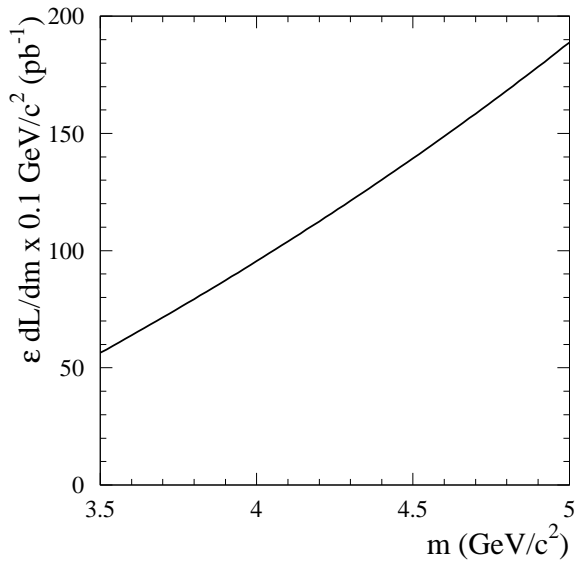}
\caption{The mass dependence of the ISR differential luminosity multiplied
by the detection efficiency for experiments at the $B$-factory 
($2E_0=10.58$ GeV, $L=500$ fb$^{-1}$, untagged ISR photon) in the charm 
production mass region.
\label{ISRLum2}}
\end{figure}
The mass dependence of the ISR differential luminosity multiplied
by the detection efficiency for the BABAR experiment is shown in
Fig.~\ref{ISRLum1} for masses below 2.2 GeV/$c^2$. The detection 
efficiency used was calculated in Sec.~\ref{taguntag} for the
process $e^+e^-\to \pi^+\pi^-\gamma$ with a tagged ISR photon.
The integrated luminosity is taken to be 500 fb$^{-1}$. The dashed
curve in Fig.~\ref{ISRLum1} shows the same quantity calculated for the
KLOE experiment with the integrated luminosity of 240 pb$^{-1}$
and detection efficiency taken for the case of an untagged
ISR photon (Fig.~\ref{fig8}). The luminosity of 240 pb$^{-1}$ was
used in the recent measurement~\cite{kloe2} of the $e^+e^-\to \pi^+\pi^-$ 
cross section in the 0.592--0.975 GeV/$c^2$ mass range. 
The total integrated luminosity collected by the KLOE is about an order of
magnitude larger, 2.5 fb$^{-1}$. The KLOE ISR luminosity is shown only up
to 0.92 GeV/$c^2$. It increases sharply and reaches 21 pb$^{-1}$ at 0.975 
GeV/$c^2$.
It should be noted that the BABAR measurement of the $e^+e^-\to \pi^+\pi^-$
cross section~\cite{babr2pi} is also based on a part of the recorded
data corresponding to 232 fb$^{-1}$. The histogram in Fig.~\ref{ISRLum1}
shows the distribution of the integrated luminosities collected in
some direct $e^+e^-$ experiments. At masses below 1.4 GeV/$c^2$ the 
statistics of the SND experiment~\cite{snd3pi}
recorded at the VEPP-2M collider is presented. This is the
largest integrated luminosity collected in this mass region in a
single experiment. The mass bin 1.0--1.1 GeV/$c^2$ does not include
about 13 pb$^{-1}$ taken by SND in vicinity of the $\phi$-meson resonance.
The significant part of the statistics from the 0.7--0.8 GeV/$c^2$
mass interval is collected in the $\omega$-meson mass window 
0.76--0.80 GeV/$c^2$. In the c.m. energy range 1.4--2.2 GeV/$c^2$ 
the experiments with 
the largest statistics are DM1 and DM2 at the Orsay DCI $e^+e^-$ collider.
The histogram at $m > 1.4$ GeV/$c^2$ shows a sum of the integrated 
luminosities collected with these detectors.

At low masses of the hadronic system the data samples of ISR events currently 
available at $B$-factories exceed the statistics collected in conventional
$e^+e^-$ experiments, especially at masses below 0.7 GeV/$c^2$ and above 
1.4 GeV/$c^2$. The ISR luminosity of the $\phi$-factory increases very
rapidly with mass. For masses below 0.8 GeV/$c^2$ the luminosity currently 
used for ISR analysis~\cite{kloe2} is comparable with that collected in 
direct $e^+e^-$ experiments. For higher masses it exceeds both BABAR and
$e^+e^-$ luminosities.

The ISR luminosity for the mass region of charm production is presented in
Fig.~\ref{ISRLum2}. It corresponds to the 500 fb$^{-1}$ integrated luminosity
collected at $2E_0=10.58$ GeV/c$^2$ and is multiplied by the detection
efficiency calculated for the case of an untagged ISR photon (Fig.~\ref{fig6}).
The ISR luminosity in this mass region significantly exceeds the 
integrated luminosity collected in direct $e^+e^-$ experiments including
the recent CLEO-c energy scan~\cite{cleoxs}, 
60 pb$^{-1}$ at twelve points between 3.97 and 4.26 GeV.

Thus, the current data samples of ISR events produced at the $B$- and 
$\phi$-factories are larger than those produced 
directly in $e^+e^-$ collisions for all masses of interest excluding the 
regions near the narrow resonances ($\omega$, $\phi$, $J/\psi$, $\psi(2S)$). 
For masses above
1.4 GeV/c$^2$ this allows to significantly improve accuracy of the measurements
of exclusive hadronic cross sections. In the mass region below 1.4 GeV/c$^2$ 
the results obtained with the ISR method are comparable to rather precise
direct $e^+e^-$ measurements. 

\subsection {Comparison with  $e^+e^-$ scan\label{isreescancomp}}
  The ISR technique offers some advantages over conventional
$e^+e^-$ measurements. One of them    
is that the entire hadronic mass range is accessible in one
experiment. This allows one to avoid relative normalization uncertainties
which inevitably arise when data from different experiments, or from different
machine settings in one experiment, are combined.   

\begin{figure}
\includegraphics[width=.4\textwidth]{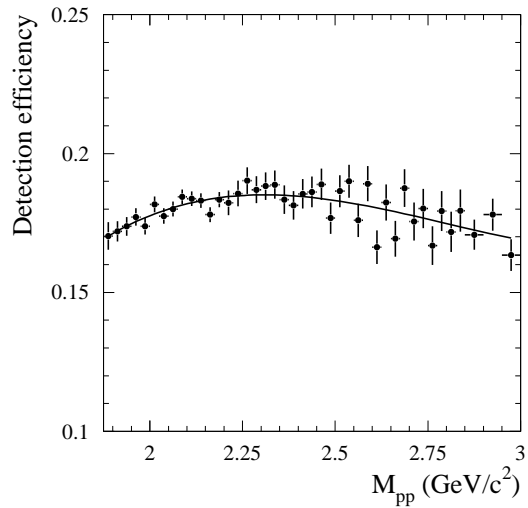}
\caption{The detection efficiency
for the process $e^+e^- \to p\bar{p}\gamma$~\cite{babrpp} as a function 
of the $p\bar{p}$ invariant mass. The curve represents the result of 
a polynomial fit.
\label{effmspp}}
\end{figure}
  The ISR measurements with a tagged photon have additional advantages.
In many cases, particularly for final states with low invariant mass
of the produced particles, the hadronic system
is collimated along the direction opposite to the
ISR photon. Therefore, the detection efficiency has low sensitivity to 
hadron angular distributions in the hadronic-system rest frame.
In Fig.\ref{fig7} the angular dependence of the detection efficiency 
is shown for the process $e^+e^- \to p\bar{p}\gamma$~\cite{babrpp}.
The angular dependence is close to uniform. This reduces the model 
dependence of the cross 
section measurement due to the unknown relation between the 
values of the proton 
electric and magnetic form factors, and significantly 
facilitates data analysis. 
Note that in conventional experiments at $e^+e^-$ or 
$p\bar{p}$ colliders the detector acceptance for the final $p\bar{p}$  or 
$e^+e^-$ systems falls to zero when  $\cos\theta_p$ approaches $\pm 1$.

For ISR events the final hadrons have non-zero momenta at the production 
threshold and are therefore  detected with full efficiency. 
In Fig.\ref{effmspp}
the detection efficiency for the process $e^+e^- \to p\bar{p}\gamma$
process~\cite{babrpp} is shown as a function of the $p\bar{p}$ invariant
mass. No strong variation of the efficiency with mass is observed, while
in direct $e^+e^-$ measurements the detection efficiency
vanishes at the threshold because of the low momenta  of the produced
particles. This feature of ISR hadron production was successfully used 
at BABAR for the measurements of the $e^+e^- \to p\bar{p}$~\cite{babrpp} and
$e^+e^- \to \pi^+\pi^-$~\cite{babr2pi} cross sections in near-threshold
mass regions. 

For the measurement of the $e^+e^- \to \pi^+\pi^-$ cross section,
particle identification plays a crucial role. In the ISR process 
$e^+e^- \to \pi^+\pi^-\gamma$ at $B$-factories most of the final pions have 
momenta larger than 1 GeV/$c$. For such pion momenta a good
$\pi$/$\mu$/$e$ separation is provided which allows one to almost completely 
remove the $e^+e^- \to e^+e^-\gamma$ and 
$e^+e^- \to \mu^+\mu^-\gamma$ backgrounds~\cite{babr2pi}.
This is in contrast with direct $e^+e^-$ 
measurements~\cite{cmd2pi1,cmd2pi2,snd2pi} in which  
it is difficult to separate $e^+e^- \to \pi^+\pi^-$ and 
$e^+e^- \to \mu^+\mu^-$ events in the most interesting $\rho$-meson mass 
region (0.60--0.95 GeV).
As a result, a sum of the cross sections is measured. The contribution 
of the process 
$e^+e^- \to \mu^+\mu^-$  is then subtracted using its theoretically
calculated cross section. This leads to an increase of statistical and 
systematic errors of the measurement.

It should be noted that the advantages of tagged ISR discussed above 
(weak mass and angular dependences of the detection efficiency) are
completely absent for untagged ISR. In this case the mass and angular 
dependences are even stronger than those for events of direct 
$e^+e^-$ annihilation. 

A disadvantage of ISR is that the mass resolution and absolute mass
scale calibration are much poorer than the beam energy spread and 
the accuracy of the beam energy setting in direct $e^+e^-$ annihilation
experiments. The influence of the resolution effects on the ISR measurement
is discussed in Sec.~\ref{masres}.

The main disadvantage of the ISR measurements is presence of a wide spectrum
of background processes different from those in direct $e^+e^-$ experiments.
For example, in $e^+e^-$ annihilation the main background process for 
$e^+e^- \to \pi^+\pi^-\pi^0$ is 
$e^+e^- \to \pi^+\pi^-\pi^0\pi^0$ with a lost $\pi^0$. For the ISR process 
$e^+e^- \to\pi^+\pi^-\pi^0\gamma$ with the $3\pi$ mass in the range 
$m_{3\pi}\pm \Delta m/2$, this background corresponds to the
contribution of the process $e^+e^- \to \pi^+\pi^-\pi^0\pi^0\gamma$  
with the $4\pi$ mass in the same range $m_{3\pi}\pm \Delta m/2$.
The presence in ISR of $4\pi\gamma$ events with arbitrary masses, which may,
in particular, be out of the $m_{3\pi}\pm \Delta m/2$ range, greatly 
increases background.

At the $\phi$-factory and in future ISR measurements at the tau-charm
factory
in Beijing the background from FSR processes should be taken into
account when the ISR photon is detected. The FSR contribution for
the $e^+e^- \to \pi^+\pi^-$ measurements at KLOE is calculated with
the PHOKHARA generator, which models FSR for pions using
scalar
QED, and also takes  into account the radiative $\phi$ decays to
$\pi^+\pi^\gamma$ via the $f_0(980)\gamma$ and $\rho\pi$ intermediate
states. The pion electromagnetic form factor used in the generator is
obtained
from a fit to the $e^+e^- \to \pi^+\pi^-$ experimental data.
In the case of the tau-charm factory,
experimental information on exclusive hadronic cross sections in the energy
range from 3.0 to 4.5 GeV obtained at B-factories can be used to estimate
the FSR contribution. Additional theoretical input is required to estimate
structure-dependent FSR.

Another background source is the non-ISR process of $e^+e^-$ annihilation
into hadrons containing a high-energy $\pi^0$. In particular, the events
of the process $e^+e^-\to X\pi^0$ with an undetected soft photon or 
merged photons from the $\pi^0$ decay may almost completely imitate the 
$e^+e^-\to X\gamma$ events. This background is usually subtracted 
statistically using for normalization selected $e^+e^-\to X\pi^0$ events with 
a reconstructed $\pi^0$. In tagged ISR measurements at 
$B$-factories the process 
$e^+e^-\to X\pi^0$  becomes the dominant background source at 
relatively high masses, about 2 GeV/$c^2$. 
It limits the mass region for ISR studies of light
hadrons to masses below 4.0--4.5 GeV/$c^2$.

In ISR measurements  with an untagged ISR photon, 
the background from $e^+e^-\to X\pi^0$ can be significantly suppressed 
by requiring that the missing
momentum in an event be directed along the beam axis. For untagged ISR,
the main sources of background are ISR processes and two-photon
processes $e^+e^-\to e^+e^-\gamma^*\gamma^* \to e^+e^-X$ in which initial
electron and positron are scattered predominantly at small angles. 
The latter background can be suppressed by a condition on the missing
mass, which should be close to zero for ISR events and has a wide 
distribution for two-photon events.

Background suppression and subtraction are the main sources of the systematic 
uncertainty on ISR measurements. 

\subsection {Colliders and detectors using ISR}
ISR processes were studied in many $e^+e^-$ experiments 
either as a source of useful physical information 
or as a source of background. For example, possibly the first
study of the process $e^+e^-\to \pi^+\pi^-\gamma$ was
performed more than 20 years ago with the ND detector at the VEPP-2M 
collider~\cite{NDpipig,NDsum}.
In this work, the FSR process $e^+e^-\to \rho \to \pi^+\pi^-\gamma$
was measured with the ISR process $e^+e^-\to \rho\gamma \to \pi^+\pi^-\gamma$
studied as a main source of background. Many interesting ISR
studies have been performed with the CLEO detector, see, e.g.,
Ref.~\cite{cleomu}.
Below we give a brief description of only three detectors: BABAR, Belle,
and KLOE, which made a great contribution both to development of the 
ISR technique and  ISR measurements of hadronic cross sections.

\subsection {PEP-II and BABAR}
The PEP-II $B$-factory at SLAC is a two-ring asymmetric-energy 
$e^+e^-$ collider   
with energies of 9 GeV for the electron  and 3.1 GeV for
the positron beam, operating at the c.m. energy of 10.58 GeV, at 
the maximum of 
the $\Upsilon (4S)$ resonance~\cite{pep2}. The maximum luminosity
achieved at PEP-II was slightly over  $10^{34}$~cm$^{-2}$~s$^{-1}$.  
The principal goal of
the PEP-II $B$-factory and the BABAR detector is studies of 
$CP$ violation in the $B$-meson system.

\begin{figure}
\includegraphics[width=.45\textwidth]{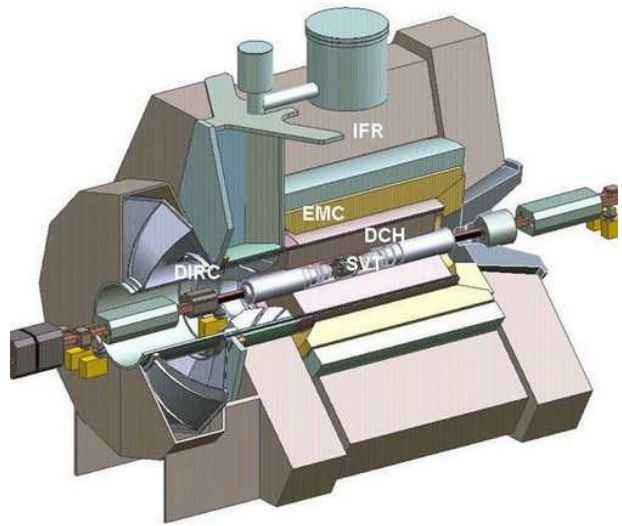}
\caption{View of the BABAR  detector.
\label{BBFig}}
\end{figure}

      The BABAR detector  (Fig.\ref{BBFig}) is described in detail 
elsewhere \cite{babar}.   Final states with charged
particles are reconstructed in the BABAR tracking system, which comprises
a  five-layer silicon vertex tracker (SVT) 
and a 40-layer  drift chamber (DCH) operating in a 1.5-T axial magnetic
field.   The vertex
position is measured by the SVT with the accuracy of  50 $\mu$m.  The
momentum resolution for 1 GeV/$c$ charged tracks is $\sigma_{p_t}/p_t=0.5$\%. 
Charged-particle identification is provided by an internally
reflecting ring-imaging Cherenkov detector (DIRC), 
and by energy loss measurements
in the SVT and DCH. 
The hard ISR photon and photons from  $\pi^0$  decays are detected in a
CsI(Tl) electromagnetic calorimeter (EMC).
The energy resolution for 1 GeV photons is about 3\%; the angles of photons are
measured with the 4 mrad accuracy. 
Muons are identified in the instrumented flux return (IFR) of
the solenoid,
which consists of iron plates interleaved with resistive plate chambers.

Experiments at the PEP-II collider with the BABAR detector were carried out
from 1999 to 2008. The total integrated luminosity is close to 530 
fb$^{-1}$. The ISR studies at BABAR started in 2001. The ISR research
program includes a study of the light hadron production with a 
tagged ISR photon
and charm and charmonium studies with an untagged photon.

\subsection {KEKB and BELLE}
The KEK B-Factory, KEK-B, is an asymmetric-energy (similar to PEP-II)
$e^+e^-$ collider with the 8-GeV
electron and 3.5-GeV positron beams and 
 the maximum luminosity of 
$2.1 \cdot 10^{34}$~cm$^{-2}$~s$^{-1}$~\cite{kekb}.  
The main physical goal of this project
is to perform a detailed study of $B$-meson properties, in particular, 
$CP$-violation.

\begin{figure}
\includegraphics[width=.48\textwidth]{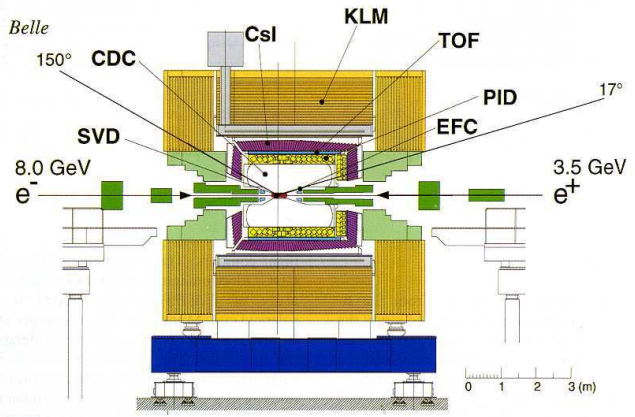}
\caption{Side view of the Belle  detector.
\label{BelFig}}
\end{figure}

   The Belle detector~\cite{BelDet} 
 (Fig.\ref{BelFig})  is configured
inside a 1.5 T superconducting solenoid. The $B$-meson vertices are measured
in a three-layer doublesided silicon vertex detector with about  50 $\mu$m
impact parameter resolution  for 1 GeV/c momentum track at
$\theta\simeq\pi/2$. Track momenta are measured in a 
50-layer wire drift chamber with a 0.4\% momentum resolution at 1 GeV/c.
Particle identification is provided by $\rm{d}E/\rm{d}x$ 
measurements in the drift
chamber, aerogel Cherenkov counters, and time-of-flight counters placed
outside the drift chamber. Electromagnetic showers are detected in
a CsI(Tl) calorimeter located inside the solenoid coil. The energy resolution
is 2\% for 1-GeV photons. An iron flux-return located outside the coil
is instrumented to detect $K_L$-mesons and identify muons.   

   Experiments with Belle started in 2000 and 
stopped in 2010. The Belle integrated luminosity reaches 1000 fb$^{-1}$.
The ISR experiments are mainly devoted to the production of charm and 
charmonium hadronic states with mass above 4 GeV/$c^2$. ISR analysis 
of light mesons is in progress.

\subsection {DA$\Phi$NE and KLOE}
DA$\Phi$NE, the Frascati $\phi$-factory \cite{DafneKloe}, 
is in operation since 1999. The main
goal of the DA$\Phi$NE project is a study of neutral and charged kaons,
intensively produced at the energy corresponding to the 
maximum of $\phi$(1020) resonance.
Similar to PEP-II and KEK-B,  DA$\Phi$NE uses two separate rings for 
storing electron and positrons, but beams have equal energies. 
The DA$\Phi$NE  design luminosity is $5\cdot 10^{32}$~cm$^{-2}$~s$^{-1}$. 
   
\begin{figure}
\includegraphics[width=.45\textwidth]{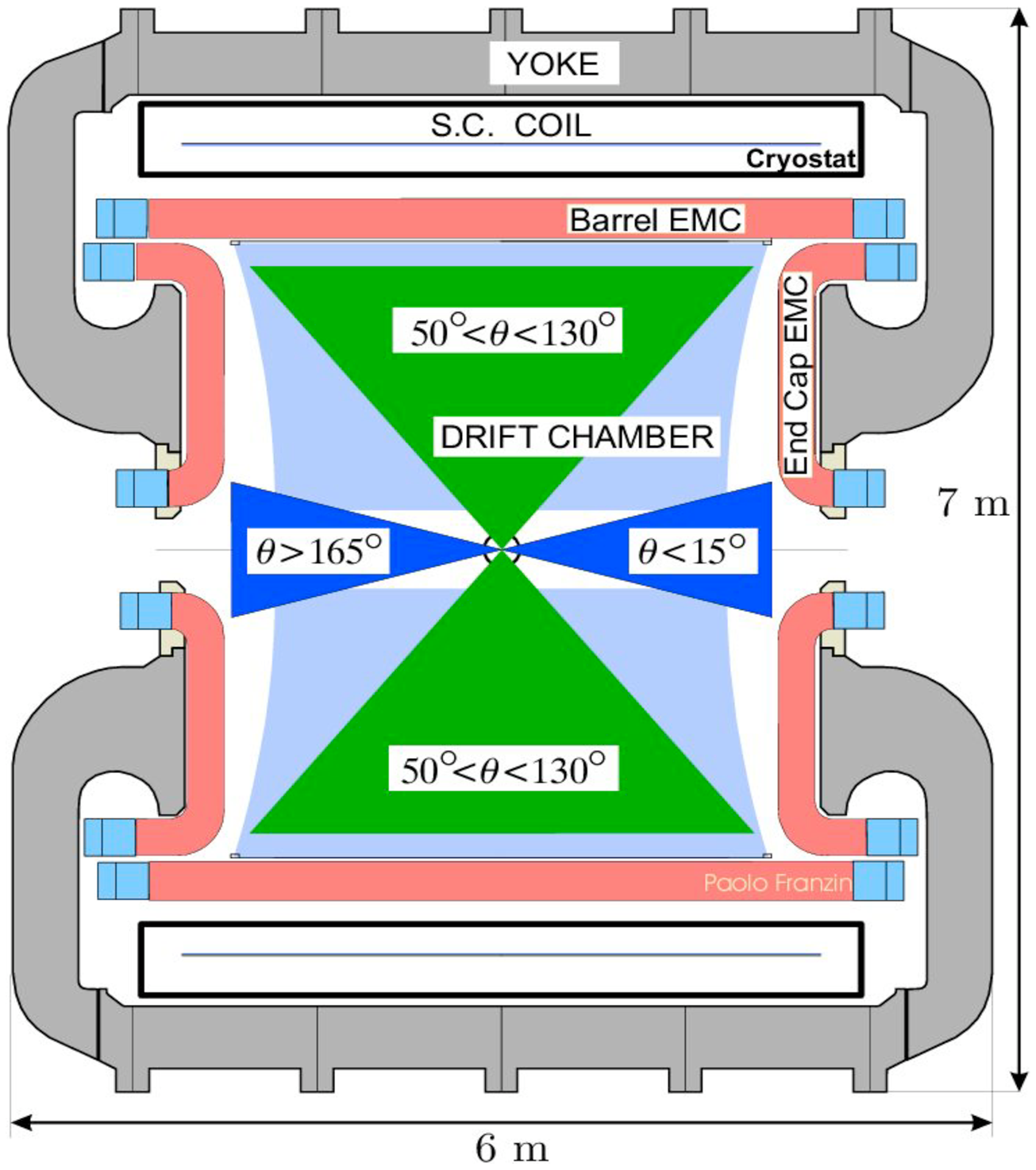}
\caption{ KLOE detector~\cite{kloe3}. The polar-angle regions used
to select tagged ($50^\circ < \theta_\gamma < 130^\circ$) and 
untagged ($\theta_\gamma < 15^\circ$ or $\theta_\gamma > 165^\circ$) ISR events 
are shown.\label{KLOE}}
\end{figure}

   KLOE \cite{DafneKloe} (Fig.\ref{KLOE}) is the main 
DA$\Phi$NE detector.
The detector consists of a large-volume drift chamber (DC) 
surrounded by a hermetic electromagnetic
calorimeter (EMC). A superconducting coil provides an axial magnetic field
of 0.52 T. In order to reduce neutral kaon regeneration and 
charged-particle multiple scattering, the 
gas mixture of 90\% helium and 10\% isobutane is used in the DC. 
Charged-track momenta are measured with the  $\sigma_{p}/p=0.4$\% accuracy. The
lead-scintillation fiber calorimeter provides the energy resolution for
electromagnetic showers of $\sigma_E/E=5.7\%/{\sqrt{E({\rm GeV})}}$, 
and the time resolution
of $\sigma_t=54{\rm ps}/{\sqrt{E({\rm GeV})}} \bigoplus 140 {\rm ps}$.

   The total integrated luminosity accumulated with KLOE is about 
3 fb$^{-1}$.
The only, but very important, ISR process studied at KLOE
is  $e^+e^-\to\pi^+\pi^-\gamma$.
\clearpage
\section{Production of light quark mesons}
\subsection{Overview}
As already mentioned in the Introduction, $e^+e^-$ annihilation into hadrons 
at c.m. energies below 2~GeV plays a very important role in 
many fundamental problems of particle physics. In particular,
knowledge of its total cross section is mandatory for   
the calculation of the muon  anomalous magnetic moment in Standard
Model. For many years  only \epem ~scan experiments  provided 
information on this reaction and determined the uncertainty 
of the SM prediction of the muon anomaly~\cite{dehz1,dehz2}.  
Main information on light vector mesons has been also obtained 
in such measurements. Unfortunately, the collected data samples
were not sufficient for a precise determination of 
parameters of excited vector mesons.
Recently, due to a very high luminosity of the \epem\  factories,
DAFNE, KEK-B, and PEP-II, the  ISR technique became a powerful tool
for an independent study of $e^+ e^-$ annihilation at low energies.

The KLOE collaboration used the ISR method at the $\phi$ meson energy
to study the reaction $e^+e^- \to \pi^+\pi^-$ and measure 
the pion electromagnetic form factor~\cite{kloe1,kloe2,kloe3,kloe3a}.
Recently, results on this process were also reported
by the BABAR collaboration~\cite{babr2pi}.

A variety of high-multiplicity final states were studied at BABAR:
$\pi^+\pi^-\pi^0$~\cite{babr3pi}, $2(\pi^+\pi^-)$, 
$\pi^+\pi^-K^+K^-$ and $2(K^+K^-)$~\cite{babr4pi},
$3(\pi^+\pi^-)$, $2(\pi^+\pi^-\pi^0)$ and
$K^+K^-2(\pi^+\pi^-)$~\cite{babr6pi},
$2(\pi^+\pi^-)\pi^0$, $2(\pi^+\pi^-)\eta$,
$K^+K^-\pi^+\pi^-\pi^0$ and $K^+K^-\pi^+\pi^-\eta$~\cite{babr5pi},
$K^+K^-\pi^+\pi^-$ and $K^+K^-\pi^0\pi^0$~\cite{babr2k2pi},
$K^+K^-\pi^+\pi^-$, $K^+K^-\pi^0\pi^0$ and 
$K^+K^-K^+K^-$~\cite{babr2k2pi}, $K^{\pm}K^0_S\pi^{\mp}$, $K^+K^-\pi^0$
and $K^+K^-\eta$~\cite{babrkkpi}. The final $K^+K^-\pi^+\pi^-$ state
was also investigated by Belle~\cite{bel2k2pi}.

Studies of the exclusive channels of \epem\ annihilation listed above
allow to determine such fundamental parameters as mass, width and 
leptonic width of various vector mesons. In addition to the low-lying
resonances, such as the $\rho$, $\omega$ and $\phi$, where ISR studies
can independently provide meaningful and competitive
information, they are indispensable for a much more precise than
before investigation of the excited vector states.

Moreover, detailed analysis of the dynamics shows that in many cases 
a multiparticle final state can be reached via different intermediate 
mechanisms. For example, four pions can be produced via $\omega\pi^0$,
$a_1(1260)\pi$, $\rho^0 f_0$, $\ldots$.
In the following sections we show a complexity of the internal substructures 
observed in some channels, which are often used  to extract  
parameters of the resonances involved in the substructures.

In general, amplitudes corresponding to different intermediate mechanisms 
interfere affecting the energy and angular distributions of the
final particles. This interference should be taken into account to
avoid additional systematic errors.

Unless otherwise stated, all cross sections in the following sections 
are corrected for effects of initial-state radiation only. Neither 
final-state radiation nor vacuum polarization corrections have been applied.

\subsection{$e^+ e^-\to \pi^+\pi^-$}
\begin{figure}
\includegraphics[width=.48\textwidth]{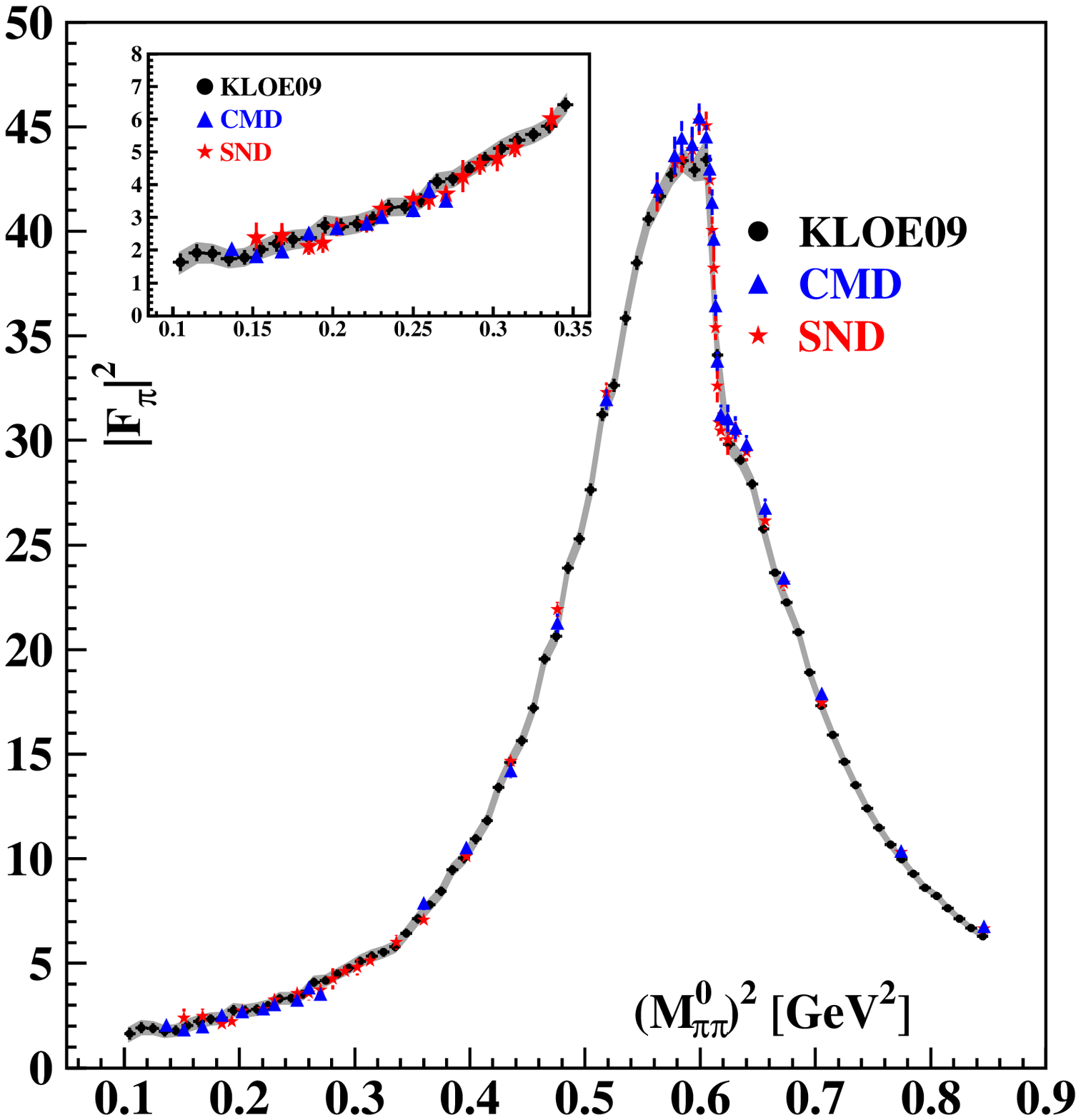}
\includegraphics[width=.48\textwidth]{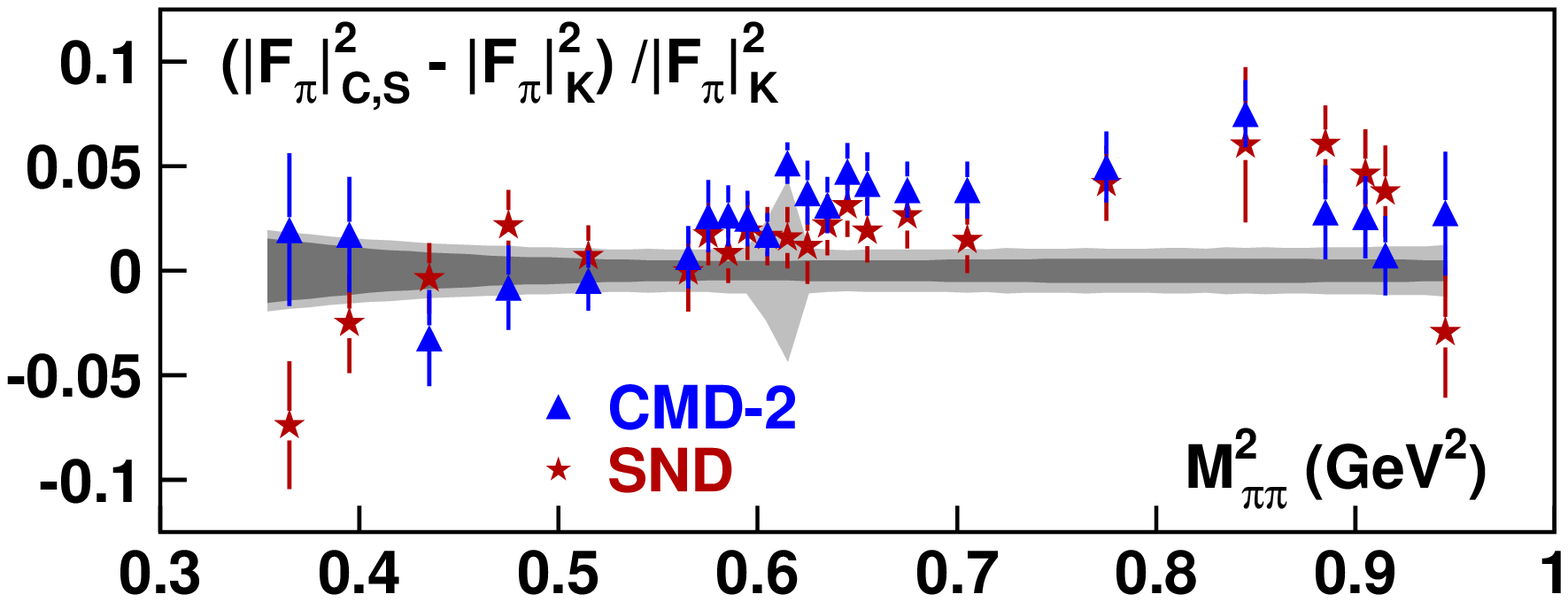}
\caption{\label{fig111} 
Top: The pion form factor obtained by KLOE in the reaction
$e^+e^-\to\pi^+\pi^-\gamma$ with a tagged ISR photon~\cite{kloe3}.
Bottom: Relative difference between the KLOE result with an untagged
ISR photon~\cite{kloe2} and the direct $e^+ e^-$ measurements
by SND~\cite{snd2pi} and CMD-2~\cite{cmd2pi2}. The dark (light) band 
indicates the KLOE uncertainty
(statistical and systematic errors combined in quadrature). For
the SND and CMD-2 data, the combined statistical and systematic 
errors are shown. 
}
\end{figure}
\begin{figure}
\includegraphics[width=.48\textwidth]{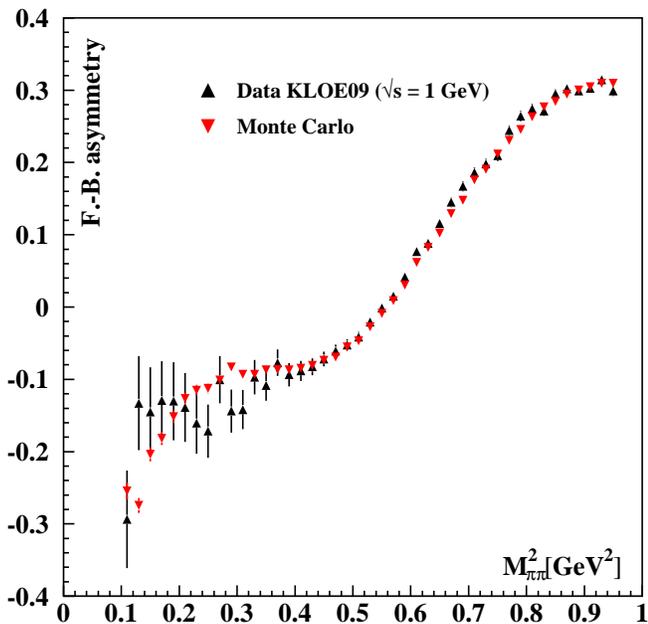}
\caption{\label{figasym} 
Forward-backward asymmetry in the reaction $e^+e^-\to\pi^+\pi^-(\gamma)$  
measured with the KLOE detector~\cite{kloe3}. }
\end{figure}
\begin{figure}
\includegraphics[width=.48\textwidth]{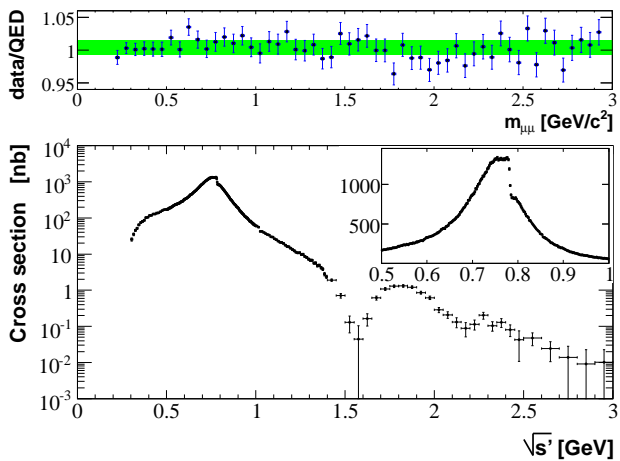}
\caption{\label{fig2} 
Top: The QED test by the ratio of the  $e^+ e^-\to\mu^+\mu^-$ 
cross section in data to the theoretical one. 
Bottom: The $e^+ e^-\to\pi^+\pi^-$ cross sections 
measured with the BABAR detector~\cite{babr2pi}. }
\end{figure}
The reaction $e^+ e^-\to \pi^+\pi^-$ was relatively well studied 
for c.m. energies up to 1.4 GeV in direct \epem~ 
experiments. The most precise measurements were performed with the 
CMD-2~\cite{cmd2pi1,cmd2pi2} and SND~\cite{snd2pi} detectors at the
VEPP-2M collider. The CMD-2 measurements have a systematic uncertainty
in the 1\% range. 

The dominant contribution to this process comes from the 
$\rho(770)$ meson production.
A measurement of the $e^+ e^-\to\pi^+\pi^-$ cross section 
in the $\rho(770)$ mass region was performed by KLOE using the 
ISR method~\cite{kloe1,kloe2,kloe3,kloe3a}.
For the first time it was demonstrated that the
cross section determined by this method could have smaller 
statistical errors than direct \epem~measurements
and could be competitive with them in a systematic uncertainty.
Both untagged~\cite{kloe1,kloe2} and
tagged~\cite{kloe3,kloe3a} ISR $\pi^+\pi^-\gamma$ events were studied with 
consistent results. While the tagged measurement has worse statistical
errors and an additional source of the systematic uncertainty due to the
FSR contribution, it  covers the region of small invariant masses
inaccessible for the untagged measurement.
The result of the tagged measurement~\cite{kloe3,kloe3a} represented as a
pion electromagnetic form factor squared is shown in Fig.~\ref{fig111}(top) 
together with the results of the direct \epem measurements with 
the CMD-2~\cite{cmd2pi2} and SND~\cite{snd2pi} detectors.
Comparison of the more precise untagged KLOE measurement~\cite{kloe2}
with the CMD-2 and SND data is given in Fig.~\ref{fig111}(bottom).
At invariant masses corresponding to the maximum of 
the $\rho$ resonance and its high-mass tail the points
from direct \epem measurements lie systematically higher than
those from KLOE. In this mass region the difference 
between the CMD-2 and KLOE measurements is definitely larger
than their combined systematic uncertainty.  
The KLOE systematic error includes the experimental
(0.6\%) and theoretical (0.6\%) uncertainties. Two main sources of 
the former are tracking and luminosity measurement.  The latter is determined
mostly by the accuracy of the radiator function calculated with 
the PHOKHARA event generator. Note that the KLOE Collaboration 
performed a dedicated study to
validate a calculation of FSR effects using forward-backward
asymmetry arising from the interference between the ISR and FSR
amplitudes (M\"uller, 2009). The study showed that the assumption of
pointlike pions works reasonably well and can be used for the 
FSR calculation, see Fig.~\ref{figasym}.

A structure seen at the top of the $\rho$-meson resonance is due to its 
interference with the much more narrow $\omega(782)$ resonance also 
decaying to $\pi^+\pi^-$. Because $\omega(782)$ mass is known precisely, the
position of this structure can be used to test the accuracy of
the mass scale calibration. Unfortunately, neither KLOE nor BABAR 
(see below) report the result of such a test.

\begin{figure}
\includegraphics[width=.48\textwidth]{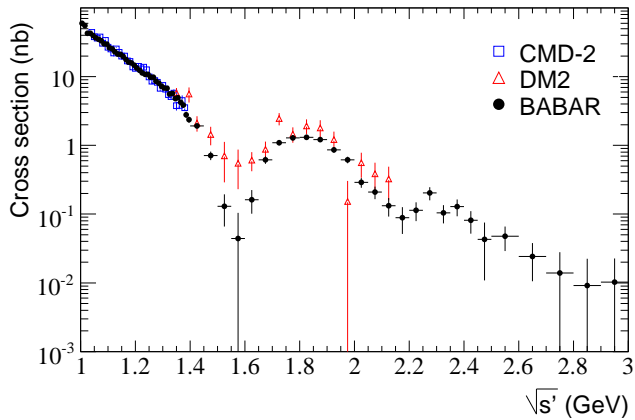}
\caption{\label{fig2a} The $e^+ e^-\to\pi^+\pi^-$ cross section 
above 1 GeV measured with the BABAR detector~\cite{babrch}. Comparison
with the CMD-2~\cite{cmd2pihigh} and DM2~\cite{DM2pihigh} 
measurements is shown. }
\end{figure}

The PEP-II B-factory also provided a large sample of the $e^+ e^-\to
\pi^+\pi^-\gamma$ events (about 530 thousand) 
and the $e^+ e^-\to\pi^+\pi^-$ cross 
section~\cite{babr2pi} was measured for the \epem\ c.m. energies
up to 3.0 GeV. In this experiment
the  $e^+ e^-\to\pi^+\pi^-$ cross section is obtained from the
ratio of the $\pi^+\pi^-$ and $\mu^+\mu^-$ mass spectra.
Due to the normalization to the cross section of the theoretically well known 
process $e^+ e^-\to\mu^+\mu^-\gamma$, 
the measurement becomes much less sensitive to the experimental
uncertainties and to the theoretical uncertainty of the radiator
function. A comparison of the measured $\mu^+\mu^-$ mass spectrum 
for the reaction $e^+e^-\to\mu^+\mu^-\gamma$ with the QED prediction 
is shown in Fig.~\ref{fig2}(top). The data and the prediction are
consistent within the estimated systematic uncertainty of 1.1\%,
dominated by the accuracy of the integrated luminosity measurement. 
Using the bin-by-bin ratio to the cross section of the process
$e^+e^-\to\mu^+\mu^-\gamma$ one  minimizes theoretical uncertainties 
and reduces  a systematic error at the $\rho$ peak to 0.5\% 
dominated by pion identification and ISR luminosity. Previously such a
test was performed in $e^+e^-$ scan experiments at the OLYA detector in
the c.m. energy range from 640~MeV to 
1400~MeV~\cite{ol2} and at the CMD-2
detector from 370~MeV to 520~MeV~\cite{cmd2pilow} 
with the achieved precision of
comparison of 3\% and 1\%, respectively.

The measured cross section is shown in 
Fig.~\ref{fig2}(bottom).   
For the first time a relatively high-statistics
measurement is performed for c.m. energies above 1~GeV. 
The cross section in this energy
range shown in Fig.~\ref{fig2a} demonstrates some statistically
significant structures which can be possibly explained by the interference
between the wide $\rho$-like excited states. Note that the cross
section shown in Fig.~\ref{fig2} is bare and includes FSR effects.

\begin{figure}
\includegraphics[width=.48\textwidth]{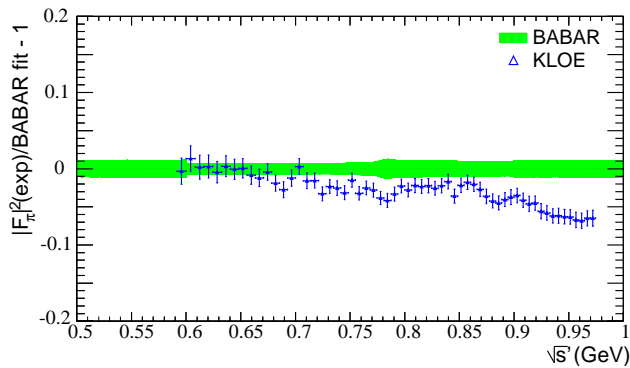}
\caption{\label{fig2b} 
The relative difference between the KLOE~\cite{kloe2} and 
BABAR~\cite{babrch} measurements. The band corresponds to
the BABAR statistical and systematic uncertainties combined in quadrature. 
}
\end{figure}
\begin{figure}
\includegraphics[width=.48\textwidth]{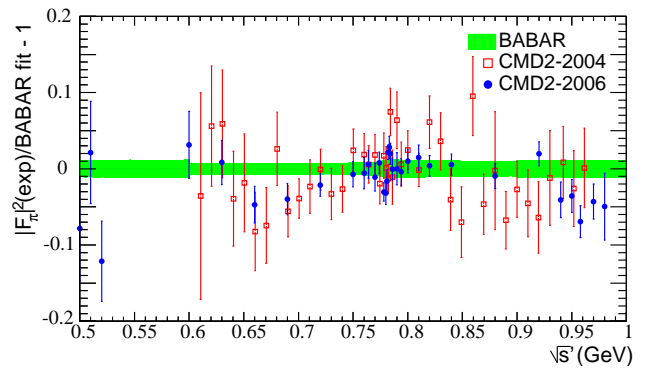}
\caption{\label{fig2c} 
The relative difference between the CMD-2~\cite{cmd2pi1,cmd2pi2} 
and BABAR~\cite{babrch} measurements. The band corresponds to
the BABAR statistical and systematic uncertainties combined in quadrature.
}
\end{figure}
The claimed sub-percent level of systematic uncertainties 
on the $e^+ e^-\to\pi^+\pi^-$ measurements can be verified by comparison 
of the results from these very different experiments. 
Above we found that the difference between the KLOE and CMD-2
measurements is larger than their combined systematic uncertainty. 
Figure~\ref{fig2b} shows a relative difference between the KLOE and BABAR 
measurements. Again the deviations larger than declared systematic 
errors are seen indicating a presence of unaccounted systematic uncertainties
in one or both experiments. 
Comparison between the CMD-2 and BABAR shown in Fig.~\ref{fig2c} also reveals
some non-statistical up to 5\% deviations both below and above the
$\rho$-resonance maximum. In the whole energy range  BABAR data are
in fair agreement with the SND~\cite{snd2pi} results  
within experimental uncertainties.

In Section~\ref{mu_g2} we discuss the impact of these measurements on
the problem of the muon anomaly.
 
\subsection{$e^+ e^-\to \pi^+\pi^-\pi^0$}
\begin{figure}
\includegraphics[width=.48\textwidth]{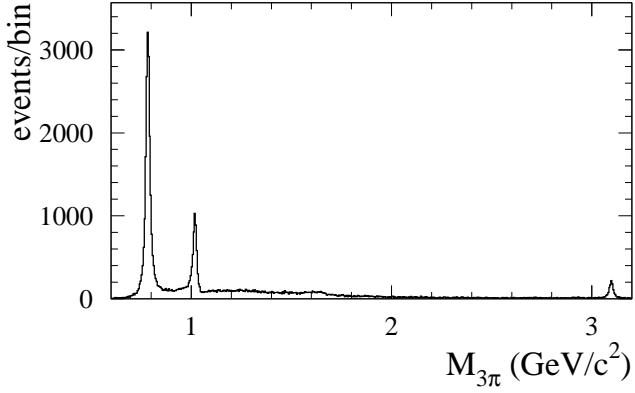}
\caption{\label{fig3a} 
The $m(\pi^+\pi^-\pi^0)$ distribution for the
$e^+ e^-\to\pi^+\pi^-\pi^0\gamma$ reaction measured with the BABAR
  detector~\cite{babr3pi}.
}
\end{figure}

\begin{figure}
\includegraphics[width=.48\textwidth]{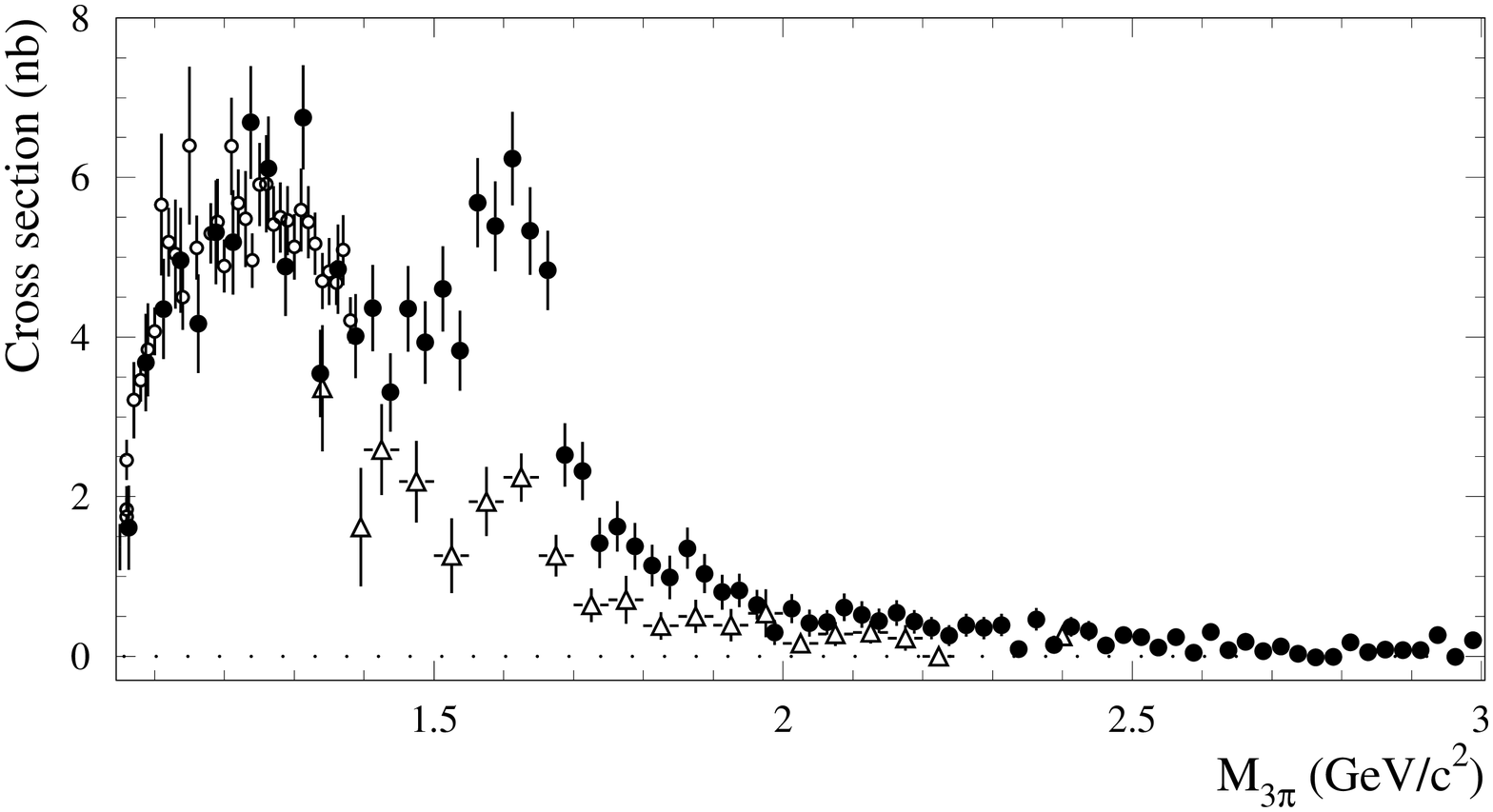}
\hspace{-3.7cm}
\includegraphics[width=3.2cm]{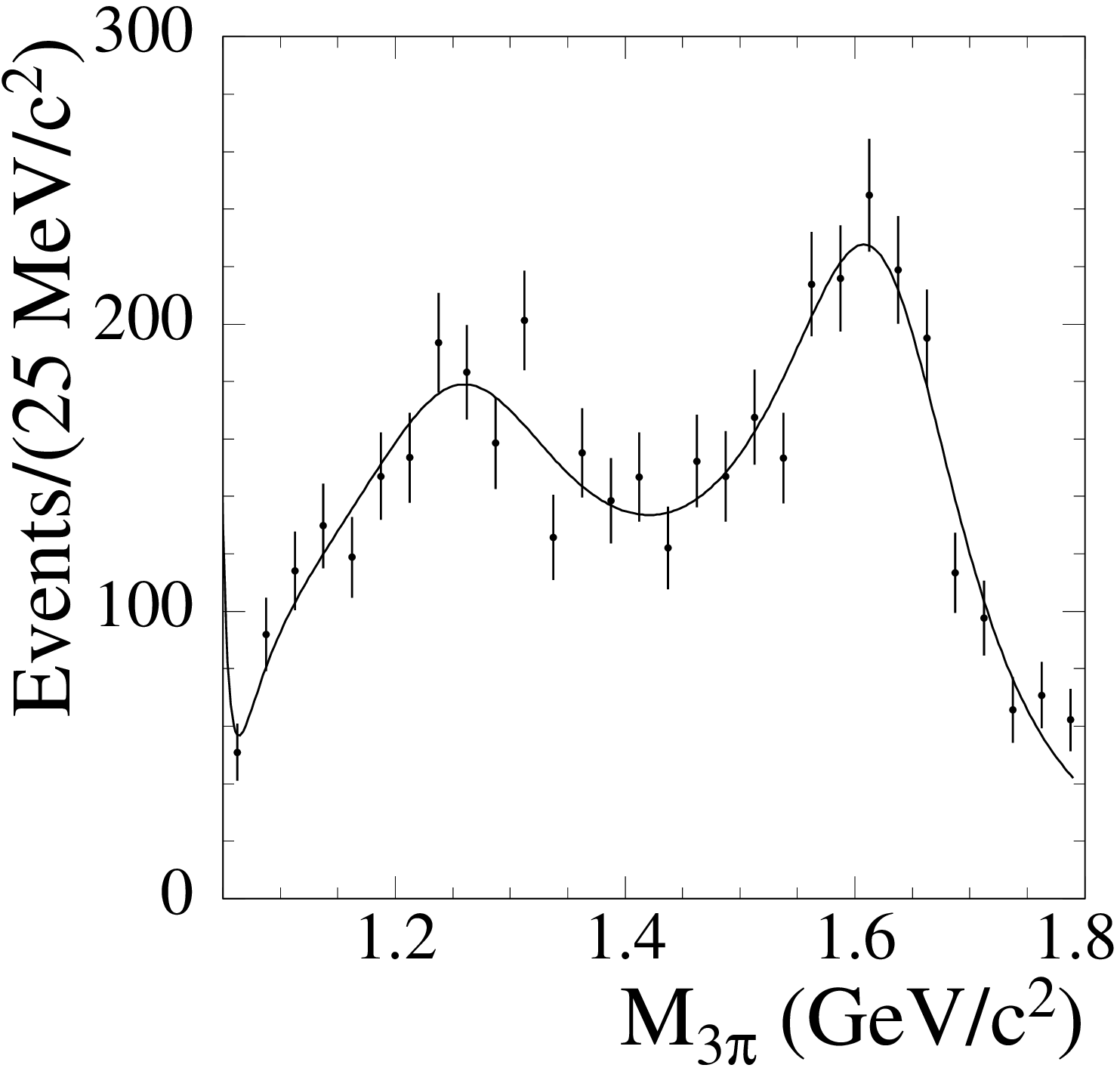}
\caption{\label{fig103} The 
$e^+ e^-\to\pi^+\pi^-\pi^0$ cross section measured with the BABAR
detector~\cite{babr3pi} in the 1-3 GeV/c$^2$ range compared with the 
SND~\cite{snd3pi} (open
circles)  and DM2~\cite{dm23pi} (triangles) data. 
The inset shows the mass distribution fitted with two resonances.
}
\end{figure}

A study of the three-pion production in the ISR process
was reported by BABAR in Ref.~\cite{babr3pi}. The 
three-pion mass distribution for the $e^+ e^-\to
\pi^+\pi^-\pi^0\gamma$ reaction shown in Fig.~\ref{fig3a} is dominated by 
the well known $\omega(782)$, $\phi(1020)$, and $J/\psi$ resonances. 
For the $\omega(782)$ and $\phi(1020)$ resonances they determine 
the product of the leptonic
width and the branching fraction to three pions consistent with
other measurements and having comparable accuracy.
Large data samples make possible
the observation of two structures in the 1--2 GeV/$c^2$
mass region (see Fig.~\ref{fig103}). The cross section below 1.4 GeV
is in agreement with the SND measurement~\cite{snd3pi}, but 
at higher energies a large deviation
from the DM2 results~\cite{dm23pi} is observed. 
The cross section in this region is fitted (see inset in Fig.~\ref{fig103}) 
assuming the presence of two excited $\omega$-like states, $\omega(1420)$ and
$\omega(1650)$~\cite{PDG08}.
The parameters of these states are still not well determined.
In this case they strongly depend on relative phases between the
corresponding amplitudes  and their phase differences 
with the $\omega(782)$ and $\phi(1020)$ amplitudes. The latter
resonances have a much larger decay rate to the $3\pi$ mode.
The obtained parameters of the $\omega(1420)$ and $\omega(1650)$ 
are summarized in Table~\ref{omfit_tab}.

The three-body final state is a relatively simple process for 
a study of hadron dynamics. Its Dalitz plot analysis shows that 
the $\rho(770)^{\pm}\pi^{\mp}$ and
$\rho(770)^0\pi^0$ intermediate states dominate at all energies.
There is also a small contribution of the $\omega\pi$ intermediate
state with $\omega$ decay to $\pi^+\pi^-$.  
\subsection{$e^+ e^-\to K^+ K^-\pi^0,\, 
K^0_S K^{\pm}\pi^{\mp},\, K^+ K^-\eta$  \label{kkpi}}
\begin{figure}
\includegraphics[width=.43\textwidth]{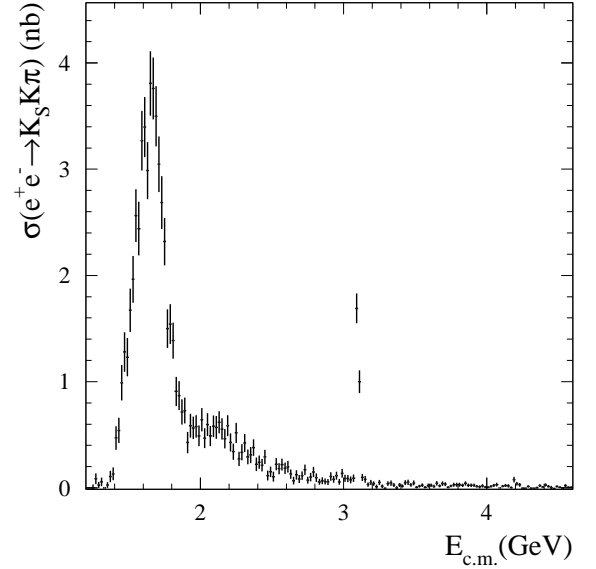}
\includegraphics[width=.43\textwidth]{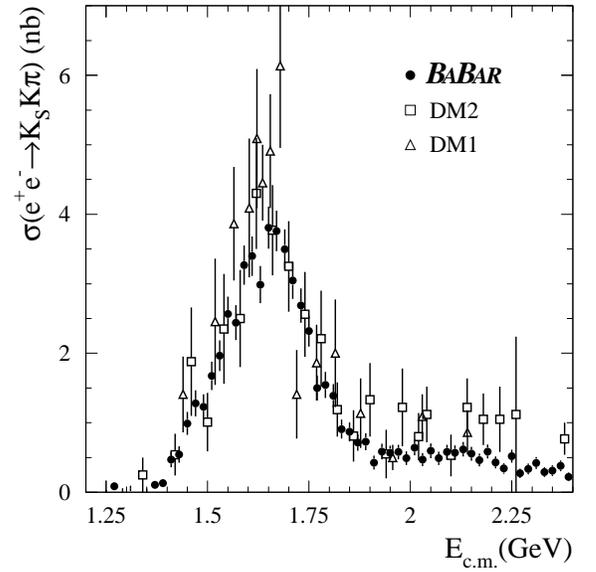}
\caption{
The $e^+ e^- \to K^0_S K^{\pm}\pi^{\mp}$ cross section 
measured by BABAR~\cite{babrkkpi} (top).
Comparison of the BABAR measurement with the results 
of the previous DM1~\cite{dm1kkpi} and DM2~\cite{dm2kkpi}
experiments (bottom).
\label{xs_kskpi}}
\end{figure}
\begin{figure}
\includegraphics[width=.43\textwidth]{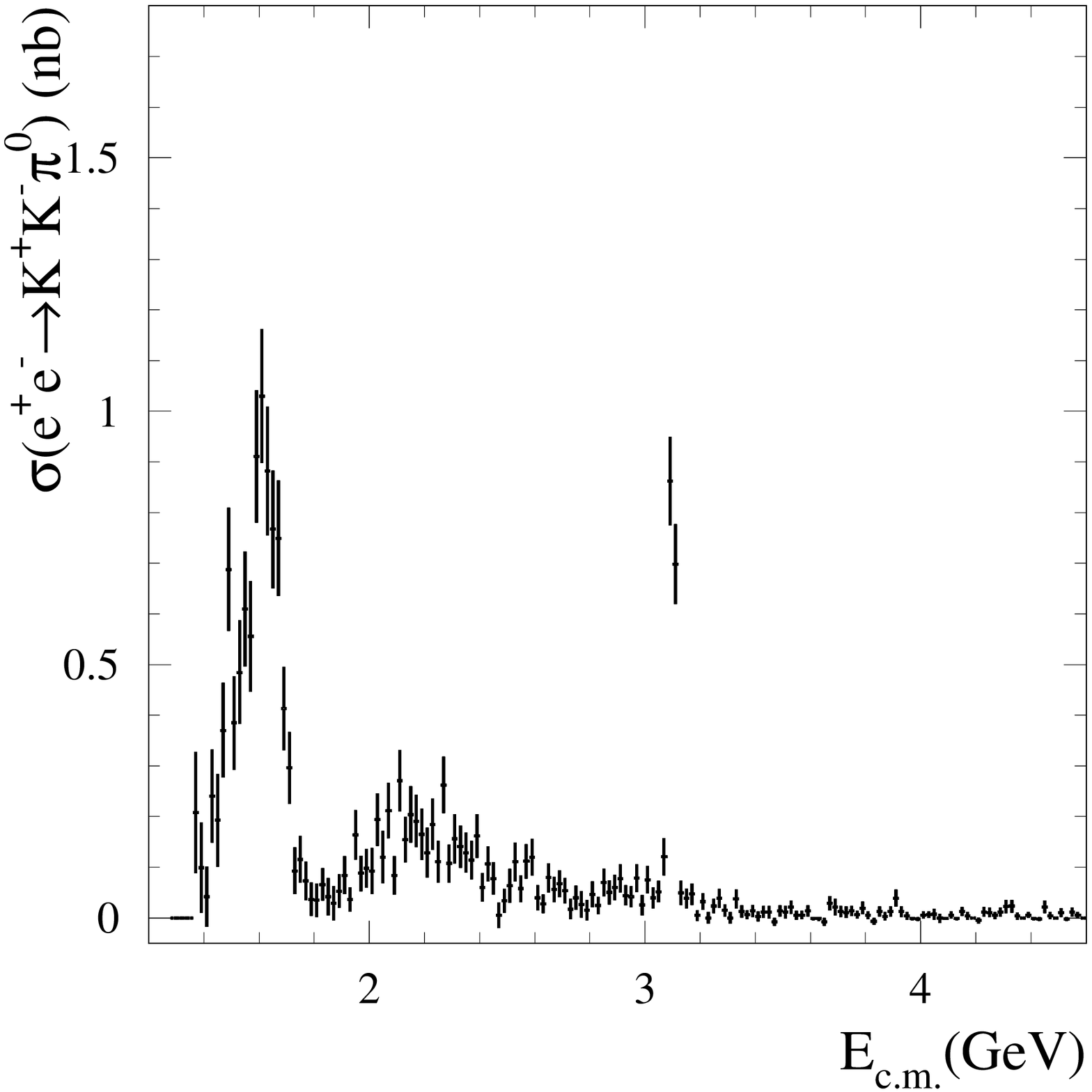}
\includegraphics[width=.43\textwidth]{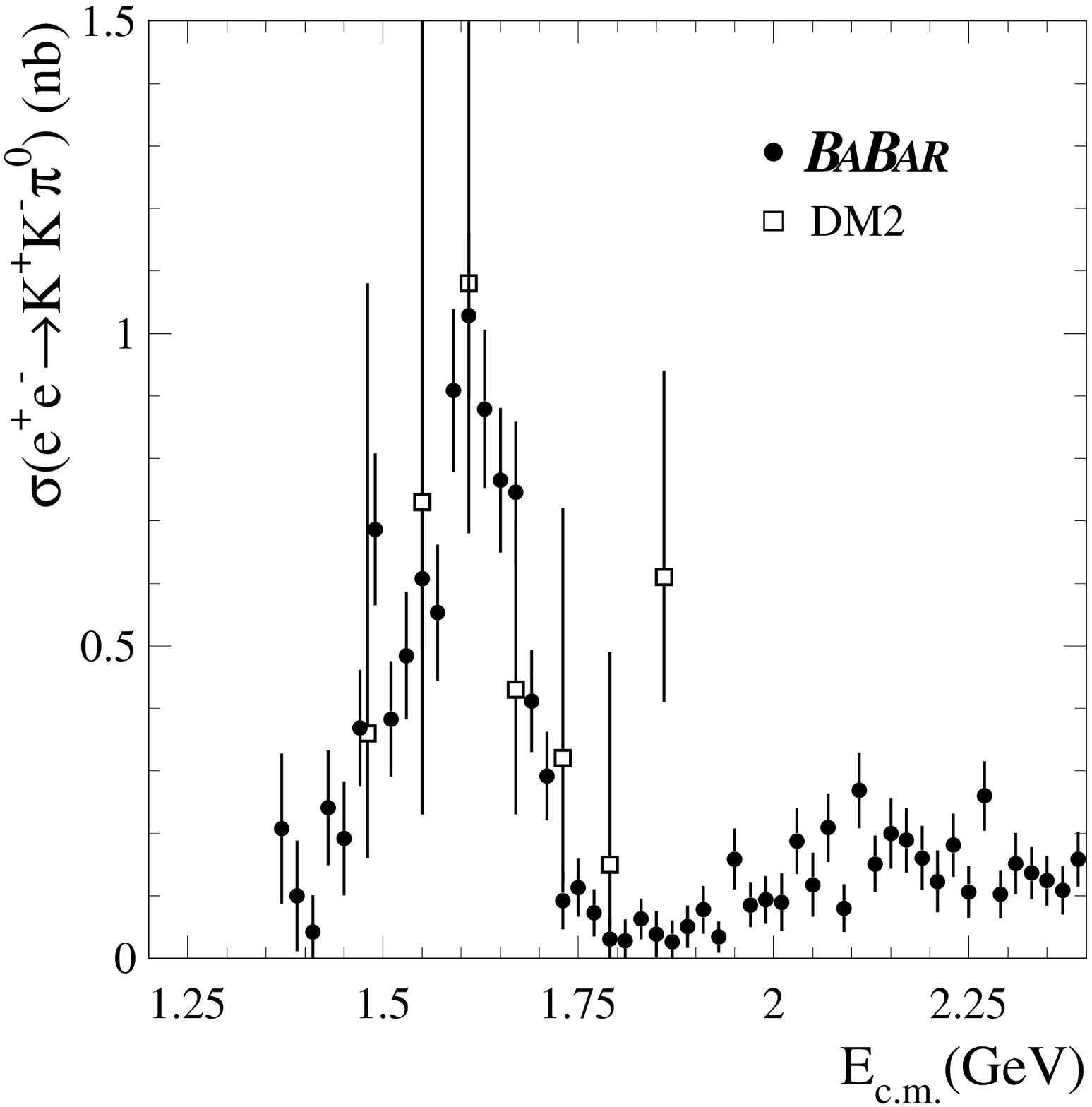}
\caption{
The $e^+ e^- \to K^+ K^-\pi^0 $ cross section measured by 
BABAR~\cite{babrkkpi}(top). Comparison of the BABAR measurement with the result
of the DM2 experiment~\cite{dm2kkpi} (bottom).
\label{xs_kkpi0}}
\end{figure}
Figures~\ref{xs_kskpi} and \ref{xs_kkpi0} show 
the $e^+ e^-\to K^+ K^-\pi^0$ and $e^+e^- \to K^0_S K^{\pm}\pi^{\mp}$
cross sections measured in the BABAR experiment~\cite{babrkkpi} (top) and 
comparison of the BABAR results with DM1~\cite{dm1kkpi} and DM2~\cite{dm2kkpi} 
measurements below 2.4 GeV where the previous data are available (bottom). 
The BABAR data are about 10 times more precise. The
``spike'' at 3.1 GeV is due to $J/\psi$ decays to these final states
and will be discussed later.

\begin{figure}
\includegraphics[width=.4\textwidth]{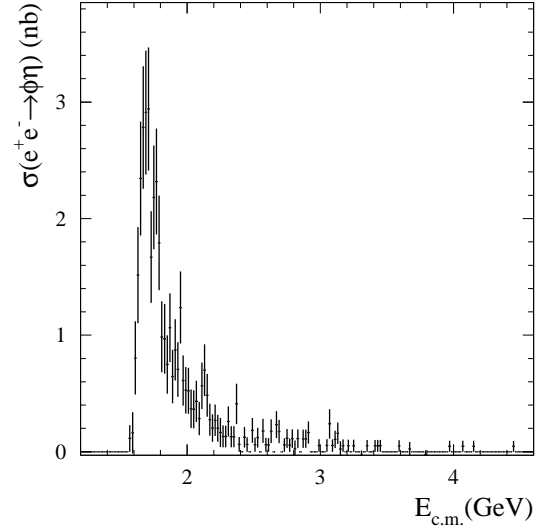}
\caption{
The $e^+ e^-\to\phi\eta$ cross sections
measured with the BABAR detector~\cite{babrkkpi}. }
\label{phieta}
\end{figure}
\begin{figure}
\includegraphics[width=.4\textwidth]{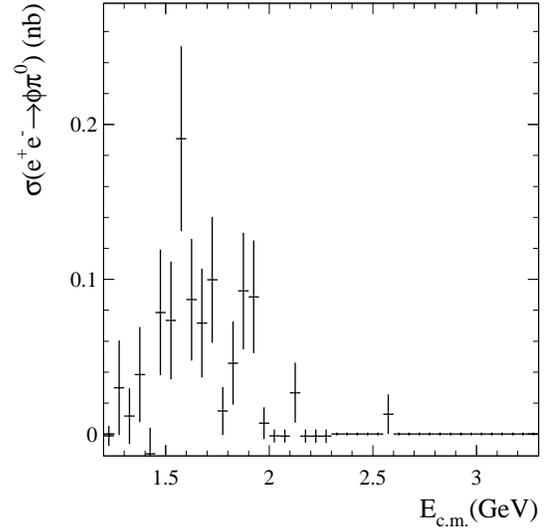}
\caption{
The $\phi\pi^0$  cross sections
measured with the BABAR detector~\cite{babrkkpi}. }
\label{phipi0}
\end{figure}
In the $K^+K^-2\gamma$ final state the $\phi(1020)\eta$ and 
$\phi(1020)\pi$ intermediate states were also observed.
The measured cross sections for these states, which were not previously studied,
are shown in Figs.~\ref{phieta} and \ref{phipi0}. 
The $e^+ e^-\to\phi(1020)\eta$ 
is the best channel for a study of the excited $\phi$ state. The contributions
of the $\omega$-like states to this channel should be 
suppressed by the OZI rule.

The reaction $e^+ e^-\to\phi(1020)\pi^0$ is promising for a search
for exotic isovector resonances. For ordinary isovector states, the
$\phi\pi^0$ decay should be suppressed by the OZI rule. 
The authors perform two fits of the cross section. In the first one
they assume a single resonance and obtain for it mass and width of
$1593 \pm 32$ MeV/$c^2$ and $203 \pm 97$ MeV, respectively. These 
parameters are compatible with those
of the  $\rho(1700)$~\cite{PDG08}. A somewhat better quality of the fit
is achieved if two resonances are assumed. The obtained parameters
of the first resonance are $1570 \pm 36 \pm 62$~MeV/$c^2$
for the mass and $144 \pm 75 \pm 43$~MeV for the width, i.e.,
consistent with those of the $C(1480)$ state observed in
Ref.~\cite{lands}. The mass and width for the second resonance are
$1909 \pm 17 \pm 25$~MeV/$c^2$ and $48 \pm 17 \pm 2$~MeV,
respectively, compatible with the dip already observed in 
other experiments, predominantly in multipion final 
states~\cite{babr6pi,DM26pi,FOCUS6pi,fenice}. 
With the limited statistics available at the moment
they cannot draw a firm conclusion: an OZI-violating decay of the
$\rho(1700)$ cannot be excluded. 

\begin{figure*}[p]
\includegraphics[width=6cm]{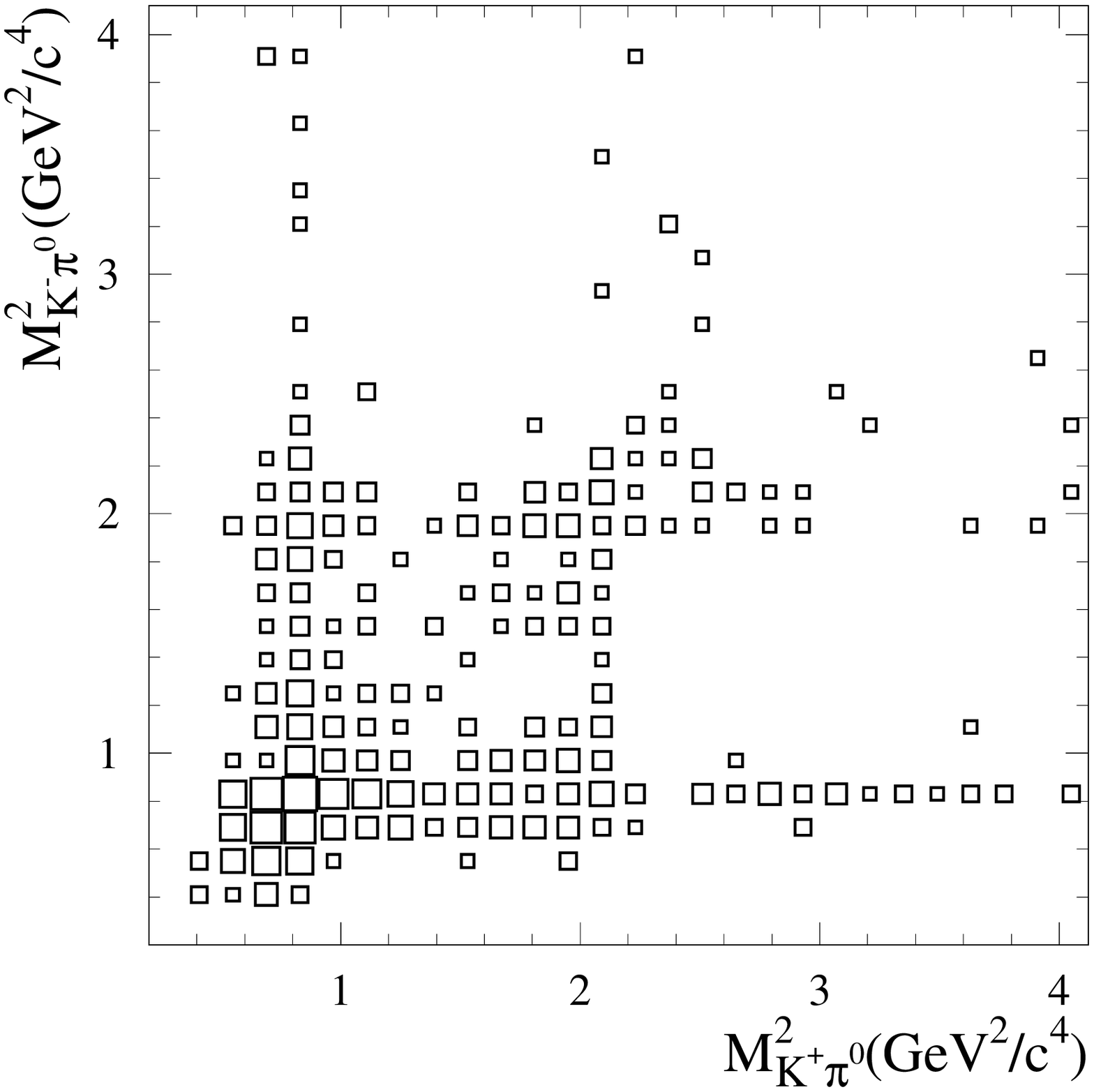}
\put(-38,150){(a)}
\includegraphics[width=6cm]{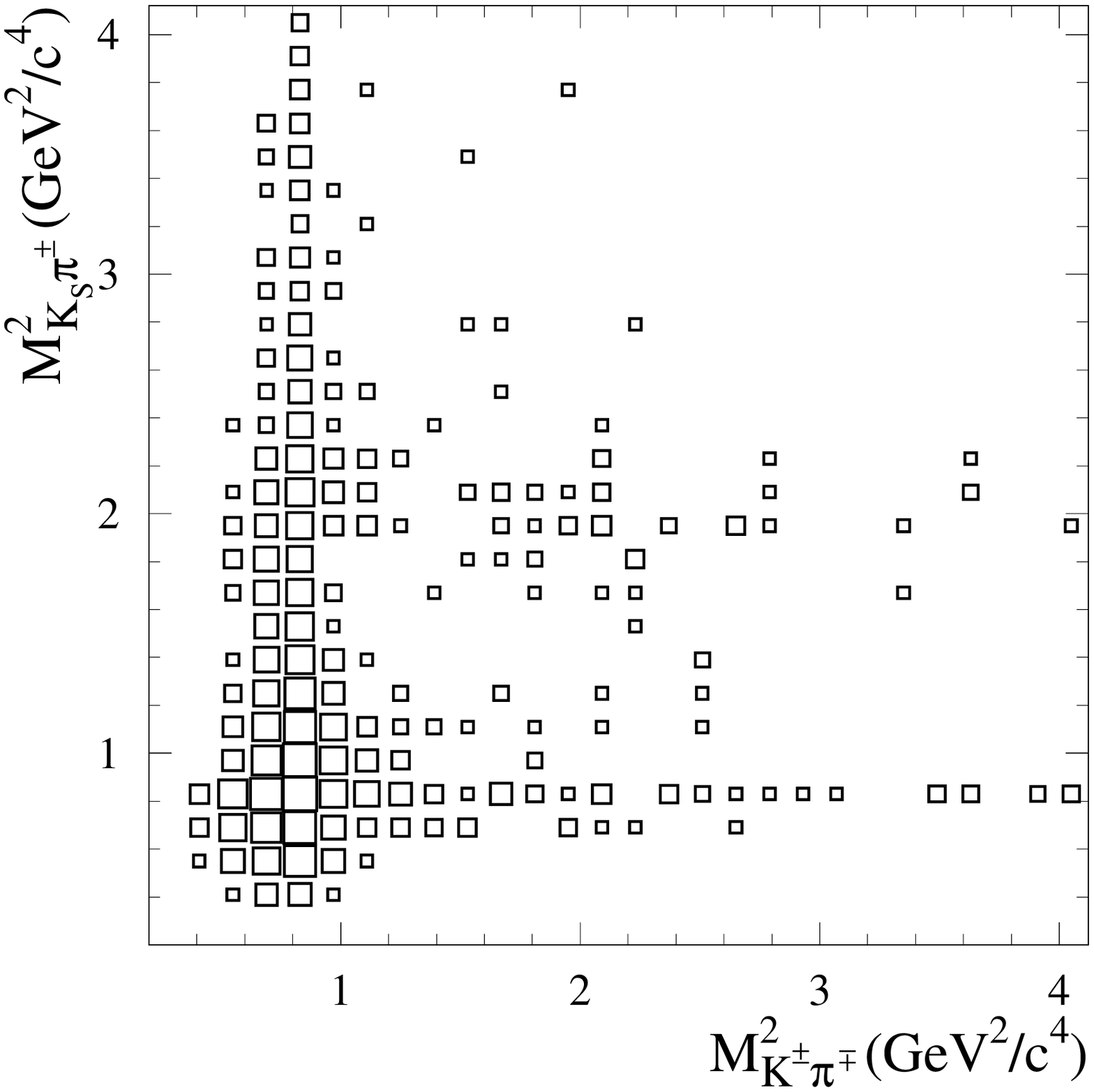}
\put(-38,150){(b)}
\caption{
The Dalitz plot distribution for the $K^+ K^-\pi^0$  (a)
and $K^0_S K^{\pm}\pi^{\mp}$ final state (b) from Ref.~\cite{babrkkpi}.
A sum over all accessible c.m. energies of the hadronic final states is given.
\label{dal-kkpi0}}
\end{figure*}
\begin{figure*}[p]
\includegraphics[width=6cm]{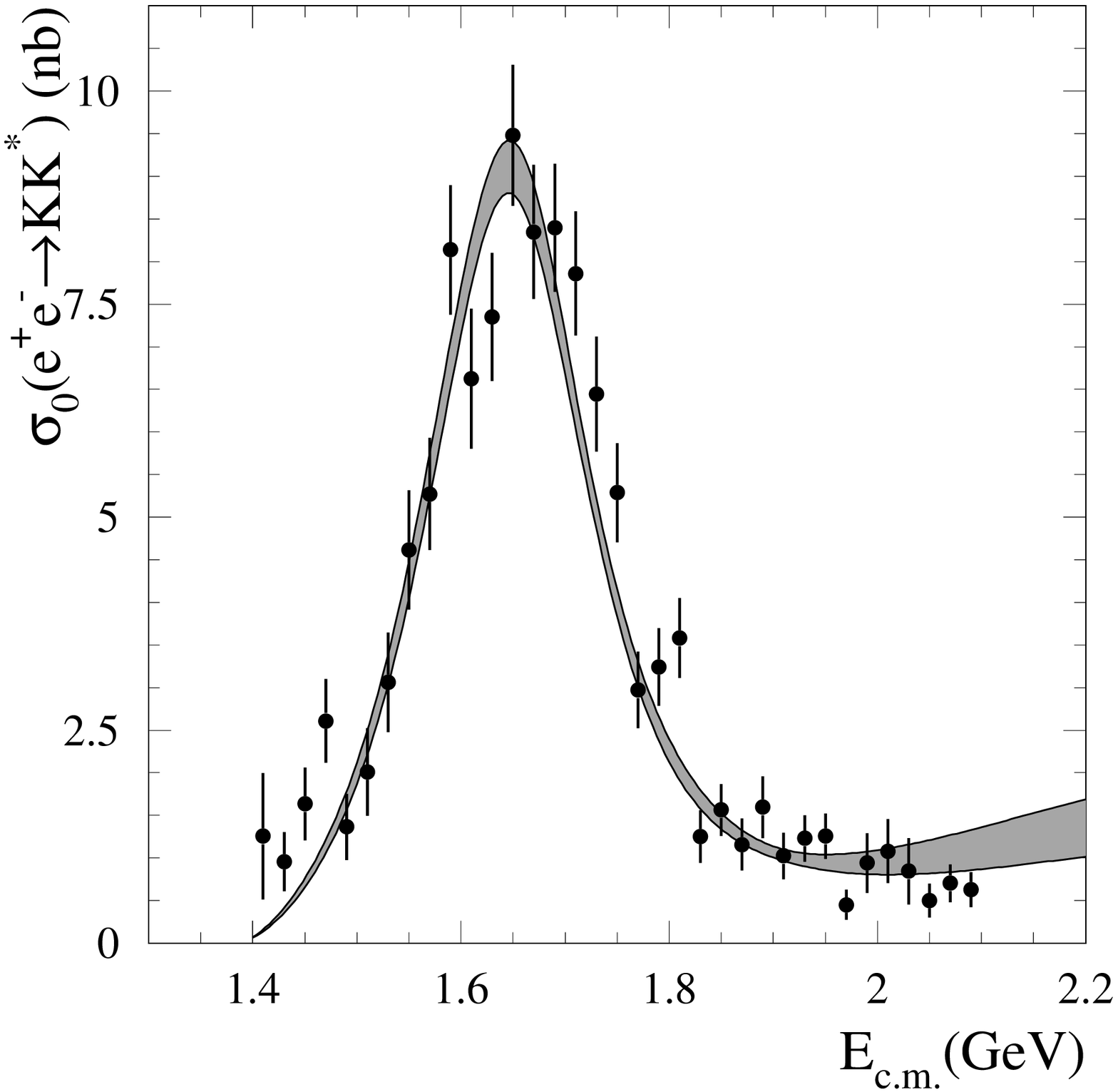}
\put(-38,150){(a)}
\includegraphics[width=6cm]{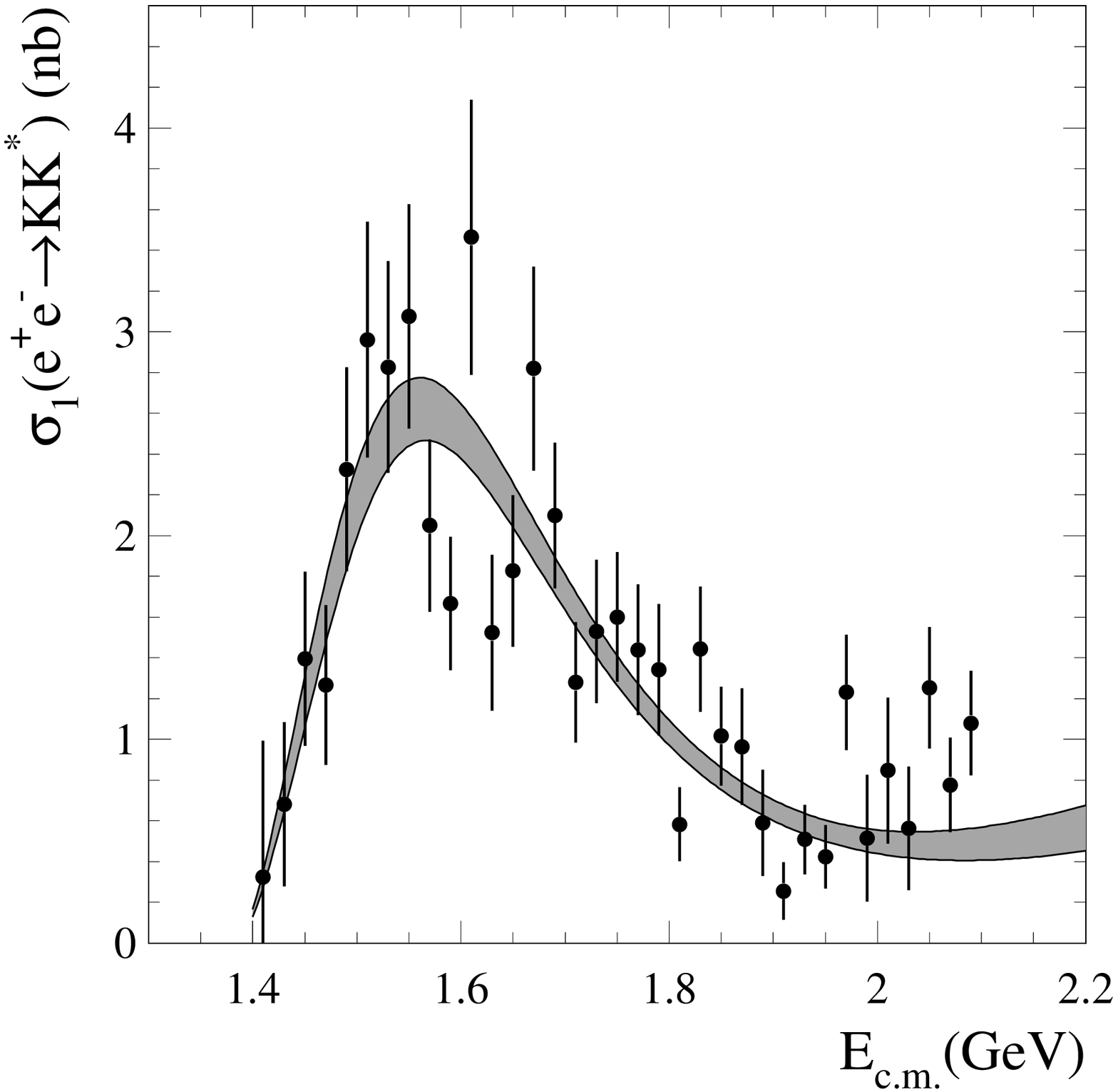}
\put(-38,150){(b)}\\
\includegraphics[width=6cm]{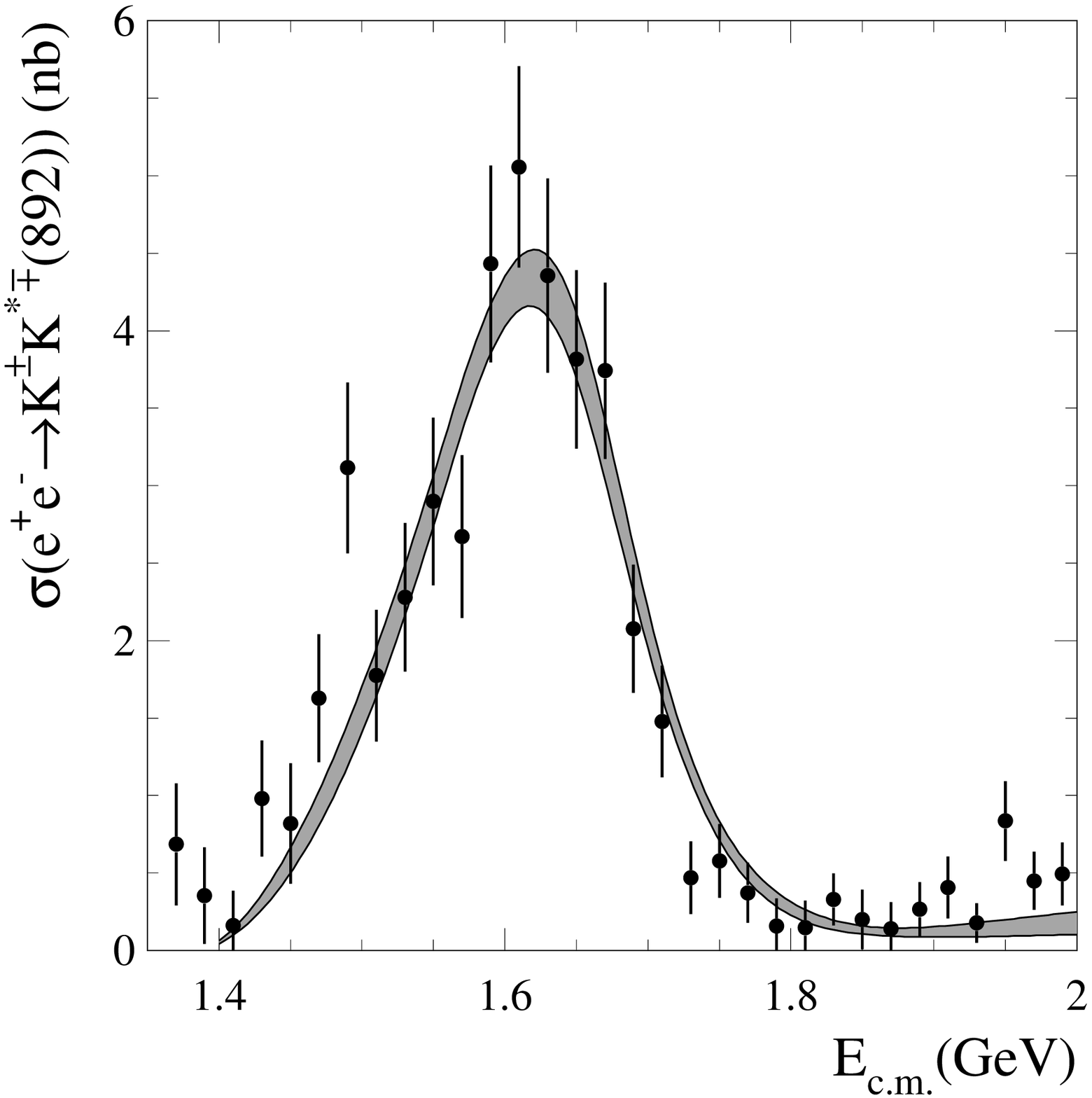}
\put(-38,150){(c)}
\includegraphics[width=6cm]{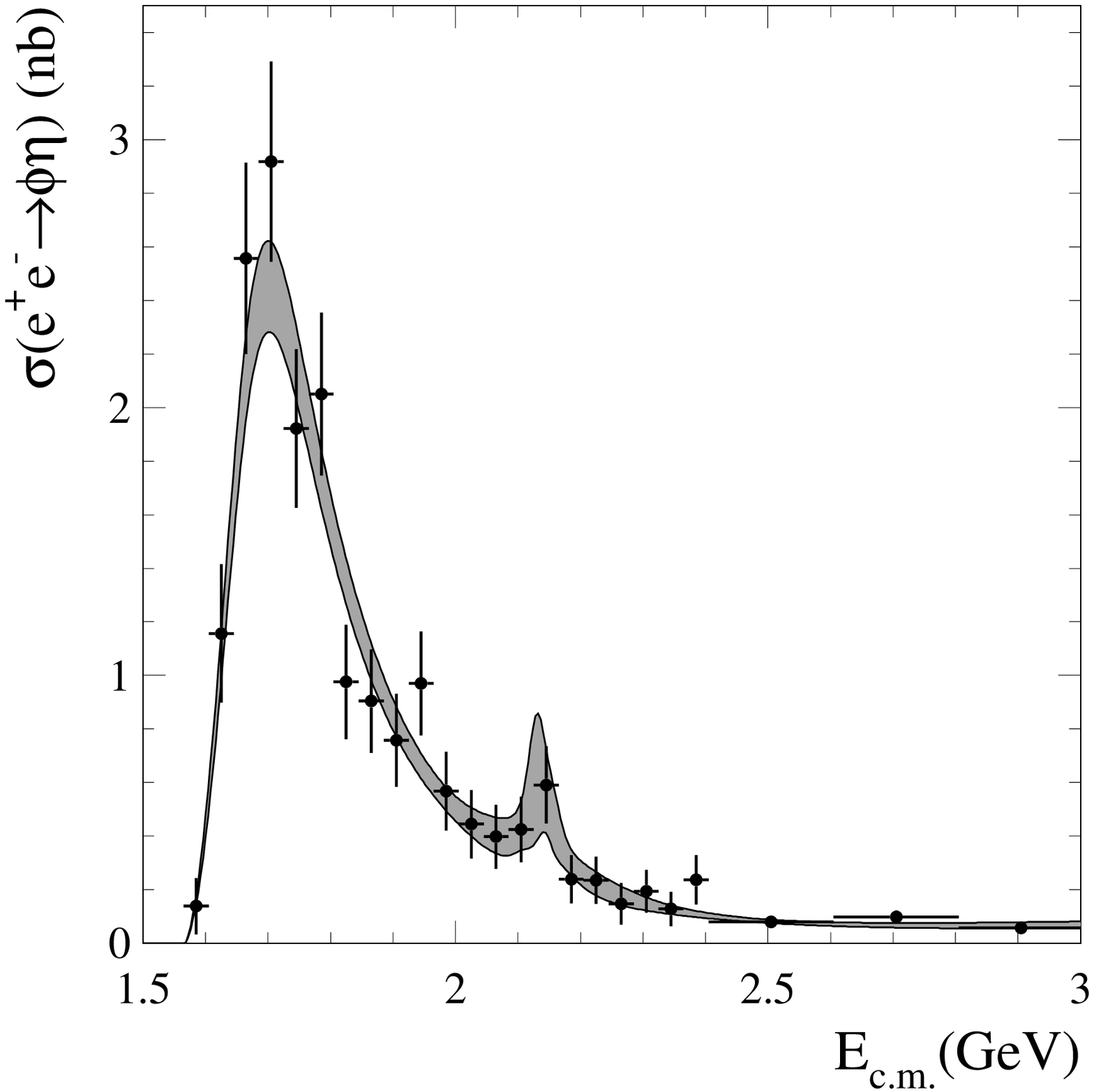}
\put(-38,150){(d)}
\caption{\label{s0-s1}
Isoscalar (a) and isovector (b) components of the $e^+e^-\to K \bar{K}\pi$ 
cross section; the $e^+e^-\to K^{\pm}K^\ast(892)^{\mp}$  cross section obtained
using $e^+e^-\to K^+K^-\pi^0$ events (c), and the $e^+e^-\to\phi\eta$ 
cross section (d)~\cite{babrkkpi}.
The points with error bars are data and the gray band represents the fit and 
its uncertainty.
}
\end{figure*}
\begin{table}\vspace{0mm}
\caption{
\label{tab:multifit}
Parameters of the isoscalar and isovector resonances 
obtained in Ref.~\cite{babrkkpi} from the global fit to
the isoscalar and isovector amplitudes using
the $e^+e^-\to K^{\pm}K^\ast(892)^{\mp}$,$K^0_S K^{\pm}\pi^{\mp}$ 
and $e^+e^-\to \phi\eta$ cross sections.
} 
\vspace{2mm}
\renewcommand{\arraystretch}{1.5}
\begin{tabular}{lcc}
\cline{1-3}
\cline{1-3}
$R$ with $I=0\hspace{10mm}$ & $\hspace{10mm}\phi'\hspace{10mm}$ & $\hspace{10mm}\phi''\hspace{10mm}$ \\
\cline{1-3}
$\Gamma^R_{ee}\mathcal{B}^{R}_{K\kst}(\rm eV)$  & $408\pm 49$& - \\
$\Gamma^R_{ee}\mathcal{B}^{R}_{\phi\eta}(\rm eV)$  & $172\pm 31$& $1.9\pm1.0$\\
$m_R(\rm MeV)$ &  $1723\pm 20 $& $ 2139 \pm 35$\\
$\Gamma_R(\rm MeV)$ & $371\pm 75$ & $76\pm62$\\ 
\cline{1-3}
$R$ with $I=1\hspace{10mm}$ & \multicolumn{2}{c}{$\hspace{10mm}\rho'\hspace{10mm}$} \\
\cline{1-3}
$\Gamma^R_{ee}\mathcal{B}^{R}_{K\kst}(\rm eV)$ & \multicolumn{2}{c}{$135\pm 12$} \\
$m_R(\rm MeV)$ &  \multicolumn{2}{c}{$1506\pm 16$} \\
$\Gamma_R(\rm MeV)$ & \multicolumn{2}{c}{$437\pm 24$} \\ 
\cline{1-3}
\end{tabular}
\end{table}
Figure~\ref{dal-kkpi0} shows the Dalitz plots for the $K^+ K^-\pi^0$
and $K^0_S K^{\pm}\pi^{\mp}$ final states. It is seen that the
$K\bar{K}^\ast(892)$ and $K\bar{K}_2^\ast(1430)$ intermediate states give
the main contribution to the $K\bar{K}\pi$ production. 
For the $K^0_S K^{\pm}\pi^{\mp}$ final state both
the neutral $K^0\bar{K}^{\ast 0}$ and charged $K^{\pm}K^{\ast\mp}$ 
combinations are involved. Since the $K^0\bar{K}^{\ast 0}$
and $K^{\pm}K^{\ast\mp}$ amplitudes are the sum and the
difference of the isovector and isoscalar amplitudes, respectively,
the Dalitz plot population for the $K^0_S K^{\pm}\pi^{\mp}$ mode
is asymmetric and strongly depends on isospin composition. 
From the Dalitz plot analysis the moduli and relative
phase of the isoscalar and isovector amplitudes
both for the $K\bar{K}^\ast(892)$ and $K\bar{K}_2^\ast(1430)$ intermediate states
were determined. The obtained
isoscalar and isovector $e^+e^- \to K\bar{K}^\ast(892)$ cross sections
are shown in  Fig.~\ref{s0-s1} (a,b). 

The global fit to the $e^+ e^-\to\phi(1020)\eta$ and
$e^+ e^-\to K^+K^-\pi^0$ cross sections, isovector
and isoscalar $K\bar{K}^\ast(892)$ amplitudes, and their
relative phase was performed to determine parameters 
of the $\phi$ and $\rho$ excitations decaying into these
final states. The fit results are shown in Fig.~\ref{s0-s1}
and summarized in Table~\ref{tab:multifit}.
The obtained mass and width of the $\phi^\prime$ and $\rho^\prime$
are in reasonable agreement with the parameters of the
$\rho(1450)$ and $\phi(1680)$ resonances measured in other experiments
(see Ref.~\cite{PDG08} for references). The parameters of the 
$\phi^{\prime\prime}$,
which is seen in the $\phi\eta$ final state, are close to those for
the $Y(2175)$ state observed in the $\phi f_0(980)$ final state.
This state will be discussed in Sec.~\ref{Sec:2k2pi}.

\subsection{$e^+ e^-\to \pi^+\pi^-\pi^+\pi^-,\, \pi^+\pi^-2\pi^0$}
\begin{figure}
\includegraphics[width=.4\textwidth]{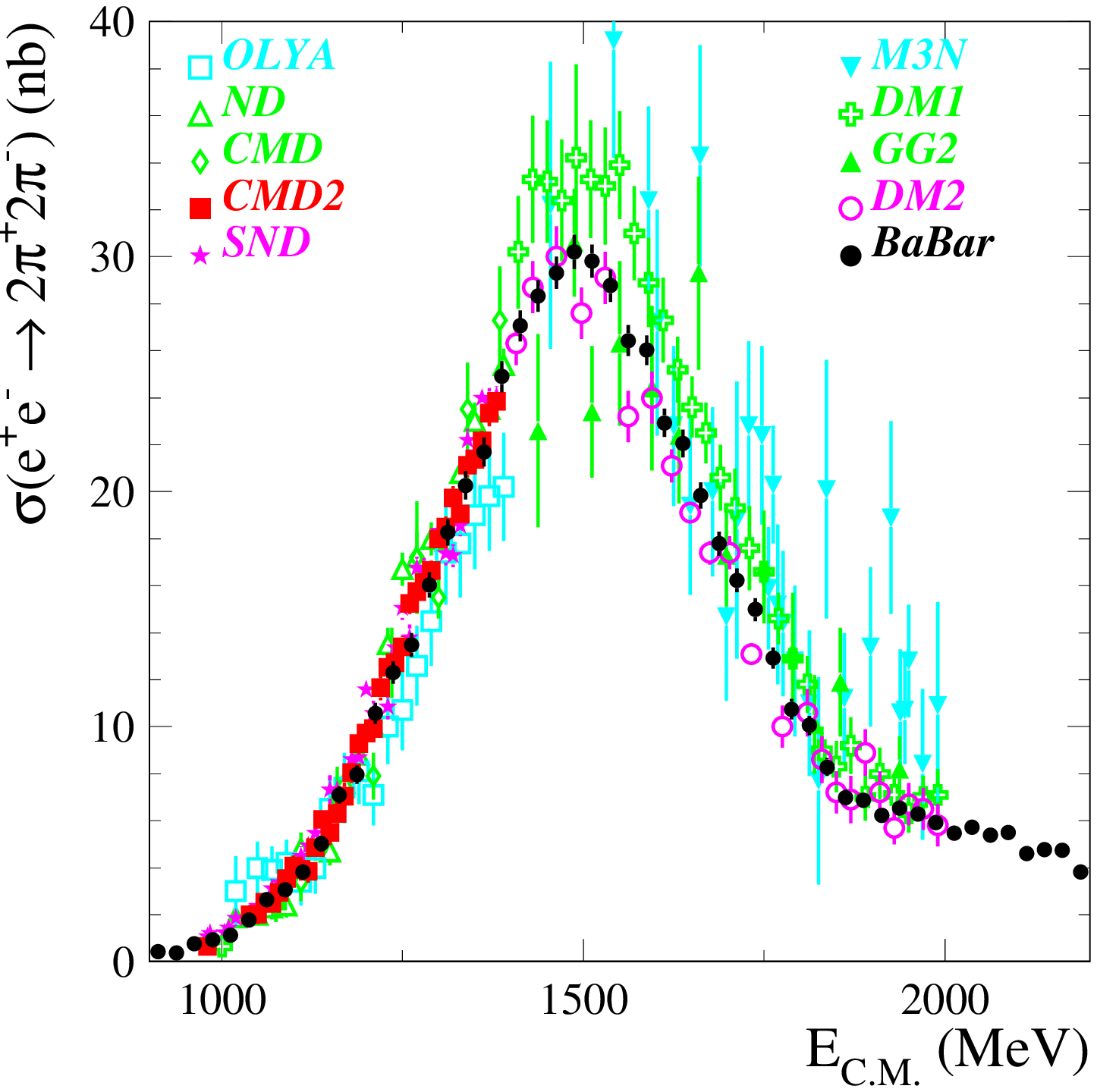}
\caption{\label{fig400a}
Comparison of the BABAR results on the $e^+ e^-\to \pi^+\pi^-\pi^+\pi^-$
cross section~\cite{babr4pi} with the previous direct \epem 
measurements~\cite{m3n4pi,gg24pi,dm14pi,olya4pi,dm24pi,NDsum,snd4pi,cmd4pi}.}
\end{figure}
\begin{figure}
\includegraphics[width=.4\textwidth]{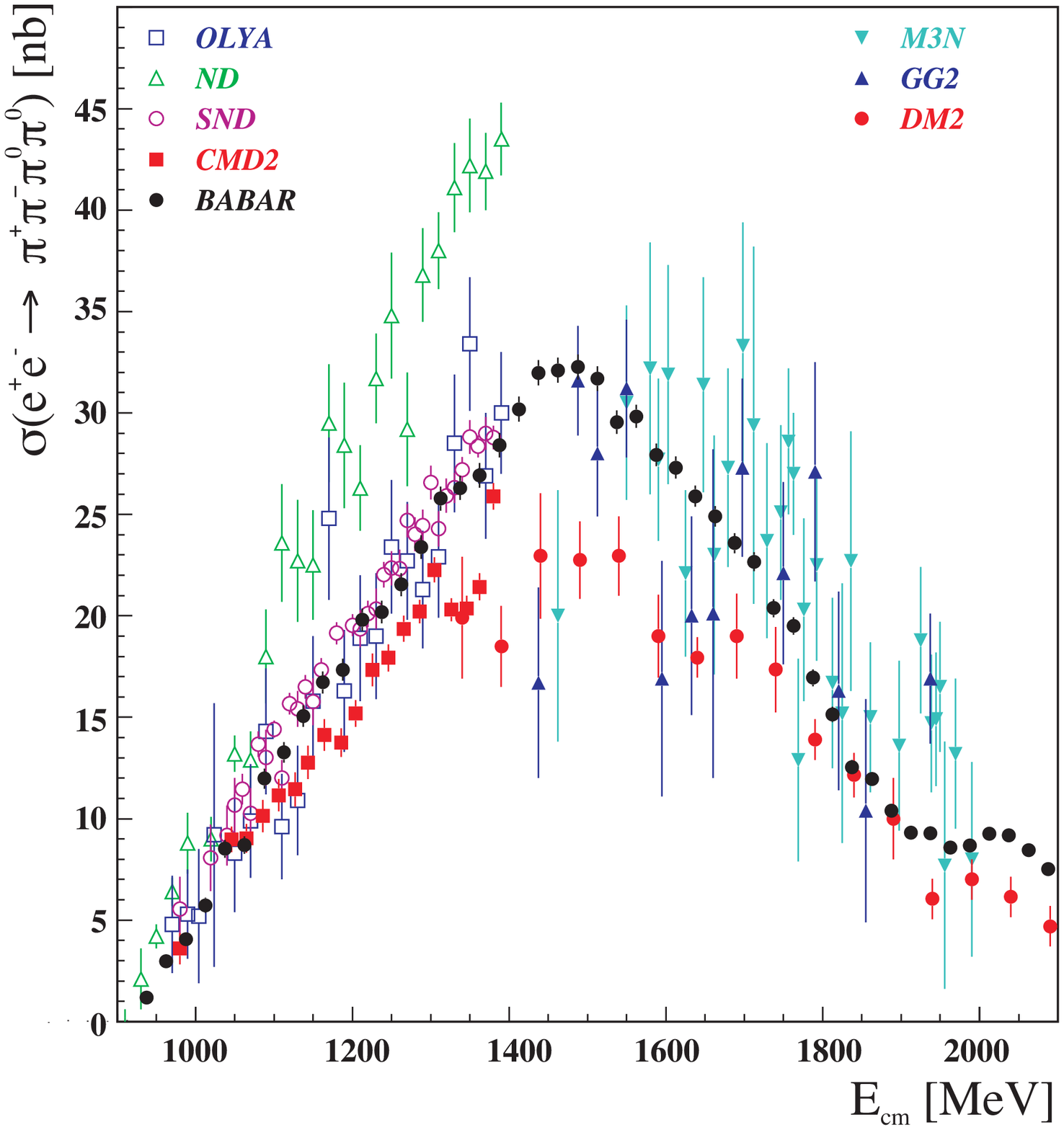}
\caption{\label{fig400b}
Comparison of the BABAR results on the $e^+ e^-\to \pi^+\pi^-2\pi^0$
cross section~\cite{dru07} with the previous direct \epem 
measurements~\cite{m3n4pi,gg24pi0,dm24pi,olya4pi0,NDsum,snd4pi,cmd4pi0}.}
\end{figure}
The reactions $e^+ e^-\to \pi^+\pi^-\pi^+\pi^-$, $\pi^+\pi^-2\pi^0$ have 
the largest cross sections in the energy region above the $\phi$-meson 
resonance. They were studied with
the BABAR detector~\cite{babr4pi,dru07} in the energy region
below 4.5 GeV.
Figures~\ref{fig400a} and \ref{fig400b} show comparison 
of the BABAR results with
the previous direct \epem\ measurements, see 
Refs.~\cite{m3n4pi,gg24pi,dm14pi,olya4pi,dm24pi,NDsum,snd4pi,cmd4pi} for
$\pi^+\pi^-\pi^+\pi^-$ and 
Refs.~\cite{m3n4pi,gg24pi0,dm24pi,olya4pi0,NDsum,snd4pi,cmd4pi0} for
$\pi^+\pi^-2\pi^0$,
in the energy range covered by these measurements. 
The large difference between the data sets from different experiments
indicates that some previous measurements had large, up to 50\%, 
unaccounted systematic errors.
The BABAR systematic uncertainty on the $e^+ e^-\to \pi^+\pi^-\pi^+\pi^-$
cross section is estimated to be about 5\% in the 1--3 GeV energy range.
For this channel the BABAR data are in good agreement with the
recent SND~\cite{snd4pi} and CMD-2~\cite{cmd4pi} measurements at the energies
below 1.4 GeV. The DM2 and BABAR data are also in reasonable agreement. 

For the $\pi^+\pi^-2\pi^0$ channel the
BABAR results are still preliminary. The estimated systematic 
uncertainty changes from 8\% in the maximum of the cross section 
to 10\% at 1 and 3 GeV. 

At energies below 1.4 GeV the BABAR cross sections agree well with the
results of the recent SND~\cite{snd4pi} and older OLYA~\cite{olya4pi0}
measurements, but not with the ND~\cite{NDsum} and 
CMD-2~\cite{cmd4pi0}
cross sections that may be affected by large unaccounted systematic 
errors as mentioned above.

\begin{figure*}
\includegraphics[width=.8\textwidth]{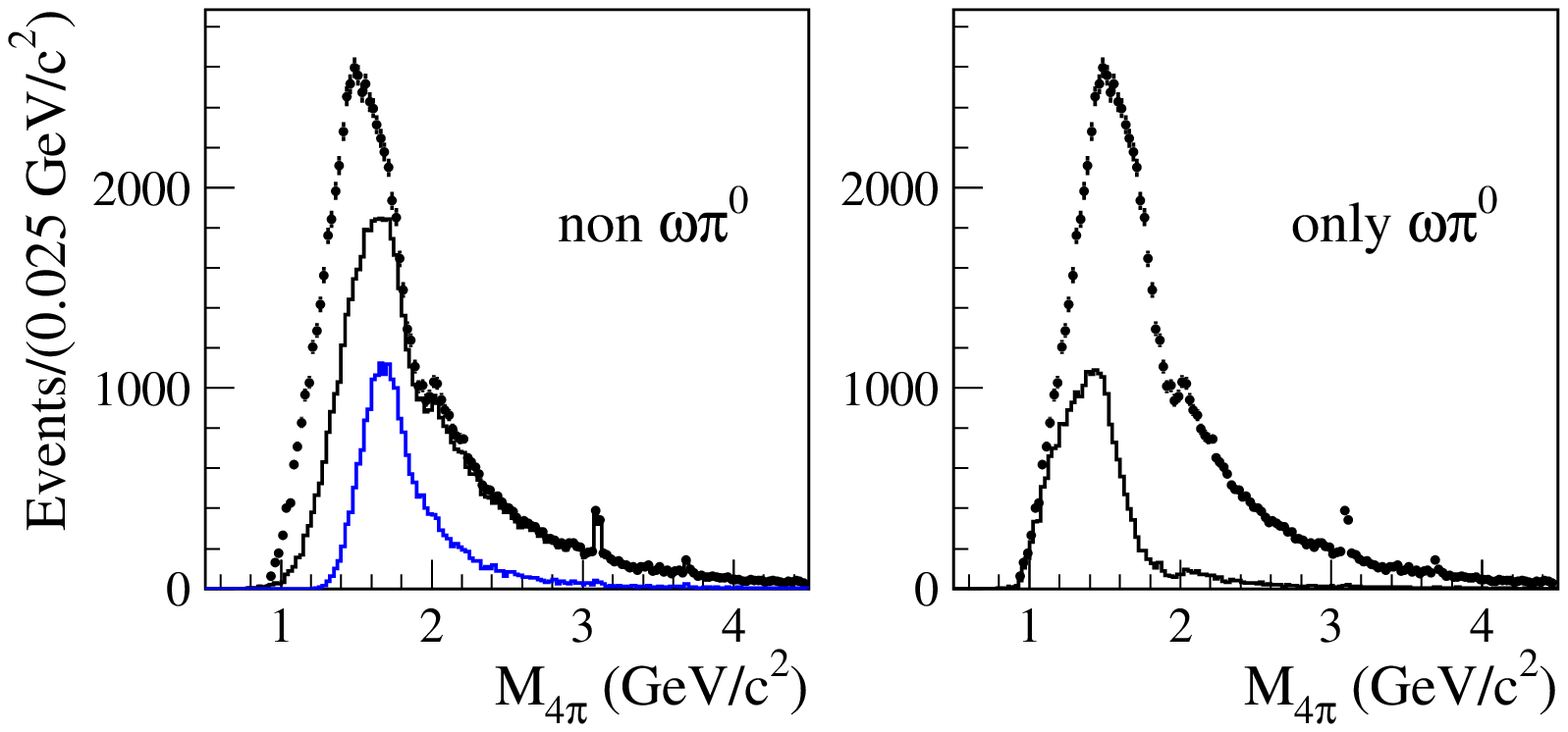}
\caption{\label{fig4a}
The $4\pi$ invariant mass spectrum for selected $e^+e^-\to \pi^+\pi^-2\pi^0$
events~\cite{dru07} (points with error bars) in comparison with the spectrum for
non-$\omega\pi^0$-events only (left) 
or with the spectrum for $\omega\pi^0$-events only (right). In the left plot
the lowest histogram shows the contribution of the 
$\rho^+\rho^-$ intermediate state.
}
\end{figure*}
The shape of the cross sections  for both reactions shows 
wide structures peaked 
at about 1.5 GeV. Different intermediate states contribute
to the $e^+ e^-\to 4\pi$ cross sections. The observed bumps are  
sums of the contributions from the $\rho(770)$, $\rho(1450)$, and $\rho(1700)$ 
decays into these intermediate states, which should be separated
for a study of the excited $\rho$ properties.
Unfortunately,
such analysis was performed at BABAR only at a qualitative level.
The two- and three-pion mass distributions for the $\pi^+\pi^-\pi^+\pi^-$
final state are relatively well described by the model of 
the $a_1(1260)\pi$
intermediate state with a small contribution of the $f_0(1300)\rho$ state. 
This agrees with the $a_1\pi$ dominance hypothesis suggested in 
Ref.~\cite{cmd4pi0} to describe the $4\pi$ dynamics at energies
below 1.4 GeV. A strong deviation from this hypothesis is observed
in the $\pi^+\pi^-2\pi^0$ channel. In addition to the expected 
$\omega\pi^0$ and $a_1\pi$ contributions, a surprisingly large 
contribution of the $\rho^+\rho^-$ intermediate
state was observed. This is demonstrated in Fig.~\ref{fig4a}, where
the $4\pi$ mass spectra for $\omega\pi$, non-$\omega\pi$, and $\rho^+\rho^-$
intermediate states are shown together with the total mass spectrum 
for the $e^+e^-\to \pi^+\pi^-2\pi^0$ reaction. 
The contributions of the different 
intermediate states were separated using simple conditions on $3\pi$ and 
$2\pi$ invariant masses. It is seen that the $\rho^+\rho^-$ cross section
is more than a half of the non-$\omega\pi$ cross section at the energy about 
1.7 GeV. For the $\pi^+\pi^-2\pi^0$ masses higher than 2.5 GeV/$c^2$ a
clear signal of the $f_0(980)$ meson  and a peak at the mass 
about 1.25 GeV/$c^2$
(probably from the $f_2(1270)$ meson) are seen in the $\pi^0\pi^0$ 
mass spectrum corresponding to the contributions of the 
$f_0(980)\rho$ and $f_2(1270)\rho$
intermediate states.

\subsection{$e^+ e^-\to K^+ K^-\pi^+\pi^-,\, K^+ K^-\pi^0\pi^0$}
\label{Sec:2k2pi}
\begin{figure}
\includegraphics[width=.4\textwidth]{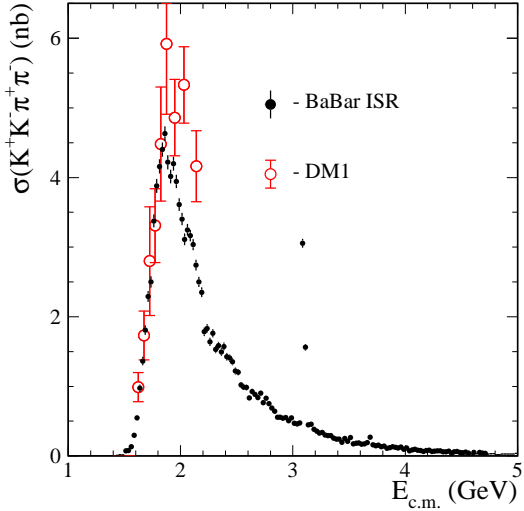}
\caption{\label{fig15a} 
The $e^+ e^-\to K^+ K^-\pi^+\pi^-$ cross section measured 
with the BABAR detector~\cite{babar2k2pi1} in comparison with the only previous 
measurement by DM1~\cite{2k2pidm1}.}
\end{figure}
\begin{figure}
\includegraphics[width=.4\textwidth]{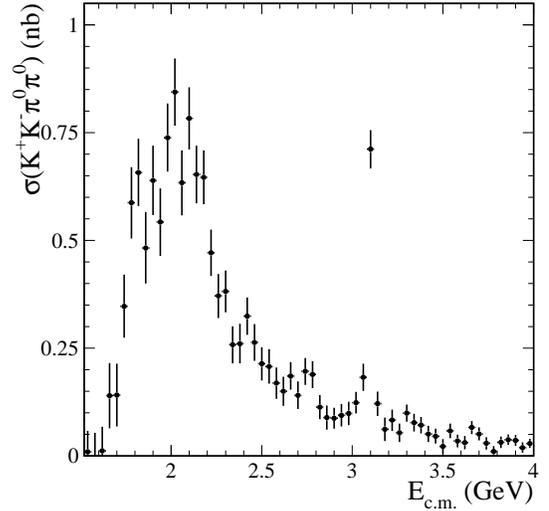}
\caption{\label{fig15b} 
The $e^+ e^-\to K^+K^-\pi^0\pi^0$ cross section measured 
with the BABAR detector~\cite{babar2k2pi1}.}
\end{figure}
The  $e^+ e^-\to K^+ K^-\pi^0\pi^0$ reaction has been never studied before the 
BABAR experiment~\cite{babr2k2pi,babar2k2pi1}, 
while the fully charged mode was previously measured with 
the DM1 detector~\cite{2k2pidm1} 
but with an about 100 times smaller data set.
The measured cross sections are shown in  Figs.~\ref{fig15a} and \ref{fig15b}.
The systematic uncertainties for these measurements are estimated to be at
the (5--9)\% level. The structures seen in the cross section energy
dependence cannot be understood without analysis of intermediate states
involved.

\begin{figure}
\includegraphics[width=.48\textwidth]{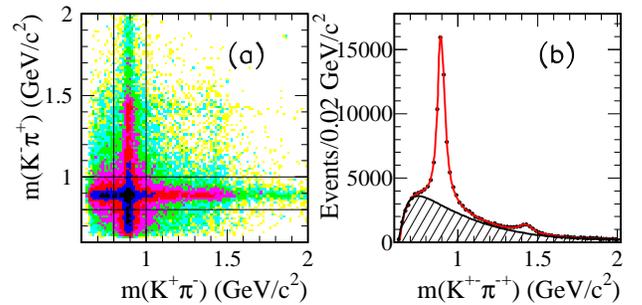}
\caption{\label{fig5a}
(a) Scatter plots $m(K^-\pi^+)$ vs. $m(K^+\pi^-)$,  
and (b) projection $m(K^{\pm}\pi^{\mp})$  plot (two entries per event) for
the  reaction $e^+ e^-\to  K^+ K^- \pi^+\pi^-$~\cite{babar2k2pi1}.
A sum over all accessible c.m. energies is given.}
\end{figure}

\begin{figure}
\includegraphics[width=.48\textwidth]{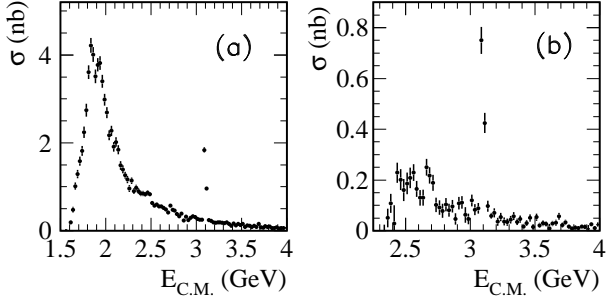}
\caption{ 
(a) The $e^+ e^-\to K^{*}(892)^0 K\pi$, 
 and (b) $K_2^{*}(1430)^0 K\pi$  cross sections~\cite{babar2k2pi1} 
  obtained from the $K^{*}(892)^0$ and $K_2^{*}(1430)^0$ signals of
Fig.~\ref{fig5a}(b), respectively.
\label{kstar_xs}}
\end{figure}

The distributions of the $K\pi$ invariant masses shown in Fig.~\ref{fig5a}
indicate that the $K^\ast(892)^0 K^{\pm}\pi^{\mp}$ and
$K^\ast(892)^{\mp} K^{\pm}\pi^0$ (similar plots are not shown)
intermediate states dominate in these reactions.
A small contribution of the $K_2^\ast(1430) K\pi$ state is also seen
(Fig.~\ref{fig5a}(b)). 
A special correlation study~\cite{babar2k2pi1} showed
that the intermediate state with two $K^\ast$, 
$K^\ast(892)\bar K^\ast(892)$, $K^\ast(892)\bar K_2^\ast(1430)$, 
and $K_2^\ast(1430) \bar K_2^\ast(1430)$, 
contributes less than 1\% to the total reaction yield (the associated 
$K^\ast(892)\bar K_2^\ast(1430)$ production is observed in
$J/\psi$ decays). Taking the numbers of 
events in the $K^\ast$ peaks for each c.m. energy interval, the
``inclusive''  $e^+ e^- \to K^\ast(892)^0  K\pi$ and
$e^+ e^-\to K_2^\ast(1430)^0  K\pi$ cross sections shown in Fig.~\ref{kstar_xs}
were extracted.
\begin{figure}
\begin{center}
\includegraphics[width=0.49\linewidth]{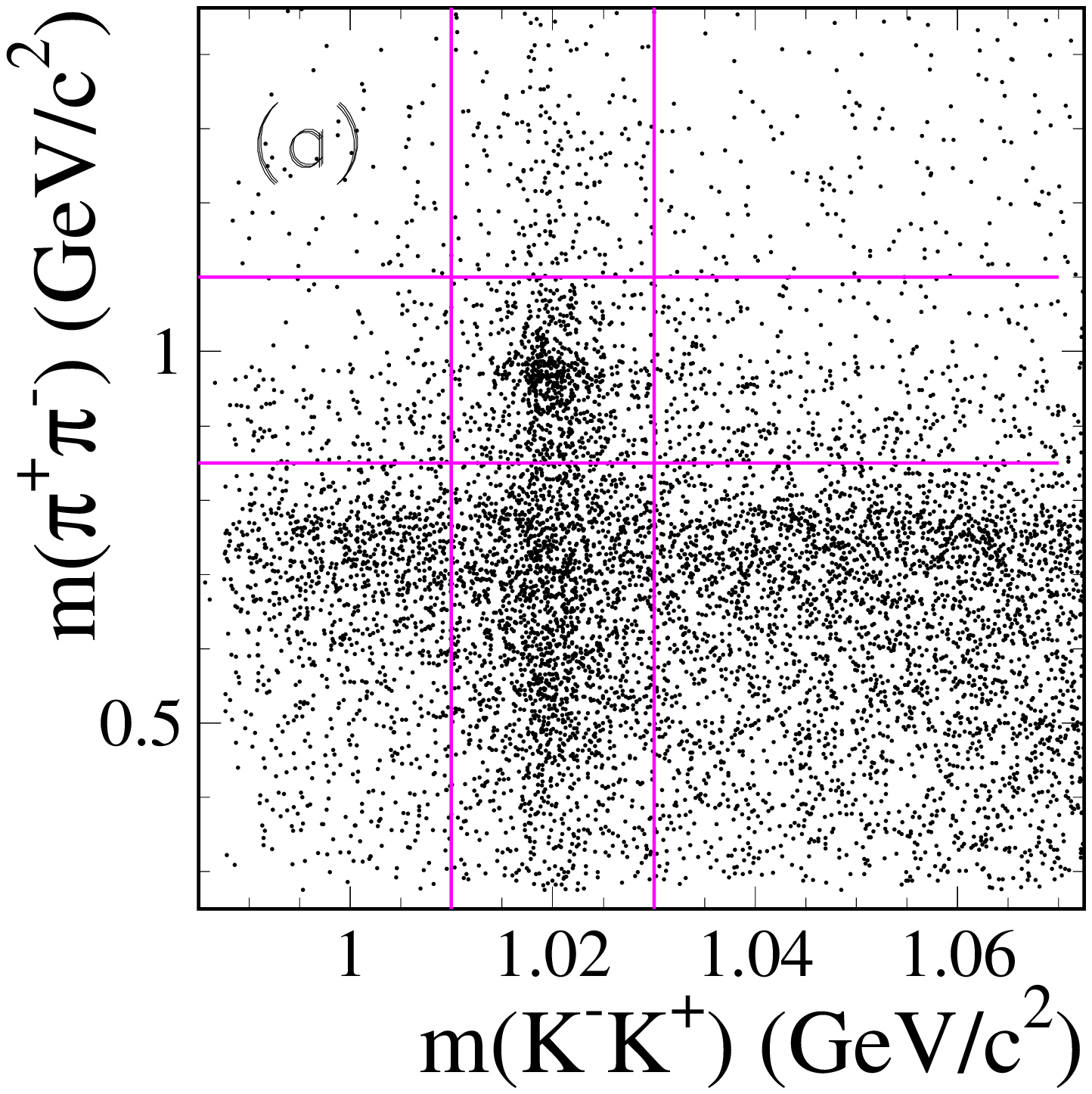}
\includegraphics[width=0.49\linewidth]{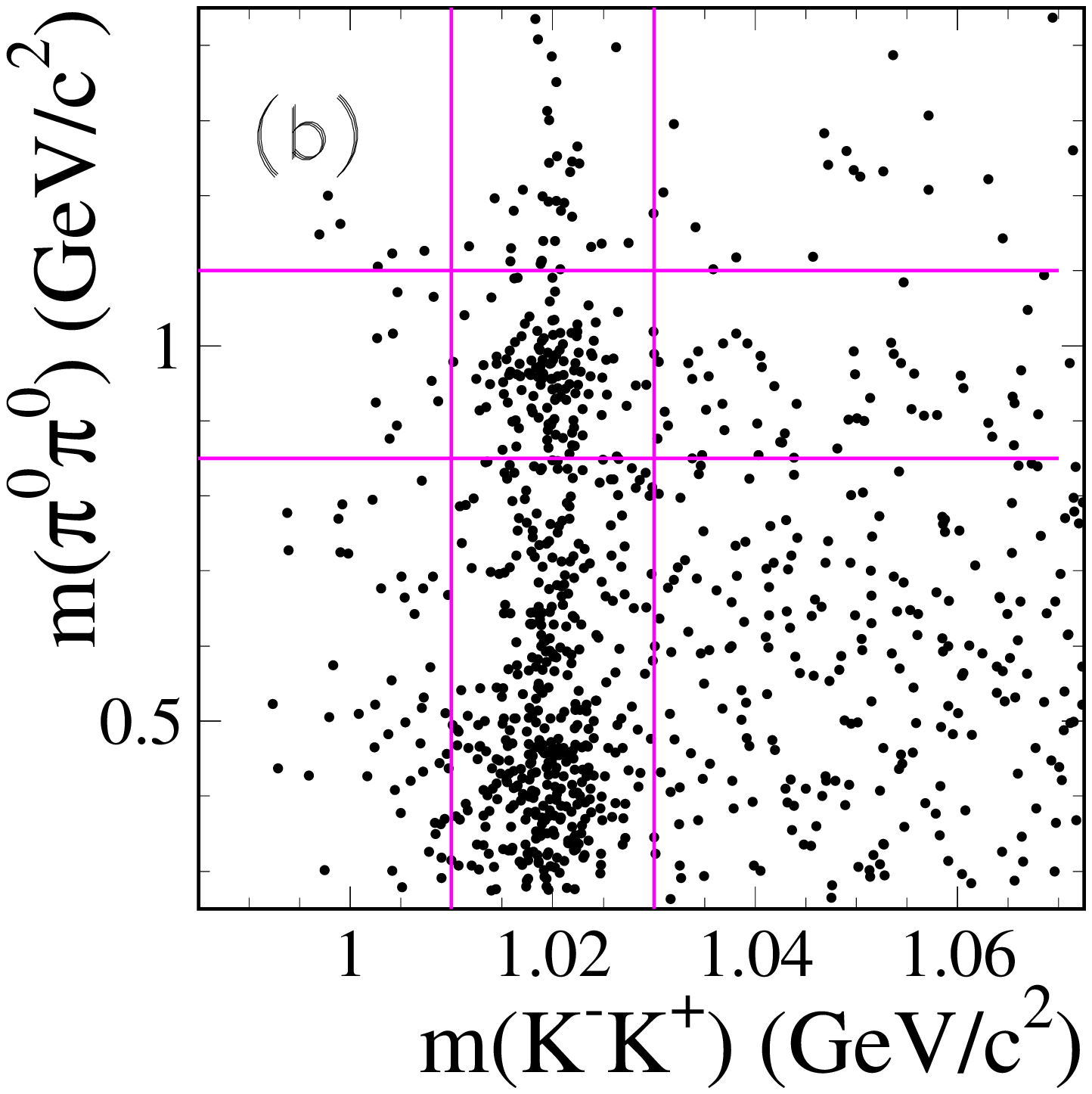}
\vspace{-0.7cm}
\caption{The
  scatter plots of the reconstructed a) $m(\pipi)$ and b)~$m(\ppz)$
  versus $m(K^+K^-)$ for selected events in the data.
  The vertical (horizontal) lines bound a
  $\phi$ ($f_0(980)$) signal region~\cite{babar2k2pi1}..
  }
\label{phif0_show1}
\vspace{-0.5cm}
\end{center}
\end{figure}
Figures~\ref{phif0_show1}(a) and~\ref{phif0_show1}(b) show scatter
plots of the reconstructed a) $m(\pipi)$ and b)~$m(\ppz)$
versus $m(K^+K^-)$ for selected events of the reactions   
$e^+ e^-\to K^+ K^-\pi^+\pi^-$  and  $e^+ e^-\to K^+ K^-\pi^0\pi^0$,
respectively. A clear $\phi(1020)$ signal is seen in the
$K^+K^-$ invariant mass in both figures and is discussed in more
detail below. A signal of the $\rho(770)$ is observed in the
$\pi^+\pi^-$ invariant mass distribution in Fig.~\ref{phif0_show1}(a).
The $\pi^+\pi^-$ invariant mass distribution for $ K^+ K^-\pi^+\pi^-$ 
events not containing the $K^\ast(892)$ meson is shown in 
Fig.~\ref{kkrho_sel}~(a). The $\rho(770)$ peak, probably
corresponding to the intermediate $K_1(1230)^{\pm}$ and 
$K_1(1400)^{\pm} K^{\mp}$ states, is clearly seen in the $\pi^+\pi^-$
mass spectrum. In Fig.~\ref{kkrho_sel}~(b) the ``inclusive'' cross section 
for the $K^+ K^- \rho(770)$ reaction is presented.
It is obtained by fitting the $\rho(770)$ signal in
the $\pi^+\pi^-$ invariant mass distributions for each c.m. energy
interval in Fig.~\ref{kkrho_sel}.
\begin{figure}
\includegraphics[width=.48\textwidth]{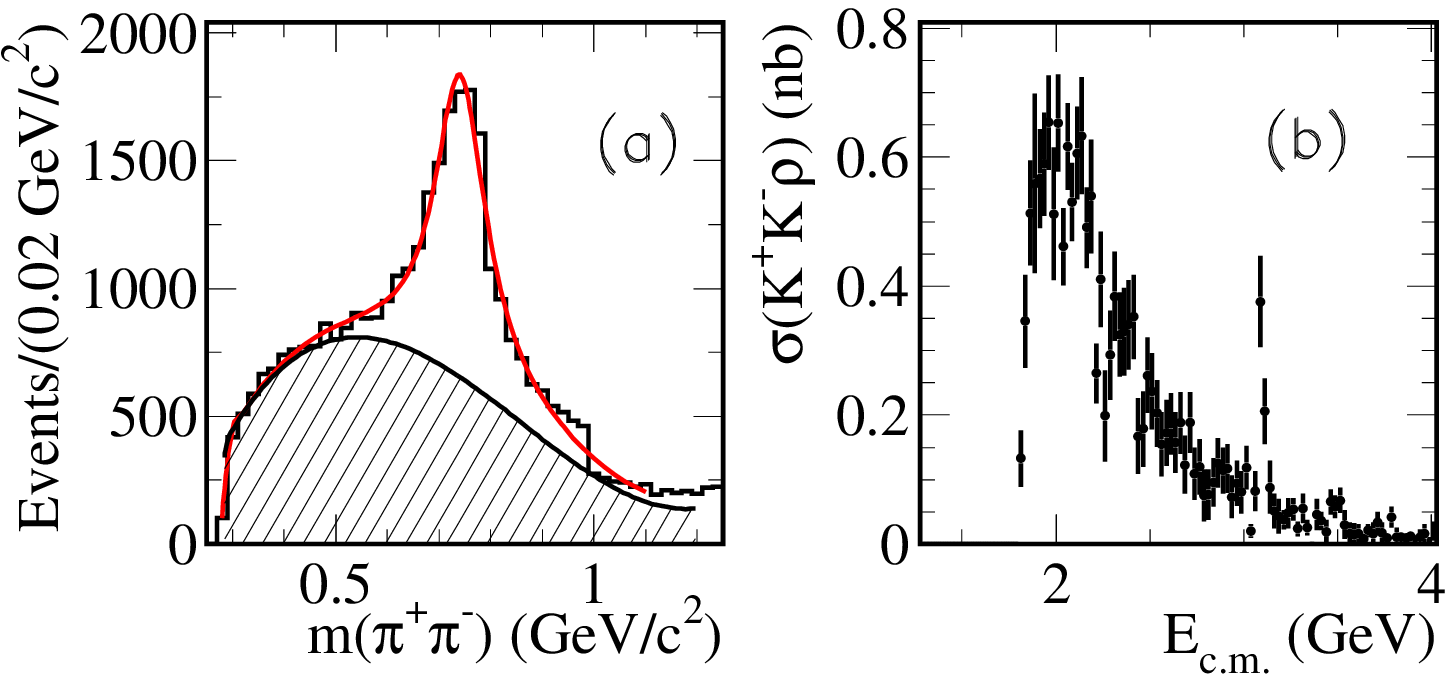}
\caption{ 
  (a) The $\pi^+\pi^-$ mass distribution for 
  $ K^+ K^-\pi^+\pi^-$ events ($K^\ast(892)K\pi$ events are excluded); 
  the solid curve represents a fit using a signal Breit-Wigner function
  with $\rho(770)$ parameters and a polynomial background (hatched area).
  (b) The $e^+ e^- \to K^+ K^- \rho(770)$  cross section 
  obtained using the fitted numbers of $\rho$-meson events in
  each 25~MeV c.m. energy interval~\cite{babar2k2pi1}.
\label{kkrho_sel}}
\end{figure}
\begin{figure}
\includegraphics[width=.4\textwidth]{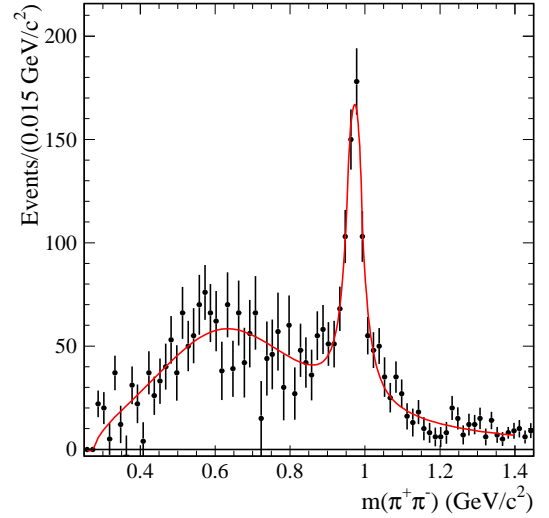}
\caption{\label{fig12a} 
The $m(\pi^+\pi^-)$ distribution for the 
$e^+e^-\to\phi(1020)\pi^+\pi^-$ reaction~\cite{babar2k2pi1}.}
\end{figure}
\begin{figure}
\includegraphics[width=.4\textwidth]{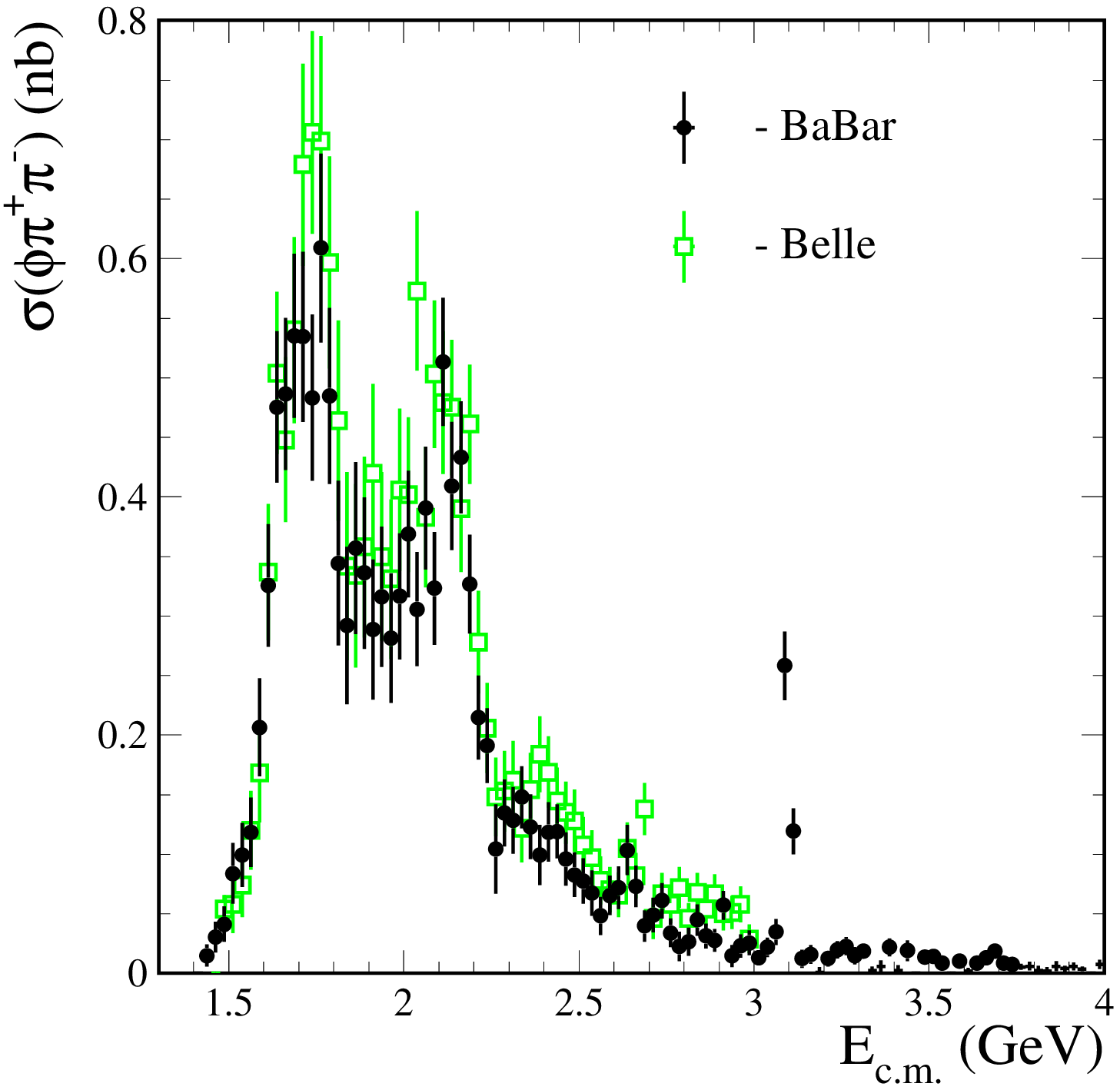}
\caption{\label{fig12b} 
The $e^+ e^- \to\phi\pi^+\pi^-$ cross sections measured with
the BABAR~\cite{babar2k2pi1} (circles) and 
Belle~\cite{bel2k2pi} (squares) detectors. }
\end{figure}

One of the interesting ISR studies performed by BABAR~\cite{babr2k2pi} 
and later reproduced by Belle~\cite{bel2k2pi} is extracting relatively small 
contributions of the $\phi(1020)\pi^+\pi^-$, $\phi(1020)\pi^0\pi^0$,
($\phi\to K^+ K^-$) intermediate states. 
Since the $\phi(1020)$ resonance is relatively narrow, the clean 
sample of $\phi\pi\pi$ events can be easily separated.
Figure~\ref{fig12a} shows the $m(\pi^+\pi^-)$ distribution for these
events demonstrating a clear signal from the $f_0(980)$ resonance 
and a bump at lower masses which can be interpreted as the $f_0(600)$ state.
A similar plot is obtained for the $\pi^0 \pi^0$ invariant mass.
These invariant mass distributions can be fitted with a superposition
of two Breit-Wigner functions for the scalar $f_0(980)$ and $f_0(600)$
resonances as shown in Fig.~\ref{fig12a}. 
The $e^+ e^- \to\phi(1020)\pi^+\pi^-$ cross section measured by BABAR 
and Belle is shown in Fig.~\ref{fig12b}. Two resonance structures
are seen at 1.7~GeV and at 2.1~GeV. The BABAR Collaboration
investigated decay mechanisms for these structures and 
concluded that the second structure decays only 
to the $\phi(1020)f_0(980)$ final state. The structure
completely disappears if events associated with the $f_0(980)$ peak
in the $m(\pi^+\pi^-)$ distribution are  removed. The first structure is
associated with the $\phi(1680)$, a radial excitation of the vector 
$s\bar s$ state. Its decays to $\phi(1020) f_0(600)$ and
$\phi(1020) f_0(980)$ are not forbidden. 
\begin{figure}
\includegraphics[width=.4\textwidth]{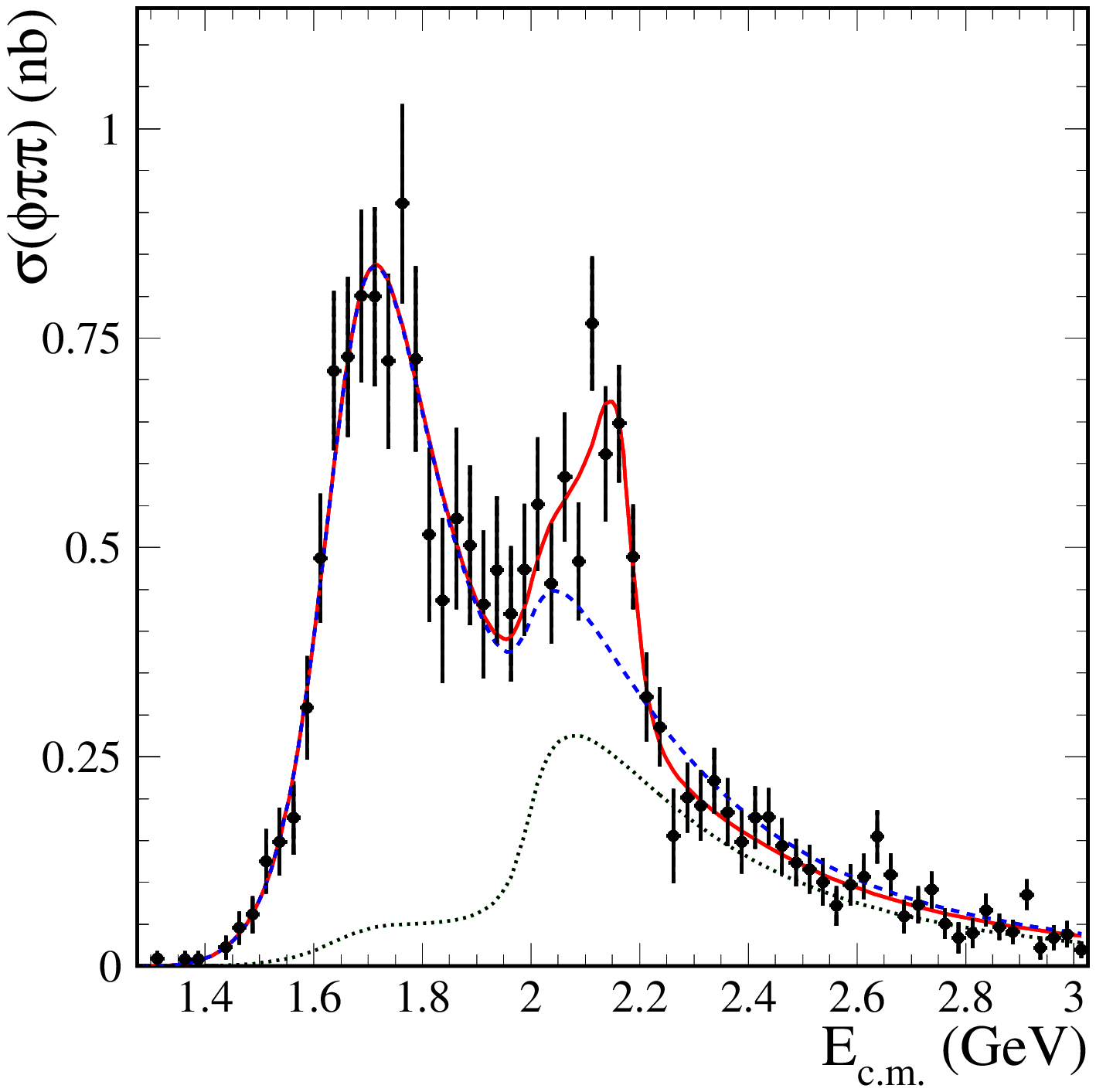}
\caption{\label{fig6aa} 
The fit to the $e^+ e^-  \to\phi\pi^+\pi^-$ cross section~\cite{babar2k2pi1} 
in the two-resonance model described in the text (solid curve). 
The contribution of the first resonance ($\phi(1680)$) is shown
by the dashed line. The dotted line shows the first resonance 
contribution in the $\phi f_0(980)$ decay mode only.}
\end{figure}
\begin{figure}
\includegraphics[width=.4\textwidth]{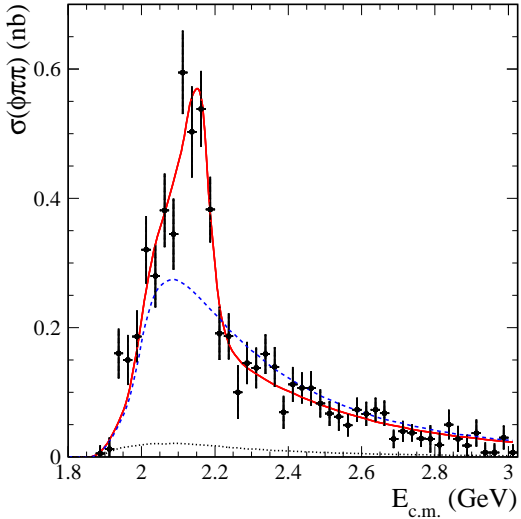}
\caption{\label{fig6ab} 
The cross section for $e^+ e^- \to\phi(1020) f_{0}(980)$ events
selected with the cut $0.85<m(\pi\pi)<1.1$~GeV/c$^2$~\cite{babar2k2pi1}.
The solid curve is the result of the  two-resonance fit.
The dashed and dotted curves are the contributions of the 
$\phi(1680) \to \phi f_{0}(980)$ and $\phi(1680) \to \phi f_{0}(600)$
decay channels, respectively.}
\end{figure}
\begin{figure}
\includegraphics[width=.4\textwidth]{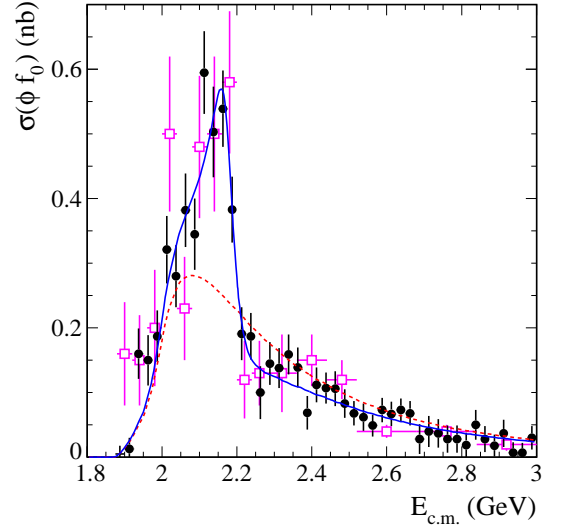}
\caption{
The $e^+ e^- \to\phi(1020) f_{0}(980)$ cross section measured 
in the $K^+ K^-\pi^+\pi^-$ (circles) and $K^+ K^-\pi^0\pi^0$
(squares) final states by BABAR~\cite{babar2k2pi1}.
The solid and dashed curves represent the results of 
the two-resonance fit described in the text. 
\label{fig6ca}}
\end{figure}
\begin{figure}
\includegraphics[width=.4\textwidth]{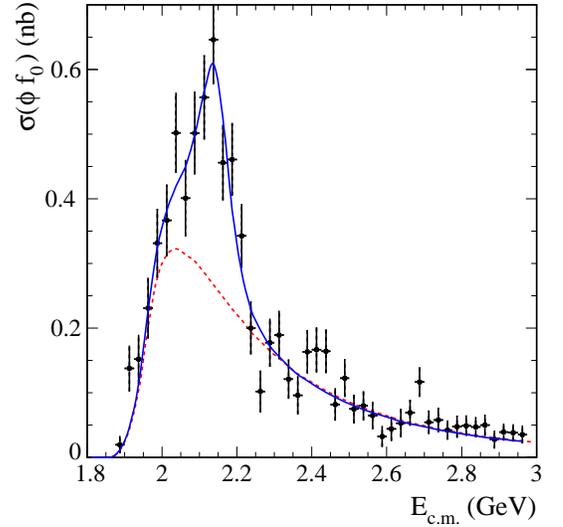}
\caption{
The $e^+ e^- \to\phi(1020) f_{0}(980)$ cross section measured 
by Belle~\cite{bel2k2pi}.
The solid and dashed curves represent the results of 
the two-resonance fit described in the text. 
\label{fig6cb}}
\end{figure}

A simple VDM based model was suggested to describe
the observed $e^+ e^- \to\phi(1020)\pi^+\pi^-$ cross section. The
model assumes that two vector mesons contribute to the cross section;
one resonance associated with the $\phi(1680)$ decays both to 
$\phi f_0(600)$ and to $\phi f_0(980)$, while
another referred to as  $Y(2175)$ decays   to $\phi f_0(980)$ only.
Since the nominal $\phi(1680)$ mass lies below the
$\phi f_0(980)$ threshold, the $\phi(1680)\to \phi f_0(980)$ decay
will reveal itself as a smooth bump in the energy dependence of 
the $e^+ e^- \to\phi f_0(980)$ cross section above 2 GeV. 
The result of the fit to the $e^+ e^- \to\phi(1020)\pi^+\pi^-$ cross section
with this model is shown in Fig.~\ref{fig6aa}. It is clearly
seen that the data above 2 GeV cannot be described with a contribution
of the $\phi(1680)$ resonance only. An additional relatively narrow resonance
$Y(2175)$ is needed to do this. The tongue below 2 GeV in the cross section for
the reaction $e^+e^- \to \phi(1680) \to \phi f_0(980)$ in Fig.~\ref{fig6aa}
is due to the finite width of the $f_0(980)$ state.

A relatively clean sample of $\phi f_0(980)$ events is selected
using the requirement $0.85 {\rm GeV}/c^2 < m(\pi\pi) < 1.1$ GeV/$c^2$. 
The cross
section for events of the $K^+ K^-\pi^+\pi^-$ mode, fitted with the model
described above, is shown in Fig.~\ref{fig6ab}. 
The contribution of the $Y(2175)$ is seen much better with this selection. 
The comparison of the $e^+ e^- \to\phi f_{0}(980)$ cross sections
measured by BABAR in two $f_{0}(980)$ decay modes, $\pi^+\pi^-$ and 
$\pi^0\pi^0$, is shown in Fig.~\ref{fig6ca}. It is seen that 
two measurements agree. 
The fit of two modes gives the peak cross section, mass and width 
of the resonance:
\begin{eqnarray*}
 \sigma_Y   & = &  0.104 \pm 0.025~{\rm nb} ,   \\
      m_Y   & = &  2.179 \pm 0.009~{\rm GeV}/c^2 ,    \\
 \Gamma_Y   & = &  0.079 \pm 0.017~{\rm GeV} .    
\end{eqnarray*}

The $e^+ e^- \to\phi f_{0}(980)$
cross section measured in the Belle experiment~\cite{bel2k2pi} is
shown in Fig.~\ref{fig6cb}, and also requires a resonance structure with
similar parameters.
Some properties of this resonance, a relatively small width and absence of
the $\phi f_0(600)$ decay, are unusual.  

The nature of this state is not clear~\cite{Y21751,Y21752}. One of the possible 
interpretations is that the $Y(2175)$ is a $s\bar s s\bar s$ 
four-quark state.
Indeed, the $f_0(600)$ does not contain strange quarks, while
the $f_0(980)$, strongly coupled with $K\bar K$, definitely contains them.
For the  $s\bar s s\bar s$ state, the observed  $Y(2175)\to \phi f_0(980)$ 
is natural decay, while the not seen $Y(2175)\to \phi f_0(600)$ transition 
is suppressed by the OZI rule. The observation of the $Y(2175)$ decay 
to the $\phi\eta$ final state containing four $s$ quarks, also supports
this hypothesis. 

\subsection{$e^+ e^- \to 2(K^+ K^-)$}
\begin{figure}
\includegraphics[width=.4\textwidth]{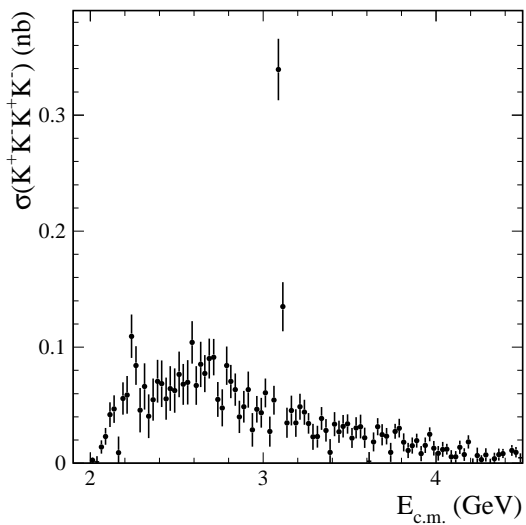}
\caption{
The $e^+ e^- \to  2(K^+ K^-)$ cross section as a function of 
  c.m. energy measured with the BABAR detector using ISR~\cite{babr2k2pi}.
\label{fig6da}}
\end{figure}

\begin{figure}
\includegraphics[width=.4\textwidth]{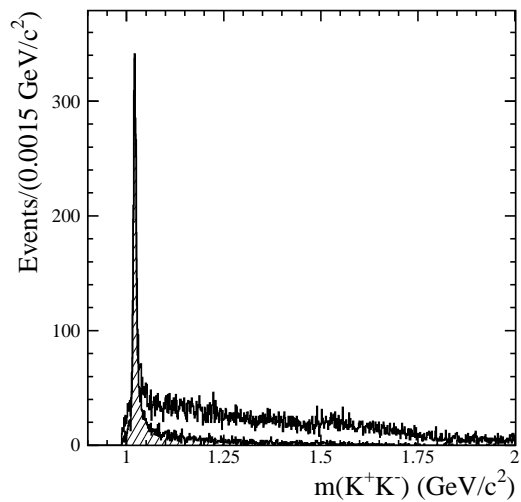}
\caption{
The $K^+ K^-$ invariant mass distribution for selected
$e^+ e^- \to 2(K^+ K^-)$ events~\cite{babr2k2pi} 
(open histogram, four entries per event),
and that for the combination in each event closest to the $\phi$-meson mass
(hatched histogram).
\label{fig6db}}
\end{figure}

The reaction $e^+ e^- \to 2(K^+ K^-)$ was studied for the 
first time by BABAR~\cite{babr2k2pi}.
The measured cross section is shown in Fig.~\ref{fig6da}. 
The most significant structure in the cross sections is due to the $J/\psi$
decay. It is natural to expect that intermediate states for this reaction
contain the $\phi(1020)$ meson which has a large decay rate to $K^+ K^-$. 
Indeed, the strong $\phi$ meson peak is seen in the $K^+ K^-$ invariant
mass distribution shown in Fig.~\ref{fig6db}. Since the $\phi$ meson 
is present in almost each four-kaon event, it is concluded that 
the reaction $e^+ e^- \to 2(K^+ K^-)$
is strongly dominated by the $\phi K^+ K^-$ production.

\begin{figure*}[p]
\includegraphics[width=6cm]{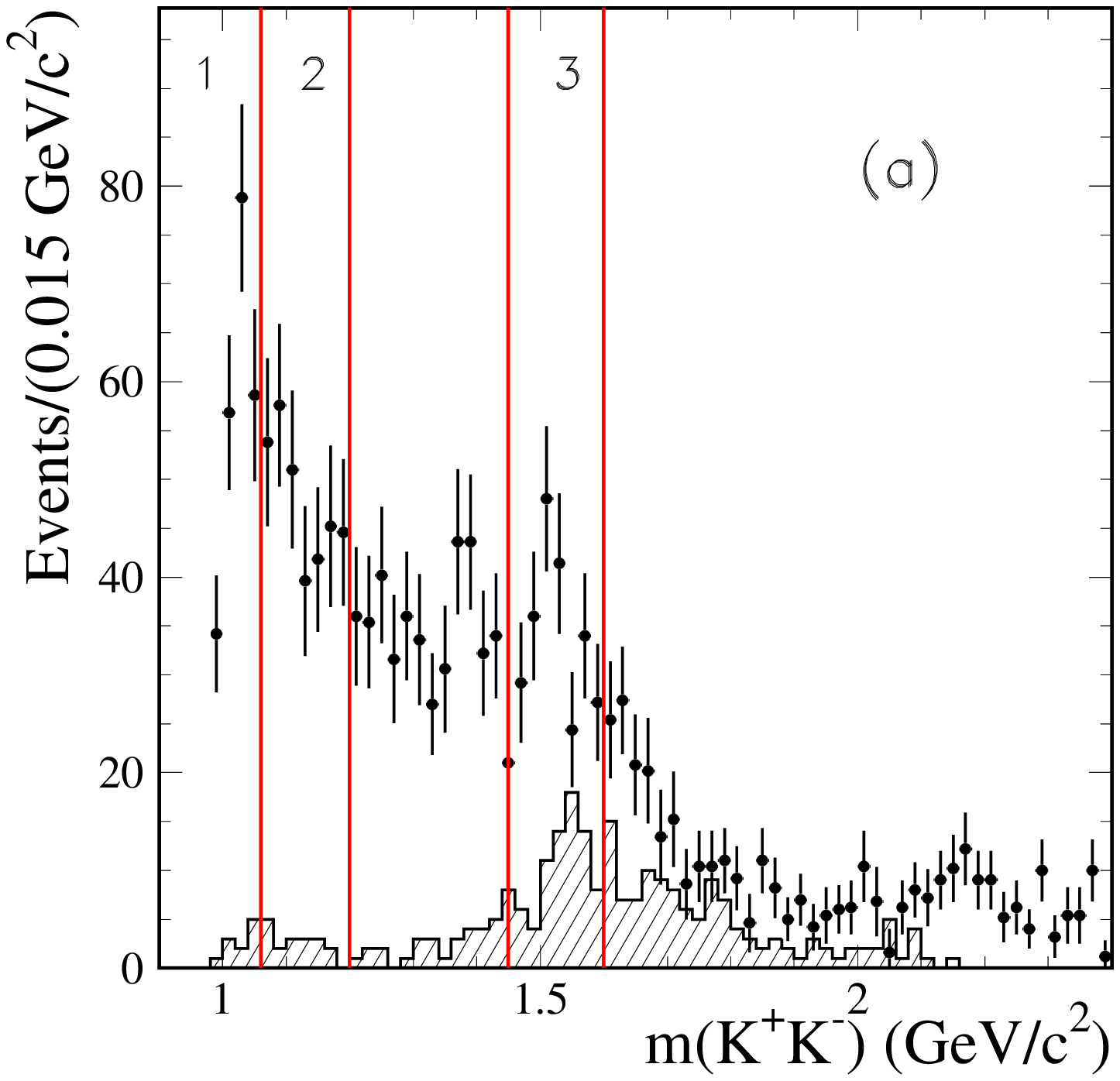}
\includegraphics[width=6cm]{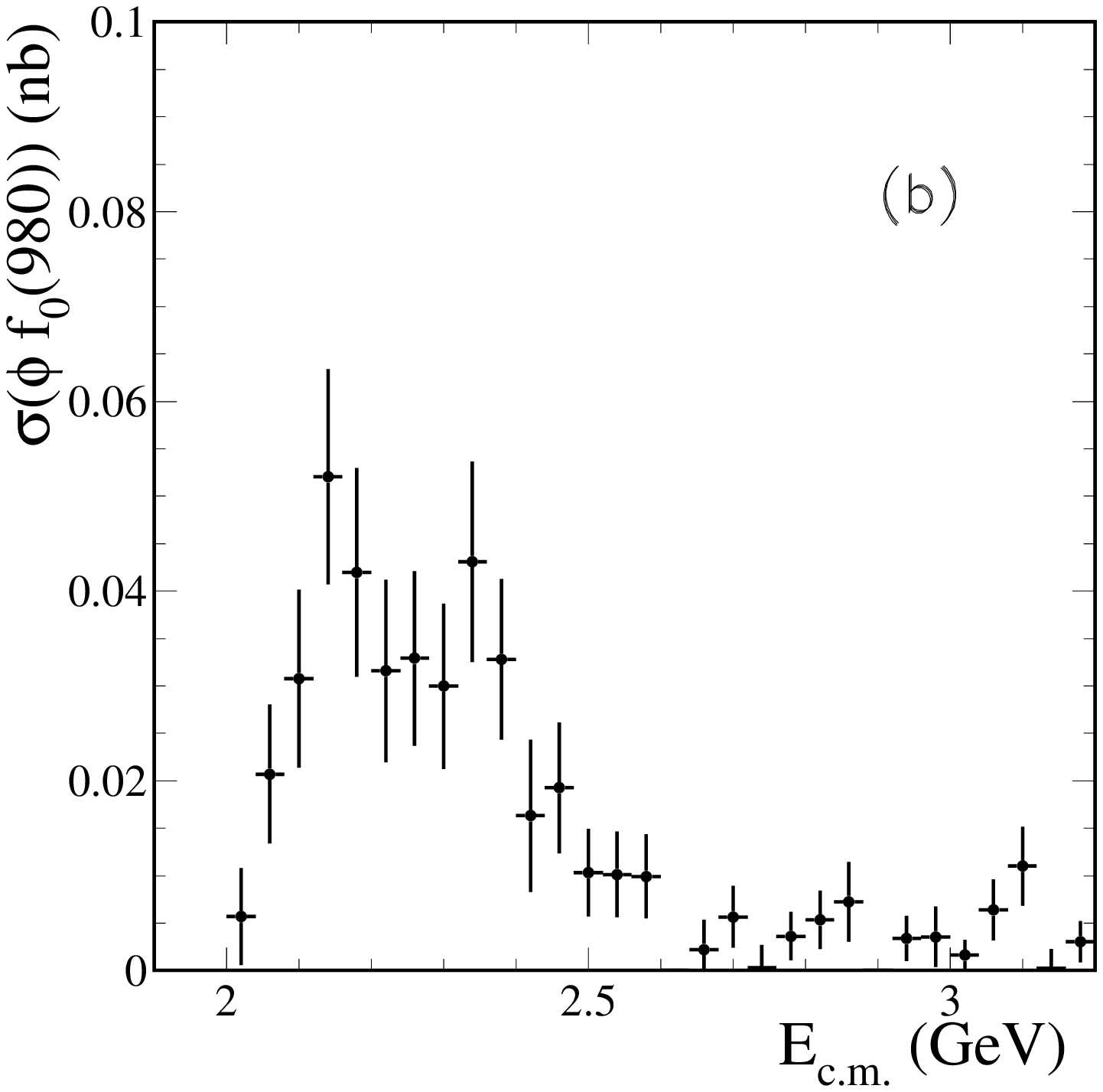}\\
\includegraphics[width=6cm]{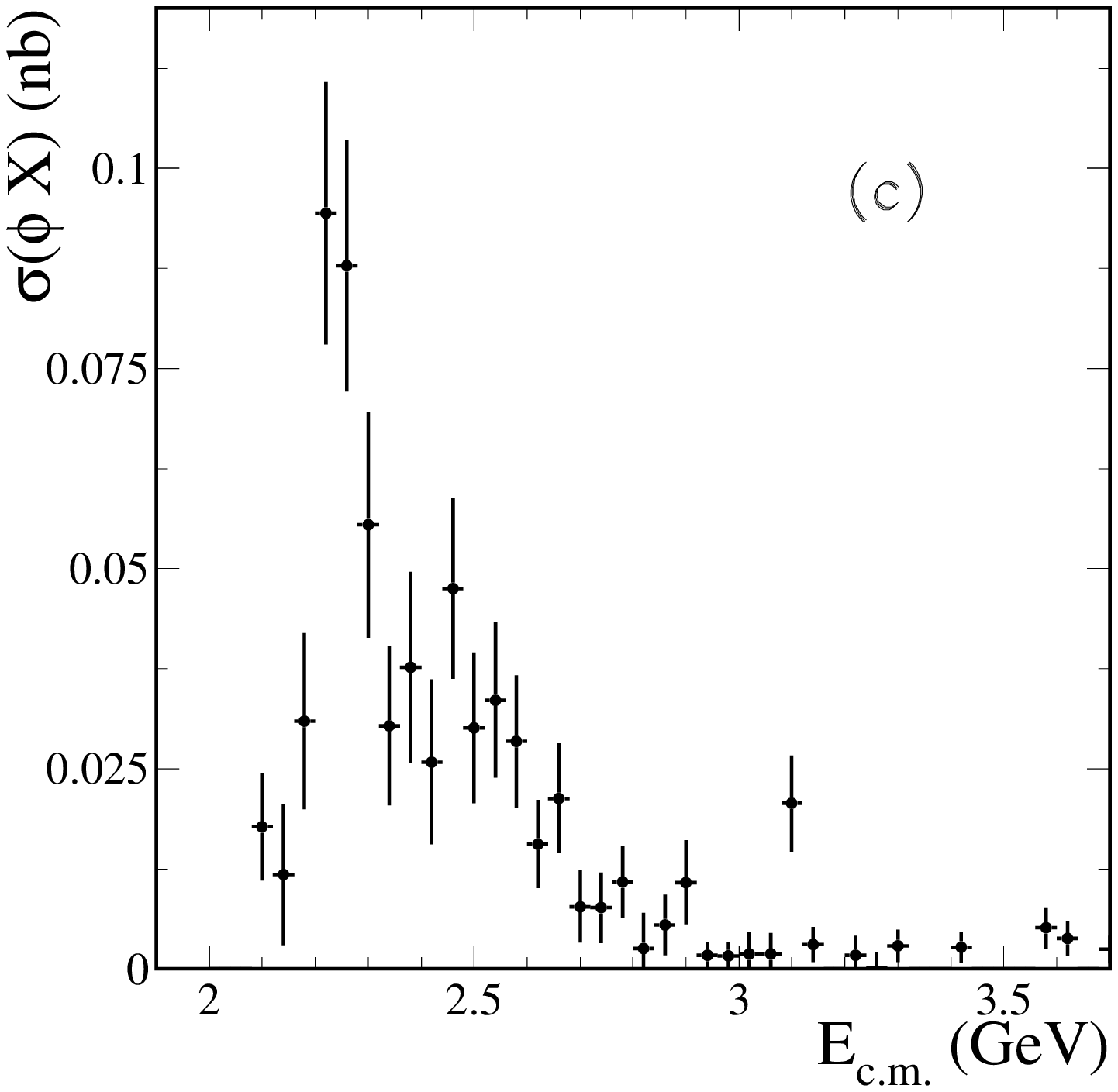}
\includegraphics[width=6cm]{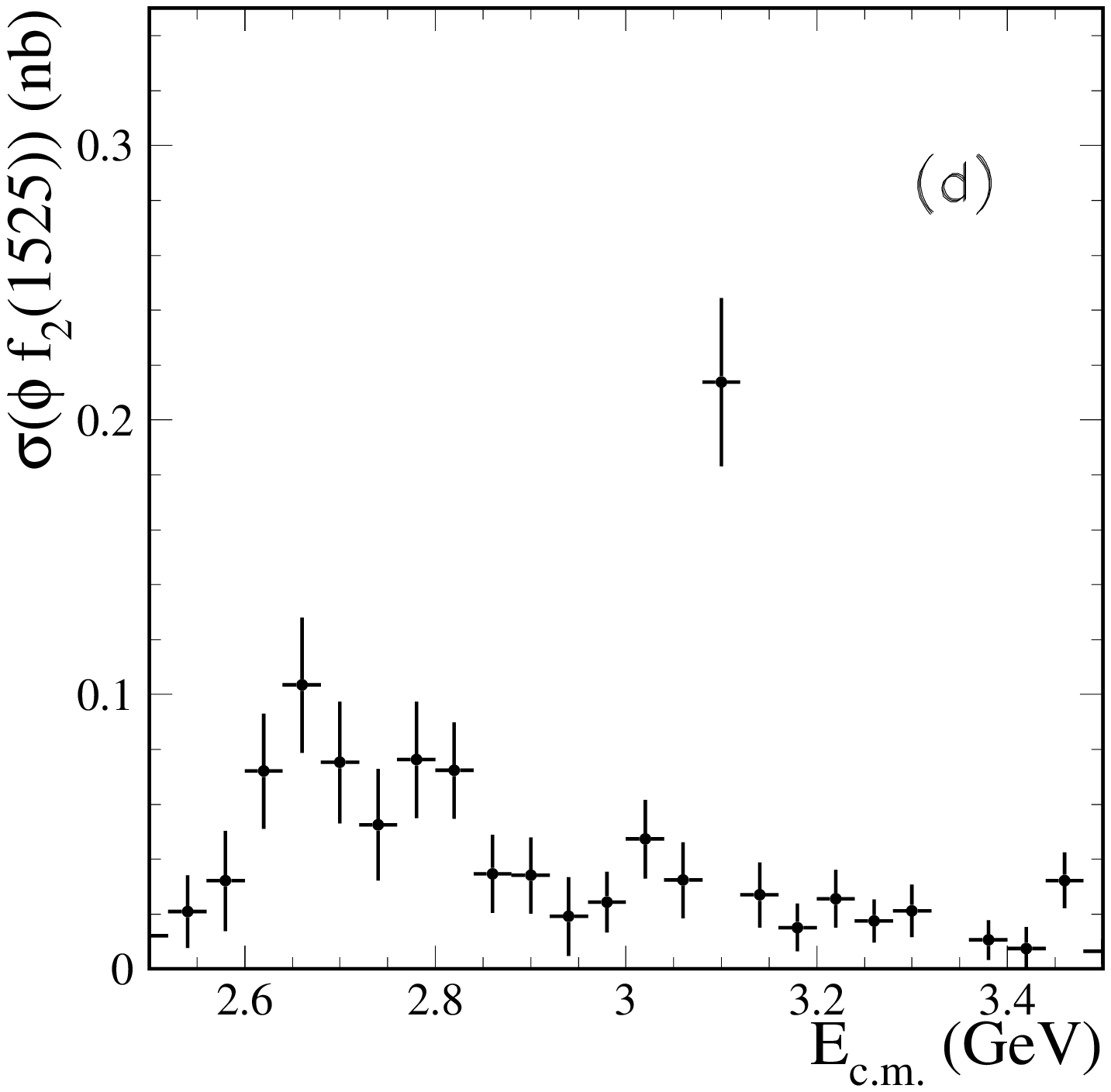}
\caption{
The $K^+ K^-$ invariant mass distribution for $\phi K^+K^-$ 
events~\cite{babr2k2pi} (a).
Events from the $J/\psi\to \phi K^+K^-$ decay are excluded from 
the spectrum shown by the open histogram.
The hatched histogram is for events from the $J/\psi$ decay.
The numbered regions of the $K^+ K^-$ mass spectrum are used to
calculate the cross sections shown in the plots (b), (c), and (d) for
the regions 1, 2, and 3, respectively.
\label{mkk_notphi}}
\end{figure*}

A study of events containing the $\phi$ meson 
was performed by BABAR. The $K^+ K^-$  pair forming the  $\phi$ meson is
selected by the requirement that its invariant mass is 
within $\pm 10$ MeV of the $\phi$ nominal mass. The invariant mass 
distribution for the second $K^+ K^-$  pair is shown in 
Fig.~\ref{mkk_notphi}(a). Figures~\ref{mkk_notphi}(b,c,d) show
the cross section for events with the $K^+ K^-$ invariant
mass in the regions 1, 2, and 3 indicated in Fig.~\ref{mkk_notphi}(a).
An enhancement in the $K^+ K^-$ invariant mass spectrum near 
the $K^+ K^-$ threshold can be interpreted as being due to
decay $f_0(980)\to K^+ K^-$. Therefore, the cross section for
the region 1 is expected to have a structure similar to that observed
in $e^+ e^- \to \phi f_0 \to K^+K^-\pi\pi$  (see Sec.~\ref{Sec:2k2pi}).
The bump at 2.175 GeV is indeed seen in the cross section shown in
Fig.~\ref{mkk_notphi}(b), however, the data sample is too low to 
perform a quantitative analysis. 

The relatively narrow region 2 with 
$1.06$ GeV/$c^2 <m(K^+ K^-)<1.2$ GeV/c$^2$ is
responsible for the spike seen at 2.25 GeV in Fig.~\ref{fig6da}.
The spike is much more significant in Fig.~\ref{mkk_notphi} (c) showing
the cross section for events from this mass region.
There is no explanation of this structure.

The peak in the $K^+ K^-$ mass spectrum near 1.5 GeV/$c^2$ 
is associated with the $f_2^\prime(1525)$. The region 3 
($1.45$ GeV/$c^2  <m(K^+ K^-)<1.6$ GeV/c$^2$) is chosen to select
$\phi f_2^\prime(1525)$ events. The cross section for
this mass region is shown in Fig.~\ref{mkk_notphi}(d) and
exhibits a broad structure at 2.7 GeV and a strong $J/\psi$
signal.
\subsection{$e^+ e^- \to 5$ mesons}
\label{5pi}
The BABAR detector studied a number of ISR reactions with five hadrons in
the final state: 
$2(\pi^+\pi^-)\pi^0$, $2(\pi^+\pi^-)\eta$, $ K^+K^-\pi^+\pi^-\pi^0$,
and $ K^+ K^-\pi^+\pi^-\eta$~\cite{babr5pi}. 

\begin{figure}[h]
\includegraphics[width=.4\textwidth]{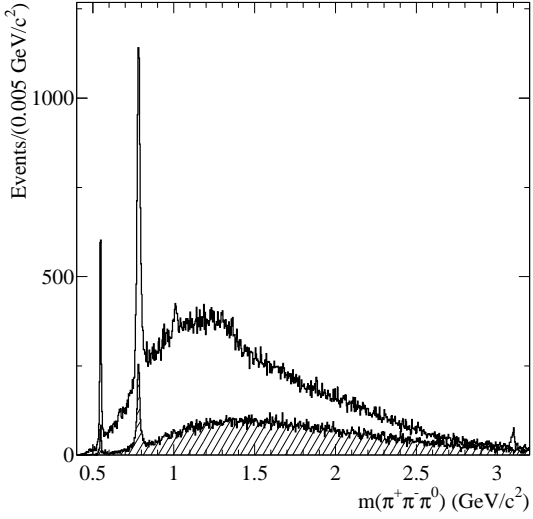}
\caption{\label{fig77a} 
The $m(\pi^+\pi^-\pi^0)$ distribution for $2(\pi^+\pi^-)\pi^0$
events~\cite{babr5pi}.}
\end{figure}

\begin{figure}[h]
\includegraphics[width=.4\textwidth]{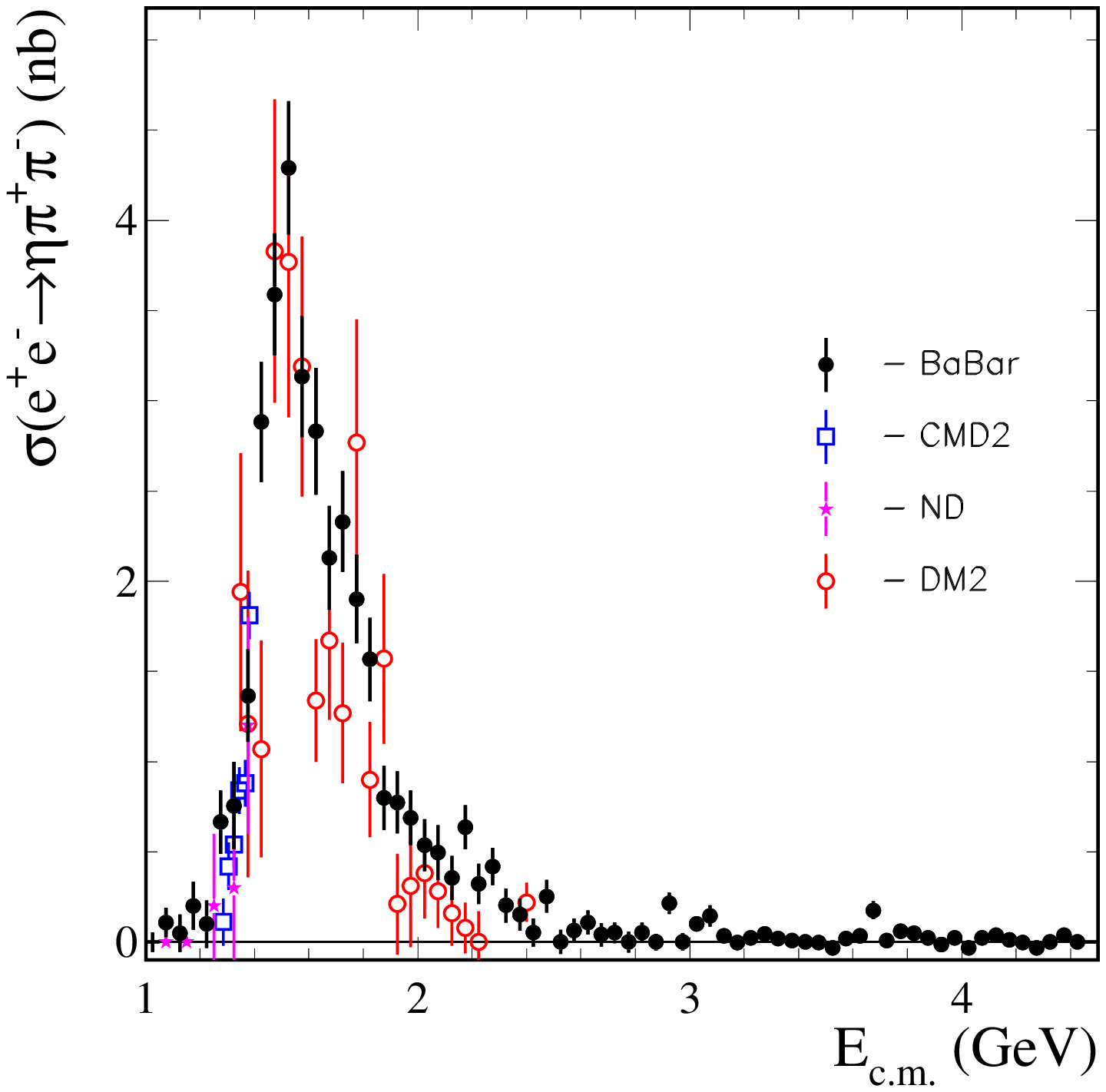}
\caption{\label{fig77b} 
The $e^+ e^-\to\eta\pipi$ cross section measured 
by  BABAR~\cite{babr5pi} in comparison
with the direct \epem measurements~\cite{dm23pi,nd5pi,cmd5pi}.}
\end{figure}

\begin{figure}[h]
\includegraphics[width=.4\textwidth]{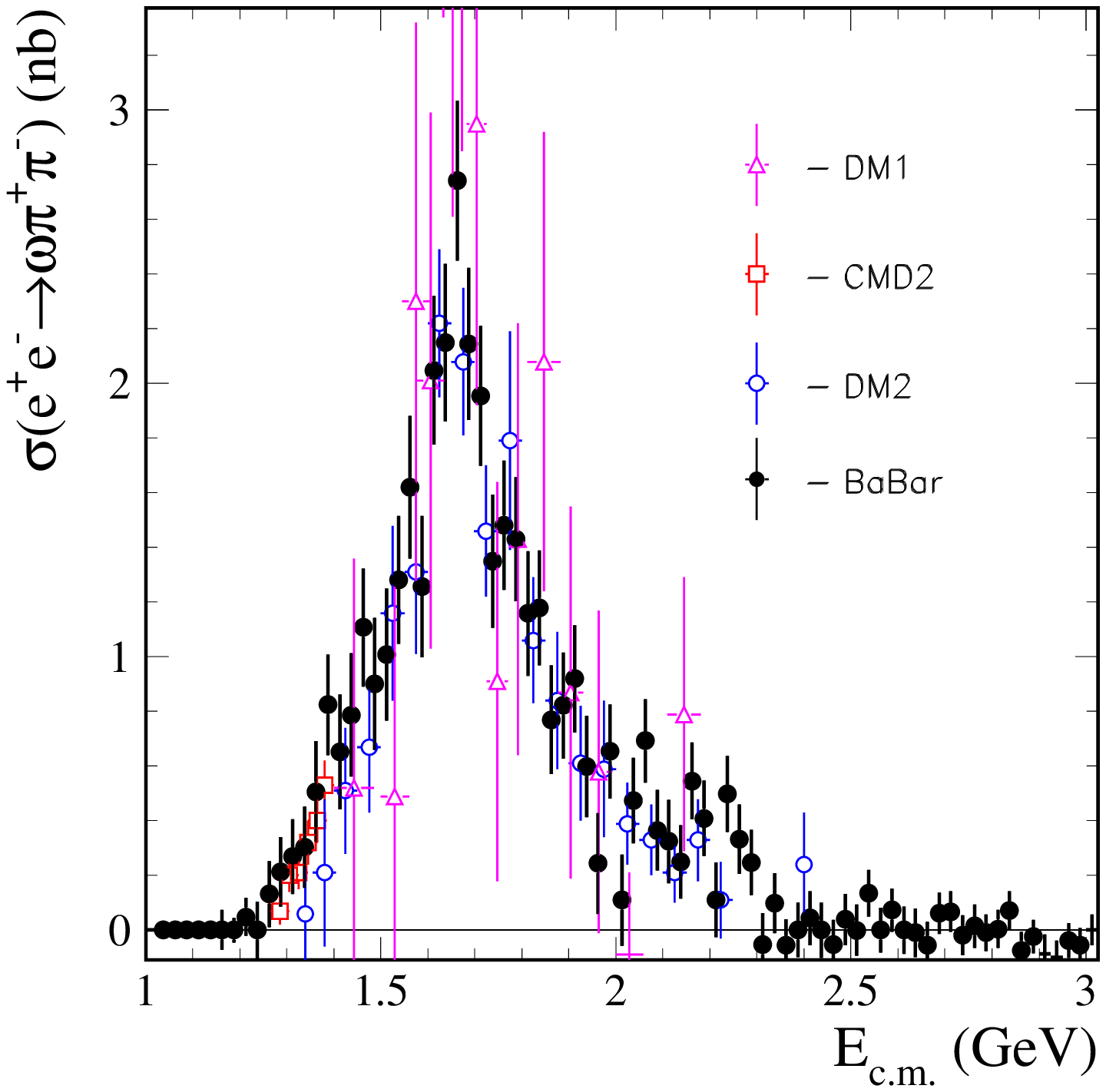}
\caption{\label{fig77c} 
The $e^+ e^-\to\omega\pipi$ cross section measured 
by BABAR~\cite{babr5pi} in comparison
with the direct \epem measurements~\cite{dm15pi,dm25pi,cmd5pi}.}
\end{figure}

The $e^+ e^-\to 2(\pi^+\pi^-)\pi^0$ reaction has the largest cross
section among the processes mentioned above. 
In the $\pi^+\pi^-\pi^0$ invariant mass spectrum for this reaction
(Fig.~\ref{fig77a}) clear signals of $\eta$ and $\omega$ mesons are seen
corresponding to the $\omega\pi^+\pi^-$ and $\eta\pi^+\pi^-$ 
intermediate states.
The cross sections for these reactions were measured in direct \epem
experiments~\cite{dm23pi,dm15pi,nd5pi,dm25pi,cmd5pi}, 
but BABAR data are significantly 
more accurate. The $e^+ e^-\to \eta\pi^+\pi^-$ and $e^+ e^-\to \omega\pi^+\pi^-$ 
cross sections measured by BABAR and in direct \epem experiments are 
shown in Fig.~\ref{fig77b} and Fig.~\ref{fig77c}, respectively.

\begin{figure}[h]
\includegraphics[width=.4\textwidth]{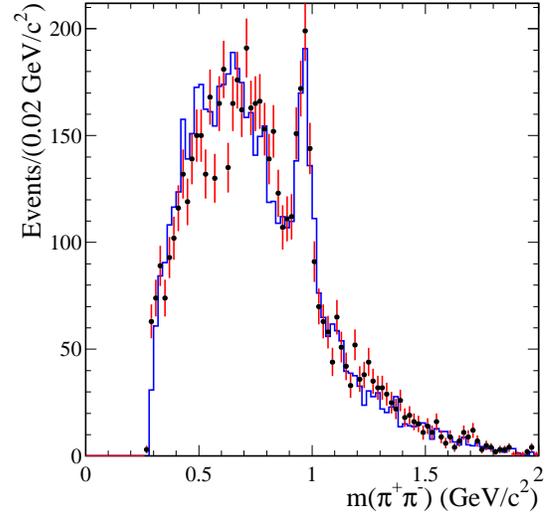}
\caption{\label{fig88a}
The $m(\pi^+\pi^-)$ distribution for selected $\omega\pi^+\pi^-$
events in data (points with error bars)
and in simulation (histogram)~\cite{babr5pi}.}
\end{figure}
\begin{figure}[h]
\includegraphics[width=.4\textwidth]{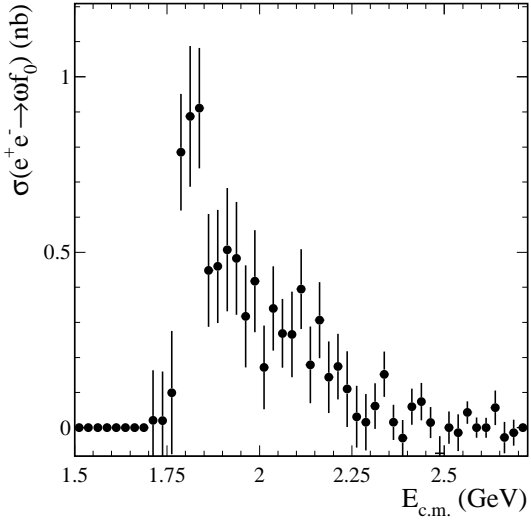}
\caption{\label{fig88b}
The $e^+ e^-\to\omega f_0(980)$ cross section measured by BABAR~\cite{babr5pi}.}
\end{figure}
\begin{figure}[h]
\includegraphics[width=.4\textwidth]{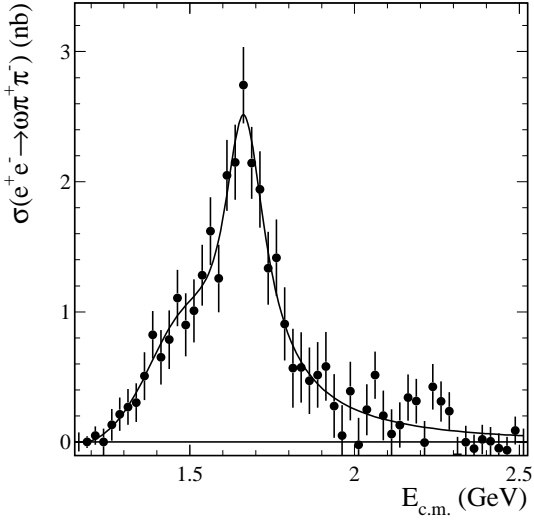}
\caption{\label{fig88c}
The fit with two Breit-Wigner functions to the $\omega\pi^+\pi^-$ cross section
with the $\omega f_0(980)$ contribution subtracted~\cite{babr5pi}.}
\end{figure}

\begin{figure}[h]
\includegraphics[width=.4\textwidth]{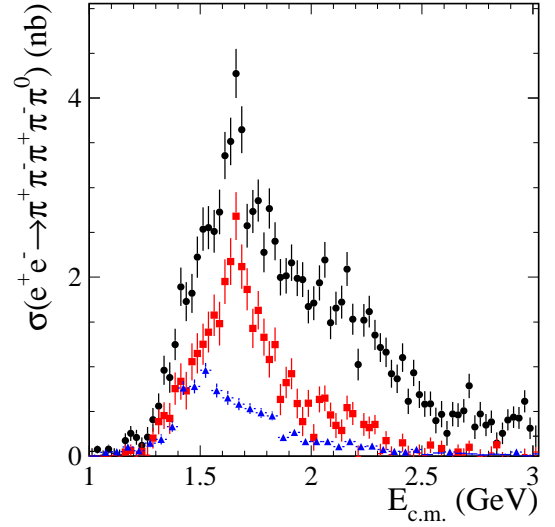}
\caption{\label{fig99a} 
The  $e^+ e^-\to 2(\pi^+\pi^-)\pi^0$ cross section~\cite{babr5pi} and
contributions from $\omega\pi^+\pi^-$(squares) and
$\eta\pi^+\pi^-$(triangles).}
\end{figure}

\begin{figure}[h]
\includegraphics[width=.4\textwidth]{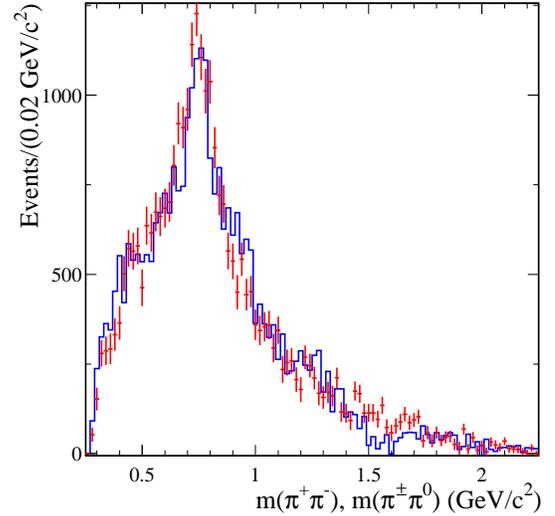}
\caption{\label{fig99b} 
The $m(\pi^+\pi^-)$ (points) and $m(\pi^{\pm}\pi^0)$ (histogram)
distributions for $2(\pi^+\pi^-)\pi^0$ events with the
$\omega\pi^+\pi^-$ and $\eta\pi^+\pi^-$ contributions excluded~\cite{babr5pi}.}
\end{figure}
\begin{figure}[h]
\includegraphics[width=.4\textwidth]{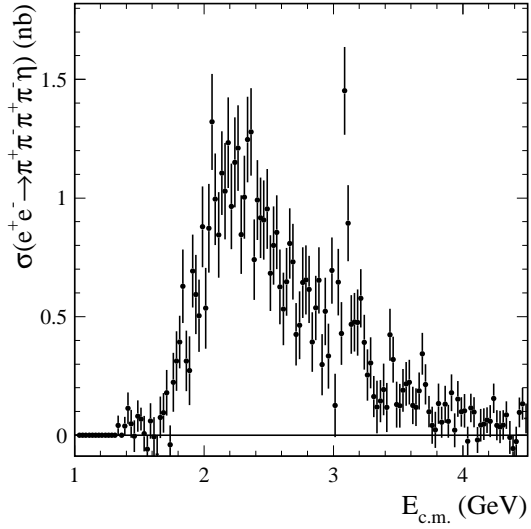}
\caption{
The $e^+ e^- \to 2(\pi^+\pi^-)\eta$ cross section
 measured by BABAR~\cite{babr5pi}.
\label{4pieta_ee_babar}}
\end{figure}
\begin{figure}[h]
\includegraphics[width=4.1cm]{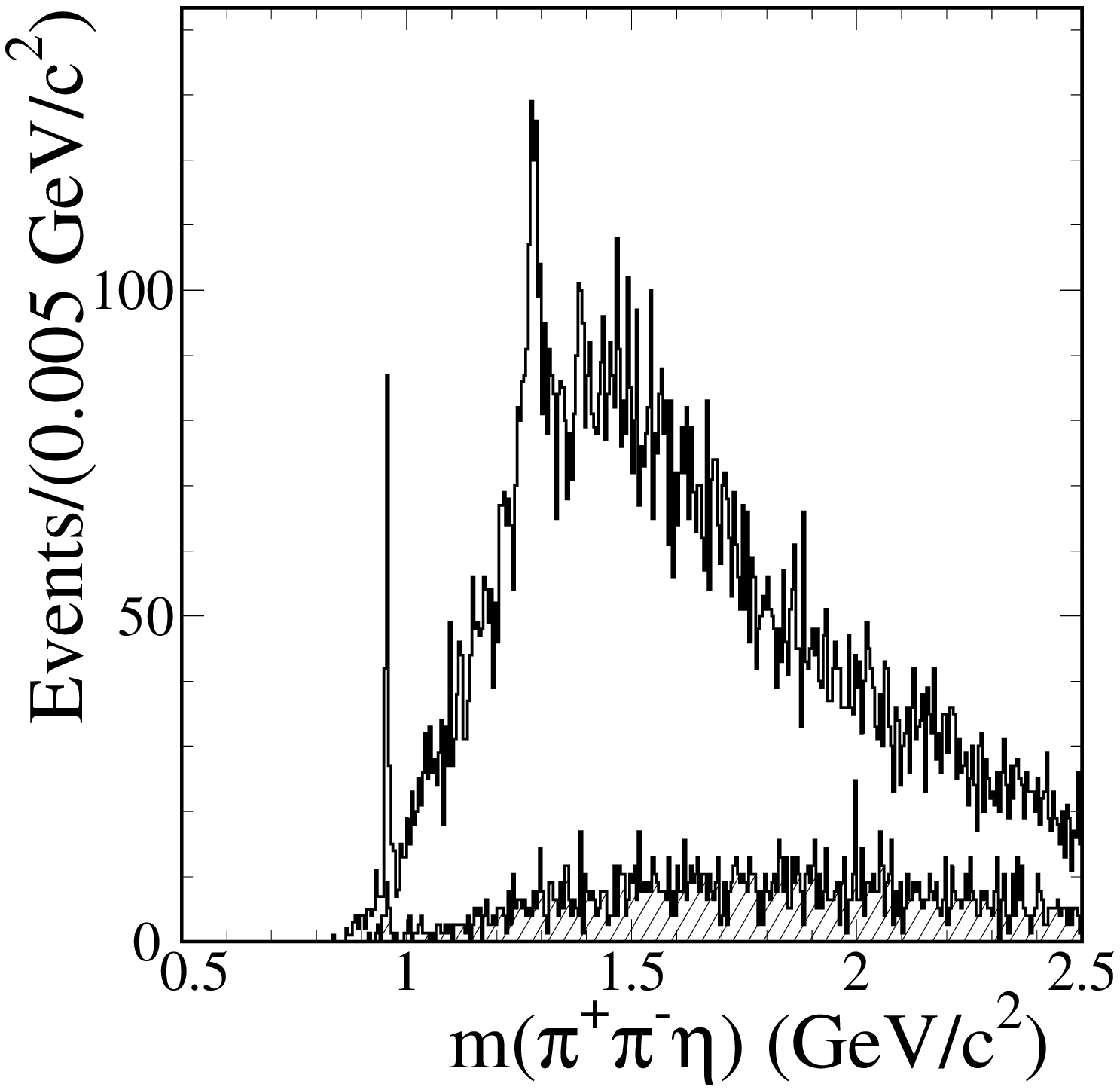}
\put(-25,100){(a)}
\includegraphics[width=4.1cm]{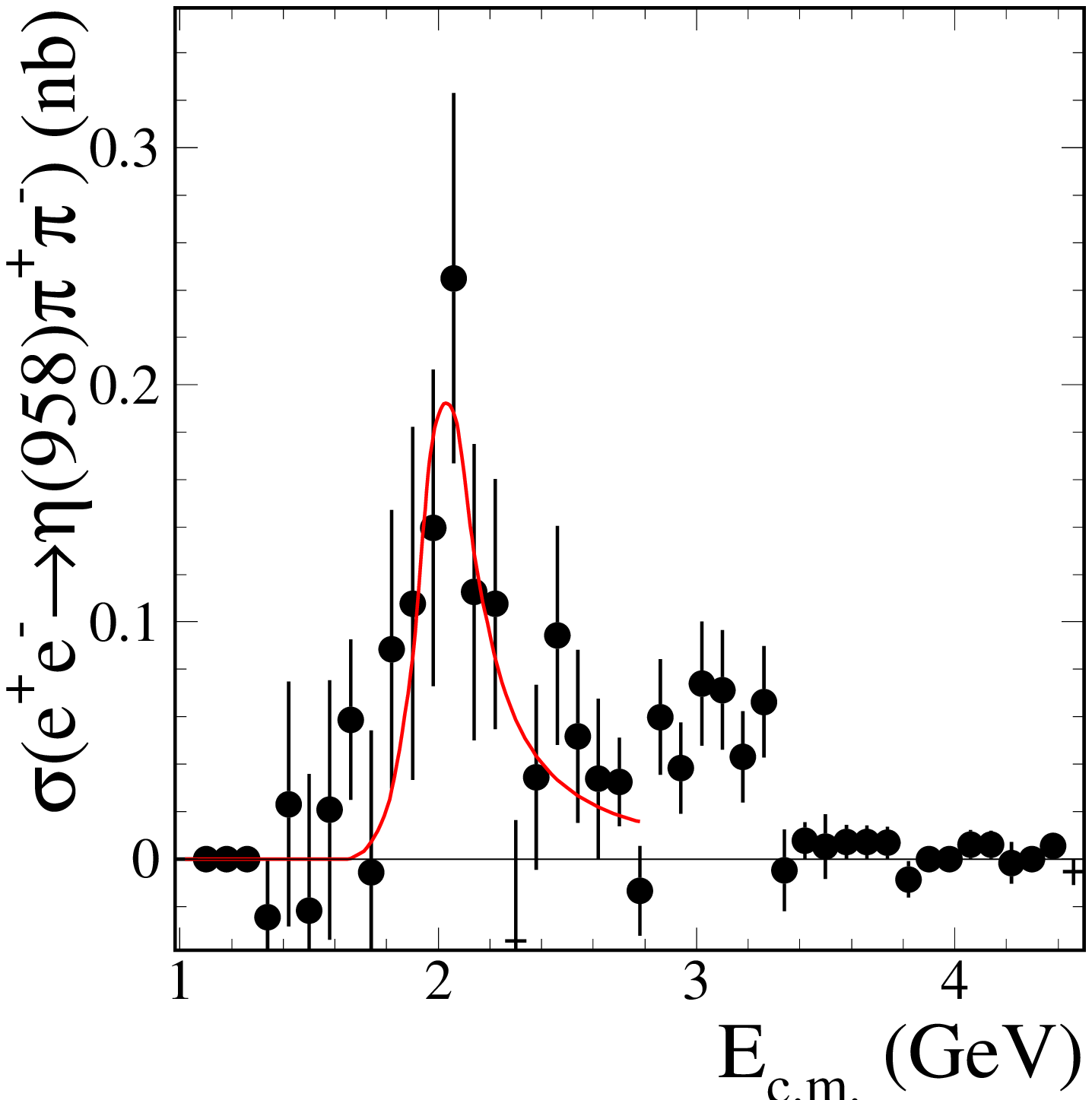}
\put(-25,100){(b)}
\caption{\label{fig10} 
(a) The $m(\eta\pi^+\pi^-)$ distribution for $ 2(\pi^+\pi^-)\eta$ events;
(b) The $e^+ e^-\to\eta(958)\pi^+\pi^- $ cross section and 
the result of the Breit-Wigner fit~\cite{babr5pi}.}
\end{figure}
\begin{figure}[h]
\includegraphics[width=4.1cm]{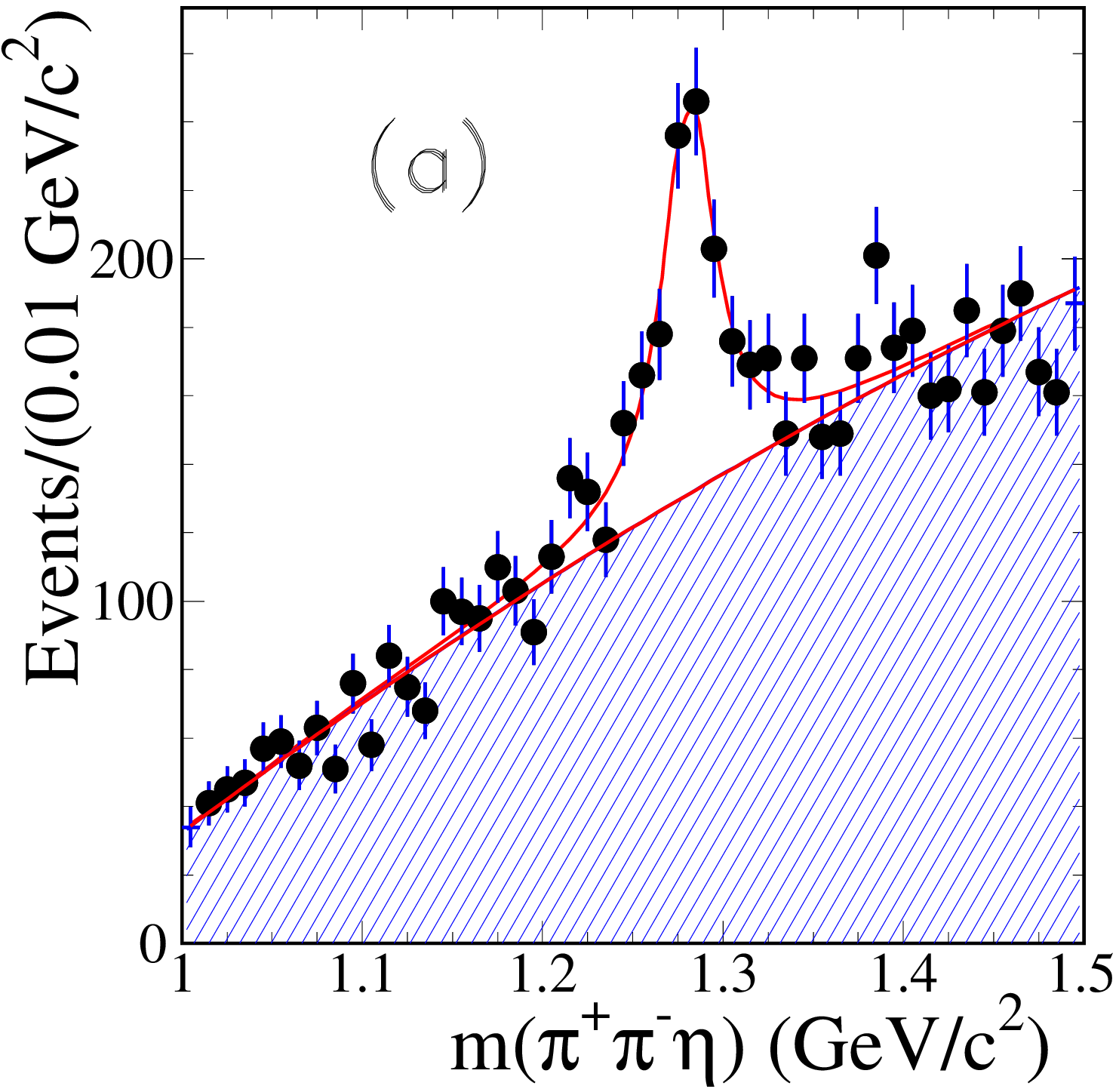}
\includegraphics[width=4.1cm]{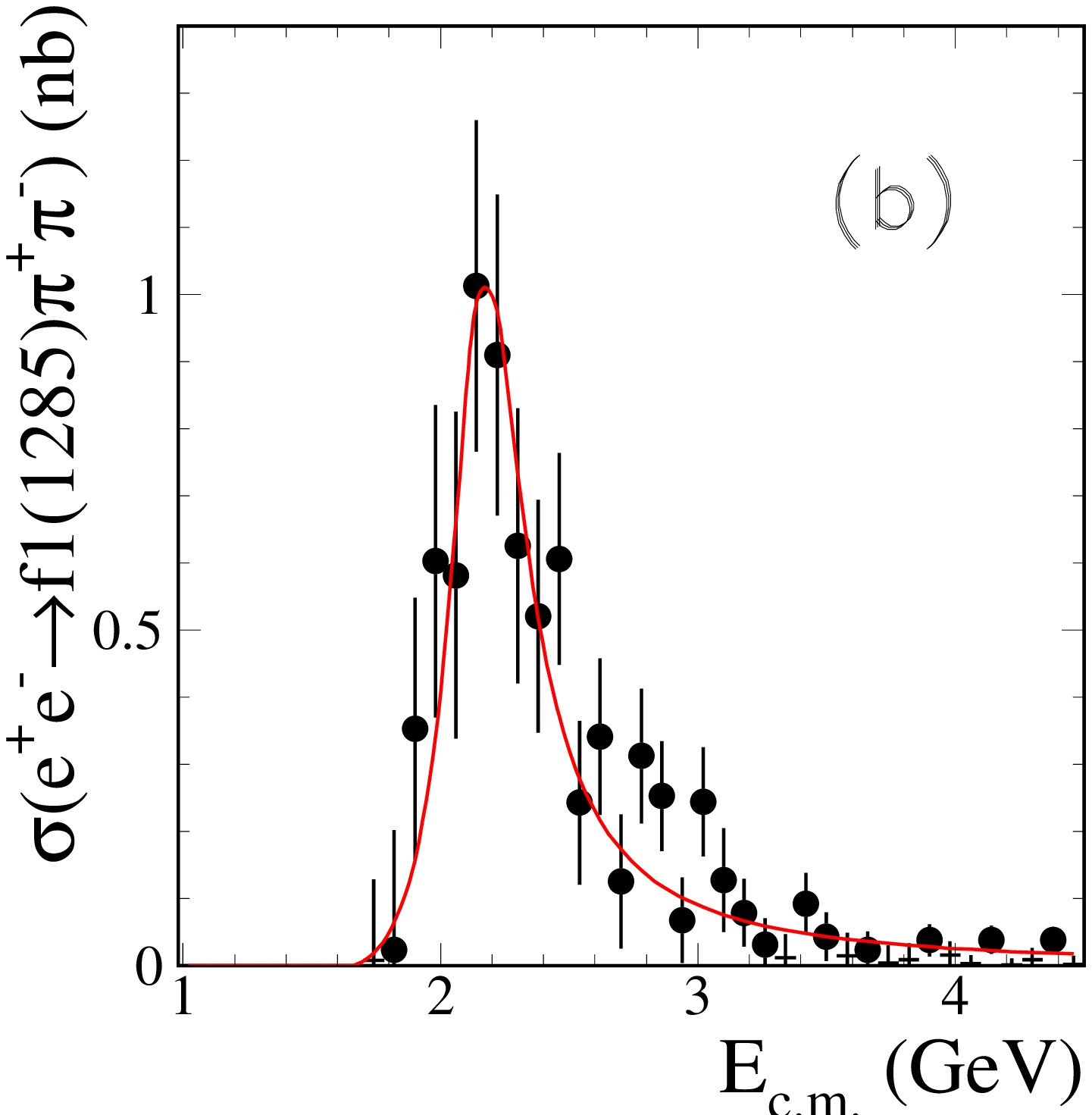}
\caption{\label{fig11} 
(a) The $m(\eta\pi^+\pi^-)$ distribution  for $ 2(\pi^+\pi^-)\eta$ events;
(b) The $e^+ e^-\to f_1(1285)\pi^+\pi^- $ cross section and the
result of the Breit-Wigner fit~\cite{babr5pi}.}
\end{figure}
The two pions from the reaction $e^+e^-\to\eta\pi^+\pi^-$ predominantly form
the $\rho(770)$. In the two-pion invariant mass spectrum for the 
$e^+e^-\to \omega\pi^+\pi^-$ reaction shown in Fig.~\ref{fig88a} 
a clear $f_0(980)$ signal is observed.
The contribution of the $\omega f_0(980)$ intermediate state was extracted,
and the cross section $e^+ e^-\to \omega f_0(980)$ was measured 
for the first time.
It is shown in Fig.~\ref{fig88b}. The $e^+ e^-\to \omega\pi^+\pi^-$ 
cross section 
after subtraction of the $\omega f_0(980)$ contribution is shown in
Fig.~\ref{fig88c}. The cross section is fitted with a sum of of two
resonances. The fit result is shown in Fig.~\ref{fig88c} and listed in
Table~\ref{omfit_tab}. The obtained parameters are in good agreement with
those obtained for the $\pi^+\pi^-\pi^0$ channel  
(see discussion in Sec.~\ref{Sec:hadsum}).

The $\omega\pi^+\pi^-$ and $\eta\pi^+\pi^-$ intermediate states
do not saturate the total 
$e^+ e^-\to 2(\pi^+\pi^-)\pi^0$ cross section as shown in  Fig.~\ref{fig99a}.
The $m(\pi^+\pi^-)$ and $m(\pi^{\pm}\pi^0)$ distributions for 
$e^+ e^-\to 2(\pi^+\pi^-)\pi^0$ events with the $\omega\pi^+\pi^-$ 
and $\eta\pi^+\pi^-$ contributions excluded are shown in 
Fig.~\ref{fig99b}. From the analysis of these two-pion mass
distributions it was concluded
that the dominant intermediate state for these events is
$\rho^0\rho^\pm\pi^\mp$. The $\rho\pi$ mass spectrum also
exhibits a resonance structure with the parameters:
\begin{eqnarray*}
      m_X & = & 1.243 \pm 0.012 \pm 0.020~{\rm\gev}/c^2; \\
 \Gamma_X & = & 0.410 \pm 0.031 \pm 0.030~{\rm\gev} . 
\end{eqnarray*}
The yield of the $X(1240)$ state is consistent with the complete dominance
of the quasi-two-body reaction 
$e^+ e^-\to \rho(770) X(1240) \to \rho^0\rho^\pm\pi^\mp$. 
The best candidates for $X(1240)$ are the $\pi(1300)$ or $a_1(1260)$ 
resonances~\cite{PDG08}.

The $e^+ e^-\to 2(\pi^+\pi^-)\eta$ reaction was studied 
for the first time by BABAR. The measured cross section is 
shown in Fig.~\ref{4pieta_ee_babar}. A rich internal structure is 
expected for the $4\pi\eta$ final state.
The four-pion mass distribution exhibits a wide resonance structure 
which can be a mixture of the known $\rho(1450)$ and $\rho(1700)$ resonances.
Figure~\ref{fig10}(a) shows the $\eta\pi^+\pi^-$ mass
distribution with two narrow peaks. The lowest mass peak corresponds
to the $\eta^{\prime}(958)$. The measured 
$e^+ e^-\to\eta^{\prime}(958)\pi^+\pi^-$ cross section is
shown in Fig.~\ref{fig10}(b). The resonance-like structure observed in
the cross section energy dependence is fitted with a single Breit-Wigner 
function. The fitted resonance parameters are 
\begin{eqnarray*}
  \sigma_0 & = & 0.18 \pm 0.07~{\rm nb}, \\
     m_x   & = & 1.99 \pm 0.08~{\rm\gev}/c^2,   \\
  \Gamma_x & = & 0.31 \pm 0.14~{\rm\gev.}
\end{eqnarray*}
There is no entry for these parameters in the current PDG 
tables~\cite{PDG08}.
Taking into account possible large systematic uncertainties on mass
and width, the observed resonance can be interpreted as the
$\rho(2150)$, extensively 
discussed in the past~\cite{PDG08}.

Another clear structure seen in the $\eta\pi^+\pi^-$ mass
distribution (Fig.~\ref{fig10}(a)) and shown in
detail  in Fig.~\ref{fig11}(a) was interpreted as 
the $f_1(1285)$ meson. The measured  $e^+ e^-\to f_1(1285)\pi^+\pi^-$ cross
section is shown in Fig.~\ref{fig11}(b). The observed resonance structure
has parameters: 
\begin{eqnarray*}
  \sigma_0 & = & 1.00 \pm 0.18 \pm 0.15~{\rm nb}, \\
     m_x   & = & 2.15 \pm 0.04 \pm 0.05~{\rm\gev}/c^2,    \\
  \Gamma_x & = & 0.35 \pm 0.04 \pm 0.05~{\rm\gev}.
\end{eqnarray*}
The mass and width are close to those measured in the reaction 
$e^+ e^-\to\eta^{\prime}(958)\pi^+\pi^-$, but the cross section is 
significantly larger. The observed structure can also  be assigned to 
the $\rho(2150)$ resonance.  
\begin{figure}[h]
\includegraphics[width=.4\textwidth]{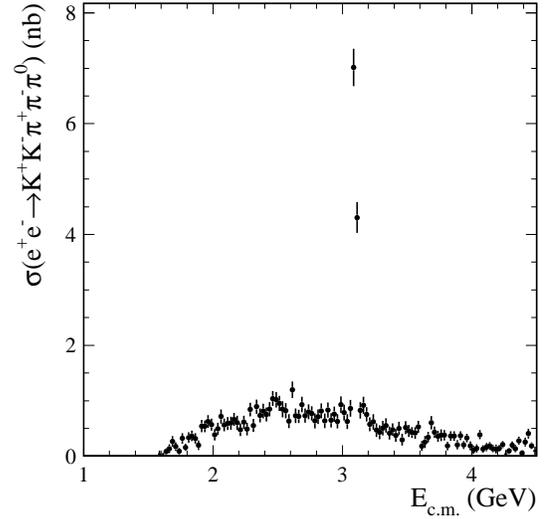}
\caption{
The $e^+ e^- \to K^+ K^- \pi^+\pi^-\pi^0$ cross section
 measured by BABAR~\cite{babr5pi}.
\label{2K3pi_ee_babar}}
\end{figure}
\begin{figure}[h]
\includegraphics[width=4.1cm]{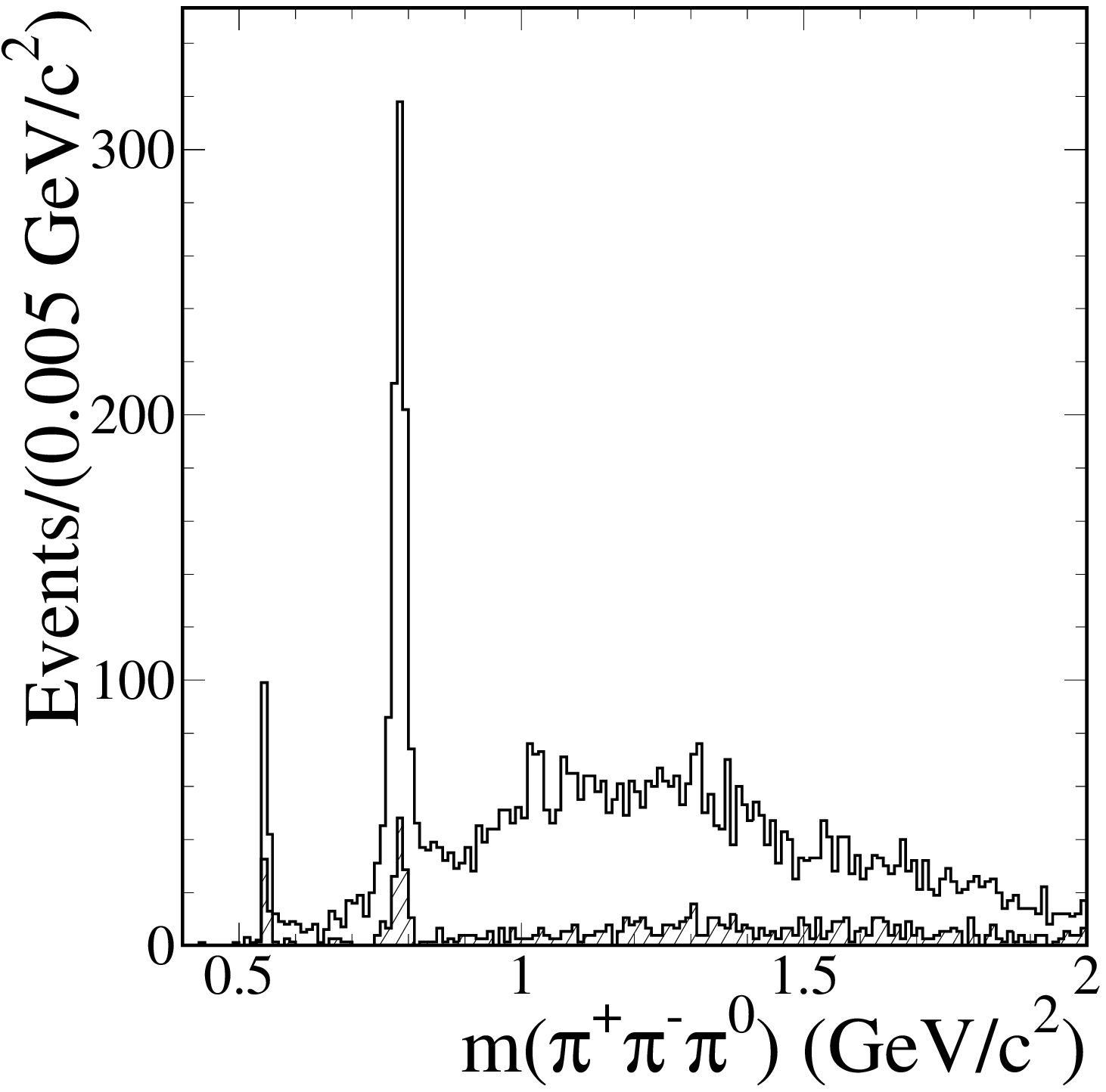}
\put(-25,100){(a)}
\includegraphics[width=4.1cm]{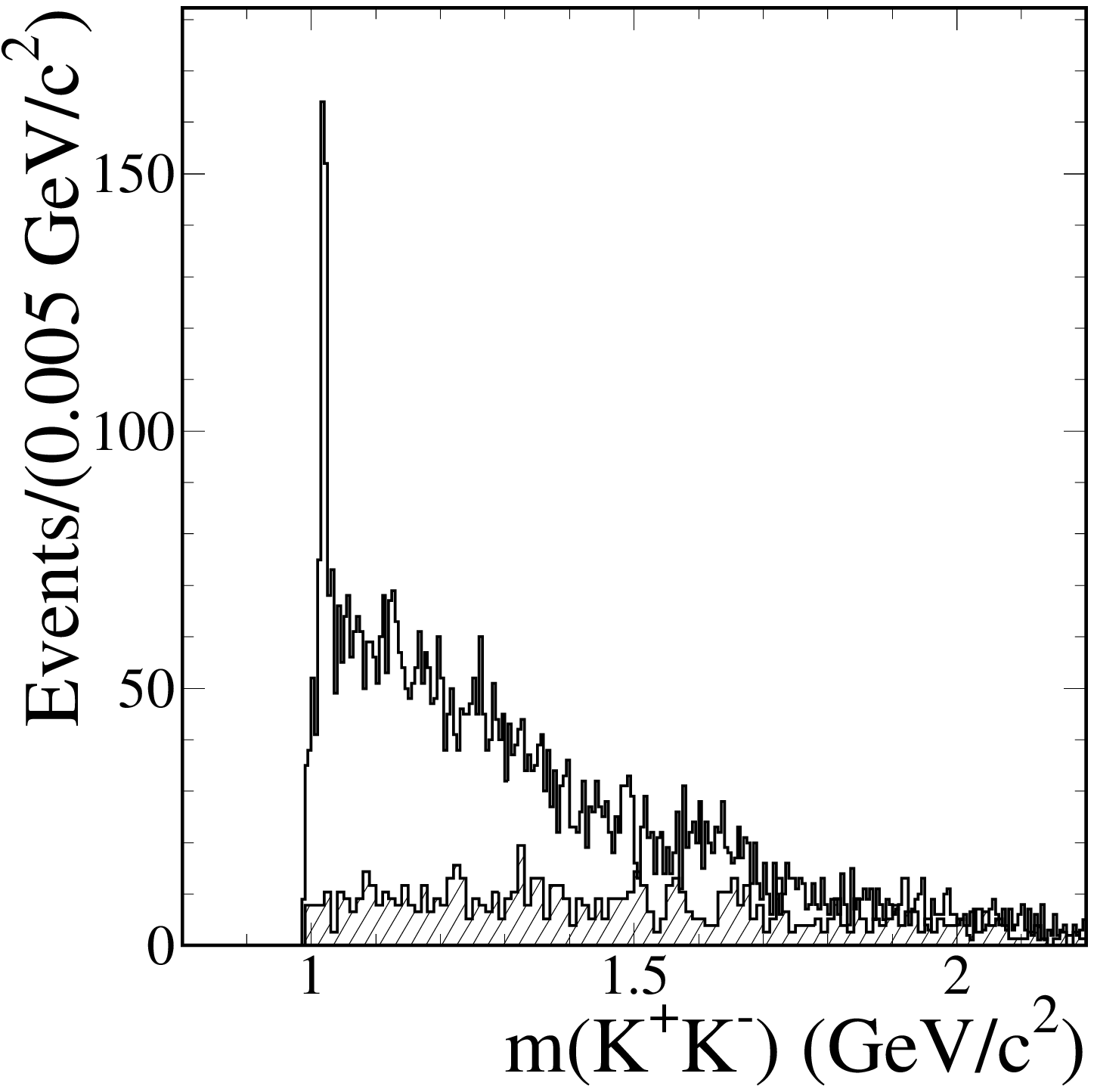}
\put(-25,100){(b)}
\caption{\label{fig121} 
The $m(\pi^+\pi^-\pi^0)$ (a)  and $m(K^+ K^-)$ (b)  distributions
for $K^+ K^- \pi^+\pi^-\pi^0$ events~\cite{babr5pi}. The hatched histogram
represents the estimated non-ISR background.}
\end{figure}
\begin{figure}
\includegraphics[width=4.1cm]{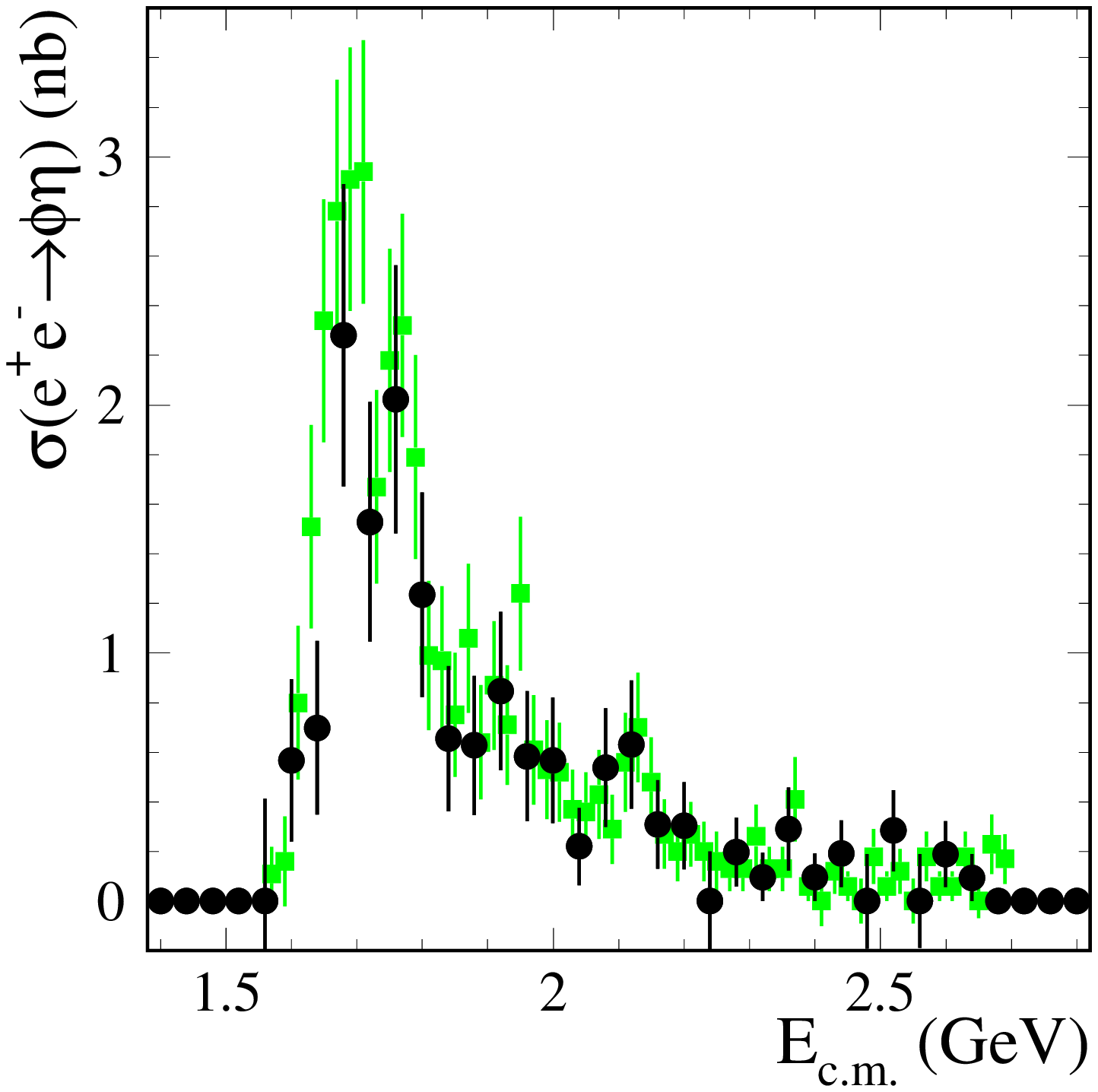}
\put(-25,100){(a)}
\includegraphics[width=4.1cm]{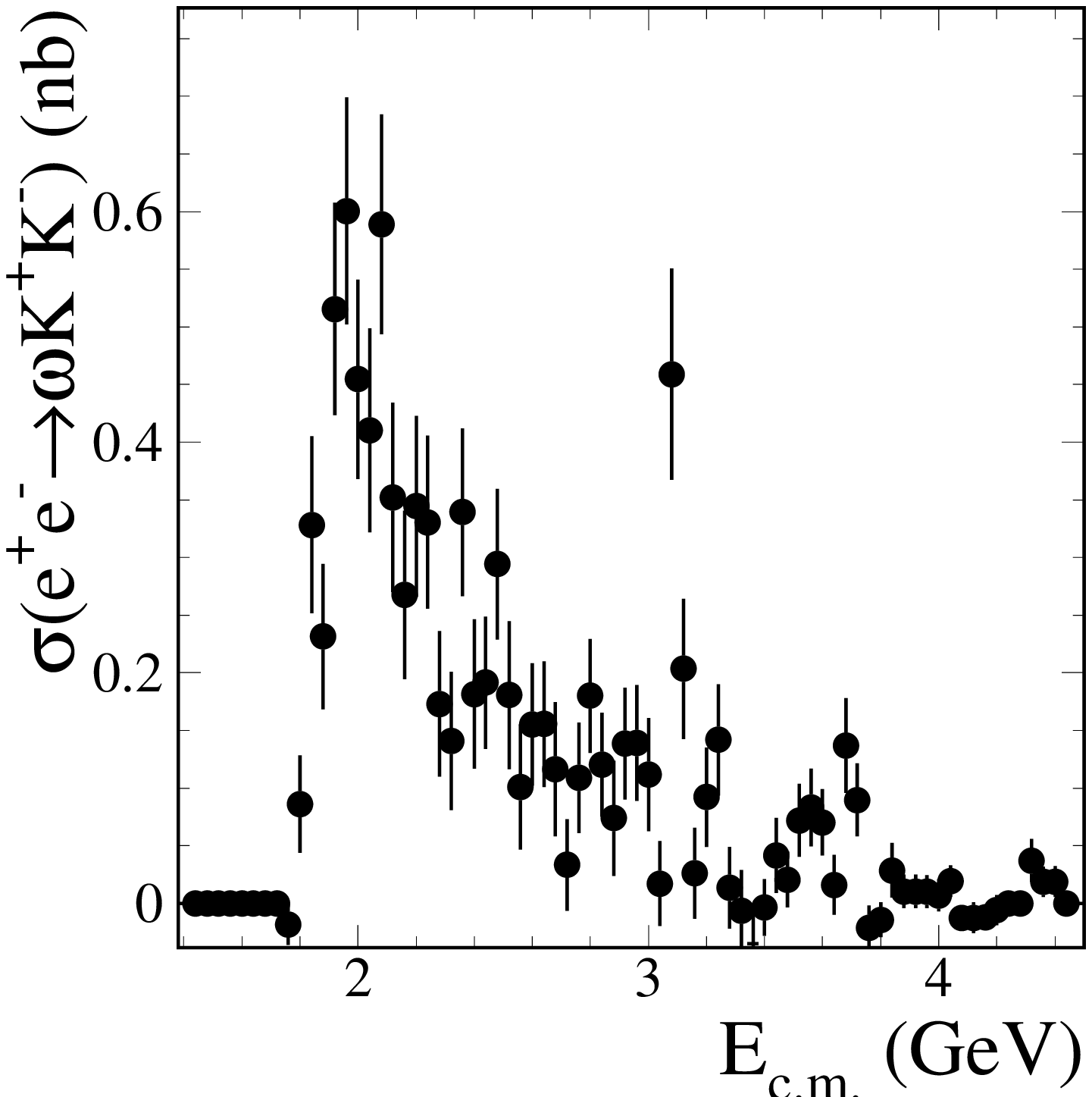}
\put(-25,100){(b)}
\caption{\label{fig13} 
(a) The $e^+ e^-\to\phi\eta$ cross section measured by BABAR in 
the $K^+ K^-\pi^+\pi^-\pi^0$~\cite{babr5pi} (circles) and $K^+ K^-\gamma\gamma$ (squares)~\cite{babrkkpi}
final states. (b) The cross sections for the $e^+ e^-\to\omega K^+ K^-$ process
measured by BABAR~\cite{babr5pi}.}
\end{figure}

The cross section for the reaction $e^+ e^-\to K^+ K^-\pi^+\pi^-\pi^0$ 
shown in Fig.~\ref{2K3pi_ee_babar} was also measured by BABAR 
for the first time. The three-pion and
two-kaon  invariant mass spectra for this reaction  
are shown in Fig.~\ref{fig121}. Clear $\eta$ and
$\omega$ signals are seen in the three-pion mass distribution
and a strong $\phi$ signal in the $K^+ K^-$ mass distribution.
Figure~\ref{fig13} shows the calculated cross sections for the 
$e^+ e^-\to\phi\eta$ (a) and $e^+ e^-\to\omega K^+ K^- $ (b)
subprocesses. The former is in good agreement with that obtained in the
$\eta\to\gamma\gamma$ mode~\cite{babrkkpi}. 
It is a first observation of the process $e^+ e^-\to\omega K^+ K^- $. 

The reaction $e^+ e^-\to K^+ K^-\pi^+\pi^-\eta$ was also studied by BABAR
in Ref.~\cite{babr5pi}. The measured cross section is small and 
rises from threshold to a maximum value of about 0.2 nb at 2.8 GeV, followed 
by a monotonic decrease with increasing energy. The clear signal of the
$\phi\eta^\prime(958)$ intermediate state is observed in the $K^+ K^-$ and
$\pi^+\pi^-\eta$ mass distributions. Unfortunately, the $\phi\eta^\prime(958)$
invariant mass spectrum is not shown in Ref.~\cite{babr5pi}. This spectrum 
is interesting since for the four-quark $Y(2175)$ resonance 
(Sec.~\ref{Sec:2k2pi}) the decay to $\phi\eta^\prime(958)$ is expected.

\begin{figure*}
\includegraphics[width=.4\textwidth]{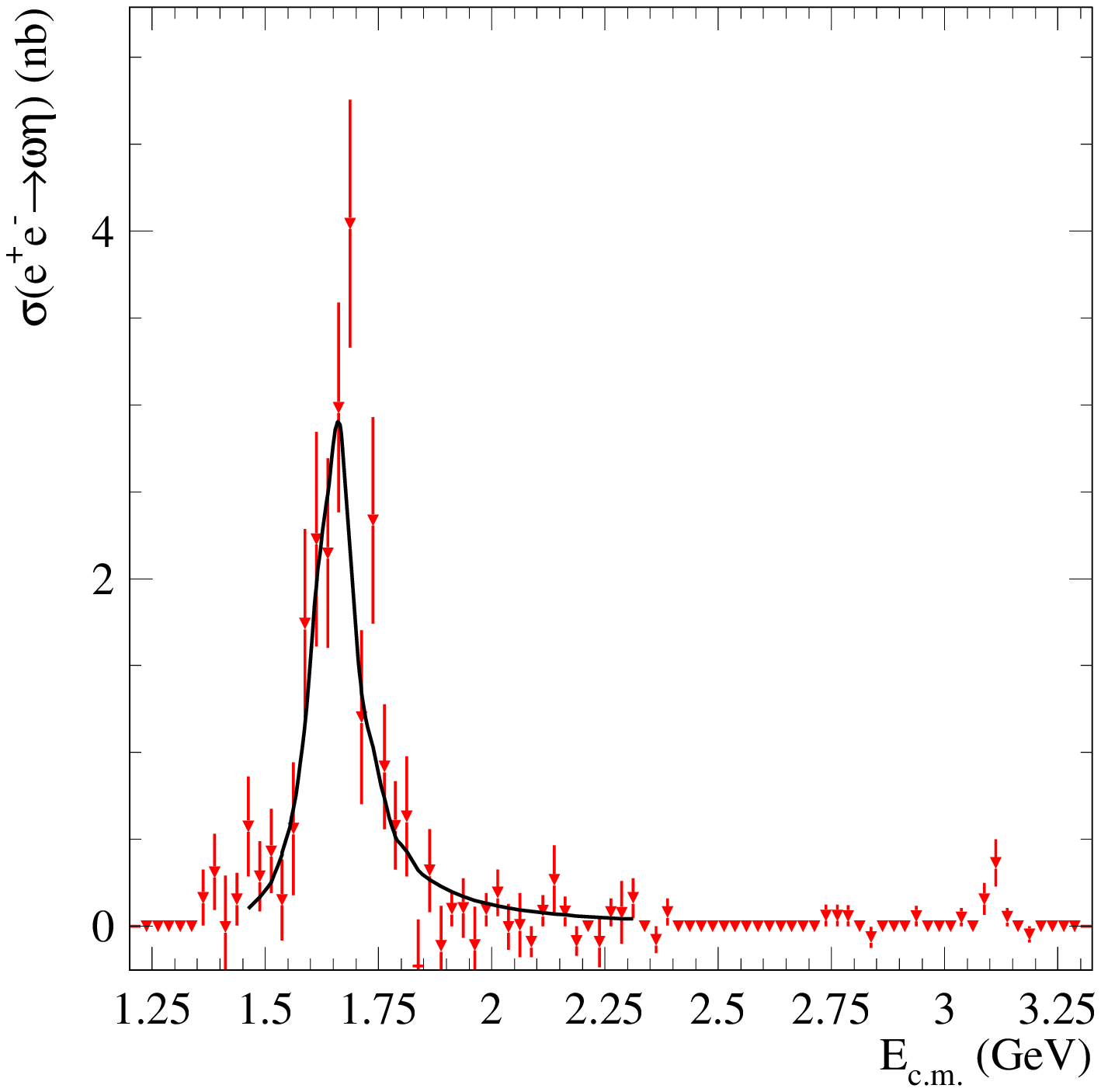}
\includegraphics[width=.4\textwidth]{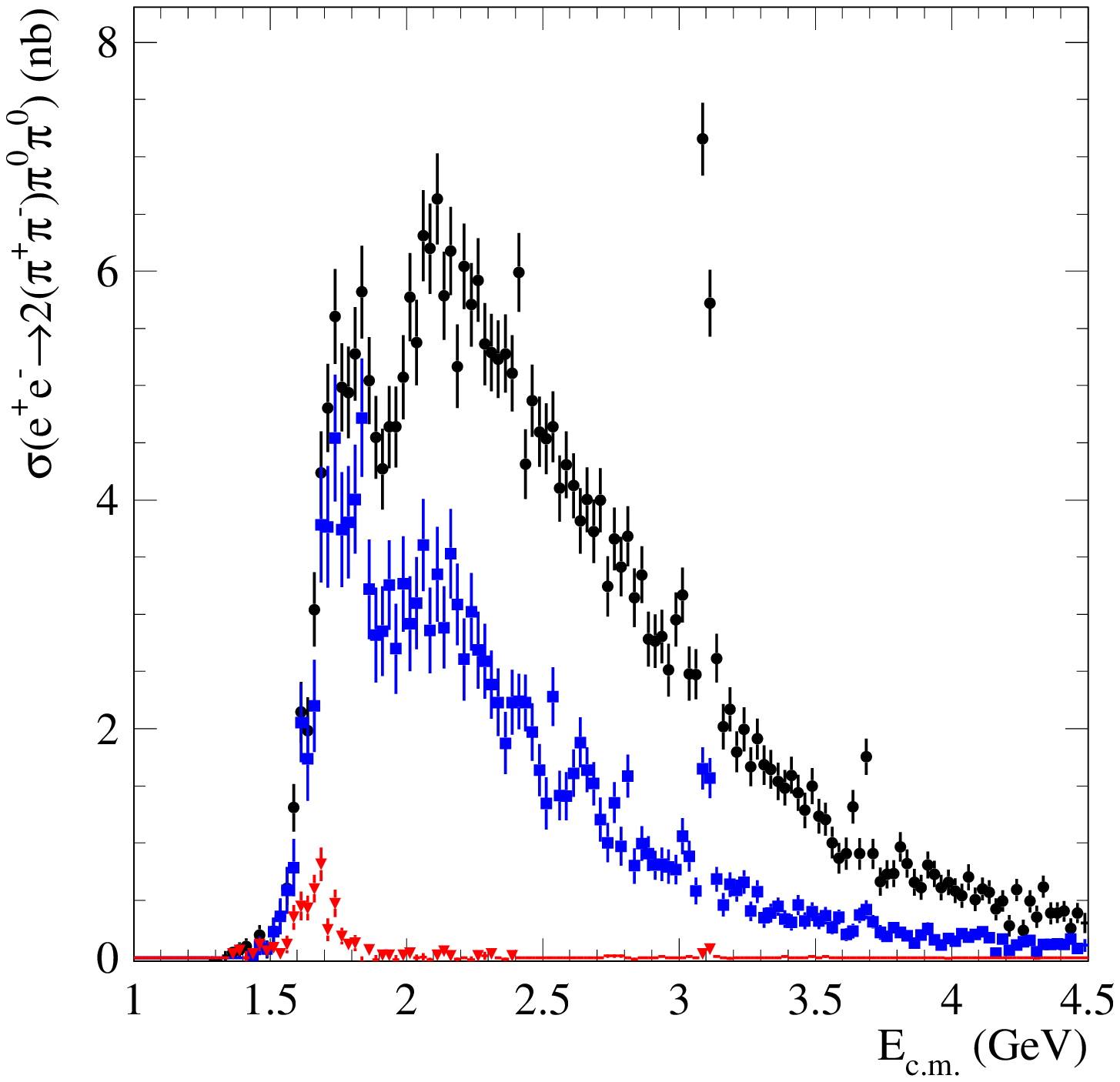}
\caption{\label{fig14} 
The $e^+ e^-\to\omega\eta$ cross section (left) and contribution of  $e^+
e^-\to\omega\pi^+\pi^-\pi^0$ (squares) and  $e^+ e^-\to\omega\eta$
(triangles) cross  sections
to all $e^+ e^-\to 2(\pi^+\pi^-\pi^0)$ events (right)~\cite{babr6pi}.}
\end{figure*}
\begin{figure*}
\includegraphics[width=8.2cm]{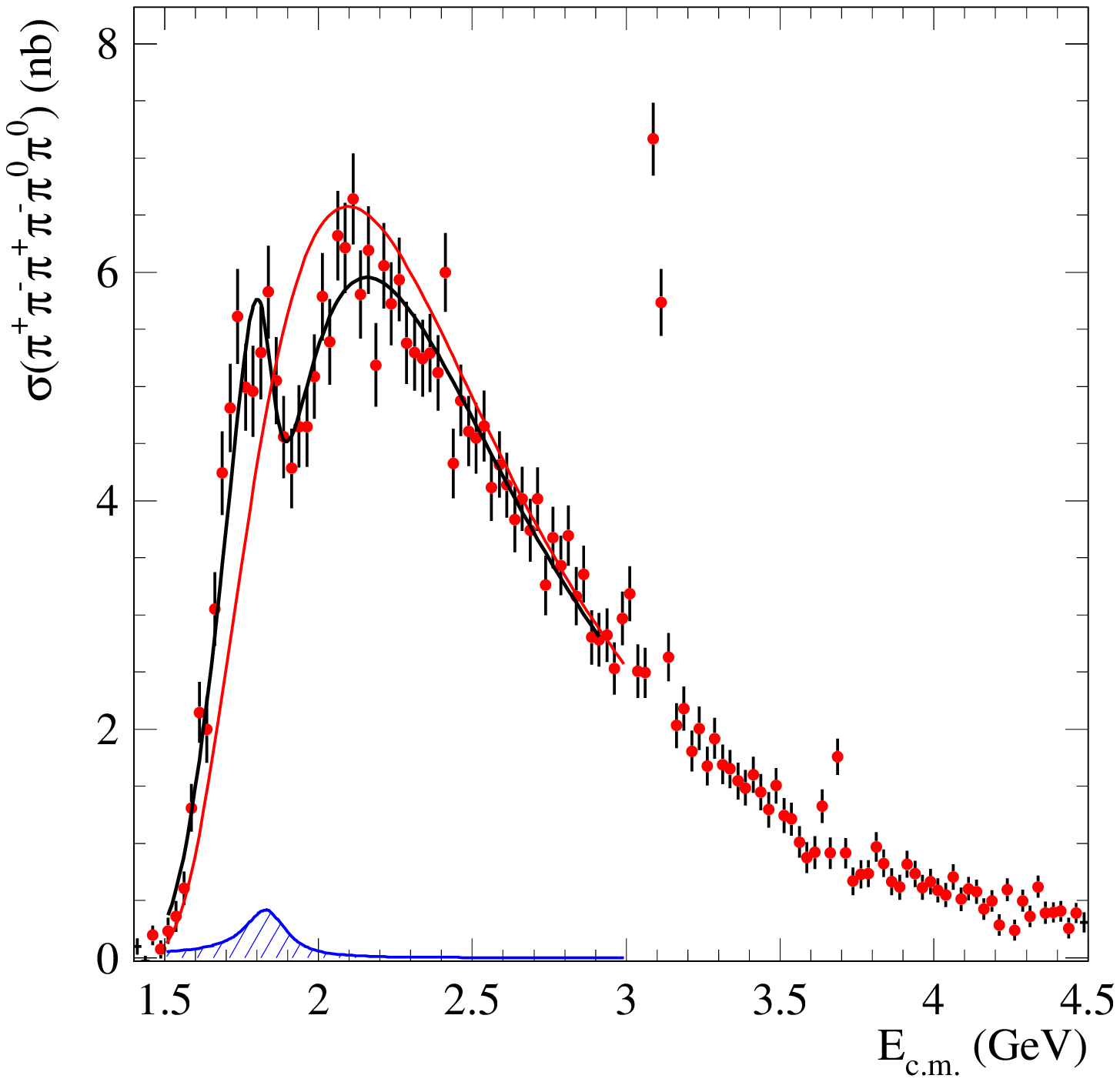}
\includegraphics[width=8.2cm]{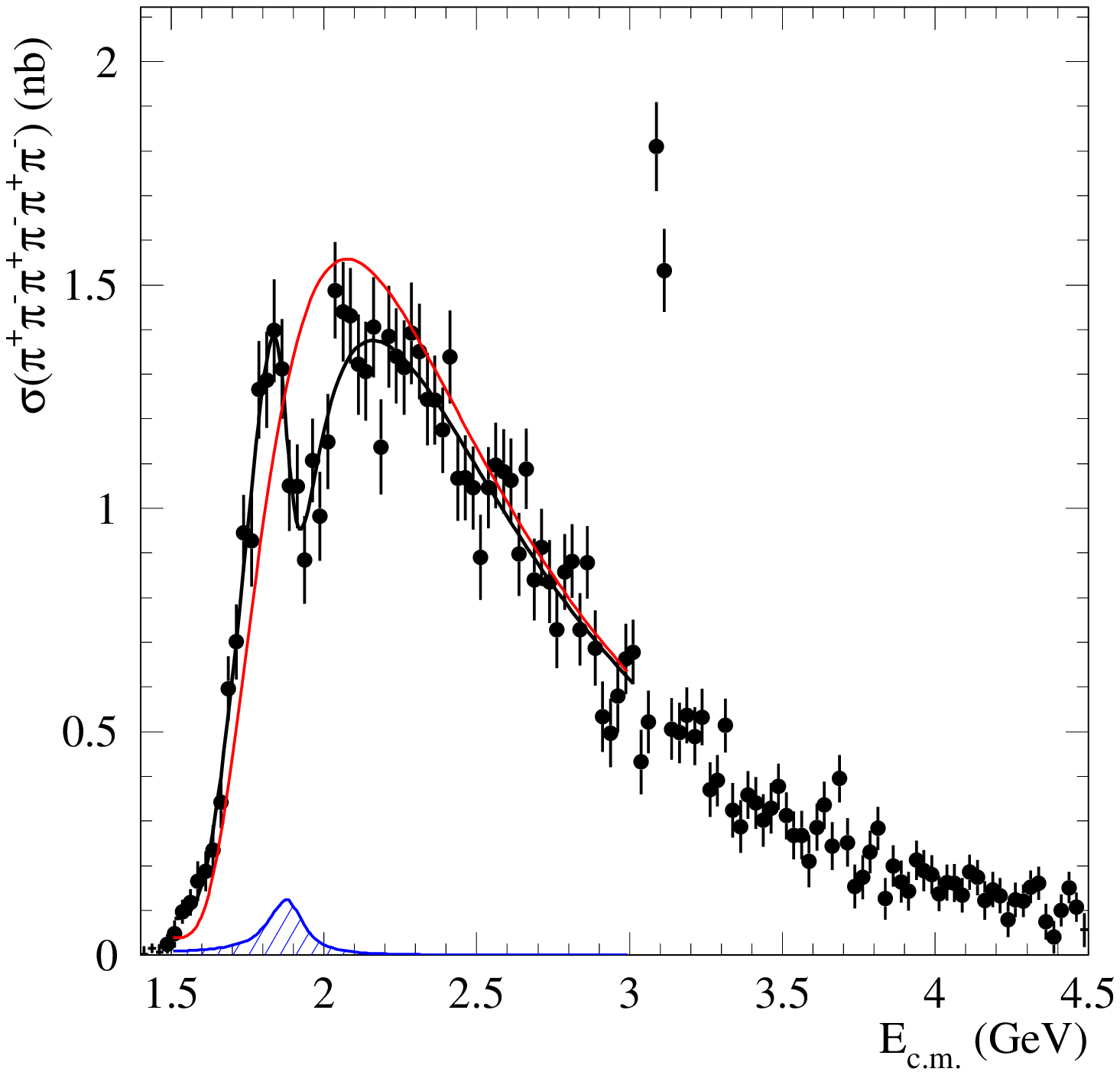}
\caption{\label{fig115} 
The $e^+ e^-\to 2(\pi^+\pi^-\pi^0) $ cross section (left) and 
the  $e^+ e^-\to 3(\pi^+\pi^-) $ cross section (right)
with a fit to the \\ (anti-)resonance function~\cite{babr6pi}.
A resonance responsible for the ``dip'' is shown shaded.}
\end{figure*}

\begin{table*}
\caption{
  Summary of the $\omega(1420)$($\omega'$) and
  $\omega(1650)$($\omega"$)  
resonance parameters obtained from the fits described in the text.
  The values without errors were fixed in the fits.
  }
\label{omfit_tab}
\begin{ruledtabular}
\begin{tabular}{l c c c c } 
Fit &  $\omega\eta$~\cite{babr6pi} & $\omega\pipi$~\cite{babr5pi} &  
$3\pi$~\cite{babr3pi} & PDG~\cite{PDG08}\\
\hline
$\sigma_{0w '}$ (\nb)     &   --    &  1.01$\pm$0.29
                                     &  --                     &   --   \\
$B_{ee}B_{w'f}\cdot 10^6$    &  --     &  0.13$\pm$0.04
                                          &  0.82$\pm$0.08    &   --   \\
$\Gamma_{ee}B_{w'f}$(\ev)   &  --       &  17.5$\pm$5.4
                                          &  369                   &   --   \\
$m_{w'}$(\gevcc)          &   --   &  1.38$\pm$0.02$\pm$0.07   
                                    &  1.350$\pm$0.030       &  1.40--1.45 \\
$\Gamma_{w'}$(\gev)    &  --         & 0.13$\pm$0.05$\pm$0.01   
                                  &   0.450$\pm$0.140      &  0.180--0.250 \\
$\phi_{w'}$ (rad.)          &  --         &     $\pi$   
                                  &      $\pi$      &   -- \\
$\sigma_{0w ''}$ (\nb)     &   3.08$\pm$0.33          & 2.47$\pm$0.18  
                                 &  --                    &   --   \\
$B_{ee}B_{w''f}\cdot 10^6$    &  --    &  0.47$\pm$0.04
                                 &  1.3$\pm$0.2                    &   --   \\
$\Gamma_{ee}B_{w''f}$(\ev)   & --   &  103.5$\pm$8.3
                                 &  286                   &   --   \\
$m_{w''}$(\gevcc)          &  1.645$\pm$0.008        &  1.667$\pm$0.013$\pm$0.006   
                                  &  1.660$\pm$0.011       & 1.670$\pm$0.030 \\
$\Gamma_{w''}$(\gev)    &  0.114$\pm$0.014         &  0.222$\pm$0.025$\pm$0.020   
                                  &  0.220$\pm$0.040       &  0.315$\pm$0.035 \\
$\phi_{w''}$ (rad.)          &   0      &       0
                                  &      0     &   -- \\
$\sigma_{0w }$ (\nb)     & 0                     &  102$\pm$67
                                 &   PDG                    &   --   \\
\chisq /n.d.f.           &   --     &     34.9/48
                         &    --      &     --     \\
\end{tabular}
\end{ruledtabular}
\end{table*}                           
\subsection{$e^+ e^-\to 3(\pi^+\pi^-),\, 2(\pi^+\pi^-\pi^0)$}
The reactions $e^+ e^-\to 3(\pi^+\pi^-)$ and $e^+ e^-\to 2(\pi^+\pi^-\pi^0)$ 
were studied before in a number of direct $e^+ e^-$ experiments,
but with limited data
samples~\cite{DM26pi,dm16pi,m3n4pi,gg24pi0,mea}. 
The BABAR detector 
studied the six-pion production using the ISR method from the threshold to  
4.5 GeV~\cite{babr6pi}. As a result,
the statistical and systematic uncertainties on the cross sections
were dramatically reduced. 

An interesting feature of the $3(\pi^+\pi^-)$ final state 
is the presence, among many $\pi^+\pi^-$ combinations, of only 
one $\rho(770)^0$ per event. No other intermediate resonance signals were
observed. For the $2(\pi^+\pi^-\pi^0)$ final state also 
one $\rho(770)$ only, neutral or charged, per event is observed
in the expected proportion $1:2$.

In the $2(\pi^+\pi^-\pi^0)$ final state, $\eta$ and
$\omega$ signals are seen in the $\pi^+\pi^-\pi^0$ invariant mass 
distribution. A small fraction of events corresponds to the associated  
production of the $\eta$ and $\omega$. Selecting these $\eta\omega$ events
the $e^+e^-\to\omega\eta$ cross section shown in Fig.~\ref{fig14}~(left) was 
measured for the first time. The observed resonance structure 
which is expected to be the $\omega(1650)$ is fitted
with a Breit-Wigner function. The fitted curve is shown in 
Fig.~\ref{fig14}~(left).
The obtained resonance parameters are listed in Table~\ref{omfit_tab} together
with the resonance parameters obtained from the fits to $e^+e^-\to 3\pi$ and
$e^+e^-\to \omega\pi\pi$ cross sections 
(see discussion in Sec.~\ref{Sec:hadsum}).
Comparison of the  $\omega3\pi$ and $\omega(782)\eta$ contributions
with the total $e^+ e^-\to 2(\pi^+\pi^-\pi^0)$ cross section
is shown in Fig.~\ref{fig14}(right). 

The total $e^+ e^-\to 2(\pi^+\pi^-\pi^0)$ and  $e^+ e^-\to 3(\pi^+\pi^-)$
cross sections shown in Fig.~\ref{fig115} have very 
similar energy dependence. 
The ratio of the cross sections is almost constant over the energy range
under study. Its average value is equal to $3.98\pm0.06\pm0.41$. 
A dip structure just below 2 GeV in the six-pion cross section
was observed in the DM2 experiment~\cite{DM26pi} and then 
confirmed in the diffractive photoproduction of six pions in 
the FOCUS experiment~\cite{FOCUS6pi}. Such a dip at 1.9~GeV 
was also observed in
the total cross section of \epem\ annihilation into hadrons 
by the FENICE detector~\cite{fenice}.
This structure in BABAR data is fitted using the Breit-Wigner function 
coherent with the smooth non-resonant background. The fitted curves for
both cross sections are shown in Fig.~\ref{fig115}. 
The following ``resonance'' parameters are obtained:
\begin{eqnarray*}
m_{6\pi}&=&1.88\pm0.03~{\rm\gev}/c^2,\\
m_{4\pi 2\pi^0}&=&1.86\pm0.02~{\rm\gev}/c^2, \\
\Gamma_{6\pi}& = &0.13 \pm 0.03~{\rm\gev},\\
\Gamma_{4\pi 2\pi^0}& = &0.16 \pm 0.02~{\rm\gev}.
\end{eqnarray*}
The parameter values seem to be essentially independent of
the final-state charge combination. These values may be also compared
with those obtained in the FOCUS experiment~\cite{FOCUS6pi}:
$m = 1.91 \pm 0.01$ \gev/$c^2$, $\Gamma = 0.037 \pm 0.013$ \gev. 
The mass values  are consistent, but the widths
obtained by BABAR are substantially larger.
Note that typical widths of known isovector resonances with 
mass near 2 GeV/$c^2$ are 200--300 MeV. 
Since the obtained mass of the resonance structure is close to 
the double proton mass, it may be interpreted as a proton-antiproton 
subthreshold bound state~\cite{dip6pi}.

\subsection{Summary}
\label{Sec:hadsum}
\begin{figure*}
\includegraphics[width=0.95\textwidth]{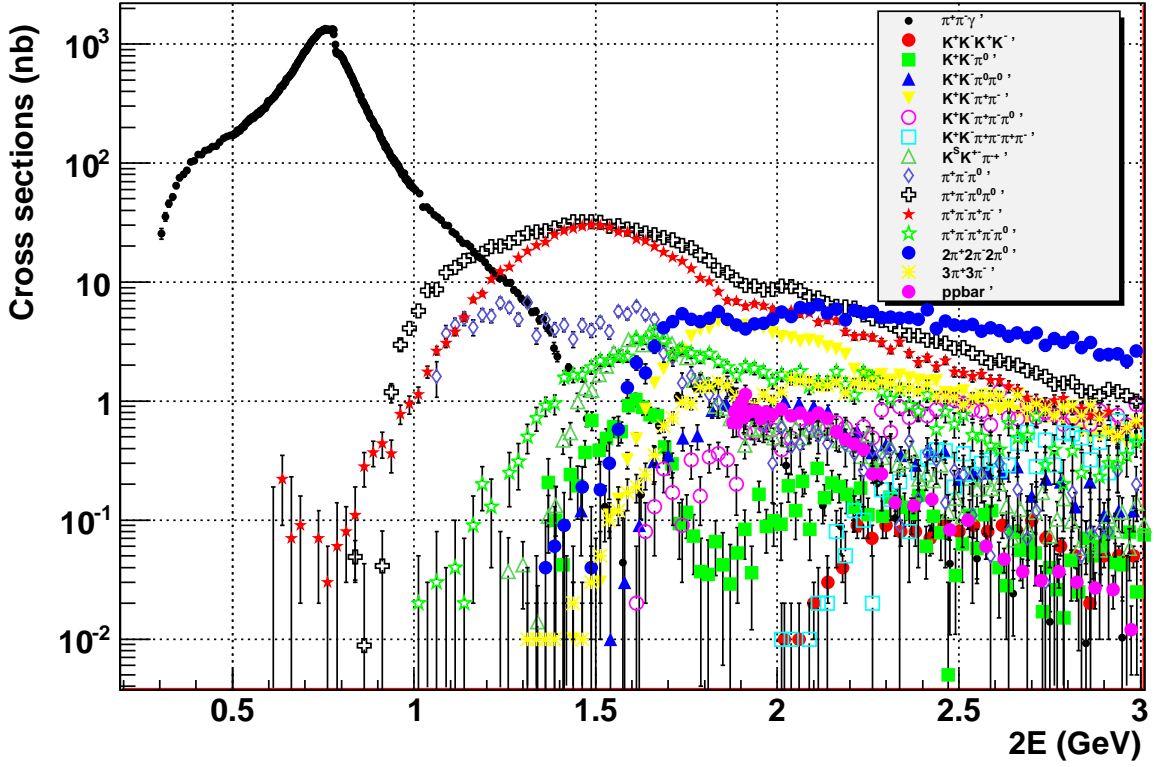}   
\caption{
The cross sections of $e^+ e^-\to$ hadrons measured with the 
BABAR detector via ISR. The results 
on $e^+e^- \to \pi^+\pi^-\pi^0\pi^0$ are preliminary. 
\label{isr_all}}
\end{figure*}

The BABAR ISR study covers the low energy range of 
$e^+ e^-$ interactions from the $2\pi$ threshold to 4.0--4.5 GeV with
exclusively measured cross sections for many processes. 
Figure~\ref{isr_all} shows all exclusive cross sections measured 
by BABAR in a single plot. One can see that in most of the cases 
cross sections strongly depend on energy and  their central
values vary by five orders of magnitude.
  
One of the purposes of the BABAR ISR program was to measure the 
total hadronic cross section in the energy range below 2 GeV
with improved accuracy~\cite{dru07}.  
To finalize this program, the cross sections at least for the 
$\pi^+\pi^-3\pi^0$, $\pi^+\pi^-4\pi^0$, $K^+K^-$, $K_S K_L$,
$K_S K_L \pi\pi$, $K_S K^\pm \pi^\mp\pi^0$ final states should
be additionally measured.

Note that the total cross section value is not the direct sum 
of the cross sections shown in Fig.~\ref{isr_all}. Each channel
has internal subprocesses which include
different resonances with different branching fractions to the
observed final states. To perform a correct summation,
each subchannel should be extracted separately and
corrected for the decay rate of an internal resonance.

The exclusive ISR study of hadron production allows to 
investigate and improve our knowledge of excited states
for light vector mesons. For most of them
the  parameters are still rather imprecise  and new
investigations are needed. 

For multihadron final states it may be difficult to isolate 
the contributions of particular vector resonances due to 
presence of many interfering intermediate states.
An example is the reaction $e^+e^-\to \pi^+\pi^-\pi^0\pi^0$,
to which the $\omega\pi^0$, $a_1\pi$, $\rho^+\rho^-$ intermediate
states give dominant contributions. The two latter states contain
wide resonances and strongly interfere.
A partial-wave analysis is required to separate the sub-processes
of the $e^+e^-\to \pi^+\pi^-\pi^0\pi^0$ and $e^+e^-\to
\pi^+\pi^-\pi^+\pi^-$ reactions.  
We hope that BABAR has enough data to perform such an analysis.
This is necessary to separate contributions of two excited $\rho$
states, $\rho(1450)$ and $\rho(1700)$, and determine their parameters.
A detailed study of intermediate mechanisms strongly benefits from the
quasi-two-body character of some final states, but is much more
difficult for multibody final states.

The BABAR ISR data on isoscalar channels already allow to improve
parameters of excited $\phi$ and $\omega$ states. A global fit
to the isovector and isoscalar components for the process
$e^+e^-\to K^\ast(892)\bar{K}$, and the $e^+e^-\to \phi\eta$
cross section~\cite{dm1kkpi} was used to determine parameters 
of the $\phi(1680)$ resonance (see Table~\ref{tab:multifit}). 

In Table~\ref{omfit_tab} we summarize the results of the fits
to the $e^+e^-\to 3\pi$, $e^+e^-\to \omega\pi\pi$,
and $e^+e^-\to \omega\eta$ cross sections 
performed by BABAR in Refs.~\cite{babr3pi,babr5pi,babr6pi},
and compare them with the corresponding PDG~\cite{PDG08} parameters. 
A simultaneous fit to all three channels could significantly 
improve the results and give additional
information on relative decay rates.

Due to numerous extensive studies of various exclusive cross sections,
we have learned a lot about the total cross section of $e^+e^-$
annihilation into hadrons and its components allowing a more precise
estimation of hadronic vacuum polarization effects to be performed 
(see also  Section~\ref{mu_g2}).

\section { Baryon form factors}
\subsection{General formulae}
The cross section of the process 
$e^+e^-\to {\cal B}\bar{\cal B}$, where ${\cal B}$ is a
spin-1/2 baryon, is given by \cite{dsigpp}
\begin{eqnarray}
\frac{{\rm d}\sigma}{{\rm d}\Omega}&=&\frac{\alpha^2\beta C}{4s}
\left [ 
(|G_M(s)|^2(1+\cos^2\theta)\right. \nonumber\\
&+&\frac{1}{\tau}|G_E(s)|^2\sin^2\theta) \\
&-&\left.\frac{1}{\tau}|G_E(s)|^2)\sin^2\theta\cos{2\varphi}
\right ],
\label{eqB1}
\end{eqnarray}
where $\beta = \sqrt{1-4m_{\cal B}^2/s}$ and $m_{\cal B}$ 
are the baryon velocity ($v/c$) and mass,
$C = y/(1-e^{-y})$ with $ y = {\pi\alpha m_{\cal B}}/{\beta \sqrt{s}}$ 
is the Coulomb correction   factor~\cite{Coulomb} for charged baryons  
($C=1$ for neutral baryons), 
$\tau = m^2/4M_{\cal B}^2$ is the inverse helicity suppression factor,
$G_M$ and $G_E$ are the baryon magnetic and  electric form factors.
The number of form factors (two) corresponds to two 
${\cal B}\bar{\cal B}$ states 
with different angular
momenta: $^3S_1$ and $^3D_1$. 
At the ${\cal B}\bar{\cal B}$ threshold 
the $D$-wave state vanishes, and $|G_{E}| = |G_{M}|$. At high $\sqrt{s}$ 
the terms containing $G_E$ are suppressed by the helicity factor $1/\tau$.
With unpolarized beams the total cross section is 
\begin{equation}            
\sigma(s) = \frac{4\pi\alpha^{2}\beta C}{3s}            
\left [|G_M(s)|^{2} + \frac{1}{2\tau}|G_E(s)|^{2}\right].            
\label{eqB2}            
\end{equation}

As discussed above (see Sections~\ref{taguntag} and 
\ref{isreescancomp}, and Fig.~\ref{fig7}), the detection 
efficiency in the ISR measurement with a tagged photon 
has weak dependence on the angular distributions of
final hadrons. In the case of the dibaryon production
this allows one to measure the total cross section [Eq.(\ref{eqB2})]
independently of the relation between
the electric and magnetic form factors. The ratio of
the form factors can then be determined from an analysis
of the baryon angular distribution. In direct $e^+e^-$
or $p\bar{p}$ experiments the range of the accessible polar
angles is limited by the detector acceptance.
In this case the cross section cannot be measured in a 
model-independent way. The detection efficiency is determined,
and the proton magnetic form factor $|G_M|$ is extracted,
usually under the assumption that $|G_M|=|G_E|$. In the BABAR
paper~\cite{babrpp} on the ISR study of the reaction $e^+e^-\to p\bar{p}$  
the effective form factor is introduced as a linear
combination of $|G_M|^2$ and  $|G_E|^2$:
\begin{equation}
|F(m)|^2=\frac{2\tau |G_M(s)|^2+|G_E(s)|^2}{2\tau +1}
\label{eqB3}
\end{equation}
With the effective form factor the total cross
section looks like 
\begin{equation}
\sigma_0(m) = \frac{4\pi\alpha^{2}\beta
C}{3m^2}(1+\frac{1}{2\tau})|F(m)|^2
\label{eqB4}
\end{equation}
The effective form factor defined in such a way allows
an easy comparison of the results of the model-independent
ISR measurement with $|G_M|$ obtained in direct
$e^+e^-$ and $p\bar{p}$ experiments under the assumption 
that $|G_M|=|G_E|$.

The modulus of the ratio of the electric and magnetic form factors
can be determined from the analysis of the baryon polar angle distribution.
This distribution can be presented as a sum of the terms proportional
to $|G_M|^2$ and $|G_E|^2$. For the $e^+e^-\to p\bar{p}\gamma$ cross section
the fully differential formula can be found in~\cite{kuhn_pp}. 
In this process the $\theta_p$ dependences of
the $G_E$ and $G_M$ terms are not strongly different from $\sin^{2}\theta_p$
and   $1+\cos^{2}\theta_p$,  describing the angular distributions for 
the electric and magnetic form factors in Eq.~(\ref{eqB1}). Note that
in direct $e^+e^-$ experiments with transversely polarized
beams a study of the proton azimuthal angle distribution can
improve $G_E/G_M$ separation (see Eq.~(\ref{eqB1})).

A nonzero relative phase between the electric and
magnetic form factors manifests itself in a polarization
of the outgoing baryons. In the reaction $e^+e^-\to {\cal B}\bar{\cal B}$ 
this polarization is perpendicular to the production plane~\cite{dub}.
For the ISR process $e^+e^-\to {\cal B}\bar{\cal B}\gamma$
the polarization observables are analyzed in Refs.~\cite{kuhn_ll,dkm}.
In the case of the $\Lambda\bar{\Lambda}$
final state the $\Lambda\to p\pi^-$ decay can be used to
measure the $\Lambda$ polarization and hence the phase
between the form factors.

\subsection{Measurement of time-like baryon form factors}
Measurements of the $e^+e^-\to p\bar{p}$ cross section
have been performed in $e^+e^-$
experiments~\cite{DM1,DM2,DM2LL,ADONE73,FENICE,BES,CLEO}
with a (20--30)\% precision.
The cross section and the proton form factor were deduced
assuming $|G_E|=|G_M|$.
More precise measurements of the proton form factor
have been performed in
$p\bar{p}\to e^+e^-$ experiments~\cite{LEAR,E760,E835}.
In the PS170 experiment~\cite{LEAR} at LEAR,
the proton form factor was measured from threshold
($p\bar{p}$ annihilation at rest) up to a  mass of 2.05~GeV/$c^2$. The
ratio  $|G_E/G_M|$ was measured using the angular dependence of the
cross section and was found to be compatible with unity.
The LEAR data show a strong dependence of the form factor
on  $p\bar{p}$ mass near threshold, and a weak dependence
in the range 1.95--2.05~GeV/$c^2$. The Fermilab experiments,
E760~\cite{E760} and E835~\cite{E835}, show that the form factor
decreases rapidly at higher masses, in agreement with the perturbative QCD
prediction $G_M\propto\alpha_s^2(m^2)/m^4$.

Experimental information on the reactions $e^+e^-\to \Lambda\bar{\Lambda}$,
$\Sigma^0\bar{\Sigma}^0$, $\Lambda\bar{\Sigma}^0$  is very scarce.
The $e^+e^-\to \Lambda\bar{\Lambda}$ cross section is measured to be
$100^{+65}_{-35}$ pb at 2.386 GeV, and at the same energy the
upper limits for $e^+e^-\to \Sigma^0\bar{\Sigma^0}$
($ <120$ pb) and $e^+e^-\to \Lambda\bar{\Sigma^0}$ ($ < 75$ pb)
cross sections have been obtained~\cite{DM2LL}.

\subsection{$e^+e^-\to p\bar{p}\gamma$}
\begin{figure}[h]
\includegraphics[width=.4\textwidth]{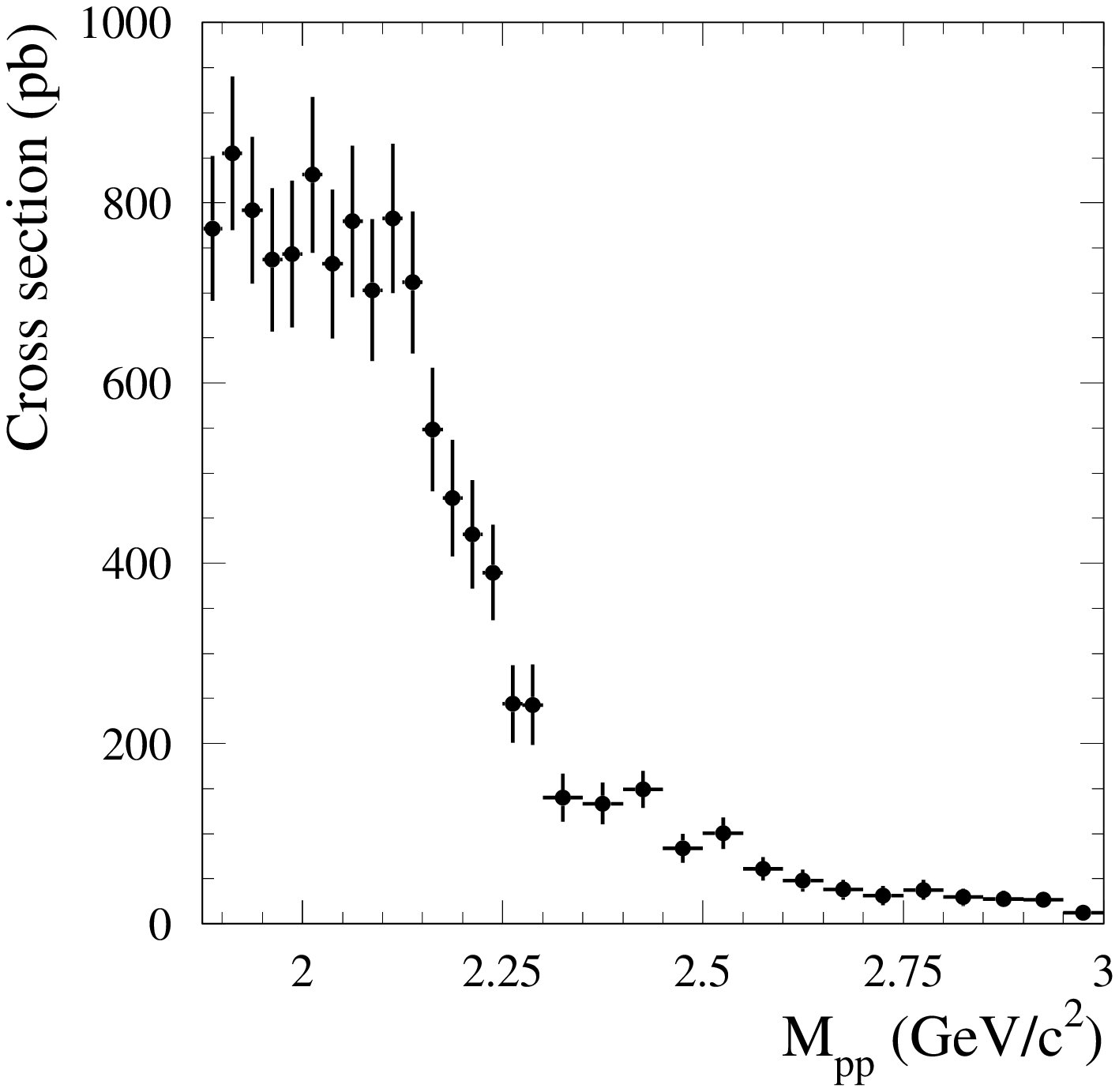}
\caption{The $e^+e^-\to p\bar{p}$ cross section measured by 
BABAR~\cite{babrpp}.
\label{prot0}}
\end{figure}

\begin{figure}[h]
\includegraphics[width=.4\textwidth]{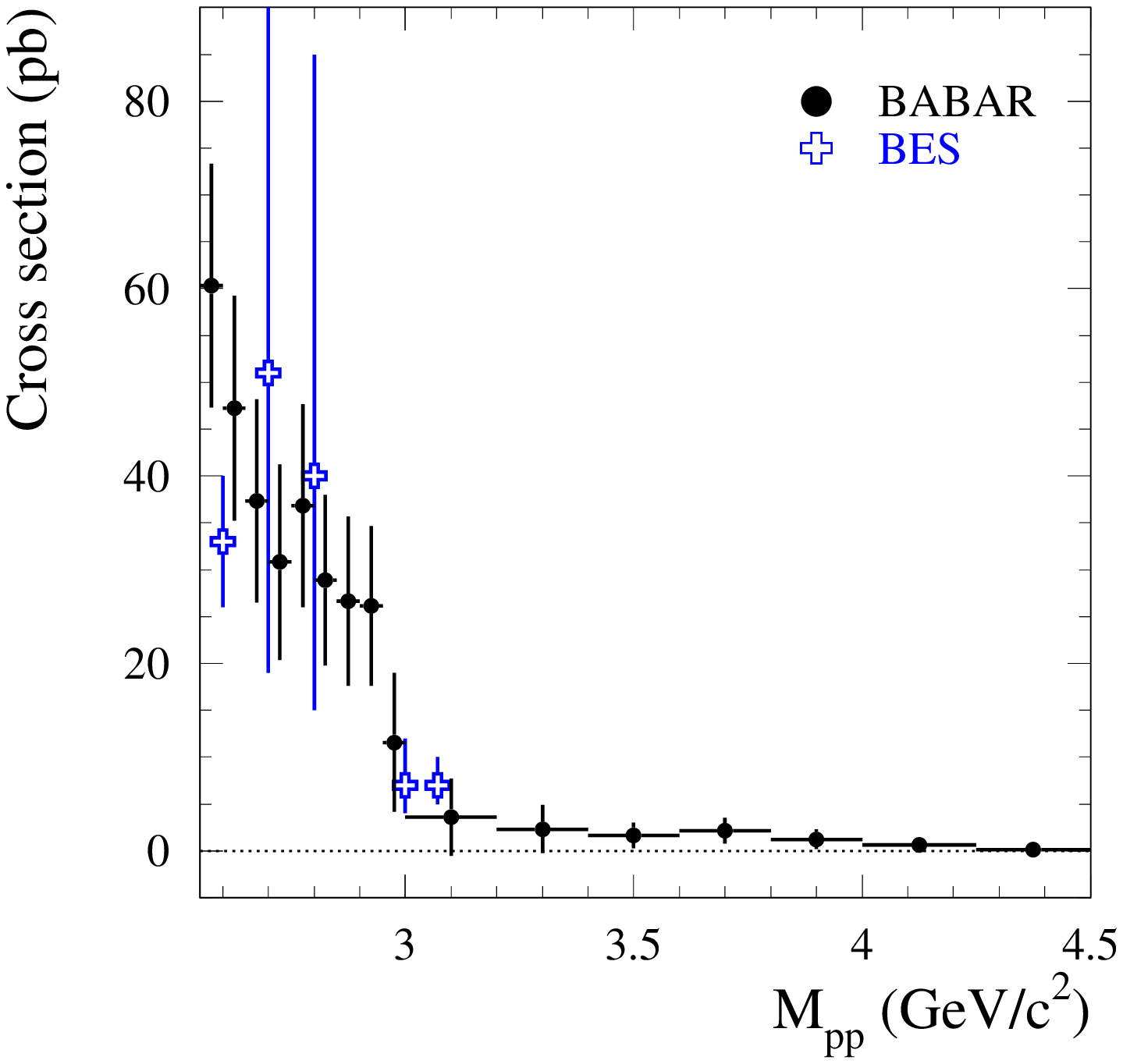}
\caption{The $e^+e^-\to p\bar{p}$ cross section measured by BABAR~\cite{babrpp} in
comparison with the BES data~\cite{BES}
\label{prot01}}
\end{figure}

\begin{figure}[h]
\includegraphics[width=.4\textwidth]{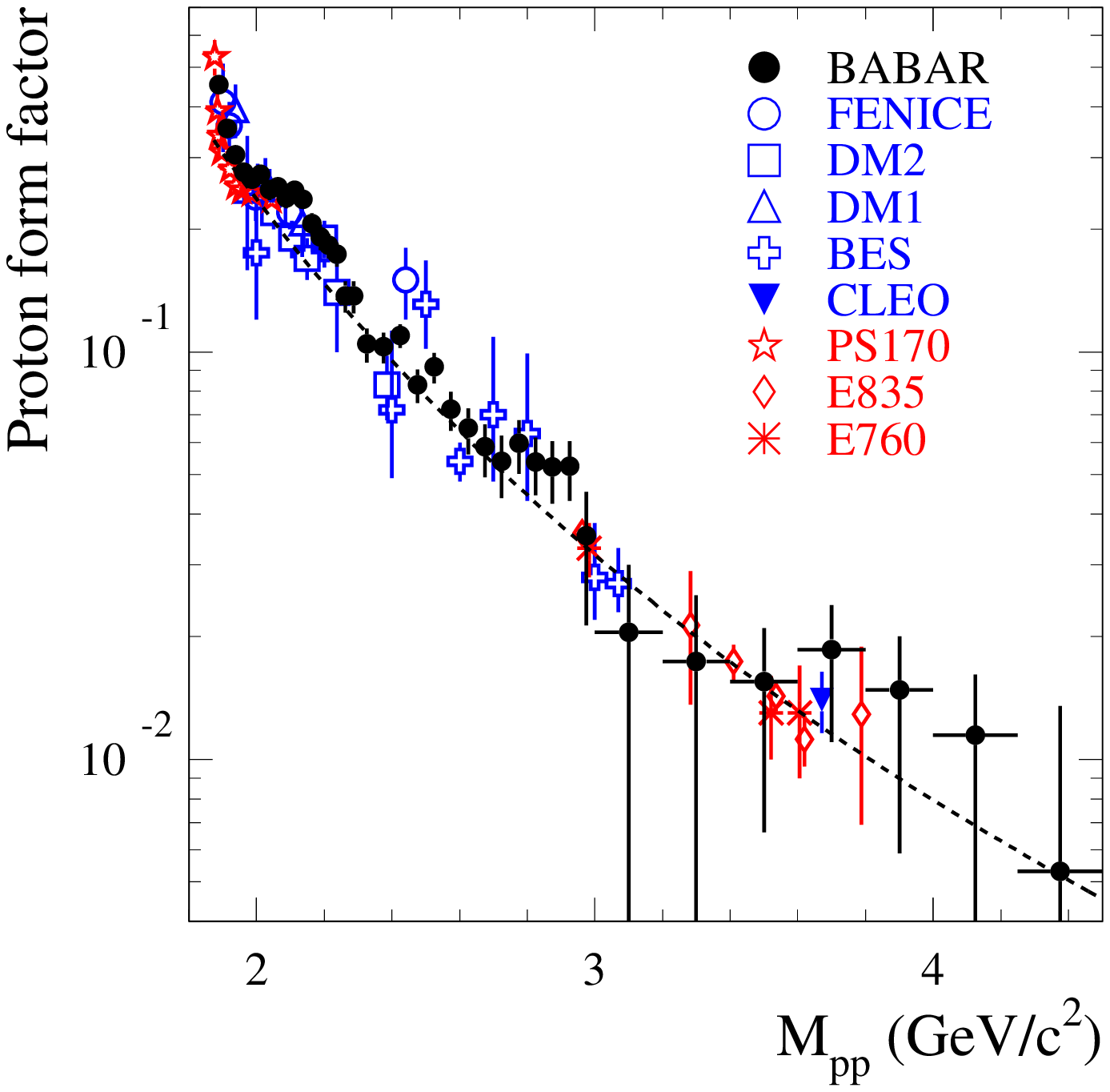}
\caption{The proton form factor measured in different 
experiments~\cite{babrpp,FENICE,DM1,DM2,BES,CLEO,LEAR,E835,E760}.
The solid line represents the QCD fit described in the text.
\label{prot1}}
\end{figure}

\begin{figure}[h]
\includegraphics[width=.4\textwidth]{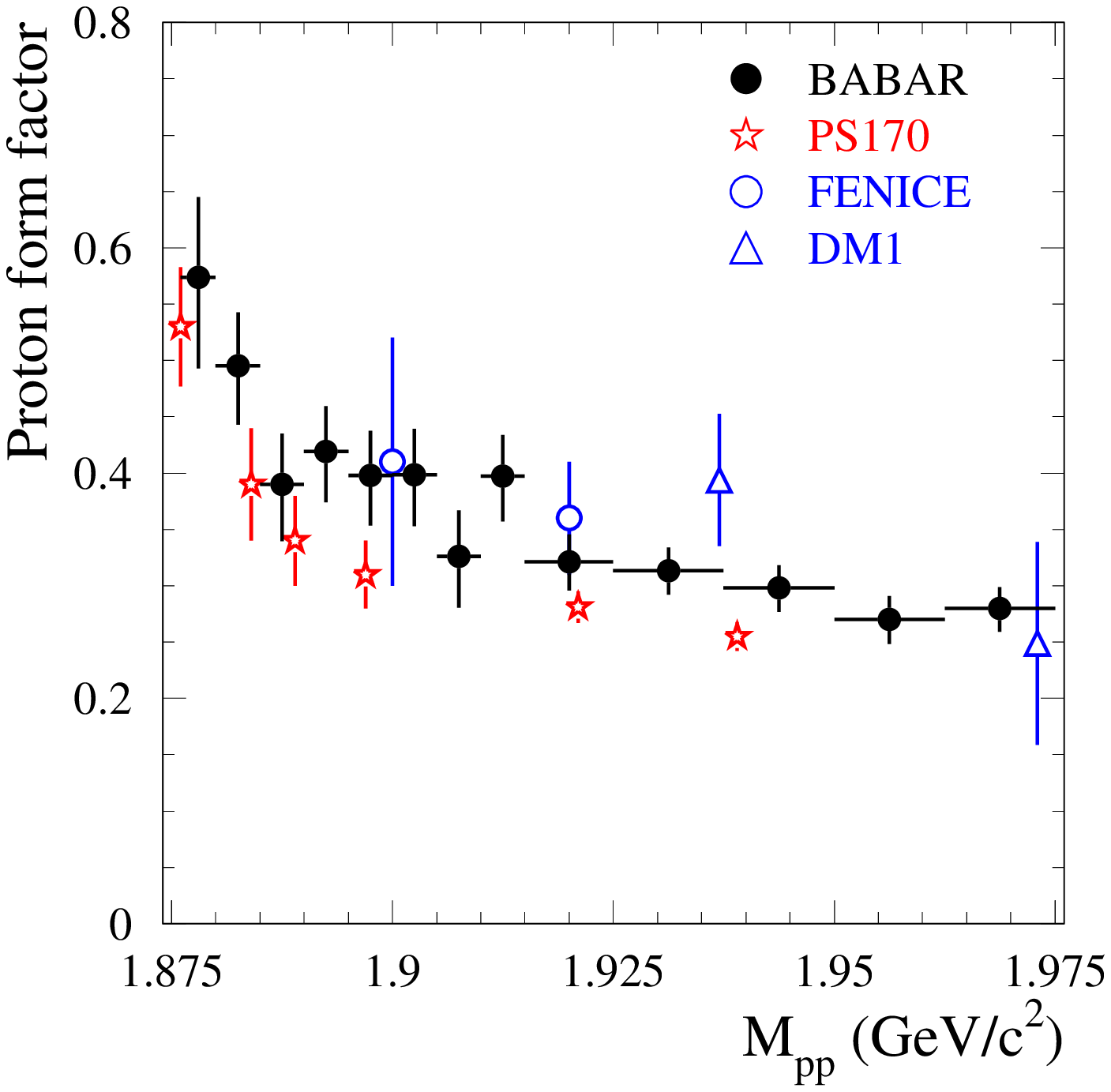}
\caption{The proton form factor in the near-threshold 
region~\cite{babrpp,LEAR,FENICE,DM1}.
\label{prot11}}
\end{figure}

The first ISR baryon experiment was the measurement of
the proton-antiproton production cross section \cite{babrpp} by BABAR.
The measured $e^+e^-\to p\bar{p}$ cross section shown
in Figs. \ref{prot0} and \ref{prot01}
is almost flat near the threshold, and then decreases from 1 nb to 
about 1 pb at 4.5 GeV. There are two rapid drops of the cross section
near 2.15 and 2.9 GeV.
The BABAR proton form factor data presented in  Fig. \ref{prot1}, in general,
agree with the previous measurements. Figure~\ref{prot11} shows an
expanded view of the near-threshold region. The BABAR measurement 
confirms the PS170 observation~\cite{LEAR} of the significant increase in the
form factor for energies approaching the $p\bar{p}$ threshold.
The proton form factor reaches about 0.6 at the threshold. 

\begin{figure}
\includegraphics[width=.4\textwidth]{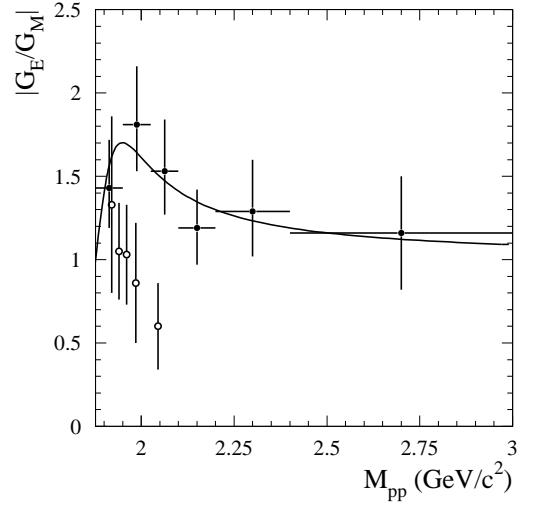}
\caption{The proton $|G_E/G_M|$ ratio measured by BABAR~\cite{babrpp}
(black points) compared with LEAR data~\cite{LEAR} (open circles). 
\label{prot2}}
\end{figure}

A study of the proton angular distribution allows one to extract the value of 
the ratio of the  electric and magnetic form factors $|G_E/G_M|$. 
The results of the BABAR $|G_E/G_M|$ measurement are shown in 
Fig.~\ref{prot2} in comparison with the data obtained at LEAR~\cite{LEAR}.
In disagreement with the LEAR result, the BABAR data indicate that $|G_E/G_M|$
significantly exceeds unity in the energy range between threshold and 2.1 GeV. 

\subsection{$e^+e^-\to \Lambda\bar{\Lambda}\gamma$}
\begin{figure}[h]
\includegraphics[width=.4\textwidth]{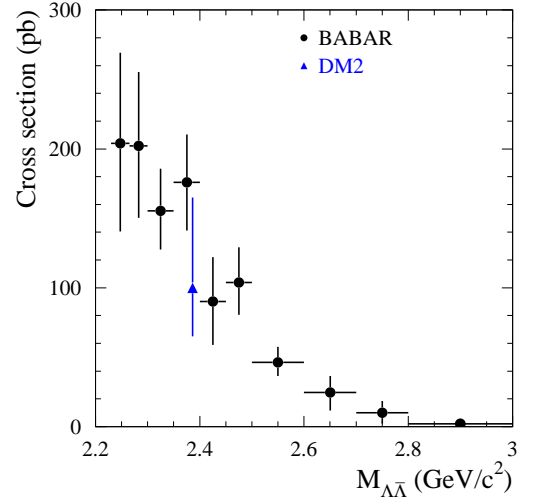}
\caption{The $e^+e^-\to \Lambda\bar{\Lambda}$ cross section
measured by BABAR~\cite{BBLS} in
comparison with the DM2 measurement~\cite{DM2LL}.
\label{crosslam}}
\end{figure}

The $e^+e^-\to \Lambda\bar{\Lambda}$ cross section measured by the BABAR 
detector~\cite{BBLS} is
shown in Fig.\ref{crosslam} in comparison with the only previous 
measurement~\cite{DM2LL}. The BABAR measurement is based on about 200 
$\Lambda\bar{\Lambda}$ events selected in the decay mode $\Lambda\to p\pi$. 

\begin{figure}[h]
\includegraphics[width=.4\textwidth]{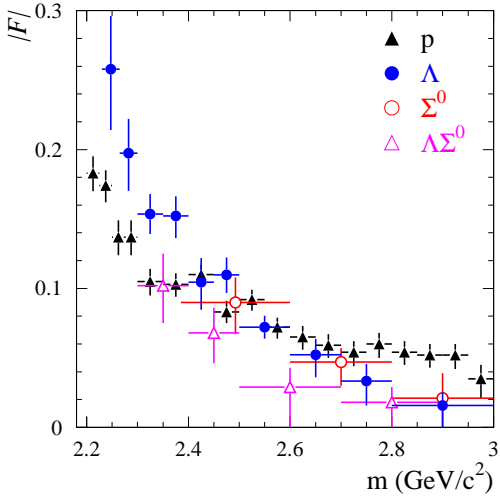}
\caption{The baryon form factors measured by BABAR~\cite{babrpp,BBLS} versus
the dibaryon invariant mass.
\label{formall}}
\end{figure}

\begin{figure}[h]
\includegraphics[width=.4\textwidth]{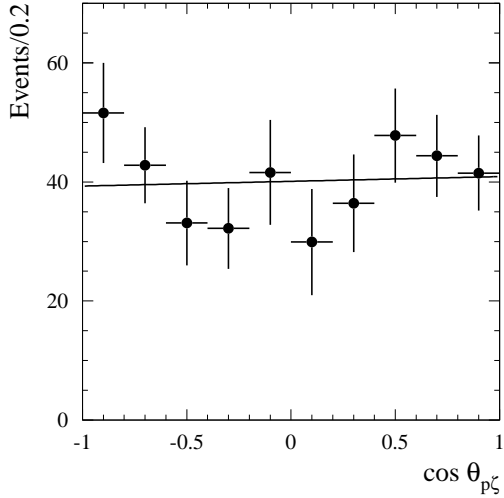}
\caption{The $\cos{\theta_{p\zeta}}$ distribution
in the $e^+e^-\to \Lambda\bar{\Lambda}$ process~\cite{BBLS}.
The line represents the result of the fit to data with a first-order
polynomial.
\label{coslmbd}}
\end{figure}

The measured $\Lambda$ effective form factor is shown in Fig.\ref{formall}.
The ratio $|G_E/G_M|$ is found to be consistent with unity. The use of the  
$\Lambda\to p\pi$ decay allows to measure the relative phase $\phi_\Lambda$
between the complex $G_E$ and $G_M$ form factors.
A non-zero $\phi_\Lambda$ leads to polarization $\zeta$ of the outgoing baryons.
The value of $\zeta$ is extracted from the analysis
of the proton angular distribution in the $\Lambda\to p\pi$ decay.
The measured $\cos\theta_{p\zeta}$ distribution,
where $\theta_{p\zeta}$ is the angle between the $\Lambda$ polarization
vector and the proton momentum in the $\Lambda$ rest frame, is shown in 
Fig.\ref{coslmbd}. No $\cos\theta_{p\zeta}$ distribution asymmetry 
corresponding to the non-zero polarization is seen. 
Because of the limited data sample  only a very
weak limit on the phase between the  $G_E$ and $G_M$ has been set
for  the $\Lambda$ hyperon:
$-0.76<\sin{\phi_\Lambda<0.98}$.

\subsection{$e^+e^-\to \Sigma^0\bar{\Sigma}^0,\,
\Lambda\bar{\Sigma}^0\,(\Sigma^0\bar{\Lambda}) $}
The BABAR measurement of the $\Sigma^0$ and $\Sigma^0\Lambda$ 
form factors is described in Ref.~\cite{BBLS}.
The decay chain $\Sigma^0\to\Lambda\gamma\to
p\pi\gamma$ is used to reconstruct $ \Sigma^0$. 
About 20 candidate events were selected
for each of the ISR reactions, $e^+e^-\to \Sigma^0\bar{\Sigma}^0\gamma$ and
$e^+e^-\to \Sigma^0\bar{\Lambda}\gamma$. 
The effective  $\Sigma^0$
and $\Sigma^0\Lambda$ form factors are shown in Fig.\ref{formall}.
The corresponding values of the $e^+e^-\to \Sigma^0\bar{\Sigma}^0$ and
$e^+e^-\to \Sigma^0\bar{\Lambda} $ cross sections are about
40 pb near the reaction thresholds.
It is  seen that the $\Lambda$, $\Sigma^0$ and $\Sigma^0\Lambda$ 
form factors are of the same order.

\subsection{Summary}
The baryon form factors are a subject of various 
phenomenological models (see Ref.~\cite{Baldi} and references therein).
QCD predicts for the baryon form factor   the asymptotic 
behavior $F(q^2)\sim \alpha_s^2(q^2)/q^4$~\cite{Chern2}. 
Comparison of this prediction with the data on the proton form factor 
is shown in Fig.~\ref{prot1}. It is seen that the asymptotic regime 
is reached at energies higher than 3 GeV.

The remarkable feature of the process $e^+e^-\to p\bar{p}$  is a
nearly flat cross section in the 200-MeV region above the $p\bar{p}$ threshold.
This feature is explained in Ref.~\cite{Baldi} by the opposite trends in the
energy dependence of the $S$-wave and $D$-wave contributions. 

A natural explanation for the sharp increase of the proton form factor 
in the vicinity of the $p\bar{p}$ threshold is the final state interaction
of the proton and antiproton (see, for example, Ref.~\cite{fsi} and references
therein). Another possibility is a contribution of the vector-meson state
located just below the $p\bar{p}$ threshold. This state is observed
in the reaction $e^+e^-\to 6\pi$~\cite{DM26pi,FOCUS6pi,babr6pi}.

The rapid drop of the cross section at 2.15 GeV may be a manifestation 
of the isovector state $\rho(2150)$, which is seen 
in the reactions $e^+e^-\to \eta^{\prime}\pi^+\pi^-$ and 
$e^+e^-\to f_1(1285)\pi^+\pi^-$~\cite{babr5pi}. 
The drop in the cross section near 2.9 GeV is still not understood. 

\begin{figure}
\includegraphics[width=.4\textwidth]{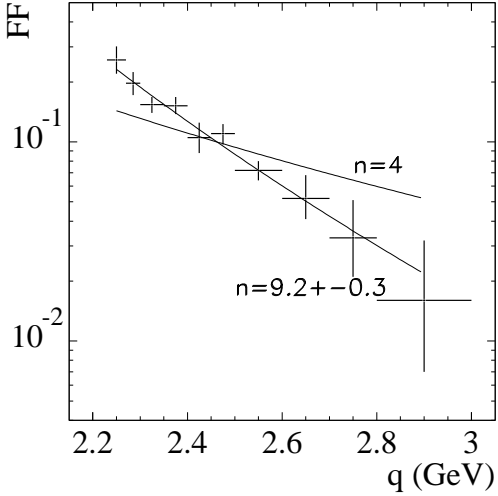}
\caption{A fit of the $\Lambda$ form factor~\cite{BBLS}  
with the power-law function
$F\sim const/q^n$ and with the QCD-inspired function $F\sim const/q^4$. 
\label{Lambfit}}
\end{figure}

 The  $e^+e^-\to\Lambda\bar{\Lambda}$ cross section has some features similar
to those for the process  $e^+e^-\to p\bar{p}$. In the energy region 
of about 200 MeV above threshold the $ \Lambda\bar{\Lambda}$ cross 
section is flat; the $G_E/G_M$ ratio is consistent with that measured
by BABAR for the  $e^+e^-\to p\bar{p}$. The attempt to explain the
unusual energy dependence of the $e^+e^-\to\Lambda\bar{\Lambda}$ cross 
section was made in Ref.~\cite{Baldi}.
A fit to the $\Lambda$ form factor with the power-law function $const/q^n$
(Fig.\ref{Lambfit}) gives $n=9.2\pm 0.3$. This $n$ value strongly differs from
the QCD asymptotic prediction $n=4$. Similarly to the $p\bar{p}$ case,
the asymptotic
regime is not reached for $e^+e^-\to\Lambda\bar{\Lambda}$ at the energies 
below 3 GeV.

   The cross sections of the processes $e^+e^-\to \Sigma^0\bar{\Sigma}^0$ and 
 $e^+e^-\to   \Lambda\bar{\Sigma}^0 (\Sigma^0\bar{\Lambda}) $
 have been measured with large errors. The corresponding 
$\Sigma^0$ and  $\Sigma^0\Lambda$ form factors  (Fig.\ref{formall})
show a monotonic decrease starting just from the  threshold. 
A fit to form factor data with the power-law function gives $n>4$
but with large errors.

It is interesting to compare the measured form factors with each other 
and with the QCD prediction for the asymptotic form factor 
ratios~\cite{ChernFFs}:
$F_p=4.1F_{\Lambda}$,
$F_{\Sigma^0}=-1.18F_{\Lambda}$, $F_{\Sigma^0\Lambda}=-2.34F_{\Lambda}$.
From comparison of the form factors in Fig.\ref{formall}, it is seen that
the prediction works (possibly accidentally) only for the ratio of the
$\Lambda$ and $\Sigma^0$ form factors. The ratio $F_{\Lambda}/F_p$
falls with energy. 
In the highest energy interval 2.8--3.00 GeV
the ratio is equal to $0.3^{+0.2}_{-0.3}$ and agrees with the 
asymptotic value 0.24. This is an indication that the asymptotic regime is
reached just above 3 GeV. 

The BABAR experiment shows that the ISR method is well suited for the
measurement of baryon form factors. Future Super $B$-factories as well
as already running BEPC $e^+e^-$ collider will
 make possible measurements of the form factors, especially for the 
proton, with unprecedented accuracy. High-precision measurements of
the proton form factor are also planned in the PANDA experiment. .  

\section{Decays of the $J/\psi$ and $\psi(2S)$}

For all the processes described in
the previous sections, clear narrow peaks are seen
in the energy dependence of the cross sections
corresponding to the $J/\psi$ and $\psi(2S)$ decays.  
The Born cross section for the ISR production of 
a narrow resonance, for example, the $J/\psi$, decaying
to the final state $h$ can be calculated using~\cite{babrmu} 
\begin{equation}
\sigma_{J/\psi}=\frac{12\pi^2\Gamma(J/\psi \to e^+e^-){\cal B}(J/\psi\to h)}
{sm_{J/\psi}}W_0(\theta_0,x_{J/\psi})
\label{gee}
\end{equation}
where $m_{J/\psi}$ and $\Gamma(J/\psi \to e^+e^-)$ are the mass and electronic
width of the $J/\psi$ meson, $x_{J/\psi} = 1-m_{J/\psi}^2/s$, and
${\cal B}(J/\psi\to h)$ is the branching fraction for the $J/\psi$ decay 
to the final state $h$. The function $W_0$ is described in 
Sec.~\ref{CalofISR} by Eq.(\ref{eq5}).
Therefore, a measurement of the number of $J/\psi\to h$
decays in the ISR process $e^+ e^-\to h\gamma$ determines the product of
the electronic width and the branching fraction:
$\Gamma(J/\psi \to e^+e^-){\cal B}(J/\psi\to h)$.

The total cross section for the process $e^+ e^-\to\gamma J/\psi$ with 
a tagged ISR photon ($\theta_0=30^\circ$) is about 3.4~pb.
With the integrated luminosity of  $\sim 500$ fb$^{-1}$
collected by the BABAR detector
it corresponds to about 1.7 million produced $J/\psi$'s. 
This number is significantly smaller than, for example, about 60 million
$J/\psi$'s produced in the BESII experiment at the BEPC 
$e^+ e^-$ collider. However, 
the general quality of the BABAR detector and its particle identification in 
particular, are much better compared to the BESII detector. 
As a result, the detector efficiency and the integrated luminosity 
are determined  with lower systematic errors. A typical systematic
uncertainty of the BABAR measurement is 3-5\%, while BESII
usually quotes 10-15\%. The lower systematic
error makes ISR results on many $J/\psi$ decays competitive with  
BESII and other previous  measurements. Practically all decays with
the rates about 10$^{-3}$ and higher can be measured via ISR with better
overall accuracy. Moreover, because of excellent particle
identification, many $J/\psi$ and $\psi(2S)$ decays with kaons
in the final state have been studied using ISR for the first time.

In the BABAR experiment the ISR method enabled to measure a few tens of  
$J/\psi$ and $\psi(2S)$ decays with the best-to-date accuracy and 
discover about 20 new decays of these resonances. 
 
\subsection{Leptonic decays}
\begin{figure}[tbh]
\begin{center}
\includegraphics[width=8.2cm]{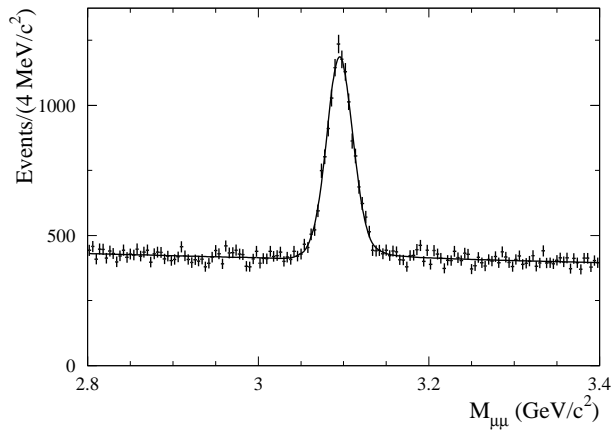}
\caption{
The $\mu^+\mu^-$ mass spectrum in the $J/\psi$ region for selected 
events of the reaction $e^+ e^- \to \mu^+\mu^-\gamma$~\cite{babrmu}.
The curve is the result of the fit described in the text.}
\label{fitmumu}
\end{center}
\end{figure}

The BABAR~\cite{babrmu} and CLEO~\cite{cleomu} collaborations performed 
a study of the $J/\psi$ production in the reaction 
$e^+e^- \to \mu^+\mu^-\gamma$.
The dimuon mass spectrum for this reaction obtained by BABAR is 
shown in Fig.~\ref{fitmumu}.
The signal of the ISR $J/\psi$ production 
is well seen in the mass spectrum. The nonresonant spectrum is due to
muon pair production in the process $e^+e^- \to \mu^+\mu^-\gamma$,
where photon can be emitted by both initial electrons and final
muons. Since the dimuon decay of the $J/\psi$ meson
proceeds through a single photon transition, 
$J/\psi \to\gamma^\ast\to\mu^+\mu^-$, the angular and momentum distributions
for events from the $J/\psi$ peak are completely identical to those
for the ISR part of nonresonant events. 
The idea of the BABAR measurement is to determine the ratio of the number of
$J/\psi$ events to the level of the nonresonant spectrum which
is well known theoretically.
The dimuon spectrum of Fig.~\ref{fitmumu} has been fit with a
function taking into account the energy dependence of the nonresonant cross
section and the experimental $J/\psi$ line shape.
The ratio
\begin{equation}
r=\frac{N_{J/\psi}}{\frac{\mathrm{d}N}{\mathrm{d}m}\cdot \Delta m}
\end{equation}
was the main fit parameter. 
After substituting cross sections for the numbers of events,
this ratio can be rewritten
\begin{equation} 
r=\frac{\sigma_{J/\psi}^{\mathrm{Born}}}
{\frac{\mathrm{d}\sigma_{\mathrm{ISR}}^{\mathrm{Born}}}{\mathrm{d}m}\cdot
\Delta m}\cdot \frac{1}{K}; \, \, 
K=\frac{\mathrm{d}\sigma_{\mathrm{Total}}^{\mathrm{vis}}/\mathrm{d}m}
{\mathrm{d}\sigma_{\mathrm{ISR}}^{\mathrm{vis}}/\mathrm{d}m} \, .
\label{ratio} 
\end{equation}
Detector acceptances and ISR radiative corrections,
which are the same for the
nonresonant ISR and $J/\psi$ contributions to the reaction
$e^+e^-\to \mu^+ \mu^-\gamma$,
cancel in the ratio.
The total nonresonant cross section includes the FSR contribution, which
is parameterized in terms of 
$K$, the ratio of the visible nonresonant total and ISR-only 
(FSR switched
off) cross sections. Since BABAR selects events with the photon
emitted at a large angle, the FSR contribution is  
relatively large.
Using simulated events, the coefficient
$ K = 1.11\pm0.01$ (statistical error only) is determined 
for the selection criteria used.

The result of the fit is shown in Fig.~\ref{fitmumu}.
The value $r = 18.94 \pm 0.44$ is found with
 $ \chi^2 / ndf=122/144$.
From the product
$r\cdot K=21.03\pm0.49\pm0.47$ the cross section
$\sigma_{J/\psi}=2124\pm49\pm47 \mbox{ fb}$ 
and the product of the $J/\psi$ parameters
$$\Gamma_{ee}\cdot B_{\mu\mu}=0.3301\pm0.0077\pm0.0073 \mbox{ keV}$$
are determined.
The main sources of the systematic error quoted are uncertainties in the
$J/\psi$ line shape and the coefficient $K$, both due to 
imperfect simulation of the detector response.

Using the values for $B_{ee}$ and $B_{\mu\mu}$~\cite{PDG04},
which are well measured in the cascade 
$\psi(2S)\to J/\psi\pi^+\pi^-$ decays~\cite{BES1},
the electronic and total widths of the $J/\psi$ meson were derived,
$$\Gamma_{ee}=5.61\pm0.20\mbox{ keV},\:\:\:
\Gamma=94.7\pm4.4 \mbox{ keV} \, . $$
These were the best-to-date measurements of the $J/\psi$ parameters.

The BABAR measurement was improved by CLEO. With the integrated luminocity 
of 281 pb$^{-1}$ collected at $\sqrt{s}=3.77$ GeV about $13\times 10^3$ ISR 
produced $J/\psi\to \mu^+\mu^-$ events were selected (compared to
$8\times 10^3$ $J/\psi\to \mu^+\mu^-$ events in the BABAR measurement
based on a 88 fb$^{-1}$ data sample). Since CLEO used the untagged approach,
the FSR contribution to the nonresonant cross section was significantly
reduced. The second important improvement of the method  was that the 
$J/\psi$ line shape was extracted from data. To do this, the data 
collected by CLEO at the $\psi(2S)$ resonance were used to select 
a clean sample of 
$\psi(2S)\to J/\psi \pi^+\pi^-,\,J/\psi\to\mu^+\mu^-$ events.
The CLEO result for the product of the $J/\psi$ parameters is
$$\Gamma_{ee}\cdot {\cal B}_{\mu\mu}=0.3384\pm0.0058\pm0.0071 \mbox{ keV}.$$
Despite the analysis improvements described above, the systematic
error of the CLEO result was not reduced compared to the BABAR measurement. 
However, the sources of the systematic uncertainties are different for the 
two measurements. So the results can be considered as completely
 independent.

The data collected at $\sqrt{s}=3.77$ GeV were used by CLEO to study
$\psi(2S)$ ISR production~\cite{cleoee}. ISR $\psi(2S)$ events
were selected in the decay modes $\psi(2S)\to \pi^+\pi^-J/\psi$,
$\pi^0\pi^0 J/\psi$, and $\eta J/\psi$ with the $J/\psi$ decaying to
the lepton pair, $e^+e^-$ or $\mu^+\mu^-$. From the number of the
$\psi(2S)$ events the products 
$\Gamma(\psi(2S)\to e^+e^-){\cal B}(\psi(2S)\to X J/\psi)$, where 
$X=\pi^+\pi^-$, $\pi^0\pi^0$, and $\eta$, were obtained.
Since the branching fractions for these decay modes are known
with 1.5--2\% accuracy~\cite{PDG08}, the measurement of the products
can by used to improve accuracy of the $\psi(2S)$ electronic
width. The CLEO result
dominates in the current PDG value $\Gamma(\psi(2S)\to e^+e^-)=2.36\pm0.04$
keV~\cite{PDG08}.
\subsection{Decays to light mesons and baryons}
\begin{figure}
\includegraphics[width=.4\textwidth]{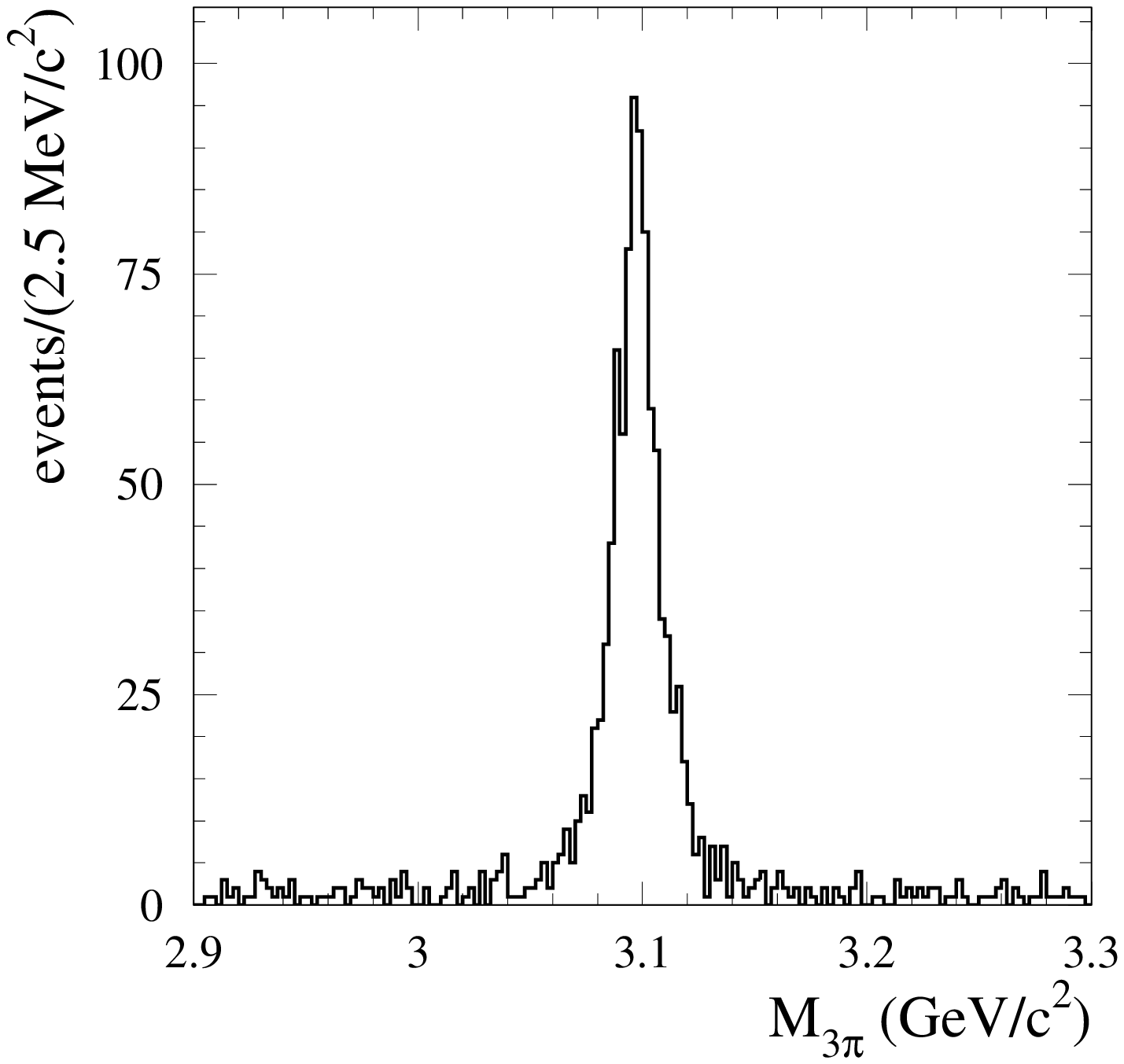}
\caption{ The $3\pi$ mass spectrum for selected 
$e^+e^-\to \pi^+\pi^-\pi^0\gamma$ data
events in the vicinity of the $J/\psi$ resonance~\cite{babr3pi}.
\label{JPsimass}}
\end{figure}

\begin{table*}[p]
\caption{
  Measurements of the $J/\psi$ and $\psi(2S)$ branching fractions via
  ISR at BABAR~\cite{babr3pi,babrpp,BBLS,babr4pi,babr6pi,babr5pi}
      compared to the current world-average values~\cite{pdg10}. 
\label{jpsitab}}
\begin{ruledtabular}
\begin{tabular}{lccc}
\multicolumn{1}{c}{Measured} & {Measured}  &  
\multicolumn{2}{c}{$J/\psi$ or $\psi(2S)$ branching fraction  (10$^{-3}$)}\\
\multicolumn{1}{c}{quantity} & {value (eV)} &
\multicolumn{1}{c}{BABAR}    & 
\multicolumn{1}{c}{PDG-2010} \\
\hline
$\Gamma^{J/\psi}_{ee}{\cal B}_{J/\psi  \to \pipi\pi^0}$  &
$122\pm 5\pm 8$ & $21.8\pm1.0\pm1.6$ &
 $20.7\pm1.2(S=1.2)$\footnote{$S$ is a PDG scale factor.} \\
$\Gamma^{J/\psi}_{ee}{\cal B}_{J/\psi  \to 2(\pipi)}$  &
$19.5\pm 1.4\pm 1.3$ & $3.70\pm0.26\pm0.37$ & $3.55\pm0.23$ \\
$\Gamma^{J/\psi}_{ee}{\cal B}_{J/\psi  \to 2(\pipi)\pi^0}$  &
$303\pm5\pm18$ &  $54.6\pm0.9\pm3.4$ & $41\pm5(S=2.4)$ \\
$\Gamma^{J/\psi}_{ee}{\cal B}_{J/\psi  \to 3(\pipi)}$  &
$23.7\pm1.6\pm1.4$    &   $4.40\pm0.29\pm0.29$  &   $4.3\pm0.4$  \\
$\Gamma^{J/\psi}_{ee}{\cal B}_{J/\psi  \to 2(\pipi\pi^0)}$  &
$89 \pm 5 \pm 10$    &   $16.5 \pm 1.0 \pm 1.8$  & $16.2\pm2.1$ \\
$\Gamma^{J/\psi}_{ee}{\cal B}_{J/\psi  \to \KKppch}$  &
$37.9\pm0.8\pm1.1$  &   $6.84\pm0.15\pm0.27$  &   $6.6\pm0.5$  \\
$\Gamma^{J/\psi}_{ee}{\cal B}_{J/\psi  \to \KKppnt}$  &
$11.8\pm0.8\pm0.9$  &   $2.12 \pm 0.15 \pm 0.18$  &   $2.45 \pm 0.32$ \\
$\Gamma^{J/\psi}_{ee}{\cal B}_{J/\psi  \to \KKKK}  $  &
$4.00\pm 0.33\pm 0.29$ &  $ 0.72 \pm 0.06 \pm 0.05$  & $0.76 \pm 0.09$ \\
$\Gamma^{J/\psi}_{ee}{\cal B}_{J/\psi  \to \Kp\Km\pipi\pi^0}  $  &
$107\pm4\pm 6$ &   $19.2 \pm 0.8 \pm 1.5$  &   $17.9 \pm 2.9(S=2.2)$\\ 
$\Gamma^{J/\psi}_{ee}{\cal B}_{J/\psi  \to \Kp\Km 2(\pipi)}  $  &
$27.5\pm2.3\pm 1.7$ &    $5.09 \pm 0.42 \pm 0.35$  & $4.7 \pm 0.7(S=1.3)$ \\
$\Gamma^{J/\psi}_{ee}{\cal B}_{J/\psi  \to \omega\pipi}{\cal B}_{\omega    \to 3\pi}$  &
$47.8 \pm3.1\pm 3.2$  &   $9.7 \pm 0.6 \pm 0.6$  & $8.6 \pm 0.7(S=1.1)$  \\
$\Gamma^{J/\psi}_{ee}{\cal B}_{J/\psi  \to \omega\pipi\pi^0}
{\cal B}_{\omega    \to 3\pi}$  &
$22\pm3\pm 2$  &   $4.1 \pm 0.6 \pm 0.4$  & $4.0\pm0.7$ \\ 
$\Gamma^{J/\psi}_{ee}{\cal B}_{J/\psi  \to \eta\pipi}{\cal B}_{\eta \to 3\pi}$  &
$0.51 \pm0.22\pm 0.03$  &  $ 0.40 \pm 0.17 \pm 0.03$  & $0.40 \pm 0.17$  \\
$\Gamma^{J/\psi}_{ee}{\cal B}_{J/\psi \to 2(\pipi)\eta}
{\cal B}_{\eta \to \gamma\gamma}$&
$  5.16 \pm0.85\pm 0.39$  & $2.35 \pm 0.39 \pm 0.20$  & $2.29 \pm 0.24$ \\
$\Gamma^{J/\psi}_{ee}{\cal B}_{J/\psi  \to \phi\eta}{\cal B}_{\phi\to\Kp\Km}
{\cal B}_{\eta\to 3\pi}$& $0.84\pm0.37\pm 0.05$ & 
$1.4 \pm 0.6 \pm 0.1$  &  $0.75 \pm 0.08(S=1.5)$\\
$\Gamma^{J/\psi}_{ee}{\cal B}_{J/\psi  \to \Kp\Km\eta}{\cal B}_{\eta\to 2\gamma} $  &
$4.8\pm0.7\pm 0.3$ &  $0.87 \pm 0.13 \pm 0.07$  &  $0.87\pm0.15$\\     
$\Gamma^{J/\psi}_{ee}{\cal B}_{J/\psi  \to \omega \Kp\Km}
{\cal B}_{\omega\to 3\pi} $  & 
$3.3\pm 1.3 \pm 0.2$ & $1.36 \pm 0.50 \pm 0.10$  &   $1.36 \pm 0.51$\\
$\Gamma^{J/\psi}_{ee}{\cal B}_{J/\psi\to\Kp\Km\pipi\eta}
{\cal B}_{\eta \to \gamma\gamma}$  &
$10.2\pm 1.3\pm 0.8$ & $4.7  \pm 0.6 \pm 0.4$  & $4.67\pm0.70$ \\
$\Gamma^{J/\psi}_{ee}{\cal B}_{J/\psi\to (K^{\ast0}\bar  K_2^{\ast0}+\mbox{c.c.})}
{\cal B}_{K^{\ast0}\to K\pi}{\cal B}_{K_2^{\ast0}\to K\pi}$ &
$8.59\pm 0.36 \pm 0.27$  &  $6.98  \pm 0.29  \pm 0.21$   &   $6.0  \pm 0.6$\\
$\Gamma^{J/\psi}_{ee}{\cal B}_{J/\psi\to (K^{0}\bar K^{\ast}(892)^{0}+\mbox{c.c.})}$ &
$26.6 \pm 2.5 \pm 1.5$ &  $4.8 \pm 0.5 \pm 0.3$  &  $4.39 \pm 0.31$  \\
$\Gamma^{J/\psi}_{ee}{\cal B}_{J/\psi\to (K^{+}\bar K^{\ast}(892)^{-}+\mbox{c.c.})}$ &
$29.0 \pm 1.7 \pm 1.3$ &  $5.2 \pm 0.3 \pm 0.2$      &  $5.12 \pm 0.30$  \\
$\Gamma^{J/\psi}_{ee}{\cal B}_{J/\psi\to K^{\ast0}\bar K^{\ast0}}
{\cal B}_{K^{\ast0}\to K^+\pi^-}{\cal B}_{\bar K^{\ast0}\to K^-\pi^+}$ &
$0.57 \pm 0.15 \pm 0.03$ &  $0.23 \pm 0.06 \pm 0.01$  &  $0.23 \pm 0.07$  \\
$\Gamma^{J/\psi}_{ee}{\cal B}_{J/\psi  \to \phi\pipi}{\cal B}_{\phi \to \Kp \Km}$  &
$2.19\pm 0.23\pm 0.07$ & $0.81 \pm 0.08 \pm 0.03$ & $0.94 \pm 0.09(S=1.2)$\\
$\Gamma^{J/\psi}_{ee}{\cal B}_{J/\psi  \to \phi\ppz}{\cal B}_{\phi\to \Kp \Km}$  &
$1.36\pm 0.27\pm 0.07$ & $0.50 \pm 0.10 \pm 0.03$  &  $0.56 \pm 0.16$\\
$\Gamma^{J/\psi}_{ee}{\cal B}_{J/\psi  \to \phi\Kp\Km}{\cal B}_{\phi    \to \Kp \Km}$  &
$2.26\pm 0.26\pm 0.16$ & 
$1.67 \pm 0.19 \pm 0.12$\footnote{${\cal B}_{J/\psi \to \phi  K\bar{K}}$}  & 
$1.83 \pm 0.24$ \\
$\Gamma^{J/\psi}_{ee}{\cal B}_{J/\psi  \to \phi f_0}
{\cal B}_{\phi  \to \Kp\Km}{\cal B}_{f_0\to \pipi}$  &
$0.69\pm 0.11\pm 0.05$ & $0.38 \pm 0.06 \pm 0.02$ &  
$0.32 \pm 0.09 (S=1.9)$ \\
$\Gamma^{J/\psi}_{ee}{\cal B}_{J/\psi  \to \phi f_0}
{\cal B}_{  \phi  \to \Kp\Km}{\cal B}_{  f_0   \to \ppz}$  &
$0.48\pm 0.12\pm 0.05$ & $0.53 \pm 0.13\pm  0.05$  &  $0.32 \pm 0.09 (S=1.9)$ \\
$\Gamma^{J/\psi}_{ee}{\cal B}_{J/\psi  \to \phi 2(\pipi)}
{\cal B}_{  \phi  \to K^+K^-}$ &
$4.7\pm 0.9\pm 0.9$ & $1.77 \pm 0.35 \pm 0.12$ & $1.66 \pm 0.23$ \\
\\
$\Gamma^{\psi(2S)}_{ee}{\cal B}_{\psi(2S) \to 2(\pipi)\pi^0} $  &
$29.7\pm 2.2\pm 1.8$ & $12.0  \pm 0.9  \pm 0.7$ & $2.9 \pm 1.0(S=4.6)$ \\
$\Gamma^{\psi(2S)}_{ee}{\cal B}_{\psi(2S) \to 2(\pipi\pi^0)}$  & 
$11.2\pm 3.3\pm 1.3$ & $5.3 \pm 1.6 \pm 0.6$ &  $5.3\pm1.7$ \\ 
$\Gamma^{\psi(2S)}_{ee}{\cal B}_{\psi(2S) \to\Kp\Km 2(\pipi)}$  & 
$4.4\pm 2.1\pm 0.3$ & $2.1 \pm 1.0 \pm 0.2$  &  $1.9\pm0.9$ \\ 
$\Gamma^{\psi(2S)}_{ee}{\cal B}_{\psi(2S) \to J/\psi\pipi}{\cal B}_{J/\psi \to 3\pi}$  &
$18.6\pm 1.2\pm 1.1$ &  $23.6  \pm 1.6  \pm 1.6$  &  
$20.7 \pm 1.2(S=1.2)$\footnote{${\cal B}_{J/\psi \to 3\pi}$.} \\
$\Gamma^{\psi(2S)}_{ee}{\cal B}_{\psi(2S) \to \omega\pipi}
{\cal B}_{\omega     \to 3\pi} $  &
$2.69\pm 0.73\pm 0.16$ & $1.22 \pm 0.33 \pm 0.07$  &  $0.73\pm 0.12$\\
$\Gamma^{\psi(2S)}_{ee}{\cal B}_{\psi(2S) \to J/\psi\eta}
{\cal B}_{\eta \to 3\pi}{\cal B}_{J/\psi \to \mumu} $  &
$1.11\pm 0.33\pm 0.07$ & $33.4  \pm 9.9  \pm 2.0$  &  $32.8 \pm 0.7$ \\
$\Gamma^{\psi(2S)}_{ee}{\cal B}_{\psi(2S) \to 2(\pipi)\eta}
{\cal B}_{\eta \to \gamma\gamma} $  &
$1.13\pm 0.55\pm 0.08$ & $1.2 \pm 0.6 \pm 0.1$  & $1.2\pm0.6$ \\
$\Gamma^{\psi(2S)}_{ee}{\cal B}_{\psi(2S) \to \Kp\Km\pipi\pi^0} $ &
$4.4\pm 1.3  \pm 0.3$ & $1.8 \pm 0.5 \pm 0.1$  & $1.26\pm 0.09(S=1.2)$\\
$\Gamma^{\psi(2S)}_{ee}{\cal B}_{\psi(2S) \to \Kp\Km\pipi\eta}
{\cal B}_{\eta \to \gamma\gamma}$ &
$1.2 \pm 0.7 \pm 0.1$  &  $1.3  \pm 0.7  \pm 0.1$   & $1.3\pm0.7$ \\ 
$\Gamma^{\psi(2S)}_{ee}{\cal B}_{\psi(2S) \to \KKppch} $  &
$1.92\pm 0.30\pm 0.06$ & $0.81 \pm 0.13 \pm 0.03$ & $0.75 \pm 0.09(S=1.9)$ \\
$\Gamma^{\psi(2S)}_{ee}{\cal B}_{\psi(2S) \to \KKppnt} $  &
$0.60\pm 0.31\pm 0.03$ & $0.25  \pm 0.13  \pm 0.02$   &  $0.25\pm0.13$  \\
$\Gamma^{\psi(2S)}_{ee}{\cal B}_{\psi(2S) \to \KKKK} $  &
$0.22\pm 0.10\pm 0.02$ & $0.09  \pm 0.04  \pm 0.01$   &  $0.060 \pm 0.014$ \\ 
$\Gamma^{\psi(2S)}_{ee}{\cal B}_{\psi(2S) \to \phi\pipi}
{\cal B}_{\phi     \to \Kp \Km} $  & $0.27\pm 0.09\pm 0.02$ & 
$0.35 \pm 0.12 \pm 0.01$  & $0.117\pm 0.029(S=1.7)$\\ 
$\Gamma^{\psi(2S)}_{ee}{\cal B}_{\psi(2S) \to \phi f_0}
{\cal B}_{\phi\to K^+K^-}{\cal B}_{ f_0     \to \pipi}   $  &
$0.17  \pm 0.06  \pm 0.02$ &
$0.22  \pm 0.08  \pm 0.02$ &
$0.068 \pm 0.024(S=1.1)$\\
\end{tabular}
\end{ruledtabular}
\end{table*}
A systematic study of the $J/\psi$ and $\psi(2S)$ decays to light hadrons was
performed in the BABAR
experiment~\cite{babr3pi,babrpp,BBLS,babr4pi,babr6pi,babr5pi}.
An example of the $J/\psi$ signal for $J/\psi\to\pi^+\pi^-\pi^0 $, 
one of the most probable 
$J/\psi$ decay modes, is shown in 
Fig.~\ref{JPsimass}~\cite{babr3pi}. It is seen that the 
nonresonant background is small. 
From the number of events at the peak, 
the product
$\Gamma(J/\psi\to e^+e^-)B(J/\psi\to 3\pi)=0.122\pm0.005\pm0.008$
keV was determined.
Using the $J/\psi$ electronic width value, known from
the ISR study of the $J/\psi\to \mu^+\mu^-$ decay, the branching fraction 
$B(J/\psi\to 3\pi)=(2.18\pm0.19)\%$ was calculated, which 
differed by about 50\% from the PDG value, $(1.47\pm0.13)\%$, 
available when the analysis~\cite{babr3pi} was carried out. 
Similar deviation was observed in the
BES experiment~\cite{bes3pi} where 
$B(J/\psi\to 3\pi)=(2.10\pm0.11)\%$ was obtained.

\begin{figure}
\includegraphics[width=.4\textwidth]{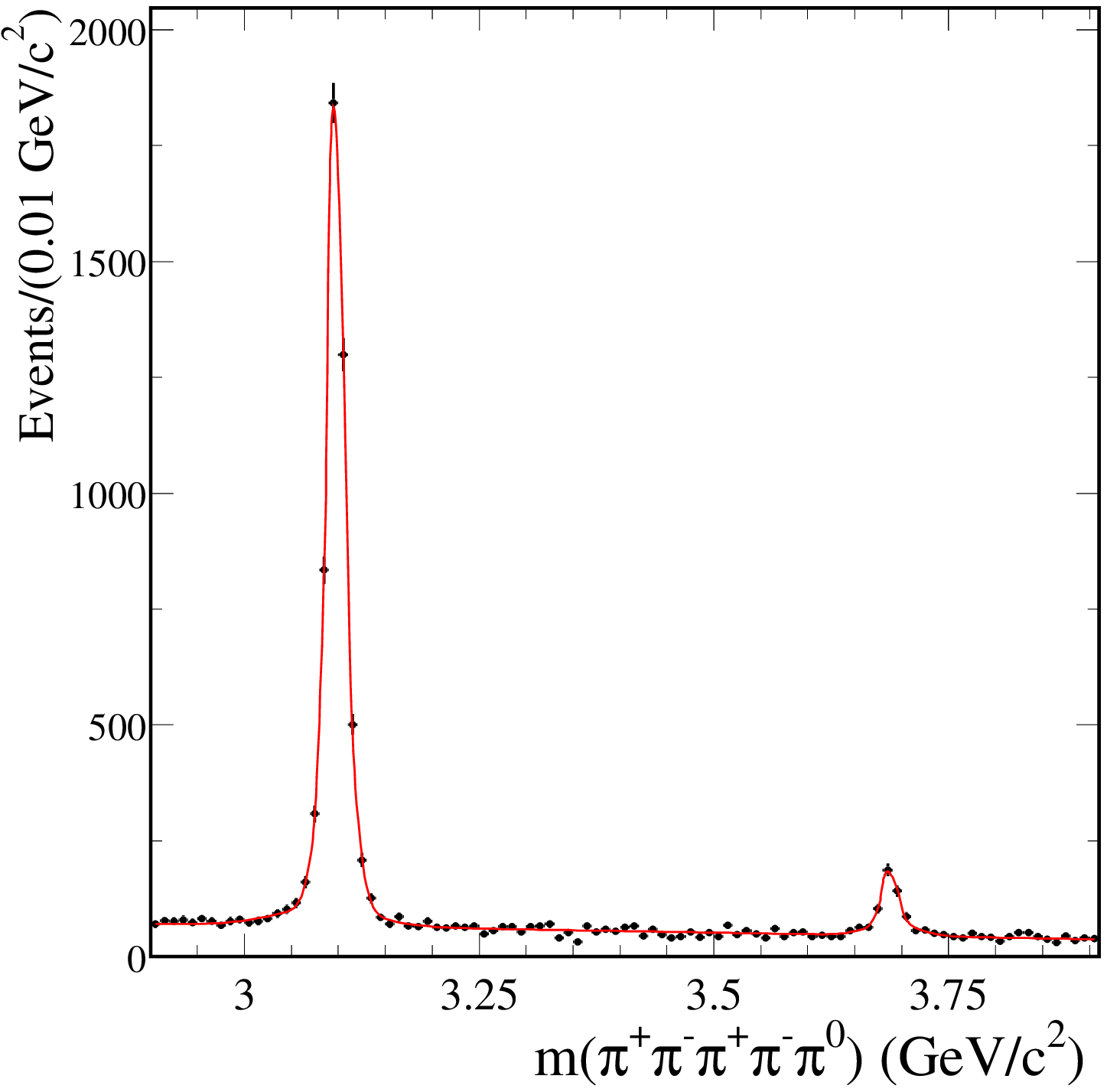}
\caption{
The $2(\pi^+\pi^-)\pi^0$ mass distribution for ISR
$e^+ e^-\to 2(\pi^+\pi^-)\pi^0\gamma$ events in the $J/\psi$--$\psi(2S)$
mass region~\cite{babr5pi}.
}
\label{jpsi5pi}
\end{figure}
Another example of a $J/\psi$ decay mode with a rather high
probability, which was studied using ISR, is shown in
Fig.~\ref{jpsi5pi}, where the signals of $J/\psi\to
2(\pi^+\pi^-)\pi^0$ and $\psi(2S)\to 2(\pi^+\pi^-)\pi^0$ are clearly
seen~\cite{babr5pi}. Again, by determining the number of peak events
over the nonresonant background and using Eq.~(\ref{gee}), the product
$\Gamma(J/\psi\to e^+e^-)B(J/\psi\to
\pi^+\pi^-\pi^+\pi^-\pi^0)=(3.03\pm0.05\pm0.18)\times 
10^{-4}$ keV was determined. The value of the $J/\psi\to
\pi^+\pi^-\pi^+\pi^-\pi^0$ branching fraction obtained from this
product, $(5.46\pm0.09\pm0.34)\%$, differed by about 5$\sigma$
from the PDG value, $(3.37\pm0.26)\%$, available when
the analysis~\cite{babr5pi} was carried out. 
As was shown in Sec.~\ref{5pi}, the five-pion final state includes
production of many intermediate resonances. All of them are seen in
the $J/\psi\to 5\pi$ decay. This may be a source of 
the systematic error unaccounted in previous measurements of
the decay. The detection efficiency in the ISR method
with a tagged photon is weakly sensitive to the dynamics of the
$J/\psi\to 5\pi$ decay. The model uncertainty in the detection efficiency
for the BABAR measurement~\cite{babr5pi} was estimeted
from the difference in efficiency values for phase-space generated
five-pion events and  events generated for the $\omega\pi^+\pi^-$ or
$\eta\pi^+\pi^-$ final states. It was found to be less than 3\%.

\begin{figure}[tbh]
\includegraphics[width=4.1cm]{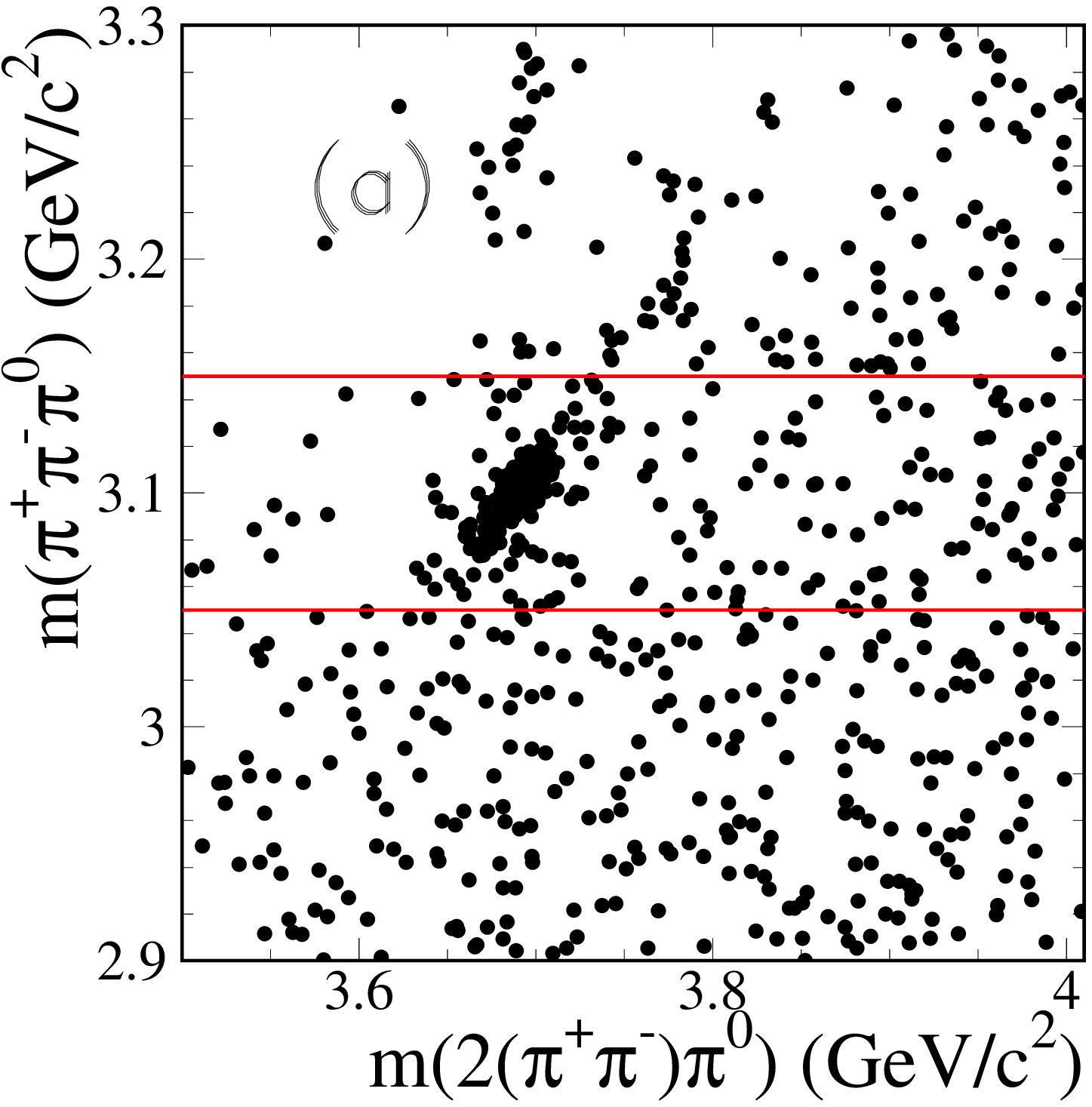}
\includegraphics[width=4.1cm]{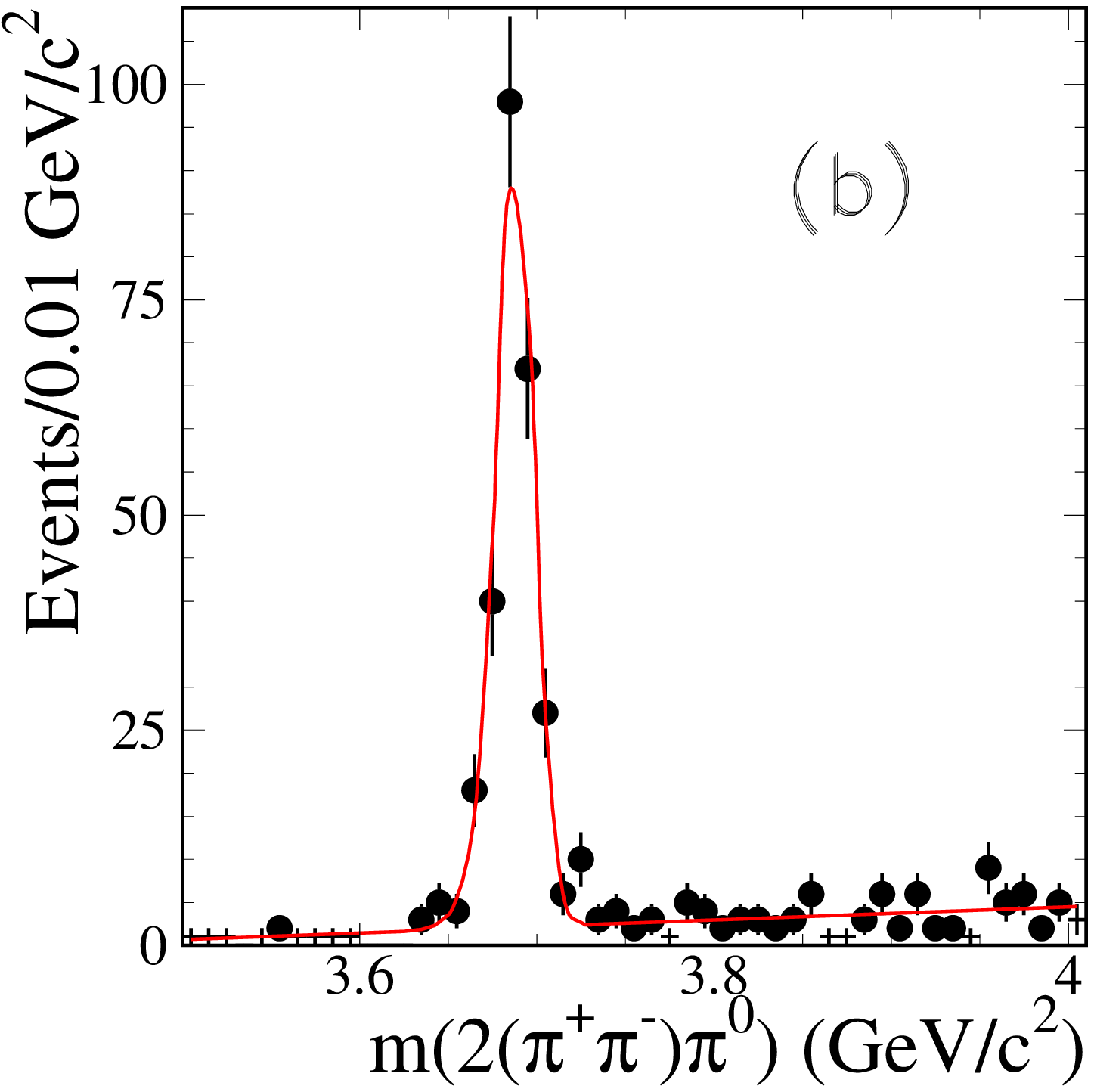}
\caption{
(a) The three-pion combination closest to $J/\psi$ mass versus
five-pion mass.
(b) The five-pion mass for
events with the three-pion mass in the $\pm 50$ MeV window around 
$J/\psi$ mass~\cite{babr5pi}.
\label{psi2s_chain}}
\end{figure}    
A part of events from the $\psi(2S)$ peak 
comes from the decay chain
$\psi(2S)\to J/\psi\pi^+\pi^-\to 2(\pi^+\pi^-)\pi^0$ with the $J/\psi$ 
decaying to three pions. To select these events, the $\pi^+\pi^-\pi^0$ 
combination with the invariant mass closest to the $J/\psi$ mass is chosen.
Figure~\ref{psi2s_chain}(a) shows the scatter plot of this three-pion mass
versus the five-pion mass. A clear signal from the above decay chain is seen. 
The five-pion mass spectrum for events with the $\pi^+\pi^-\pi^0$ mass within 
the $\pm 0.05$~\gevcc\ window around the $J/\psi$ mass is shown 
in Fig.~\ref{psi2s_chain}(b). From the fit to the mass spectrum with a
double-Gaussian function the number of detected 
$\psi(2S)\to J/\psi\pi^+\pi^-\to 2(\pi^+\pi^-)\pi^0$ events was determined to
be $256\pm 17$, and the triple product
\begin{eqnarray*}
  {\cal B}(\psi(2S)\to J/\psi\pi^+\pi^-) 
{\cal B}(J/\psi\to\pi^+\pi^-\pi^0)\\
\times\Gamma(\psi(2S)\to e^+e^-)=(1.86\pm0.12\pm0.11)\times 10^{-2}\mbox{ keV}
\end{eqnarray*}
was obtained. 
By using the world-average $\Gamma(\psi(2S)\to e^+e^-)$ and 
${\cal B}(\psi(2S)\to J/\psi\pi^+\pi^-)$ values~\cite{PDG08}, 
the branching fraction ${\cal B}(J/\psi\to\pi^+\pi^-\pi^0) =
(2.36\pm 0.16\pm 0.16)\%$ was obtained, which is in good agreement 
with the BABAR measurement in the $3\pi$ final state:
${\cal B}(J/\psi\to\pi^+\pi^-\pi^0) = (2.18\pm 0.19)\%$~\cite{babr3pi}.
This, in particular, confirms the correctness of the normalization procedure
used for the measurement of ${\cal B}(J/\psi\to 5\pi)$.

 Table~\ref{jpsitab} presents 
measurements of the $J/\psi$ and $\psi(2S)$ decay rates performed 
with the BABAR
detector via ISR for many multihadron final states. 
The current PDG values are shown in the last 
column~\cite{pdg10}. In most of the
cases these values are close to those of BABAR emphasizing   
their importance. Note also that in a few cases the scale factor is
significantly higher than one indicating a large difference between
the BABAR measurement and previous results.

\begin{figure}[tbh]
\begin{center}
\includegraphics[width=8.2cm]{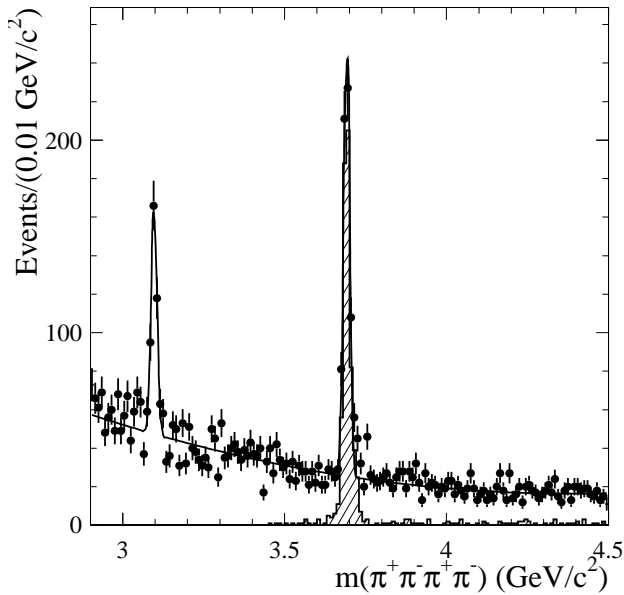}
\caption{
The $2(\pipi)$ mass distribution for ISR-produced
$e^+ e^-\to 2(\pipi)$ events  
in the mass region around the $J/\psi$ and $\psi(2S)$~\cite{babr4pi};
there are clear signals at the $J/\psi$ and 
        $\psi(2S)$ mass positions. The latter is dominates\d by the 
$\psi(2S)\to J/\psi\pipi\to\mumu\pipi$ transition;
selected events with two muons from the $J/\psi$ decays are shown 
by the shaded histogram.  
}
\label{jpsi4pi}
\end{center}
\end{figure}
\begin{figure}[tbh]
\begin{center}
\includegraphics[width=8.2cm]{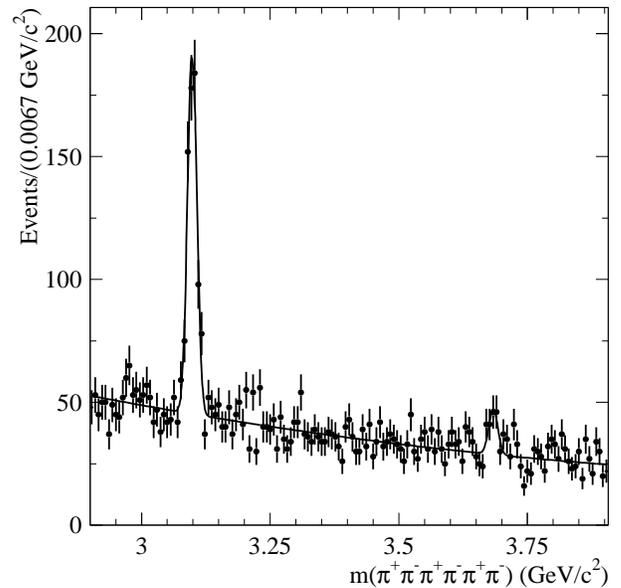}
\caption{
The $3(\pipi)$ mass distribution for ISR-produced
$e^+ e^-\to 3(\pipi)$ events  
in the mass region around the $J/\psi$ and $\psi(2S)$~\cite{babr6pi};
there are clear signals at the $J/\psi$ and 
        $\psi(2S)$ mass positions. 
}
\label{jpsi6pi}
\end{center}
\end{figure}
As can be seen from Table~\ref{jpsitab}, the $J/\psi$ decay rates to
even numbers of pions ($4\pi, 6\pi$ {\ldots}) are much smaller compared to
the decays to odd numbers of pions.    
Indeed, a strong decay of the $J/\psi$ to an even number of pions is
forbidden by G-parity conservation. 
It is expected that this decay is dominated by a single photon transition,
$J/\psi\to\gamma^\ast\to n\pi$.
No such suppression occurs for the strong $J/\psi$ decays
to other modes, such as to three or five pions, which mainly 
proceed through three gluons.
The $2(\pipi)$ and $3(\pipi)$ mass spectra for events of the ISR processes
$e^+ e^-\to 2(\pipi)\gamma$ and $e^+ e^-\to 3(\pipi)\gamma$,
in the mass regions of
the $J/\psi$ and $\psi(2S)$ resonances, are shown in Figs.~\ref{jpsi4pi}
and \ref{jpsi6pi}, respectively. From the fits to the mass spectra the
numbers of $J/\psi$ and $\psi(2S)$ events and also the level of the nonresonant
background are determined. The latter is proportional to the value
of the nonresonant $e^+ e^-\to 2(\pipi)$ or $e^+ e^-\to 3(\pipi)$
cross section. In the BABAR paper~\cite{babr6pi} the ratio
\begin{equation}
R_{J/\psi}=\frac{6\pi^2\Gamma(J/\psi \to e^+e^-){\cal B}(J/\psi\to
f)/m_{J/\psi}^2}{\sigma_{e^+ e^-\to f}(m_{J/\psi})}
\label{rjpsi}
\end{equation}
is calculated, where $\sigma_{e^+ e^-\to f}$ is the value of 
the nonresonant cross section to the final state $f$ at the $J/\psi$ mass.
The numerator of the ratio represents the integral over the $J/\psi$ 
excitation curve. The $R_{J/\psi}$ values for the $4\pi$, $6\pi$, $2K2\pi$,
$2K4\pi$, and $4K$ final states are listed in Table~\ref{jpsi_to_xs}
together with the $R_{J/\psi}$ value obtained for the $\mumu$ final
state. The $R_{J/\psi}$ 
values for the $4\pi$ and $6\pi$ final states are closer to that for
$\mumu$ compared to the final states with kaons and  indicate
that the single-photon exchange dominates for the $J/\psi$ decays into these 
modes. For the $J/\psi$ decays to the final states with kaons, which can
contain a sizeable isoscalar component, the single-photon transition 
is expected to be less dominant, as indicated by the 
larger central values  of  the ratios.
\begin{table}[tbh]
\caption{
Ratios of the $J/\psi$ partial production rates to continuum cross
sections $R_{J/\psi}$ (see Eq.(~\ref{rjpsi})). The result for $\mumu$ 
is from Ref.~\cite{babrmu}.
The result for $3(\pipi)$ is from Ref.~\cite{babr6pi} and
the results for $2(\pipi)$, $K^+ K^- \pipi$ and $K^+ K^- K^+ K^-$ are
from  Ref.~\cite{babr4pi}.
}
\label{jpsi_to_xs}
\begin{ruledtabular}
\begin{tabular}{cc}
Final state & 
$R_{J/\psi}$
(MeV) \\ \hline
$2(\pipi)$ & $85.1 \pm 7.9$ \\
$3(\pipi)$ & $106 \pm 10$ \\
$2(\pipi\pi^0)$ & $99.1 \pm 6.5$ \\
$K^+ K^- 2(\pipi)$ & $122 \pm 10$ \\
$K^+ K^- \pipi$ & $166 \pm 19$ \\ 
$K^+ K^- K^+ K^-$ & $138 \pm 32$ \\
$\mumu$ & $84.12 \pm 2.69$ 
\end{tabular}
\end{ruledtabular}
\end{table}

\section{ISR studies in the charmonium region}
In this chapter we will discuss recent progress in the charmonium
spectroscopy mainly achieved due to the application of the ISR method,
see also recent reviews~\cite{ufn,qwg}.
We will start with the description of the open charm final states 
addressing later so-called charmonium-like states, presumably
states with hidden charm.
 
\subsection{Final states with open charm}
For a quarter of a century our knowledge of the vector charmonia
above the threshold of open charm production
(throughout this section referred to as $\psi$ states) was based on 
the pioneer experiments of Mark-I~\cite{mk1} and DASP~\cite{dasp}.
Even such basic parameters of the $\psi$ mesons as mass, width and 
leptonic width were known with large uncertainties mainly determined
by low statistics of the old experiments.
In Ref.~\cite{seth} an attempt was made to use the updated information
on the $R$ values from Crystal Ball~\cite{cbl} and
BES~\cite{besr} to improve these parameters. Finally, the 
BES Collaboration performed a
global fit of the data on $R$ collected by BES in the energy range 
from 3.7 to 5~GeV~\cite{besfit}. In some cases the obtained values of mass,
width and leptonic width for the $\psi$ states differ significantly 
from the older values and still suffer from big uncertainties caused
by insufficient statistics and model dependence primarily due to
numerous thresholds of charm production opening in this energy region.  
It became clear that serious progress would be possible after
tedious exclusive studies, which recently became possible due
to ISR analyses of BABAR and Belle based on very large integrated
luminosities.
 
Exclusive $e^+e^-$ cross sections for hadronic final states containing
charm mesons in the $\sqrt{s}$=3.7-5~GeV/$c^2$ energy range were 
measured by BABAR~\cite{babrdd1,babrdd2} and  
Belle~\cite{beldd,beldst,bel4415,beldstd,belamc} using 
ISR to reach the charmonium region. Note that in these analyses
Belle systematically employs a partial reconstruction technique
to increase the detection efficiency and suppress background.  

\begin{figure}[h]
\includegraphics[width=0.49\textwidth]{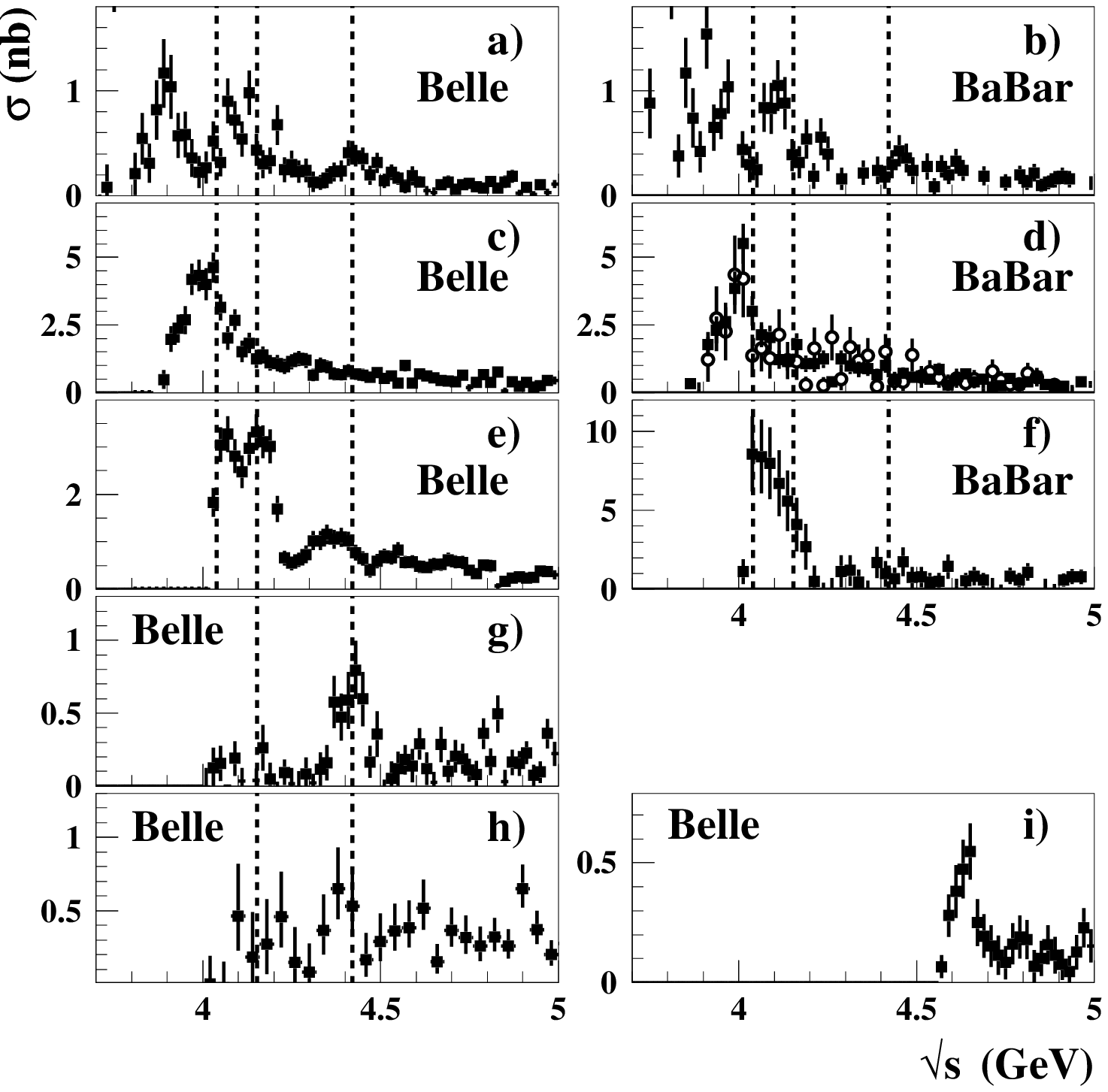}
\caption{Measured $e^+e^-$ exclusive open-charm cross sections for
  $\sqrt{s}$=3.7-5.0~GeV/$c^2$ 
  from Belle and BABAR, showing
  (a)~$D\bar{D}$~\cite{beldd},
  (b)~$D\bar{D}$~\cite{babrdd1} , 
  (c)~$D^+D^{*-}$~\cite{beldst}; 
  (d)~$D\bar{D}^*$ for $D$=$D^0$ (solid squares) 
  and $D$=$D^{\pm}$ (open circles)~\cite{babrdd2}; 
  (e)~$D^{*+}D^{*-}$~\cite{beldst}; 
  (f)~$D^*\bar{D}^*$~\cite{babrdd2}; 
  (g)~$D^0D^-\pi^+$~\cite{bel4415};
  (h)~$D^0D^{*-}\pi^+$~\cite{beldstd}; 
  (i)~$\Lambda_c\bar{\Lambda}_c$~\cite{belamc}. 
  Vertical dashed lines indicate  $\psi$ masses
  in the region.
\label{fig:BelleBaBar}}
\end{figure}

The $D\bar{D}$ cross sections in the entire charm energy range from
Belle~\cite{beldd} and BABAR~\cite{babrdd1}
are shown  in Figs.~\ref{fig:BelleBaBar}(a),(b) and are consistent
with each other.
Both exhibit clear evidence of structures near 4.1 and 4.4~GeV/$c^2$. 
They also observe a structure 
(Figs.~\ref{fig:BelleBaBar}(a) and (b))
at 3900 MeV
which must be taken into account to describe the
 $D\bar{D}$ cross section and $R$
in the region between the $\psi(3770)$ and $\psi(4040)$.
This enhancement is not considered as a new $c\bar{c}$ resonance,
as it is qualitatively consistent with the energy dependence of the
sum of the cross sections for various channels opening in this 
energy range predicted in
a coupled-channel model~\cite{eichten80}. 
The $D\bar{D}^*$ cross sections from Belle~\cite{beldst}
and BABAR~\cite{babrdd2} shown in Figs.~\ref{fig:BelleBaBar}(c),(d)
exhibit a single broad peak near threshold (close to the
$\psi(4040)$ position),
whereas the $D^*\bar{D}^*$   results from Belle~\cite{beldst}
and BABAR~\cite{babrdd2} (Figs.~\ref{fig:BelleBaBar}(e),(f)) 
feature several local maxima and minima in this energy range.

BABAR~\cite{babrdd2}
performed unbinned maximum likelihood fits to the 
 $D\bar{D}$, $D\bar{D}^*$, and $D^*\bar{D}^*$ spectra. The expected 
$\psi$ signals were
parameterized by $p$-wave relativistic Breit-Wigner (RBW) functions
with their parameters fixed to the PDG08
values~\cite{PDG08}. 
An interference between the resonances and
the non-resonant contributions was required in the fit. 
The computed ratios of the branching
fractions for the $\psi$ resonances and the quark model predictions
are presented in Table~\ref{tab:psir}. The BABAR results 
deviate from some of the theoretical expectations, which
often differ from each other. 
\begin{table}
\caption{Ratios of branching fractions for the 
$\psi(4040)$, $\psi(4160)$ and $\psi(4415)$ resonances from 
BABAR~\cite{babrdd2}. 
  Theoretical expectations are from models denoted
  $^3P_0$~\cite{barnes05}, 
  $C^3$~\cite{eichten05},
  and $\rho K \rho$~\cite{swanson06}.
\label{tab:psir}}
\setlength{\tabcolsep}{0.30pc}
\begin{center}
\begin{tabular}{c|r|cccc}
\hline\hline
\rule[10pt]{-1mm}{0mm}
State & Ratio\ \ \ \ \ & Measured & $^3P_0$ & $C^3$ & $\rho K \rho$ \\
\hline
\rule[10pt]{-1mm}{0mm}
$\psi(4040)$ & $D\bar{D}/D\bar{D}^*$ & 0.24$\pm$0.05$\pm$0.12 & 0.003 
& & 0.14 \\
& $D^*\bar{D}^*/D\bar{D}^*$ & 0.18$\pm$0.14$\pm$0.03 & 1.0   &      & 0.29 \\
$\psi(4160)$ & $ D\bar{D}/ D^*\bar{D}^*$   & 0.02$\pm$0.03$\pm$0.02 
& 0.46  & 0.08 &      \\
& $D\bar{D}^*/D^*\bar{D}^*$ & 0.34$\pm$0.14$\pm$0.05 & 0.011 & 0.16 &      \\
$\psi(4415)$ &  $D\bar{D}/ D^*\bar{D}^*$ & 0.14$\pm$0.12$\pm$0.03 
& 0.025 &      &      \\
& $D\bar{D}^*/D^*\bar{D}^*$ & 0.17$\pm$0.25$\pm$0.03 & 0.14  &      &      \\
\hline\hline
\end{tabular}
\end{center}
\end{table}

The $e^+e^- \to D^0D^-\pi^+$ cross section measured by
Belle~\cite{bel4415} 
is shown in Fig.~\ref{fig:BelleBaBar}(g)
and exhibits an unambiguous $\psi(4415)$ signal. 
A study of the resonant structure 
shows clear signals for the $\bar{D}^*_2(2460)^0$ and ${D}^*_2(2460)^+$
mesons and constructive interference between the neutral
 $D^0\bar{D}^*_2(2460)^0$
and the charged   $D^-{D}^*_2(2460)^+$ decay amplitudes.
Belle performed a likelihood fit to the  $D^0D^-\pi^+$  
mass distribution with a 
$\psi(4415)$ signal parameterized by an $S$-wave RBW function. The
significance for the signal is $\sim$10$\sigma$ and the peak mass
and total width are in good agreement
with the PDG06~\cite{PDG06} values and the BES fit
results~\cite{besfit}. 
The product of the  branching fractions
${\cal B}(\psi(4415) \to D\bar{D}^*_2(2460)) \times
{\cal B}(\bar{D}^*_2(2460) \to  D\pi^+)$ 
was found to be between 10\% and 20\% depending
on the $\psi(4415)$ parameterization. The
non-resonant  $D^0D^-\pi^+$ branching fraction  was found to be $<$22\% of 
\linebreak
${\cal B}(\psi(4415) \to D\bar{D}^*_2(2460) \to D^0D^-\pi^+)$. 
Similarly, the energy dependence of the cross section of the
$D^0D^{*-}\pi^+$ final state, shown in
Fig.~\ref{fig:BelleBaBar}(h), has
been measured by Belle~\cite{beldstd};
a marginal signal of the $\psi(4415)$ is found (3.1$\sigma$), and
its branching fraction was limited to $<$10.6\%.
Very recently BABAR~\cite{babrdd3} and Belle~\cite{belds} reported
consistent results on the cross sections of 
$D^+_sD^-_s,~D^+_sD^{*+}_s$ and $D^{*+}_sD^{*-}_s$.

The Belle collaboration has also measured the cross section of the
process \epem\ $\to \Lambda_c^+\Lambda_c^-$~\cite{belamc}.
Because of the large number of the $\Lambda_c$ decay channels 
with small branching fractions 
full reconstruction of  both $\Lambda_c$ is not
effective. The strategy of a search for 
$\Lambda_c^+\Lambda_c^-\gamma$ events at Belle \cite{belamc} was 
the following: one of the $\Lambda_c$ baryons was reconstructed using three
decay modes $pK^0_S,~pK^-\pi^+,~\Lambda\pi^+$. Then in the spectrum of masses
recoiling against the $\Lambda_c^+\gamma$ system, a peak at the $\Lambda_c^-$
mass was searched for. This peak presumably corresponded 
to the process $e^+e^-\to
\Lambda_c^+\Lambda_c^-\gamma$. 
The resulting exclusive cross section of the process
$e^+e^-\to \Lambda_c^+\Lambda_c^-$ is shown in
 Fig.\ref{fig:BelleBaBar}(i). 
The cross section is nearly flat from the threshold up to 5.4 GeV/c$^2$
except the region just above threshold, where a peak with the mass 
$M=4634^{+10}_{-30}$  MeV/c$^2$, width $\Gamma=92^{+40}_{-27}$  MeV/c$^2$
and significance of 8.2 $\sigma$ is observed. The state is denoted as
$X(4630)$ and the product of the branching fractions measured for it is
${\cal B}(e^+e^-)\times{\cal B}(\Lambda_c\bar{\Lambda}_c)
=(0.68\pm0.33)\times 10^{-6}$. 
 The nature of this enhancement remains unclear. Although both mass and
width of the $X(4630)$ 
are consistent within errors with those of another Belle state $Y(4660)$,
that
was found in $\psi(2S)\pi\pi$ decays via ISR and is described in the
next section~\cite{bely431},
this could be coincidence and does not exclude
other interpretations.

Although in general the energy behavior of the exclusive
cross sections from BABAR and Belle qualitatively follows the
expectations of the coupled-channel model~\cite{eichten80},
some features are not reproduced by theory. This is confirmed 
by the measurement  of CLEO~\cite{cleoxs}, which scanned the energy
range
between 3.97 and 4.26~GeV and
reported the cross sections for final states consisting of two charm
mesons
($D\bar{D},~D^*\bar{D},~D^*\bar{D}^*,~D^+_sD^-_s,~D^{*+}_sD^-_s$,
 and $D^{*+}_sD^{*-}_s$) as well as for those in which the charm-meson
pair is accompanied with a pion. 
The updated potential model predictions of 
Eichten~\cite{eichten80,eichten05}
fail to describe many features of the data.

\subsection{New charmonium-like states}
\begin{figure}[h]
\includegraphics[width=0.49\textwidth]{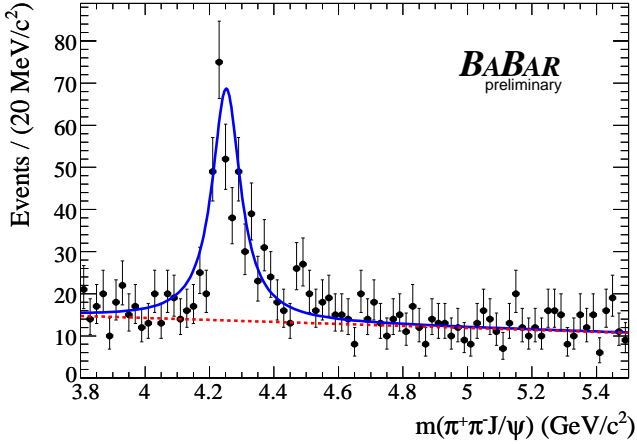}
\caption{The invariant mass of $J/\psi\pi^+\pi^-$ 
         candidates produced in initial state radiation,
         $e^+e^-\to\gamma_{ISR}\,J/\psi\pi^+\pi^-$.
         Points with error bars represent data,
         and the curves show the fit (solid) to
         a signal plus a linear background (dashed)~\cite{babry422}.
         \label{fig:y4260_babar}}
\end{figure}

\begin{figure}[h]
\includegraphics[width=0.38\textwidth,angle=-90]{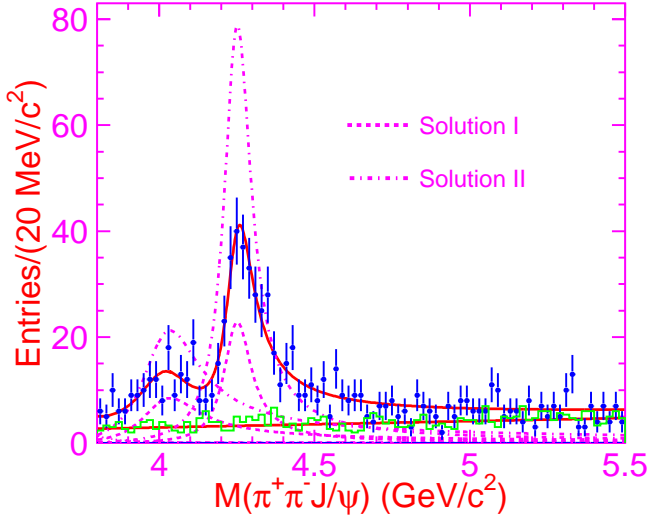}
\caption{The invariant mass of $J/\psi\pi^+\pi^-$ 
         candidates produced in initial state radiation
         studied by Belle~\cite{bely421},
         with $J/\psi$-sidebands already subtracted,
         unlike Fig.~\ref{fig:y4260_babar}.
         Points with error bars represent data,
         the solid curve shows the best fits to the data
         to two resonances including interference 
         with a floating phase, and the dashed and dashed-dot curves show
         the two pair of individual resonance contributions for
         the two equally probable best-fit phases.
         \label{fig:y4260_belle}}
\end{figure}

\begin{figure}[h]
\includegraphics[width=0.485\textwidth]{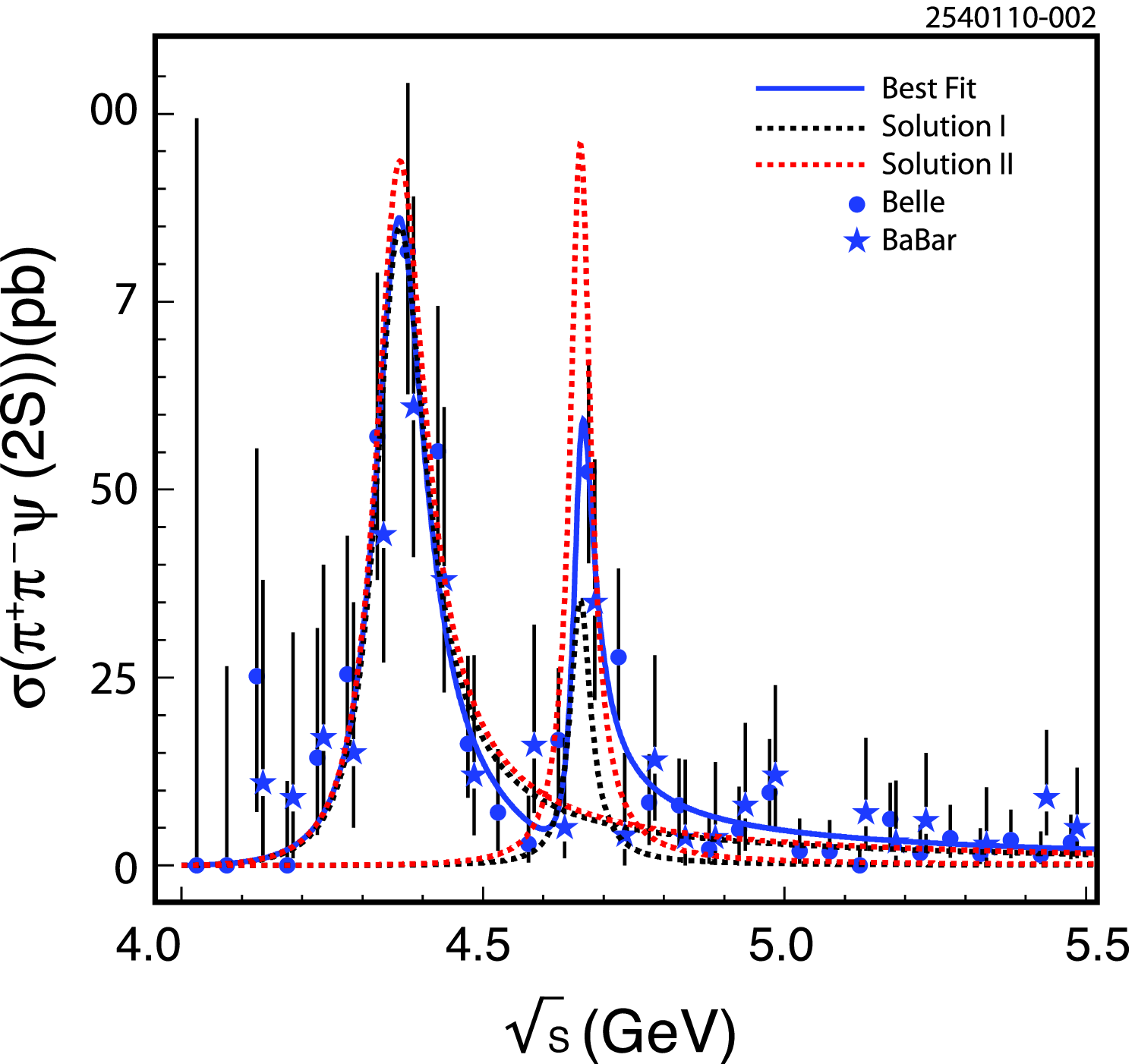}
\caption{From a binned maximum likelihood fit~\cite{bely432} 
         of combined Belle and BABAR data,
         the $\psi(2S)\pi^+\pi^-$ invariant-mass cross section as
         a function of $\sqrt{s}$. The solid circles
         and stars show the Belle and BABAR data, respectively. 
         The solid curve shows the best fits to the data
         to two resonances including interference 
         with a floating phase, and the dashed curves show
         the contributions of two pairs of individual resonances  for
         the two equally probable best-fit phases.
\label{fig:fit_liu}}
\end{figure}

The first observation of an unexpected vector charmonium-like
state was made by BABAR~\cite{babry421} in ISR production 
of $Y(4260) \to J/\psi\pi^+\pi^-$, which was later updated~\cite{babry422} 
with twice the data, as shown in Fig.~\ref{fig:y4260_babar}.
CLEO~\cite{cleoy421} and Belle~\cite{bely421} confirmed the 
BABAR result, but Belle also found a smaller, broader structure at 
4008~MeV/$c^2$, as seen in 
Fig.~\ref{fig:y4260_belle}. Aside 
from the lower mass state, for which the updated BABAR~\cite{babry422} 
analysis placed an upper limit, the three sets of measurements were 
quite consistent in mass and width, as shown in Table~\ref{tab:y4260},
but only roughly so in strength.
BABAR~\cite{babry431} found one more apparent enhancement
$Y(4360)$ in $\psi(2S)\pi^+\pi^-$, which Belle~\cite{bely431} measured
with somewhat larger mass and smaller width,
as seen in Table~\ref{tab:y4360}. 
Belle also found a second structure near 
4660~MeV/$c^2$ in the same final state, as seen in Fig.~\ref{fig:fit_liu}.
(A combined fit~\cite{bely432} to Belle and BABAR  $\psi(2S)\pi^+\pi^-$
 data found consistency between them.)
Because dipion transitions between vector quarkonia are commonplace
for charmonium and bottomonium, it was natural 
to ascribe the $Y$'s to excited
vector charmonia. 
A number of additional features of these states
are in conflict with this hypothesis. 
Only one, $Y(4660)$, is remotely near a predicted 
$1^{--}$ $c\bar{c}$ state ($1\,^3D_1$). The $Y(4260)$ and
$Y(4360)$ did not show up in inclusive hadronic cross 
section ($R$) measurements~\cite{besr},
as would be expected
of such states (there is no fine-grained $R$-scan data near
$Y(4660)$).
\begin{table}
\caption{Measured properties of the $Y(4260) \to J/\psi\pi^+\pi^-$.
The Belle~\cite{bely431} single-resonance fit result is quoted to
allow for comparison to the other two.
\label{tab:y4260}}
\renewcommand{\tabcolsep}{1mm}
\begin{center}
\begin{tabular}{c|cc}
\hline\hline
\rule[10pt]{-1mm}{0mm}
Quantity & Value & From ($\chi^2$/ndf)\\
\hline
\rule[10pt]{-1mm}{0mm}
$M$  & 4259$\pm$8$^{+2}_{-6}$ & BABAR~\cite{babry422} \\
(MeV/$c^2$)     & 4263$\pm$6             & Belle~\cite{bely421}  \\
     & 4284$^{+17}_{-16}$$\pm$4 & CLEO~\cite{cleoy421}   \\
     & 4263$\pm$5             & Avg (1.8/2)\\
\hline
\rule[10pt]{-1mm}{0mm}
$\Gamma$  & 88$\pm$23$^{+6}_{-4}$ & BABAR~\cite{babry422} \\
 (MeV)         &126$\pm$18             & Belle~\cite{bely421}  \\
          & 73$^{+39}_{-25}$$\pm$5& CLEO~\cite{cleoy421}   \\
          & 108$\pm$15            & Avg (2.4/2)\\
\hline
\rule[10pt]{-1mm}{0mm}
${\cal B}\times\Gamma_{ee}$  &5.5$\pm$1.0$^{+0.8}_{-0.7}$ & 
BABAR~\cite{babry422} \\
(eV)   &9.7$\pm$1.1                  & Belle~\cite{bely421}  \\
   &8.9$^{+3.9}_{-3.1}\pm1.8$   & CLEO~\cite{cleoy421}   \\
                             &8.0$\pm$1.4                  & Avg (6.1/2)\\
\hline\hline
\end{tabular}
\end{center}
\end{table}

\begin{table}[t]
\caption{Measured properties of the two enhancements
found in the $\psi(2S)\pi^+\pi^-$ mass distribution, 
the $Y(4360)$ and $Y(4660)$.
Liu {\it et al.}~\cite{bely432} performed a binned
maximum likelihood fit to the combined Belle and BABAR
cross section distributions (Fig.~\ref{fig:fit_liu}).
\label{tab:y4360}}
\renewcommand{\tabcolsep}{1mm}
\begin{center}
\begin{tabular}{c|cc}
\hline\hline
\rule[10pt]{-1mm}{0mm}
Quantity & Value & From ($\chi^2$/d.o.f.)\\
\hline
\rule[10pt]{-1mm}{0mm}
$M$  & 4324$\pm$24 & BABAR~\cite{babry431} \\
(MeV/$c^2$)& 4361$\pm$9$\pm$9 & Belle~\cite{bely431}  \\
     & 4353$\pm$15      & Avg (1.8/1)\\
     & 4355$^{+~9}_{-10}\pm$9 & Liu~\cite{bely432}\\
\hline
\rule[10pt]{-1mm}{0mm}
$\Gamma$  & 172$\pm$33 & BABAR~\cite{babry431} \\
   (MeV)  &  74$\pm$15$\pm$10 & Belle~\cite{bely431}  \\
          &  96$\pm$42        & Avg (6.8/1)\\
          & 103$^{+17}_{-15}\pm$11 & Liu~\cite{bely432}\\
\hline
\rule[10pt]{-1mm}{0mm}
$M$  & 4664$\pm$11$\pm$5 & Belle~\cite{bely431}  \\
(MeV/$c^2$)& 4661$^{+9}_{-8}\pm$6 & Liu~\cite{bely432}\\
\hline
\rule[10pt]{-1mm}{0mm}
$\Gamma$  &  48$\pm$15$\pm$3 & Belle~\cite{bely431}  \\
   (MeV)  &  42$^{+17}_{-12}\pm$6 & Liu~\cite{bely432}\\
\hline\hline
\end{tabular}
\end{center}
\end{table}
A comparison of the measured $J/\psi\pi^+\pi^-$ and total 
hadronic cross sections 
in the $\sqrt{s}\simeq 4260$~MeV region
yields a lower bound for $\Gamma(Y \to J/\psi\pi^+\pi^-)$$>$508~keV 
at 90\%~C.L., 
an order of magnitude higher than expectations for conventional 
vector charmonium states~\cite{mo06}. 
Charmonium would also feature dominant open charm decays,  
exceeding those of dipion transitions
by a factor expected to be $\gtrsim$100,
because such is the case for $\psi(3770)$ and $\psi(4160)$.
As summarized in Table~\ref{tab:opencharm}, 
no such evidence has been found, significantly
narrowing any window for either charmonia
or, in some cases, quark-gluon hybrid interpretations.
CLEO~\cite{cleoy422} studied direct production of
the $Y(4260)$ in $e^+e^-$ collisions
and identified the only non-$J/\psi\pi^+\pi^-$ decay modes seen so far, 
$J/\psi\pi^0\pi^0$ and
$J/\psi K^+K^-$, occuring at roughly half and one-sixth,
respectively, of the $J/\psi\pi^+\pi^-$ rate. The $J/\psi K^+K^-$
decay mode was also observed by Belle~\cite{bely422}.

\begin{table}
\caption{Upper limits at 90\% C.L. on the ratios 
$\sigma(e^+e^- \to
  Y \to T) / \sigma(e^+e^- \to Y \to J/\psi\pi^+\pi^-)$ at $\sqrt{s} =
  4.26$~GeV/$c^2$ (CLEO~\cite{cleoxs}) and 
  ${\cal B}(Y \to T) / {\cal B}(Y \to J/\psi\pi^+\pi^-)$ (for  $Y(4260)$) 
  (BABAR~\cite{babrdd1,babrdd2} and Belle~\cite{beldst}), 
  where $T$ is an open charm final state.\label{tab:opencharm}}
\setlength{\tabcolsep}{0.80pc}
\begin{center}
\begin{tabular}{l|l}
\hline\hline
\rule[10pt]{-1mm}{0mm}
  \ \ \ \ $T$  & $Y(4260)$  \\
\hline
\rule[10pt]{-1mm}{0mm}
$D\bar{D}$          & 4.0~\cite{cleoxs}      \\
                    & 7.6~\cite{babrdd1}     \\
$D\bar{D}^*$        & 45~\cite{cleoxs}       \\
                    & 34~\cite{babrdd2}      \\
$D^*\bar{D}^*$      & 11~\cite{cleoxs}       \\
                    & 40~\cite{babrdd2}      \\
$D\bar{D}^*\pi$     & 15~\cite{cleoxs}       \\
                    & 9~\cite{beldst}   \\
$D^*\bar{D}^*\pi$   & 8.2~\cite{cleoxs}   \\
$D_s\bar{D}_s$      & 1.3~\cite{cleoxs}   \\ 
$D_s\bar{D}_s^*$    & 0.8~\cite{cleoxs}   \\ 
$D_s^*\bar{D}_s^*$  & 9.5~\cite{cleoxs}   \\
\hline
\hline
\end{tabular}
\end{center}
\end{table}

\begin{figure}[h]
\includegraphics[width=0.4\textwidth]{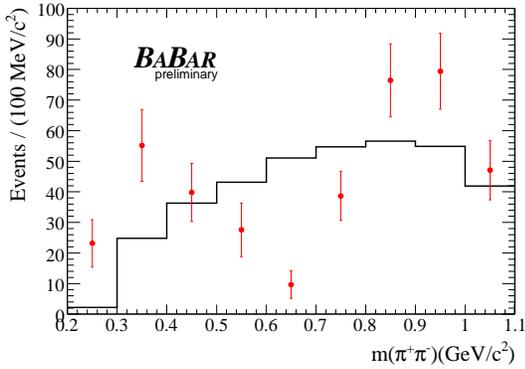}
\caption{The dipion invariant
         mass distribution on ISR-produced $Y(4260) \to
         J/\psi\pi^+\pi^-$ decays, where points represent data 
and the line histogram phase-space MC simulation~\cite{babry422}.
\label{fig:dipion_babar}}
\end{figure}
\begin{figure}[h]
\includegraphics[width=0.45\textwidth]{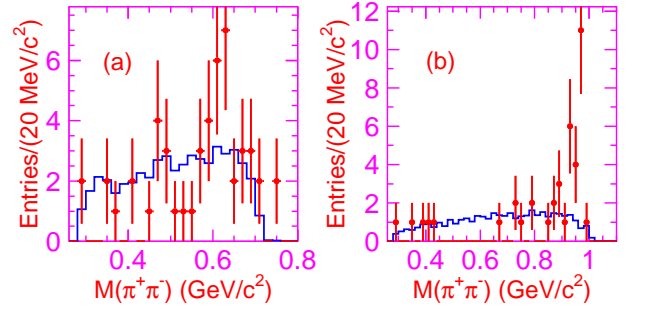}
\caption{\label{fig:dipion_belle} The dipion
         invariant mass distribution on ISR-produced (a)
         $Y(4360) \to \psi(2S)\pi^+\pi^-$ and (b) 
$Y(4660) \to \psi(2S)\pi^+\pi^-$,
         where points represent data and the line histogram phase-space
         MC simulation~\cite{bely431}.
}
\end{figure}

Any interpretation for these vector states will not only 
have to explain their masses, widths, and manifest
reluctance to materialize in open charm or unflavored
light meson final states. The dipion invariant mass
spectra exhibit curious structures, as seen for the $Y(4260)$ in 
Fig.~\ref{fig:dipion_babar}~\cite{babry422},
for the $Y(4360)$ in Fig.~\ref{fig:dipion_belle}(a)~\cite{bely431},
and for the $Y(4660)$ in Fig.~\ref{fig:dipion_belle}(b)~\cite{bely431}.
The first shows a distinctly non-phase-space double-hump
structure which is qualitatively confirmed by Belle~\cite{bely421},
the second exhibits a majority of events at higher masses,
and the third indicates a quite dominant $f_0(980)$ component.

\section{Some implications for theory and perspectives}
\label{mu_g2}

The progress in precision of the low energy data on $e^+e^- \to $~hadrons
achieved recently due to ISR studies allows an update of the 
estimation of the hadronic contribution to the muon anomalous 
magnetic moment to be performed. It is
well known that the precision of the
Standard Model  prediction of this quantity is 
limited by the contributions from strong interactions. 
These are conventionally separated into a theory-driven
light-by-light contribution, see a recent review in \cite{lbl},
and two experiment-driven vacuum polarization contributions,  
the dominant lowest-order and higher-order parts. The lowest-order term can be 
calculated from a dispersion integral~\cite{muonold,gourdin} 
in which the integrand 
contains a combination of experimental data on cross sections 
of $e^+e^- \to $~hadrons and perturbative QCD. The integral ranges
from the threshold of hadron production, i.e., from the 
$\pi^0\gamma$ threshold to infinity:
\begin{equation}
{a}^{\rm had,LO}_{\mu}=
\left( \frac{\alpha m_\mu}{3\pi} \right)^2
\int_{m^2_\pi}^{\infty} ds\: \frac{R(s)\:\hat{K}(s)}{s^2}.
\label{eq:glo}
\end{equation}
The function $\hat{K}(s)$ in the integration kernel is rather smooth,
whereas a factor $1/{s}^2$ emphasizes the low-energy part of the
spectrum. Of particular importance is the process
$e^+e^- \to \pi^+\pi^-(\gamma)$, which provides
about 73\% of the lowest-order hadronic contribution and about 62\% 
of its total quadratic error.

\begin{figure}[h]
\includegraphics[width=0.45\textwidth]{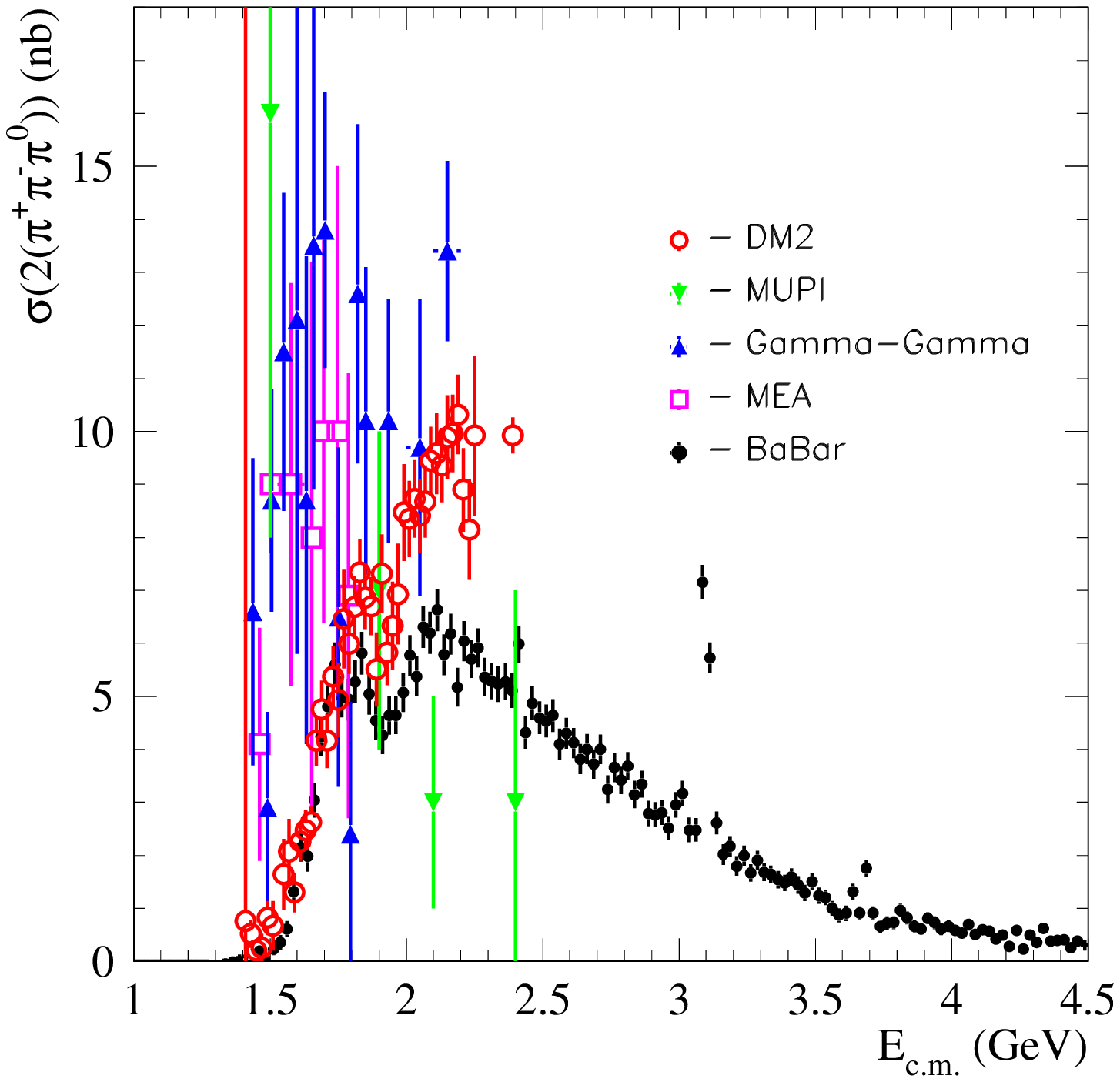}
\caption{The cross section of the process
$e^+e^-\to 2\pi^+2\pi^-2\pi^0$~\cite{DM26pi,m3n4pi,gg24pi0,mea,babr6pi}.
\label{fig:6p1}}
\end{figure}

\begin{figure}[h]
\includegraphics[width=0.45\textwidth]{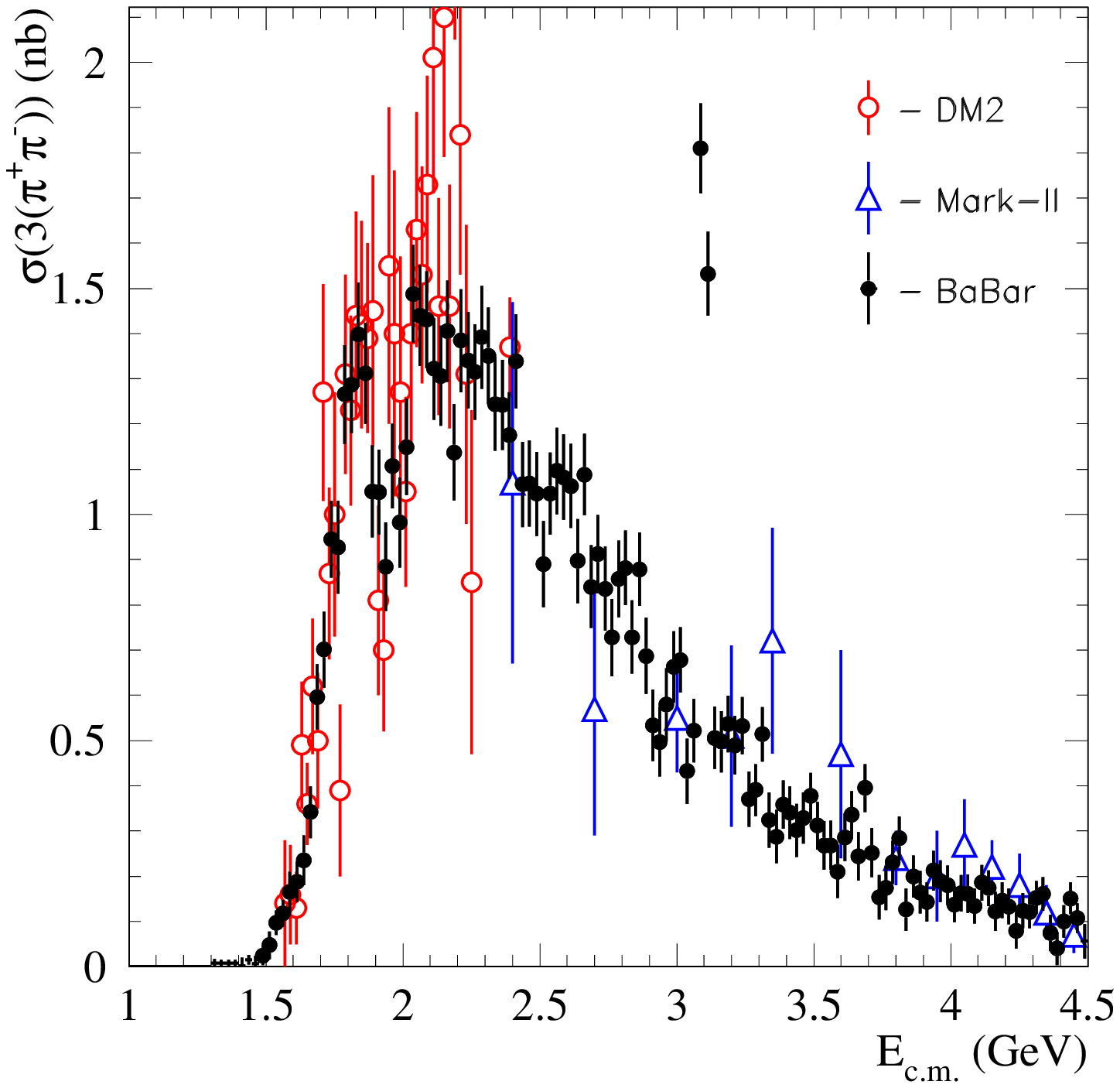}
\caption{The cross section of the process 
$e^+e^-\to 3\pi^+3\pi^-$~\cite{DM26pi,mark2,babr6pi}.
\label{fig:6p2}}
\end{figure}

In most cases new ISR results from BABAR are consistent with 
previous measurements and have comparable or better accuracy.
However, not always these results agree with the corresponding 
old datasets. For example, from Fig.~\ref{fig103} discussed in Ch.~3 
it is clear that the cross section
of the process \epem $\to \pi^+\pi^-\pi^0$ obtained by BABAR~\cite{babr3pi}
is consistent with that of SND~\cite{snd3pi} below $\sqrt{s}=1.4$~GeV,
but is  much higher than that at DM2~\cite{dm23pi} above this energy. The
energy dependence of the cross section observed by DM2 is also
inconsistent with other measurements (see the discussion of this
problem in Ref.~\cite{snd3pi}) and the existence of the rather well 
established $\omega(1420)$ and  $\omega(1650)$ resonances. The
contribution of this process to $\amulo$, which was equal to
$(2.45 \pm 0.26) \cdot 10^{-10}$ before BABAR~\cite{dehz2}, becomes    
$(3.25 \pm 0.09) \cdot 10^{-10}$ after the new results are taken 
into account~\cite{dav07}. For the process 
\epem $\to 2\pi^+2\pi^-$, which 
cross section is one of the largest above 1~GeV, the new BABAR
measurement~\cite{babr4pi} is in good agreement with the older results and 
after taking them into account the precision of the corresponding 
contribution improves by a factor of two.
Another example is the measurement of two six-pion final 
states~\cite{babr6pi}. In Figs.~\ref{fig:6p1} and \ref{fig:6p2}
we compare the cross sections from BABAR with those in older measurements. 
It is clear that the improvement is dramatical because 
older measurements were too imprecise to make a reasonable 
prediction. We summarize the discussed contributions to
$\amulo$ integrated from threshold to 1.8~GeV for the 
measurements before BABAR (see the references in Ref.~\cite{dehz2}) and
with BABAR in Table~\ref{tab:oldnew}~\cite{dav07}. 

\begin{table}[t]
\begin{center}
\begin{tabular}{ccc}
\hline
\hline
Process & Before BABAR & With BABAR \\
\hline
$\pi^+\pi^-\pi^0$ & $2.45 \pm 0.26$ & $3.25 \pm 0.09$ \\
\hline
$2\pi^+2\pi^-$ & $14.20 \pm 0.90$ & $13.09 \pm 0.44$ \\
\hline
$3\pi^+3\pi^-$ &$0.10 \pm 0.10$ & $0.11 \pm 0.02$ \\
\hline
$2\pi^+2\pi^-2\pi^0$ & $1.42 \pm 0.30$ & $0.89 \pm 0.09$ \\
\hline
\hline
\end{tabular}
\caption{The contribution of some multipion processes to
$\amulo$ integrated from threshold to 1.8~GeV for the 
measurements before BABAR and
including the new BABAR results. For the  $\pi^+\pi^-\pi^0$
final state the contribution of the $\omega$ and $\phi$ mesons is
excluded. All values are in units of $10^{-10}$. \label{tab:oldnew}} 
\end{center}
\end{table}

The calculation using all multibody modes measured by 
BABAR~\cite{babr3pi,babr4pi,babr5pi,babr6pi} together with the 
relevant older measurements (for a complete list of references 
see Ref.~\cite{dehz2}) gives for the contribution of hadronic
continuum from threshold to 1.8~GeV the value 
$(54.2 \pm 1.9) \cdot 10^{-10}$. It is consistent with the result
of the older calculation~\cite{dehz2},
 $(55.0 \pm 2.6) \cdot 10^{-10}$, and more precise.
 
A very important step in the calculations of ${a}^{\rm had,LO}_{\mu}$
has been recently made in Ref.~\cite{mi09}, which for the first time
took into account the BABAR measurement of the 
reaction \epem $\to \pi^+\pi^-$~\cite{babr2pi} and finally in 
Ref.~\cite{mi10}, which also accounted for the KLOE results
on the $\pi^+\pi^-$ final state~\cite{kloe2,kloe3a}.
It uses the whole set
of experimental data in the energy range up
to 1.8~GeV, the region dominated by hadronic resonances, and
perturbative QCD for the contribution of the quark continuum beyond
that energy. In particular, they modify the treatment of the
$\omega(782)$ and $\phi(1020)$ resonances using the data from 
Ref.~\cite{babr3pi}, and include preliminary data of BABAR on 
$e^+e^- \to \pi^+\pi^-2\pi^0$~\cite{dru07}. They also use isospin
invariance to estimate the contributions from several unmeasured 
final states with six pions and $K\bar{K}(n\pi)$ relating them to 
those of known channels.  For other channels they 
refer to earlier calculations~\cite{dehz1,dehz2,dav07}.
  
Contributions to ${a}^{\rm had,LO}_{\mu}[\pi\pi]$  
from the individual $\pi^+\pi^-$ cross sections 
measured at BABAR~\cite{babr2pi}, KLOE~\cite{kloe2,kloe3a}, 
CMD2~\cite{cmd2pi1,cmd2pi2}, and SND~\cite{snd2pi}
are given in the middle column of Table~\ref{tab:resultspipi}~\cite{mi10}.
In the right column we list  the corresponding CVC predictions
for the $\tau^{\pm} \to \pi^{\pm}\pi^0\nu_{\tau}$ branching fraction
corrected for isospin-breaking  effects~\cite{eetaunew}. 
Here the first error is experimental and the second estimates the 
uncertainty in the isospin-breaking corrections. 
 For each experiment, all available data in the energy range from threshold 
   to $1.8$~GeV ($m_\tau$ for ${\cal B}_{\rm CVC}$) are used, 
and the missing part is 
   completed by the combined $e^+e^-$ data.
\begin{table}[t]
\setlength{\tabcolsep}{0.0pc}
\begin{tabularx}{\columnwidth}{@{\extracolsep{\fill}}lrr} 
\hline\noalign{\smallskip}
Experiment & \multicolumn{1}{c}{${a}^{\rm had,LO}_{\mu}[\pi\pi]~[10^{-10}]$} & 
\multicolumn{1}{c}{${\cal B}_{\rm CVC}~[\%]$} \\
\noalign{\smallskip}\hline\noalign{\smallskip}
BABAR & $514.1\pm 3.8~ (1.00)$  & $ 25.15 \pm 0.18 \pm 0.22~ (1.00)$ \\
KLOE  & $503.1\pm 7.1~ (0.97)$  & $ 24.56 \pm 0.26 \pm 0.22~ (0.92)$   \\
CMD2  & $506.6\pm 3.9~ (0.89)$  & $ 24.96 \pm 0.21 \pm 0.22~ (0.96)$   \\
SND   & $505.1\pm 6.7~ (0.94)$  & $ 24.82 \pm 0.30 \pm 0.22~ (0.91)$  \\
\noalign{\smallskip}\hline
\end{tabularx}
\caption[.]{\label{tab:resultspipi}Estimated ${a}^{\rm
    had,LO}_{\mu}[\pi\pi]$ 
and ${\cal B}_{\rm CVC}$ contributions from the $e^+e^-$ data~\cite{mi10}.
}
\end{table}

The average of the four separate results gives
${a}^{\rm had,LO}_{\mu}[\pi\pi]=(507.8\pm2.8_{\rm tot})\cdot 10^{-10}$~\cite{mi10}. 
The comparison with their previous result~\cite{eetaunew}, 
$  {a}^{\rm had,LO}_{\mu}[\pi\pi]=(503.5 \pm 3.5_{\rm tot}) \cdot 10^{-10}$, 
shows that the inclusion of the new BABAR data significantly 
increases the central value 
of the integral.
For the higher-order hadronic contributions to $\amulo$,
which are also estimated based on \epem data,
 there is 
a slight gain in accuracy from 
$-9.79 \pm 0.08_{\rm exp} \pm 0.03_{\rm rad}$~\cite{teub07} to
$-9.79 \pm 0.06_{\rm exp} \pm 0.03_{\rm rad}$~\cite{teub10}.

\begin{figure}
\includegraphics[width=0.45\textwidth]{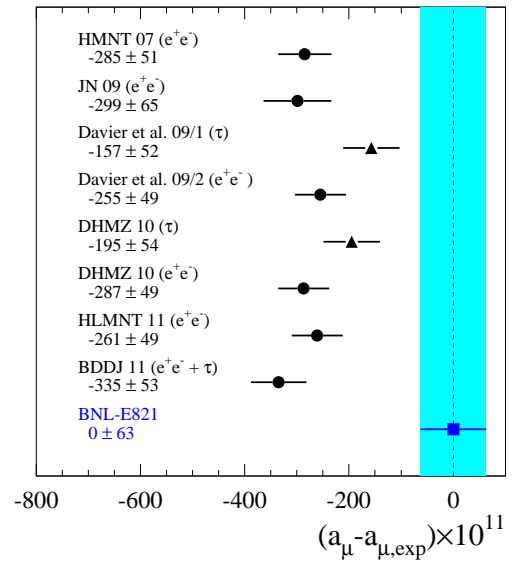}
\caption{Compilation of recent results for $a_{\mu}$.
\label{fig:amures}}
\end{figure}
A compilation of recent results for $a_{\mu}$, from which
the central value of the experimental 
average~\cite{brook} has been subtracted, is given in Fig.~\ref{fig:amures}.
The shaded vertical band indicates the experimental error. 
The SM predictions are taken from: HMNT 07~\cite{teub07}, JN 09~\cite{jeger},
Davier et al. 09/1 ($\tau$-based)~\cite{eetaunew},
Davier et al. 09/2 ($e^+e^-$-based, with BABAR $\pi^+\pi-$ data)~\cite{mi09},
HLMNT 11 ($e^+e^-$-based, with BABAR and KLOE $\pi^+\pi^-$ data)~\cite{teub10},
DHMZ 10 ($\tau$ and $e^+e^-$)~\cite{mi10}, 
BDDJ 11 (($\tau$ and $e^+e^-$)~\cite{bddj11}.
  
There have been only a few tests of CVC based on the new data. The most
interesting final state is of course that with two pions, where a
serious discrepancy between the \epem and $\tau$ data was 
reported~\cite{dehz1,dehz2}. The results
of the most recent tests in this channel~\cite{eetaunew,mi10} are based on the
reevaluation of isospin-breaking corrections. The predictions       
for the branching fraction of $\tau^{\pm} \to \pi^{\pm}\pi^0\nu_{\tau}$
shown in  Table~\ref{tab:resultspipi}. can be compared to the world
average value of $(25.51 \pm 0.09)$\%~\cite{pdg10}. Although the
difference between the CVC prediction and the experimental value
is less significant than previously~\cite{dehz2,dav07}, it is still
substantial for all groups but BABAR. A new approach to the problem
was suggested very recently in Ref.~\cite{jeger11}, where the
$\rho-\gamma$ mixing was properly taken into account. The CVC
prediction for the branching fraction to two pions is 
$(25.20 \pm 0.17 \pm 0.28)$\% in good agreement with the direct measurement. 
Finally, Ref.~\cite{bddj11} reports consistent results on $e^+e^-$ and
$\tau$ in the 2$\pi$ channel and obtains a theoretical prediction
for $a_{\mu}$,   which is lower than the experimental value by 4.1$\sigma$.
   
One more test of CVC that included recent ISR results from BABAR has been 
performed in Ref.~\cite{cher09}. The authors use CVC together with the
data on $e^+e^- \to \eta\pi^+\pi^-$ and $e^+e^- \to \eta^{\prime}\pi^+\pi^-$
to estimate the branching fraction of the corresponding $\tau$ decays.
For the former final state the estimate based on the older
data~\cite{etadm1,nd5pi,dm25pi,cmd5pi} predicts for the branching
fraction $(0.132 \pm 0.016)\%$, somewhat smaller but not incompatible 
with $(0.165 \pm 0.015)\%$ obtained from the BABAR data~\cite{babr5pi}.
The average of the two gives the CVC prediction of $(0.150 \pm 0.016)\%$,
in good agreement with the new world average
${\cal B}(\tau^- \to \eta\pi^-\pi^0\nu_{\tau})=(0.139 \pm 0.010)\%$
that uses the new presise measurement at Belle~\cite{etabel}.
For the  ${\cal B}(\tau^- \to \eta^{\prime}\pi^-\pi^0\nu_{\tau})$ 
they give an upper limit of $<3.2 \cdot 10^{-5}$ at 90\% CL,
which is a factor of 2.5 more restrictive than the upper limit based
on the only existing measurement by CLEO: $< 8 \cdot 10^{-5}$~\cite{cleoetap}.

There are two recent evaluations
of $\Delta{\alpha}^{(5)}_{\rm had}(M^2_Z)$, the
hadronic contribution to the running $\alpha$ from five 
flavors~\cite{mi10,teub10}. In Ref.~\cite{teub10}  
a data set of \epem cross sections includes multibody data from
BABAR and $2\pi$ data from KLOE and the calculation gives the value
$0.02760 \pm 0.00015$, slightly higher and 
significantly more accurate than the previously accepted value    
$0.02758 \pm 0.00035$~\cite{bp05}. The estimation performed in
Ref.~\cite{mi10} additionally uses the BABAR data on the $\pi\pi$
final state and perturbative QCD between 1.8 and 3.7 GeV and gives an even
more precise value $0.02749 \pm 0.00010$.

\section{Conclusions}
Successful experiments at high-luminosity \epem\ colliders
($\phi$- and $B$-factories) opened a new era in a study of  
$e^+e^-$ annihilation
into hadrons at low energies using a novel method of initial-state
radiation
usually referred to as ISR  or radiative return.

Modern detectors operating at these factories which collected
unprecedentally high integrated luminosity allow this method to
compete
with direct \epem\ experiments.

A lot of new data on the cross sections of  $e^+e^-$ annihilation
into hadrons were obtained using ISR, first of all, on the production
of mesons from threshold of their production 
$\sim 2m_{\pi}$ to the c.m. energy of
about 4-5 GeV. More than 30 processes have been studied in which mesons and
hadronic resonances were produced, many of them for the first time.

Valuable information on the particles with mass of about a few GeV
has been obtained, primarily on excited vector mesons, radial and/or
orbital excitations. 
Parameters of vector charmonia were investigated, new data
on more than 40 decay channels were obtained, many decays observed for the
first time.

New data on production cross sections were obtained for various baryons:
proton, $\Lambda$, $\Sigma^0$ and $\Lambda_c$ hyperon, opening new 
possibilities for testing form factor models.

New states ($\rho(1900)$, $Y(2175)$, $Y(4260)$, $Y(4320)$ {$\ldots$}), 
some of them with presumably exotic quark structure, have been discovered. 
Their nature is not yet established and widely discussed.

New values of the cross sections obtained using ISR can be used for more
precise predictions of the muon anomalous magnetic moment, running fine-
structure constant at the Z boson mass, tests of CVC and many other
theoretical models.

Only part of the available ISR data sample has been processed, e.g., for
BABAR it is about 1/3, analysis is in progress. Belle has only started
a corresponding data processing.

If existing projects of Super B-Factories are approved, prospects
of reaching an integrated luminosity by a factor of 30-100 exceeding
that today appear. Such experiments will improve accuracy for many processes
which studies are now statistically limited.

\section*{Acknowledgments}
The authors are grateful to Vera Luth for the idea
of this review. We are indebted to our colleagues 
from the BABAR and Belle Collaborations for many years of 
fruitful work. Thanks are also due to Brian Heltsley from
CLEO and Graziano Venanzoni from KLOE for valuable comments.  

This work was supported in part by the  grants RFBR
10-02-00695, 11-02-00112, 11-02-00558, 
and the DFG grant GZ 436 RUS 113/769/0-3.

\bibliographystyle{apsrmp}

\end{document}